\documentclass[%
reprint,
nofootinbib,
 amsmath,amssymb,
 aps, 
 twocolmun,
]{revtex4-2}
\usepackage{hyperref}
\usepackage{subfig,graphicx}
\usepackage{dcolumn}
\usepackage{bm}
\usepackage{comment}
\usepackage{cleveref}
\usepackage{multirow}
\usepackage{newtxtext,newtxmath}
\usepackage{ragged2e}

\newcommand{\beq}{\begin{equation}}
\newcommand{\eeq}{\end{equation}}
\newcommand{\beqn}{\begin{eqnarray}}
\newcommand{\eeqn}{\end{eqnarray}}
\newcommand{\beqns}{\begin{eqnarray*}}
\newcommand{\eeqns}{\end{eqnarray*}}
\newcommand{\bei}{\begin{itemize}}
\newcommand{\eei}{\end{itemize}}

\newcommand{\rar}{\rightarrow}

\newcommand{\mumu}{\mu^+\mu^-}
\newcommand{\pipi}{\pi^+\pi^-}

\def\MeVM{~${\rm MeV}/c^2$}

\newcommand{\costhpi}{$\cos\theta^*_\pi$}

\newcommand{\acosthpi}{$|\cos\theta^*_\pi|$}

\newcommand{\mpipi}{$m_{\pi\pi}$}
\newcommand{\mmumu}{$m_{\mu\mu}$}
\newcommand{\mumug}{$\mu\mu\gamma$}
\newcommand{\pipig}{$\pi\pi\gamma$}
\newcommand{\KKg}{$KK\gamma$}
\newcommand{\eeg}{$ee\gamma$}
\newcommand{\vxy}{$V_{xy}$}

\input babarsym

\begin{document}
\normalsize
\begin{flushleft}
\begin{minipage}{5cm}
\mbox{BABAR-PUB-26/01}\\
\mbox{SLAC-PUB-260811}\\
\end{minipage}
\end{flushleft}
 


\title{\textbf{A New Precise Measurement of the \boldmath$e^+e^-\rar\pi^+\pi^-(\gamma)$ Cross Section with the Initial-State Radiation Method at \babar}}

\author{J.~P.~Lees}
\author{V.~Poireau}
\author{V.~Tisserand}
\author{E.~Grauges}
\author{A.~Palano}
\author{G.~Eigen}
\author{D.~N.~Brown}
\author{Yu.~G.~Kolomensky}
\author{M.~Fritsch}
\author{H.~Koch}\thanks{Deceased}
% ...
\collaboration{The \babar\ Collaboration}
\noaffiliation



 
\begin{abstract}
We report precise results for the 
$e^+e^-\rightarrow \pi^+\pi^-(\gamma)$ cross section in the energy range between threshold and 1.4\,GeV, relevant for the hadronic vacuum polarization contribution to the muon $g-2$ prediction. Using the full 460\,fb$^{-1}$ of on- and off-$\Upsilon(4S)$ data,
the cross section is derived from the 
ratio of $\pi^+\pi^-$ to $\mu^+\mu^-$ initial-state radiation (ISR) events, in which many systematic uncertainties cancel.
The pion/muon separation does not rely on particle identification as in the 2009 \babar\
analysis, but rather makes use of the well-known, different angular distributions in the 
two-particle center-of-mass system. The compatibility and correlation between the 2026 and 2009 measurements are discussed.
The new measurement is also compared with other two-pion results.
The leading-order hadronic contribution to the muon magnetic anomaly calculated using the new 2026 and combined 2009 and 2026 \babar\ measurements in the energy range of 0.3--1.4\gev are $525.9\pm 3.5_\mathrm{stat}\pm3.8_\mathrm{syst}$ and $517.8\pm 2.4_\mathrm{stat}\pm3.3_\mathrm{syst}$ in units of $10^{-10}$, respectively.


\end{abstract}

\maketitle

\clearpage
\tableofcontents

\clearpage
\section{ Introduction}

Precise measurements of the $e^+e^-\to{\rm hadrons}$ cross section 
are necessary to evaluate dispersion integrals~\cite{Bouchiat:1961lbg} for calculations of 
hadronic vacuum polarization (HVP). Of particular interest is the
contribution $a_\mu^\textrm{had}$ to the muon magnetic moment anomaly
$a_\mu$, which requires data in a region 
dominated by the process $e^+e^-\to\pi^+\pi^-(\gamma)$.
Over the years, the theoretical prediction for $a_\mu$ has been linked to the 
progress of  $e^+e^-$ data. Unfortunately the situation is rather confusing with conflicting measurements in the dominant (73\%) $\pi^+\pi^-$ channel, preventing their combined use~\cite{Aliberti:2025beg}. A specific study by \babar\ of radiative processes~\cite{BaBar:2023xiy} has raised issues on the treatment of radiative corrections in experiments such as KLOE relying heavily on the next-to-leading order (NLO)-only \textsc{Phokhara} generator~\cite{Davier:2023fpl}. The photon-inclusive \babar\ analysis is not sensitive to this problem and the fact that there is a fair agreement of the first \babar\ measurement~\cite{BaBar:2009wpw,BaBar:2012bdw} with CMD-3~\cite{CMD-3:2023alj}, $\tau$~\cite{Davier:2023fpl,Castro:2024prg} and lattice-based~\cite{Aliberti:2025beg,Boccaletti:2024guq} results is comforting. Therefore new independent investigations are needed in order to clarify the situation and improve the Standard Model prediction. This is the purpose of the present study.

This paper is organized as follows. After the general features described in the 
Section~\ref{sec:generalfeatures}, the new method deployed in this study is presented in Section~\ref{sec:method}. 
The analysis steps toward the mass spectra determination are detailed in Sections~\ref{sec:outline} to ~\ref{sec:pionFF}, including event selection and studies of related efficiencies and data/MC corrections.
The mass spectra determination, which is the core of the analysis based on angular distributions, is explained in full detail in Sections~\ref{sec:spectra} and \ref{sec:unfolding}. The results on the muon channel, including the QED test that validates the angular method, and the determination of the effective luminosity, are reported in Section~\ref{sec:qed-test}. The measurement of the $\pi^+\pi^-(\gamma)$ cross section and its combination with the 2009 \babar\ measurement are described in Sections~\ref{sec:pionXsect} and \ref{sec:combination}, respectively. Finally the contribution to the anomalous magnetic moment of the muon derived from the present $\pi^+\pi^-(\gamma)$ cross section is given in Section~\ref{sec:amu}.  

\section{General features of this analysis} 
\label{sec:generalfeatures}

\subsection{The ISR approach}

\begin{figure}[tb] 
\centering
  \includegraphics[width=0.45\linewidth]{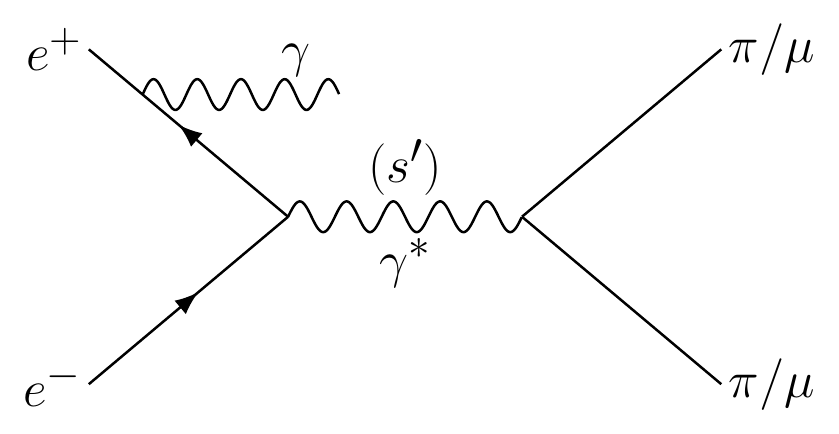}
  \includegraphics[width=0.45\linewidth]{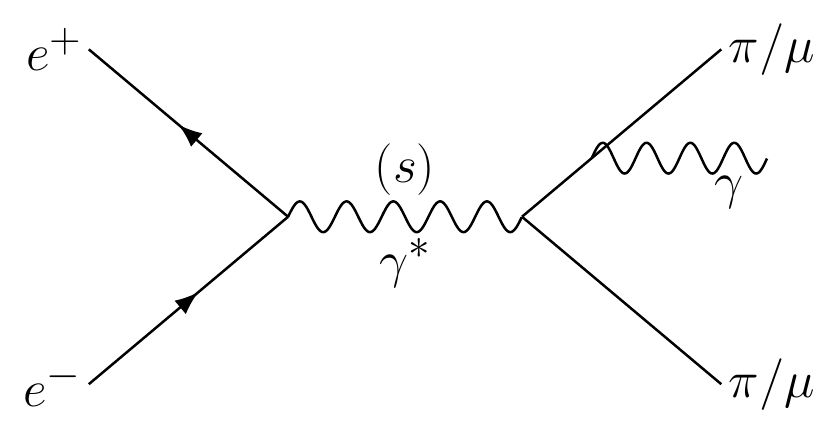}
  \includegraphics[width=0.45\linewidth]{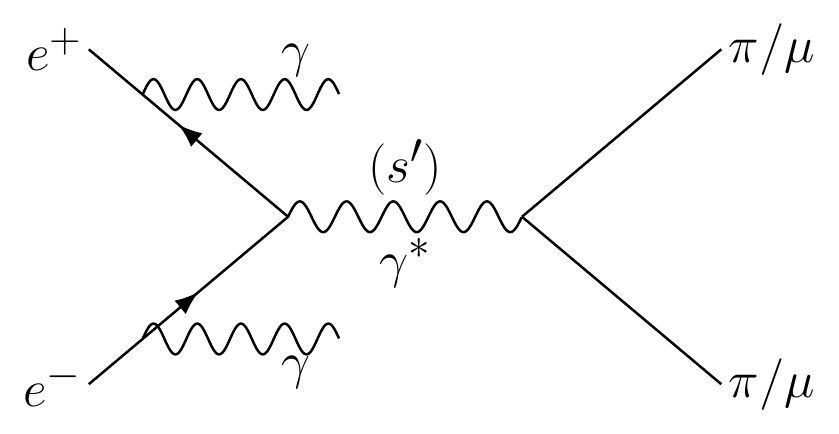}
  \includegraphics[width=0.45\linewidth]{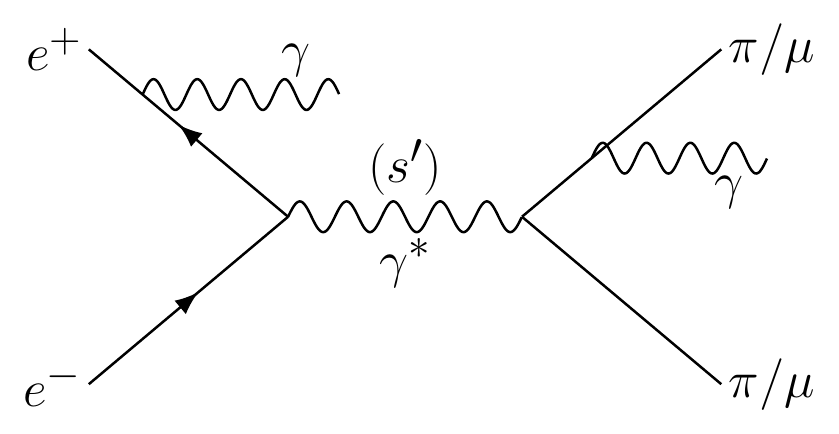}
  \caption{\label{diagrams} \small \justifying The Feynman diagrams for the processes most
relevant to this study, from left to right, top to bottom: lowest-order (LO) ISR $\mu\mu$ or 
$\pi\pi$, LO FSR $\mu\mu$ or $\pi\pi$ (very small process for $\pi\pi$),  
next-to-leading order (NLO) ISR with additional ISR, NLO with additional FSR. 
}
\end{figure}

In the initial-state radiation (ISR) method, the cross section for $e^+e^-\to X$ at the reduced energy 
$\sqrt{s'}=m_X$, where $X$ can be any final state, e.g.\ $\mu^+\mu^-$ or $\pi^+\pi^-$, 
is deduced from a measurement of the radiative process 
$e^+e^-\to X\gamma$ where the photon is emitted by the initial 
$e^+$ or $e^-$ particles. 
The reduced energy is related to the energy $E_\gamma^*$ 
of the ISR photon in the $e^+e^-$ center-of-mass (CM) frame
by $s'=s(1-2E_\gamma^*/\sqrt{s})$, 
where $s$ the square of the $e^+e^-$ CM
energy. 
Since the ISR photon in this analysis is detected at a large angle to the beams, the system recoiling against it is well contained in the detector acceptance and boosted such that the reconstruction efficiency is high even down to its threshold. Higher-order radiation must also be considered with additional photons at ISR and final-state radiation (FSR) levels.
Figure~\ref{diagrams} shows the four 
important Feynman diagrams relevant to this study. We have considered and measured the contributions from all diagrams with up to three photons.

\subsection{Cross-section measurement through the $\pi\pi/\mu\mu$ ratio}
\label{pimuratio}

The cross section for the process $e^+e^-\to X$ 
is related to the $\sqrt{s'}$ spectrum of $e^+e^-\to X\gamma$ events through
\begin{equation}
\label{def-lumi}
  \frac {dN_{X\gamma}}{d\sqrt{s'}}~=
   ~\frac {dL_\textrm{ISR}^\textrm{eff}}{d\sqrt{s'}}~
    \varepsilon_{X\gamma}(\sqrt{s'})~\sigma_{X}^0(\sqrt{s'})~,
\end{equation}
where $dL_\textrm{ISR}^\textrm{eff}/d\sqrt{s'}$ is the effective luminosity, 
$\varepsilon_{X\gamma}$ is the full acceptance for the event sample,
determined by simulation with corrections obtained from data, 
and $\sigma_X^0$ is the bare cross section (excluding vacuum polarization)
for the process 
$e^+e^- \to X$ (including additional FSR photons).  
The $\sqrt{s'}$ spectrum is obtained from the observed 
mass $m_{X}$ distributions through an unfolding procedure. 

Equation~(\ref{def-lumi}) applies equally to $X=\pi\pi(\gamma)$ and $X=\mu\mu(\gamma)$ final states, so
that the ratio of cross sections is directly related to the ratio of the pion to muon spectra
as a function of $\sqrt{s'}$. The details of the procedure will be given later. 

Several advantages follow from taking the $\pi\pi$ to $\mu\mu$ ratio as a function of $(\sqrt{s'})$:
\begin{itemize}
\item the result is independent of the \babar\ luminosity $L_{ee}$;
\item the determination of the ISR luminosity comes from the muon data,
independently of the number of additional ISR photons, thus not depending
on a theoretical calculation;
\item the ISR photon efficiency cancels out;
\item vacuum polarization also cancels out.
\end{itemize}
Furthermore the Monte Carlo (MC) generator and the detector simulation are only used to 
compute the acceptance of the studied $X\gamma_\textrm{ISR}$ processes.

 \subsection{Measurement of additional radiation}

 Since the original 2009 analysis of the $e^+e^-\rar\pi^+\pi^-$
cross section~\cite{BaBar:2009wpw,BaBar:2012bdw}, \babar\ has adopted a loose approach for the event selection which includes photons in addition to the main ISR photon. These additional photons can be emitted either from the initial or final states. Additional ISR photons are concentrated at small angles to the beam, in which case they are found through kinematical fits, or directly detected if emitted in the detector acceptance. Additional FSR photons are directly detected and separated from large-angle ISR photons through their angular distributions thanks to the high CM energy available at \babar. So far, \babar\ is the only experiment measuring additional radiation, ensuring a large insensitivity to the description of radiative effects in the MC generators.

 \babar's results extending the range of radiation measurements from NLO with one additional photon to next-to-next-to-leading order (NNLO) with two additional photons, both for ISR and FSR, were published recently in Ref.~\cite{BaBar:2023xiy}. The comparison to the prediction of the \textsc{Phokhara} generator reveals clear disagreements: first, the predicted NLO rate of additional ISR photons at small angles to the beam is about 25\% larger than the measured one; and second, observed NNLO photons have no counterpart in the generator since \textsc{Phokhara} is limited to NLO. These shortcomings affect the calculation of the total event acceptance, but only to a very small extent because of (1) the inclusiveness of the event selection regarding additional radiation, (2) the approximate cancellation between the NLO excess and the NNLO deficit in the generator, and (3) the dominant additional radiation from ISR cancels in the $\pi\pi$/$\mu\mu$ ratio.

 \subsection{Relation to the first \babar\ measurement}

The 2009 analysis~\cite{BaBar:2009wpw,BaBar:2012bdw} was based on about half of the full data sample and used particle identification (PID) to separate muons and pions. This required tracks with momenta above $1\gevc$, reducing the achievable statistical precision, and the uncertainties related to the PID efficiencies were among the largest systematics. Here, we take a complementary approach, exploiting the different angular distributions to statistically separate pions and muons (and kaons and residual electrons), which allows the use of all well measured tracks with a much looser cut of transverse momentum $p_\mathrm{T}>0.1\gevc$. Another difference is the use of the NLO \textsc{Phokhara} event generator for simulation of the signal channels whereas the 2009 analysis used \textsc{AfkQed} (see Section~\ref{sec:mc}).

In addition to the mass-dependent efficiencies reevaluated with the full data sample and updated reconstruction and simulation, we have determined their angular dependence. Furthermore we have improved our understanding of higher order photon radiation and FSR from pions~\cite{BaBar:2023xiy}.

\subsection{Blind approach}

For any precision analysis involving corrections of systematic effects, it is
mandatory to proceed in a blind way, in order not to be influenced by results
from other experiments or by any theoretical prejudice. 
This is critical here because of the existing discrepancies between the results on the $e^+e^-\rar\pi^+\pi^-$
cross section. 

Blinding is achieved by hiding two important individual data/MC corrections, namely for trigger and tracking efficiencies, by applying offsets known to only one person for each of these corrections.
These multiplicative offsets are applied to MC efficiencies during the full analysis up to final unblinding. Also independent hidden offsets, known only by a third person, are applied when $\pi\pi$ and $\mu\mu$ contributions are separated. These independent blinding factors differ from unity by up to several percent, positive or negative. Only when all the studies have been successfully completed with internal consistency checks are the six offsets released and absolute cross sections computed. The unblinding is performed in two steps, first with the $e^+e^- \rightarrow \mu^+\mu^-(\gamma)$ cross section which is compared to the QED prediction as an important check~\cite{BaBar:2009wpw,BaBar:2012bdw}, then with the $e^+e^- \rightarrow \pi^+\pi^-(\gamma)$ cross section as the final goal of the analysis.
The analysis was unblinded for the preliminary results in 2025~\cite{Zhang:2026jvb,Polat:2026ysh} and was reblinded for the further improvements made to obtain the final results presented here (Appendix~\ref{app:changes}).

The relevance of the multiple blinding method is enhanced 
as it guarantees an unbiased comparison of the results of the present analysis with the previously published 2009 results from \babar~\cite{BaBar:2009wpw,BaBar:2012bdw}. 
\section{The new $\pi\pi/\mu\mu$ separation method}
\label{sec:method}
\subsection{Motivation}
 
The idea~\cite{davier-ti-2017}
is to fit in each mass bin the angular distributions in the $X^+X^-$ CM frame in order to obtain the respective contributions of the different 
components $X=\pi/\mu/K/e$ of the generic process
$e^+e^-\to X^+ X^- \gamma$. We define $\theta^*_\X$ 
as the $X^-$ angle in the $X^+X^-$ CM frame with respect to the ISR photon 
direction. The $\theta^*_\X$ distributions in the ISR process 
follow the original polar angular distributions of the corresponding Born 
processes:
\beqn
  \frac{d\sigma(e^+e^-\to \pi^+\pi^-)}{d\cos\theta_\pi} &\sim&  \sin^2\theta_\pi\nonumber \\
  \frac{d\sigma(e^+e^-\to K^+K^-)}{d\cos\theta_K} &\sim&  \sin^2\theta_K \label{eq-angle}\\
  \frac{d\sigma(e^+e^-\to \mu^+\mu^-)}{d\cos\theta_\mu} &\sim&  \beta_\mu[1+\cos^2\theta_\mu+\nonumber \\
   & & (1-\beta_\mu^2)\sin^2\theta_\mu],\nonumber   
\eeqn
where $\beta_\mu$ is the muon velocity in the $e^+e^-$ frame. 
The muon distribution is flat near threshold, whereas it behaves as $1+\cos^2\theta^*_\mu$ at
large energies. 

\begin{figure}[tbp] \centering
\includegraphics[width=\columnwidth]{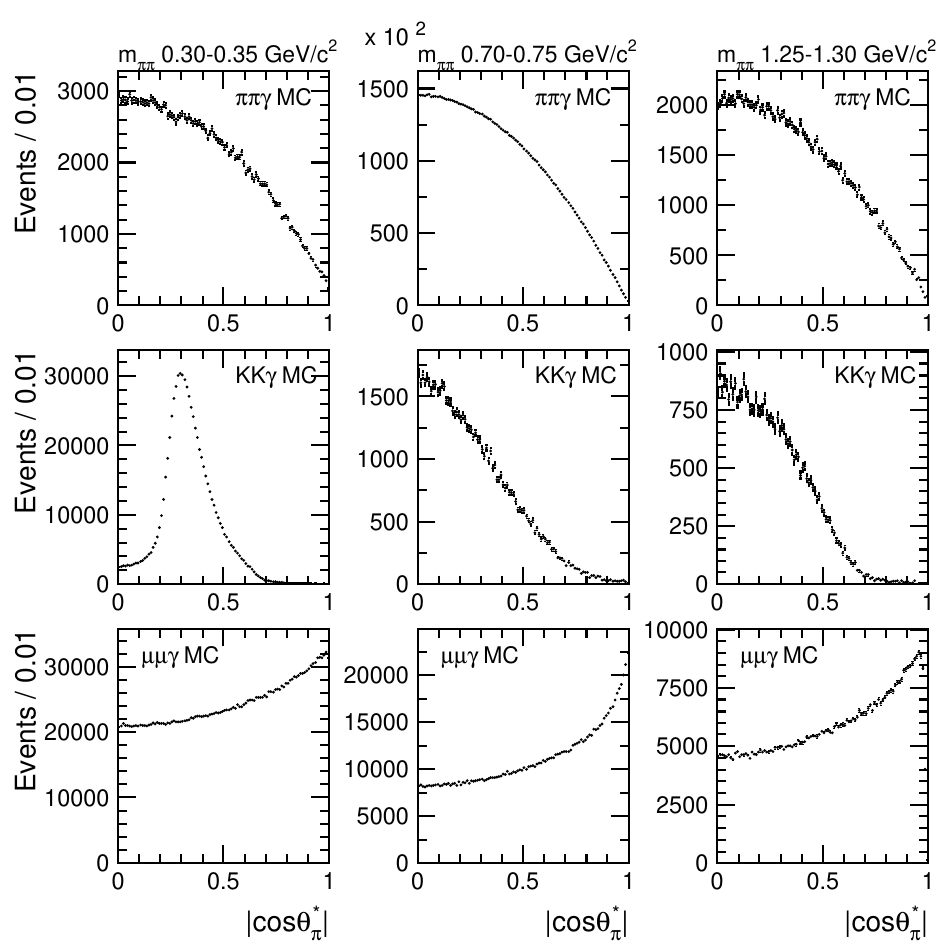}
    \caption{\label{mc-cosths}   \small \justifying
Some $|\cos\theta^*_\pi|$ distributions in different mass intervals for 
$\pi\pi\gamma$, $KK\gamma$, and $\mu\mu\gamma$ processes in the detector 
geometrical acceptance, as generated by \textsc{Phokhara}. 
  } 
\end{figure}

In the data, since the nature of the charged particle is a priori unknown, experimental angular distributions and two-track invariant mass values are computed 
under a track mass hypothesis.
Thus, with a pion mass, the distribution using $\cos\theta^*_\pi$ in bins of $m_{\pi\pi}$ corresponds to the true distribution for pions while it is 
not the case for particles with a different mass. However, as seen in
Fig.~\ref{mc-cosths} for MC samples within the analysis acceptance (see Section~\ref{evt-sel}), 
the relative distributions are markedly different, reflecting
their proper physical distributions and kinematics.
The special shape of the $KK\gamma$ process and its strong variation with $|\cos\theta^\ast_\pi|$ at low mass values are due to the choice of the pion mass hypothesis.

In each mass bin, the angular distributions for each $X=\pi/\mu/K$ process are generated from simulation and corrected using data. They are used as templates for the angular fits of the data components.
The electron case is more complicated because the selected sample contains not only radiative Bhabha events in the far tails of the discriminating variables (\dedx\ and calorimeter energy deposit) that are not reliably simulated,
but also events from the $e^+e^-\to\gamma\gamma$ process followed by photon conversion in the detector material. 
Thus, the templates for electron events are obtained directly from data (Section~\ref{sec:eeg}). 
The contribution from the $p\overline{p}\gamma$ process is treated differently because it is very small and well measured~\cite{BaBar:2013ves}. 
It is subtracted before the angular fits. 

It should be emphasized that the ISR method with an energetic large-angle 
photon in the detector acceptance is ideal for observing the full 
$\cos\theta^*_X$ distribution since the $X^+X^-$ pair is boosted
into the detector acceptance in a direction opposite to the photon. 
This is in contrast to usual `scan' $e^+e^-$ experiments 
where generally the small-angle region is not covered by the detector, thus 
strongly limiting the ability to separate pion and muon pairs in this way.
 
\subsection{Requirements}
 
\begin{figure}[tbp] \centering
\includegraphics[width=\columnwidth]{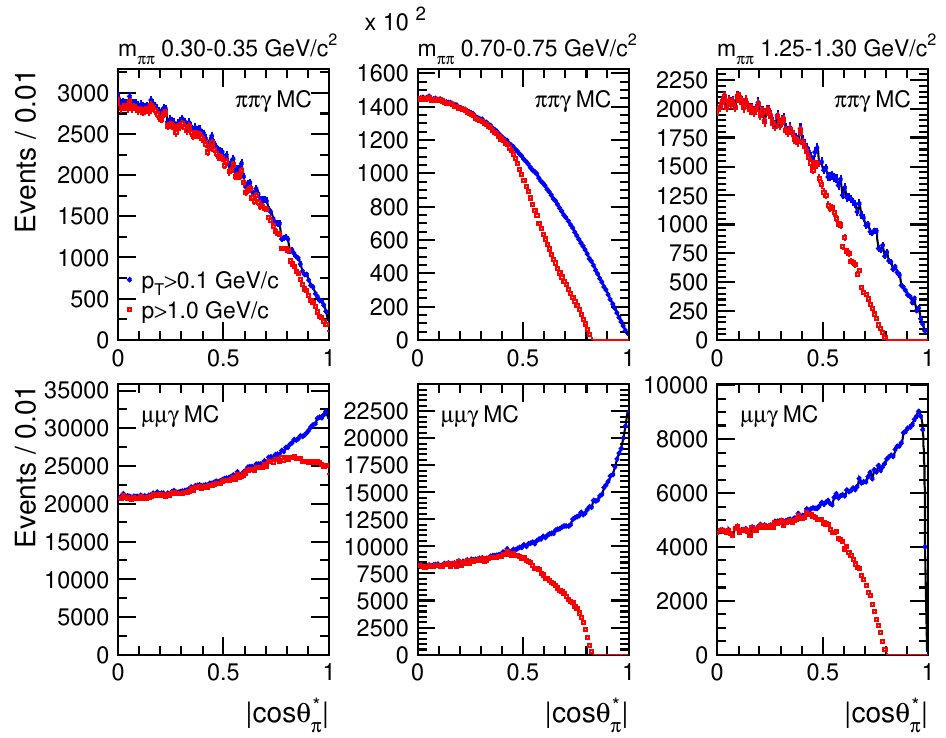}
    \caption{\label{low-p}   \small \justifying
The $|\cos\theta^*_\pi|$ distributions in three mass intervals for $\pi\pi\gamma$ 
(top) and $\mu\mu\gamma$ (bottom) MC events within the geometrical acceptance 
selections, without momentum selection (black histogram),
and different threshold values: $p>1.0\gevc$ (red, 2009 analysis) and $p_\mathrm{T}>0.1\gevc$ (blue, this analysis).
  } 
\end{figure}

It is clear from Eq.~(\ref{eq-angle}) that separation of the pion and muon 
processes through their angular distribution requires a good sensitivity 
at large $|\cos\theta^*_\pi|$ values, corresponding to momentum-asymmetric 
topologies. Thus, it is necessary to be able to detect particles at low 
momenta. For $|\cos\theta^*_\pi|$ values near 1, the $\pi^+\pi^-\gamma$ distribution is 
strongly suppressed while the $\mu^+\mu^-\gamma$ one is maximal. 
The importance of low-momentum detection is illustrated in Fig.~\ref{low-p}. 
The momentum selection at $1\gevc$ used in the 2009 analysis was necessary 
to ensure reliable PID. Dropping the 
need for PID makes this selection irrelevant, allowing one to reach large
$|\cos\theta^*_\pi|$ values as seen in Fig.~\ref{low-p}, and to perform a statistical pion-muon separation
with a large increase in statistical precision, especially for muons. 
As tracks in \babar\ are reconstructed 
down to $p_\mathrm{T}=100$\,MeV/c, this is fully consistent with the analysis 
requirements.
 
Since $\mu\mu\gamma$ events strongly dominate over $\pi\pi\gamma$ near
threshold and at large masses, this analysis is restricted to the $m_{\pi\pi}$ mass interval between 0.3 and $1.4\gevcc$, fully matched to what is needed for the computation of the HVP contribution to the muon magnetic moment.

Note that, although PID is not used in this analysis for $\pi\pi/\mu\mu$ separation, it is still necessary for specific efficiency studies that have to be done separately for pion and muon samples. 
\section{Analysis Outline}
\label{sec:outline}

\subsection{The \babar\ detector and data samples}

The analysis is based on  $424.2\invfb$ of data~\cite{BaBar:2013agn} collected with the \babar\ 
detector at the SLAC PEP-II asymmetric-energy $e^+e^-$ storage 
rings operated at the $\FourS$ resonance.
In addition, 36.2\,fb$^{-1}$ of data\footnote{The Run~6 off-resonance sample of 7.8\,fb$^{-1}$ was not used for this analysis.} 
were collected 40\mev below the resonance. 
The \babar\ detector is described in detail elsewhere~\cite{BaBar:2001yhh,BaBar:2013byz}.
Charged-particle tracks are measured with a five-layer double-sided 
silicon vertex tracker (SVT) together with a 40-layer drift chamber (DCH) 
inside a 1.5\,T superconducting solenoidal magnet. Photons are assumed to 
originate from the average beam crossing point
and their energy is measured in a CsI(Tl) electromagnetic calorimeter (EMC).
Charged-PID uses the ionization losses $\dedx$
in SVT and DCH, the Cherenkov radiation detected in a ring-imaging
device (DIRC), the shower energy deposit in the EMC ($E_\mathrm{cal}$), and the
shower shape in the instrumented flux return (IFR) of the magnet. The IFR 
system is made of modules of resistive plate chambers (RPC), then in a later 
stage modules of limited streamer tubes (LST), interleaved 
with iron slabs, arranged in a barrel and two endcaps. 

\subsection{Monte Carlo generators and simulation}
\label{sec:mc}
Signal ISR processes $e^+e^- \to X\gamma_{\rm ISR}$ with $X=\mu\mu$, $\pi\pi$ or 
$KK$ 
are simulated with the full NLO \textsc{Phokhara9.1} event generator~\cite{phokhara-v9.1,Czyz:2003hc,Czyz:2004rj}. 
All NLO diagrams for virtual and real photon production are included,  
when radiation may occur in the initial or the final state. 
Photons are generated with an angular distribution according to the full NLO matrix element.  
For muons, matrix elements are given by QED for ISR, FSR, and interference.
For pions, ISR radiation is given by QED while FSR is described by a point-like model of the pion.
No NNLO diagram is included, and no more than one additional ISR or FSR photon is generated.
The main ISR photon $\gamma_{\rm ISR}$ is generated in the $e^+e^-$ CM frame with a polar angle in the range $20^\circ-160^\circ$, 
safely bracketing the detector acceptance. A minimal ISR photon energy of 4\gev in the
CM frame is required at production level, 
which restricts the generated mass range to the region relevant for this study.
Simulated samples corresponding to about 10 times the number of data events are generated for the signal channels.

For systematic studies\footnote{For the 2009 \babar\ measurement, while \textsc{Phokhara} was still under development, the full samples of signal events were generated with \textsc{AfkQed}.}, small samples of signal ISR processes are simulated with the 
\textsc{AfkQed} MC event generator, which is based on the formalism of 
Refs.~\cite{Czyz:2000wh,Binner:1999bt}.  
Additional ISR photons are generated collinear to the $e^+$ or $e^-$ beam with
the structure function method~\cite{Caffo:1997yy}, and additional FSR photons 
with {\small PHOTOS}~\cite{Barberio:1990ms}. 
As the structure function method relies on a resummation of leading logarithms, 
it takes into account real and virtual higher order contributions, although in an approximate way. 
 
Background ISR processes $e^+e^- \to X\gamma_{\rm ISR}$ are simulated with the 
\textsc{AfkQed} event generator. Background processes $e^+e^-\to\qqbar$ 
($q=u,d,s,c$) are generated with {\small JETSET}~\cite{Sjostrand:1993yb}, and $e^+e^-\to\tau^+\tau^-$ 
with {\small KK2f}~\cite{Jadach:1999vf}. 
The response 
of the \babar\ detector is simulated with {\small GEANT4}~\cite{GEANT4:2002zbu}.

\subsection{Event selection}
\label{evt-sel}
Two-body ISR events are selected 
by requiring an ISR photon candidate with $E_\gamma^*>4\gev$ and 
laboratory polar angle $\theta$ in the range $0.35-2.4\rad$, 
and exactly two 
`good' tracks of opposite charge, each with transverse
momentum $p_\textsc{T}>0.1\gevc$ and within the polar angle range $0.40-2.45\rad$.
The `good' tracks are required to have at least 
15 hits in the DCH, and must originate from within
$5\mm$ of the collision axis (distance of closest approach in the transverse 
plane $\mathrm{doca}_{xy}<5\mm$) and within $6\cm$ from the beam spot along the beam direction
($|\Delta_\mathrm{z}|<6\cm$).
 If several ISR photon candidates are detected, the one with the highest $E_\gamma^*$ is chosen, resulting in an incorrectly assigned ISR photon in less than $\sim10^{-4}$ of the events.

\sloppy
A veto based on a combination of $E_\mathrm{cal}$ and $\dedx$, $((E_\mathrm{cal}/p-1)/0.15)^2~+~((\dedx_\mathrm{DCH}-690)/150)^2~<~1$),
reduces electron contamination. Remaining background from photon conversions are efficiently vetoed by requiring the distance $V_{xy}$ in the transverse plane 
between the two-track vertex and the beam collision spot to satisfy $V_{xy}<0.5$\,cm. 
Events may be accompanied by any number of standard \babar\ tracks, not satisfying the above criteria, 
and any number of additional photons.

Events are subjected to a set of kinematic fits, described in Section~\ref{sec:kin-fit}, and selected using an optimized combination of $\chi^2$ values.

\subsection{Corrections to the detector simulation}

The simulation
is initially used in order to incorporate in a consistent way all 
effects entering the final event acceptance. Then corrections between data and
MC are obtained for each contribution to the efficiency using dedicated studies performed
on data and simulation samples. The main corrections originate from the trigger conditions, tracking, $V_{xy}$ selection, momentum calibration,
and the $\chi^2$ of the kinematic fits. Corrections are applied in two steps: 
(1) data/MC shape corrections as a function of 
$|\cos\theta^*_{\pi}|$ are made
to the MC reference distributions before performing the template fits, and (2)
absolute data/MC corrections are applied to the $\pi\pi$ and $\mu\mu$ mass spectra obtained from the fits.

Care is taken to avoid correlations between efficiency measurements. Thus, the corrected efficiency is
\begin{equation}
\label{eff-corr}
   \varepsilon = \varepsilon_{\rm MC}\left (
   \frac {\varepsilon_\textrm{trigger}^{\rm data}}{\varepsilon_\textrm{trigger}^{\rm MC}}
      \right )\left (
   \frac {\varepsilon_\textrm{tracking}^{\rm data}}{\varepsilon_\textrm{tracking}^{\rm MC}}
         \right )\left (
   \frac {\varepsilon_{V_{xy}}^{\rm data}}{\varepsilon_{V_{xy}}^{\rm MC}}
      \right )\left (
   \frac {\varepsilon_{\chi^2}^{\rm data}}{\varepsilon_{\chi^2}^{\rm MC}}
      \right ).
\end{equation}
The corrections $C_i=\left (
   \frac {\varepsilon_i^{\rm data}}{\varepsilon_i^{\rm MC}}
      \right )$ are reviewed in turn in the following sections. 
They are applied as $|\cos{\theta^*_\pi}|$ and mass dependent corrections to the MC efficiency. 
\section{Kinematic fitting}
\label{sec:kin-fit}

\subsection{Fit description}

As described in Section~\ref{evt-sel}, the event definition is 
enlarged to include the radiation of one photon in addition to the already 
required ISR photon. Two types of fits to $e^+e^-\to \pi\pi\gamma{_{\rm{ISR}}\gamma}$ are considered, according to the 
following situations:
\begin{itemize}
\item The additional photon is detected in the EMC, in which case its energy 
and angles are used in the fit: we call this a large angle (LA) fit, while 
the extra photon can be either from FSR or from ISR at LA to the 
beams. The threshold for
the additional photon is kept low (50\mev). In the case of several extra detected photons, LA fits are performed using
each photon in turn and the fit with the best $\chi^2_\mathrm{LA}$ is retained.
\item The additional photon is assumed to be from ISR at small angle (SA) to the 
beams. Since no more information 
is available, it is postulated that the extra photon is perfectly aligned with 
either the $e^+$ or $e^-$ beams. 
The corresponding 3-constraint fit (so-called SA fit) ignores additional photons measured in the EMC and returns the
energy of the fitted collinear ISR photon. 
Between the two possibilities associated with each beam, the one with a better $\chi^2_\mathrm{SA}$ value is retained. 
\end{itemize}

In both cases, the constrained fit procedure uses the energy and 
direction of the measured ISR photon (plus those of the LA photon in the LA fit), and the measured momenta and 
angles of the two tracks with their covariance matrix in order to satisfy the
four energy-momentum conservation equations. The pion mass is assumed for the two charged particles. 
Since the 
energy response of the EM calorimeter is asymmetric with a longer tail when the 
measured energy is smaller than the true energy, the photon energy variable is
transformed into a Gaussian-distributed symmetric variable using the so-called
Novosibirsk function (a log-normal distribution)~\cite{hafner-thesis}. Calibration of the photon energy and energy-dependent parameters of the Novosibirsk function is performed beforehand through a dedicated study, using  PID-selected data and MC, subjected to a 3C fit to $e^+e^-\to \mu\mu\gamma_{\rm{ISR}}$ that ignores the $\gamma_{\rm{ISR}}$  energy.

Each event is characterized by two $\chi^2$ values, $\chi^2_\mathrm{LA}$ and
$\chi^2_\mathrm{SA}$ from the LA and SA fits, respectively, which are reported 
on a 2d plot.
In practice, the quantities $\ln{(\chi^2+1)}$ are used so that the long tails
can be properly visualized. Events without any extra measured
photons have only the $\chi^2_\mathrm{SA}$ value and they are plotted separately.

\begin{figure}[tbp] \centering
      \includegraphics[width=\linewidth]{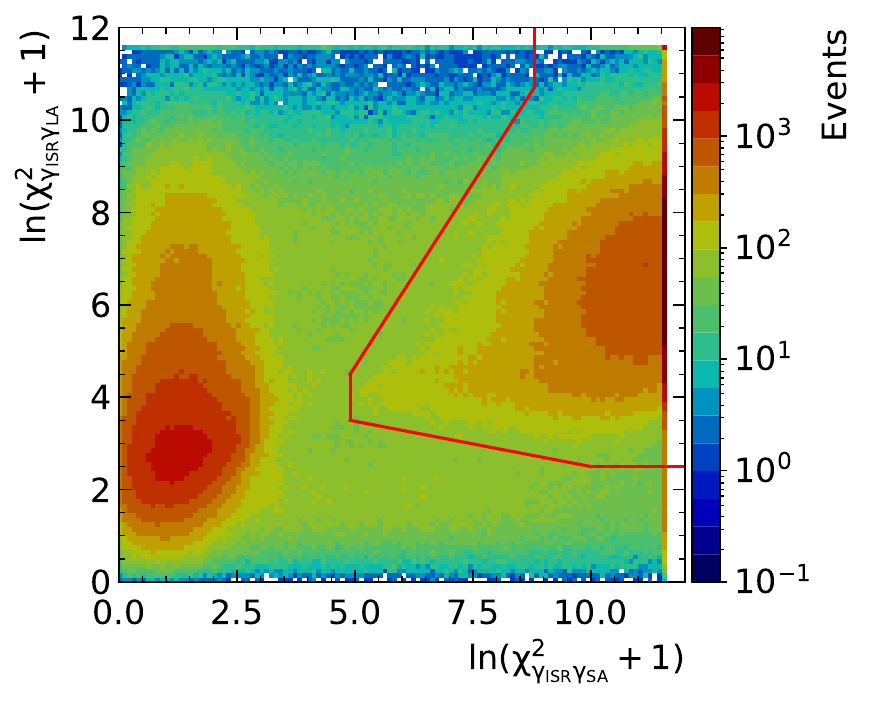}
      \includegraphics[width=\linewidth]{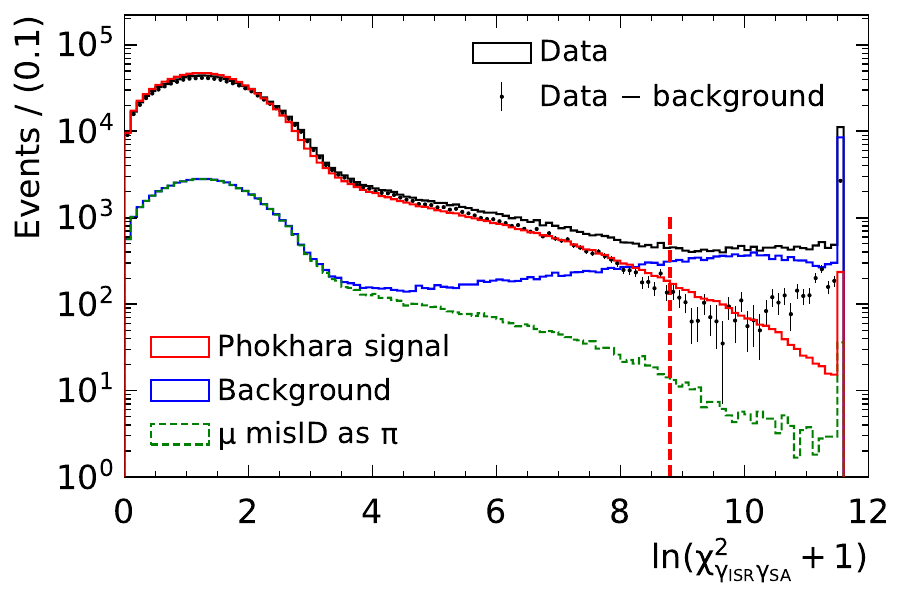}
    \caption{\label{fig:2d-chi2}   \small \justifying
The 2d-$\chi^2$ distribution for $\pi\pi\gamma_\mathrm{ISR}(\gamma_\mathrm{SA/LA})$ in data (top) and a comparison of $\chi^2_{\gamma_{\textrm{ISR}}\gamma_{\textrm{SA}}}$
between data and MC for events that do not have an additional measured photon (bottom) in the mass range of $0.5<m_{\pi\pi}<1$\gevcc. The zones delimited with red solid lines (top) and to the right of the vertical dashed lines (bottom) are rejected according to the optimized selection.
}
\end{figure}
It is easy to visualize the different interesting regions in the 2d-$\chi^2$
plane, as outlined in Fig.~\ref{fig:2d-chi2} for $\pi\pi\gamma$ data with both $\pi$ selected with a tight selector. 
Most of the events peak at small values of both $\chi^2$, but the tails 
along the axes clearly indicate events with additional radiation: 
small-angle ISR along the $\chi^2_\mathrm{LA}$ axis (with large ISR energies at large
values of $\chi^2_\mathrm{LA}$), and FSR or large-angle ISR along the $\chi^2_\mathrm{SA}$ 
axis (with large additional radiation energies at large values of $\chi^2_\mathrm{SA}$). 
Events along the diagonal at large $\chi^2$ values do not satisfy either hypothesis and 
result from resolution effects for the pion tracks (also secondary interactions) or for the 
primary ISR photon, or possibly from additional radiation of more than one photon. 
There are significant fractions of events without any extra measured photons
with energies above the 50\mev threshold, corresponding to 67\% for $\pi\pi$ and 40\% for $\mu\mu$ of the selected event samples.
The $\chi^2_\mathrm{SA}$ distribution of these $\pi\pi$ events is also shown in Fig.~\ref{fig:2d-chi2}.
The signal distribution (red curve in Fig.~\ref{fig:2d-chi2} bottom) is quite similar to the distribution of the $\mu\mu$ dominant background (green curve), while other backgrounds populate the tail of the distribution ($\ln{(\chi^2_\mathrm{SA}+1)}>4.$).  

\subsection{Optimization of $\chi^2$ selection}

Multibody background populates the region where both $\chi^2$ 
are large and consequently a background region is defined in the 2d-$\chi^2$
plane where signal events are rejected.
A Boosted Decision Tree (BDT) optimization is performed separately in three mass windows, low ($<0.5$\gevcc), intermediate (0.5$-$1.0\gevcc), and high (1.0$-$1.4\gevcc), using the two variables $\chi^2_\mathrm{SA}$ and $\chi^2_\mathrm{LA}$, treating the $\pi\pi\gamma$, $\mu\mu\gamma$ and $KK\gamma$ processes as signal and all the other processes as background. 
The contours of the signal MC events selected using the best signal over background BDT selection in the three mass windows are similar. The contour for the intermediate mass region is indicated in Fig.~\ref{fig:2d-chi2}.

\subsection{Strategy to determine the $\chi^2$ efficiency in data}

The background contributions to the $\mu\mu\gamma(\gamma)$
sample are significant only in the
background region of the 2d-$\chi^2$ plane, and 
the determination of the 2d-$\chi^2$ selection efficiency is relatively 
straightforward in the muon channel.
The situation for the pion channel is very different, as it is not possible to 
directly measure the efficiency of the 2d-$\chi^2$ selection in data because of 
an overwhelming background in the rejected regions (see Fig.~\ref{fig:2d-chi2}).

The rejected signal events with large $\chi^2$ have several sources. Some are common to pions and muons: mis-reconstructed ISR photon and/or charged tracks, tails of the $\chi^2$ distributions of events with additional ISR or FSR, and more than one additional photon (mostly ISR). In addition, pions are affected by secondary interactions and pion-induced fake photons.   
The strategy is thus to rely on the $\chi^2$-selection studies performed on muon data
to account for the common losses and to further investigate the losses 
specific to pions. 

\subsubsection{Determination of the $\chi^2$ selection efficiency for $\mu\mu\gamma$}

\begin{figure*}[tbp]
  \begin{minipage}{0.49\textwidth}
  \centering
  \vspace{-5mm}
  \includegraphics[width=\linewidth]{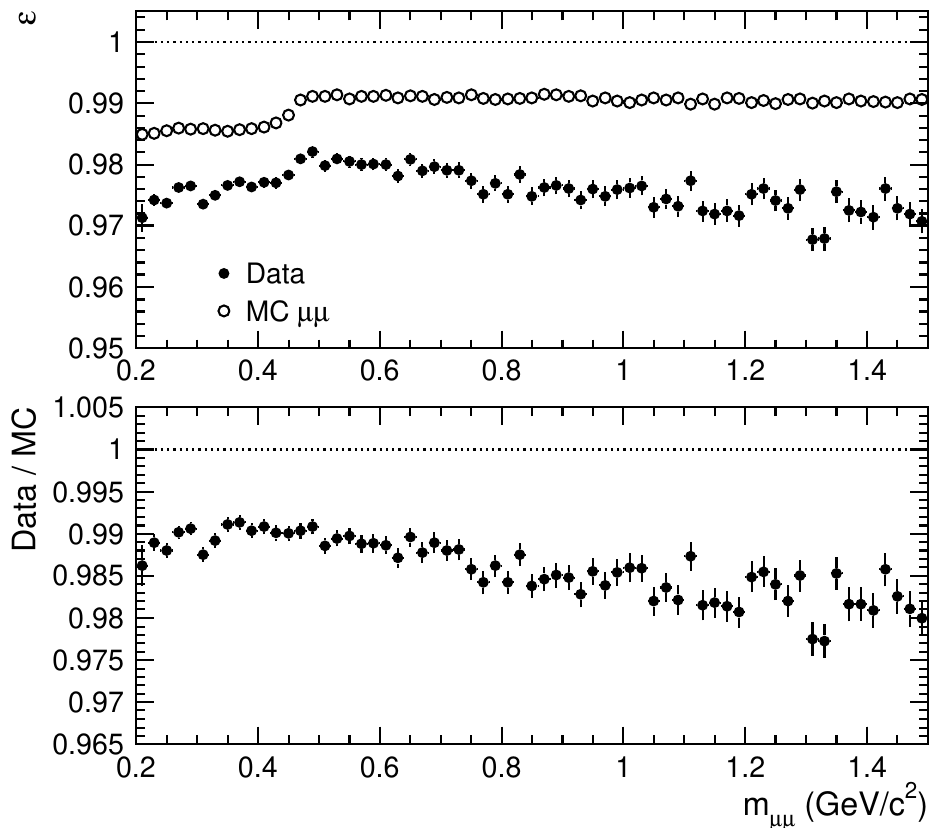}
  \caption{\small \justifying 
  The efficiency of the $\chi^2$ selection for $\mu\mu\gamma(\gamma)$ data (after background
  subtraction) and $\mu\mu\gamma(\gamma)$ MC simulation as a function of mass.}
  \label{fig:effchi2_massmumu}
  \end{minipage}
  \begin{minipage}{0.49\textwidth}
  \centering
  \includegraphics[width=\linewidth]{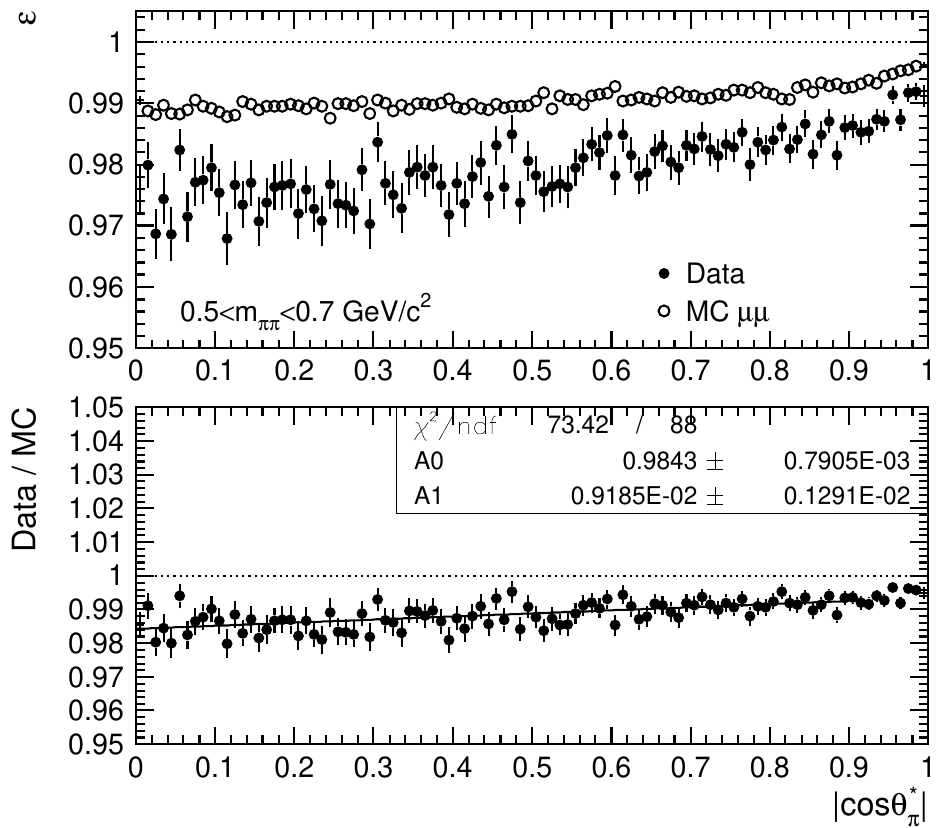}
  \caption{\small \justifying The efficiency of the $\chi^2$ selection for $\mu\mu\gamma(\gamma)$ data (after background
  subtraction) and $\mu\mu\gamma(\gamma)$ MC simulation as a function of $|\cos\theta^\ast_{\pi}|$ in the mass range between 0.5 and 0.7\gevcc with the pion mass hypothesis.}
  \label{fig:effchi2_costhpi}
  \end{minipage}
\end{figure*}

The efficiency of the 2d-$\chi^2$ selection for muons is obtained using events with two muons identified by a tight muon selector with background estimated using MC at the per mil level or below. The  efficiency obtained in background-subtracted data and $\mu\mu\gamma$ MC is rather insensitive to the stringency of muon ID requirement and the variation is taken as systematic uncertainty. The result as a function of mass is presented in Fig.~\ref{fig:effchi2_massmumu}.  The efficiency in data is typically 1\% lower than in MC and decreases slightly with mass.
The higher-order radiation observed in data but missing in \textsc{Phokhara} reported in Ref.~\cite{BaBar:2023xiy} may be part of the difference.
The discrepancy between data and simulation is due to different shapes of the $\ln(\chi^2_\textrm{LA/SA}+1)$ distributions, but the difference is directly measured and taken into account.

The 2d-$\chi^2$ selection efficiency for muons as a function of $|\cos\theta^\ast_{\pi}|$ is measured in mass bins of typically 200\,MeV/c$^2$. The data/MC efficiency ratio displays small slopes with $|\cos\theta^\ast_\pi|$ with little variation between neighboring bins so that several mass bins are merged together to provide template corrections with reduced statistical fluctuations. One example is shown in Fig.~\ref{fig:effchi2_costhpi}.
The fit performed is illustrative and the non-fitted corrections are applied directly to the templates predicted by MC simulations in the corresponding mass range. 

\subsubsection{Determination of the $\chi^2$ selection efficiency for $\pi\pi\gamma$}
\label{sec:eff-chi2-pipi}

The efficiency of the $\chi^2$ selection for the $\pi\pi\gamma$ data sample is derived using the relation:
\begin{equation}
\epsilon_{\chi^2}^{\pi\pi\gamma,\mathrm{data}}=\epsilon_{\chi^2}^{\mu\mu\gamma,\mathrm{data}}+\epsilon_{\chi^2}^{\pi\pi\gamma,\mathrm{MC}}-\epsilon_{\chi^2}^{\mu\mu\gamma,\mathrm{MC}},\label{eq:eff-derived}
\end{equation}
where $\epsilon_{\chi^2}^{\mu\mu\gamma,\mathrm{data}}$ and $\epsilon_{\chi^2}^{\mu\mu\gamma,\mathrm{MC}}$ are the 2d-$\chi^2$ selection efficiencies of data and MC events for the $\mu\mu\gamma$ samples from the previous subsection, and $\epsilon_{\chi^2}^{\pi\pi\gamma,\mathrm{MC}}$ is the corresponding efficiency for the $\pi\pi\gamma$ MC sample without any particle identification.
This relation provides a first approximation of the data-to-MC correction to the $\chi^2$ selection efficiency in pions, where it is assumed that the efficiency difference between pions and muons is given by the simulation. One expected difference is related to the FSR rates. However it was shown in Ref.~\cite{BaBar:2023xiy} that the latter are well simulated in \textsc{Phokhara} in both channels, and so is the FSR component of the pion to muon efficiency difference. Pion specific inefficiencies related to interactions are not so well simulated, so that additional corrections are subsequently applied (Section~\ref{subsec:eff-chi2-specpi}). 

\begin{figure*}[tbp]
  \begin{minipage}{0.49\textwidth}
  \centering
  \includegraphics[width=\linewidth]{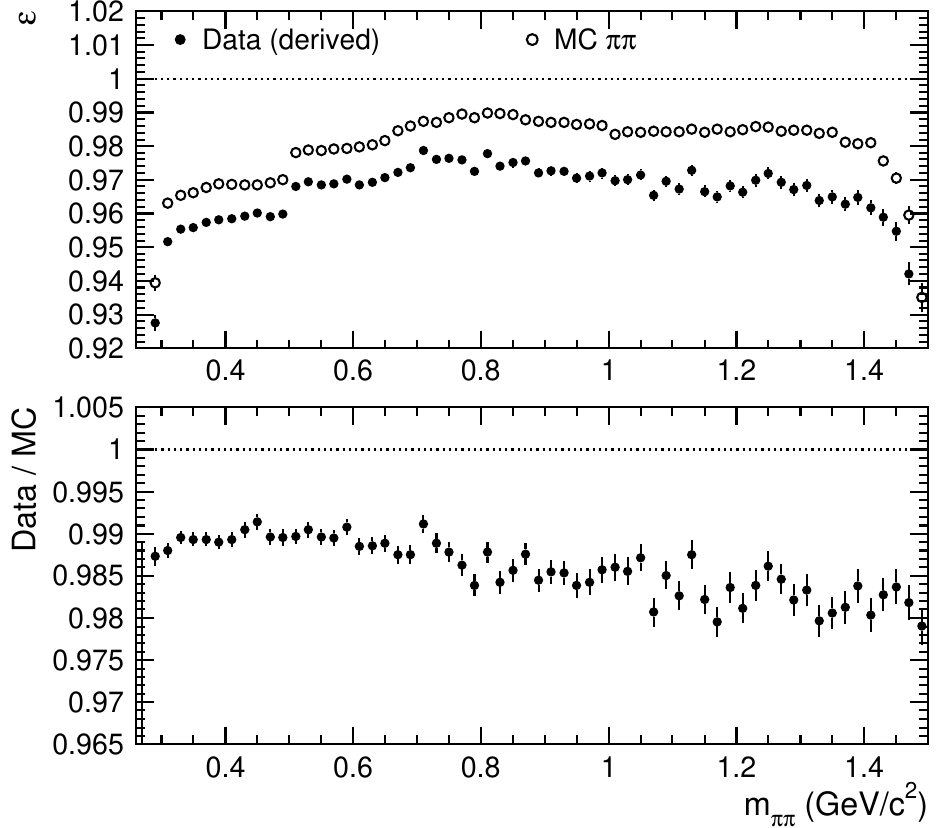}
  \caption{\small \justifying The derived $\chi^2$ efficiency for $\pi\pi\gamma(\gamma)$ data and $\pi\pi\gamma(\gamma)$ MC simulation as a function of mass.}
  \label{fig:effchi2_masspipi}
  \end{minipage}
  \begin{minipage}{0.49\textwidth}
  \centering
  \vspace{+6mm}
  \includegraphics[width=\linewidth]{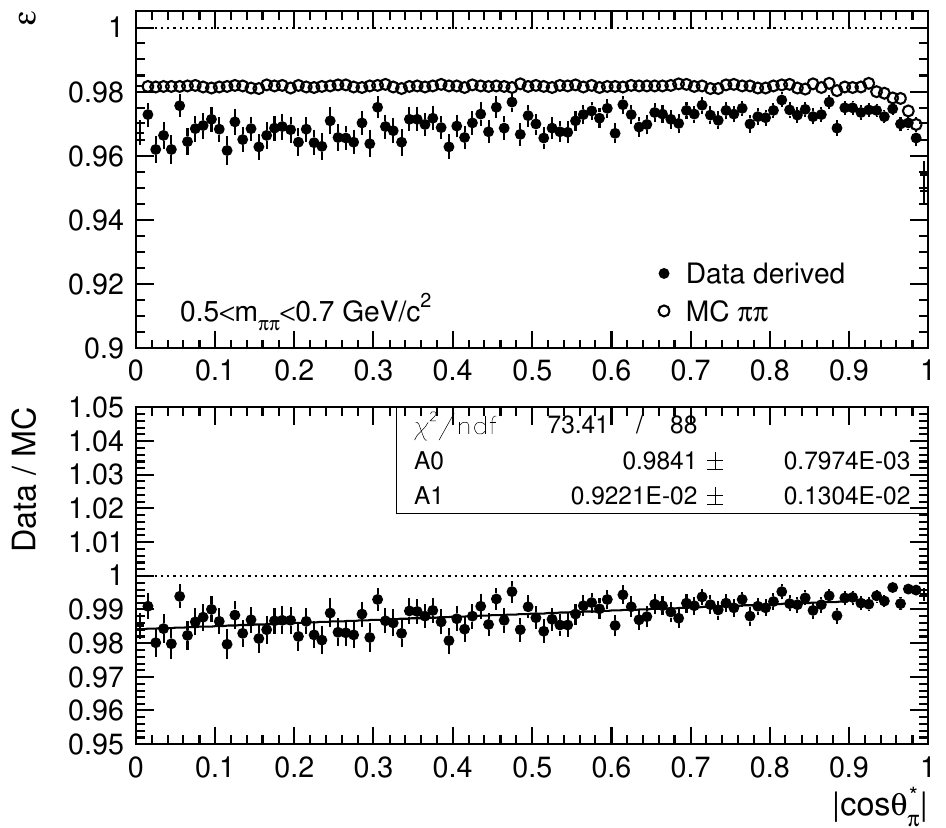}
  \caption{\small \justifying The derived $\chi^2$ efficiency for $\pi\pi\gamma(\gamma)$ data and $\pi\pi\gamma(\gamma)$ MC simulation as a function of $|\cos\theta^\ast_\pi|$ in the mass range between 0.5 and 0.7\gevcc with the pion mass hypothesis.}
  \label{fig:effchi2_costhpi_data2pi}
  \end{minipage}
\end{figure*}

The efficiency derived from Eq.~(\ref{eq:eff-derived}) as a function of mass in the pion mass hypothesis is presented in Fig.~\ref{fig:effchi2_masspipi}. In comparison with Fig.~\ref{fig:effchi2_massmumu}, the efficiency for the $\pi\pi\gamma$ samples is lower and shows a stronger mass dependence. The steps in the MC efficiency distribution at 0.5\gevcc and 1.0\gevcc correspond to the mass boundaries of the different $\chi^2$ selections. Given the high efficiency values of the selection, the data/MC correction derived from Eq.~(\ref{eq:eff-derived}) and shown in the bottom panel of Fig.~\ref{fig:effchi2_masspipi} is expected to be comparable with that of the $\mu\mu\gamma$ samples.  

The 
efficiency as a function of $|\cos\theta^\ast_\pi|$ 
is 
derived separately in different mass bins. Neighboring bins having a similar data/MC dependence on $|\cos\theta^\ast_\pi|$ are again merged. One example is presented in Fig.~\ref{fig:effchi2_costhpi_data2pi} in the same mass interval as in Fig.~\ref{fig:effchi2_costhpi}. 

\subsection{Corrections to simulation for specific pion effects}
\label{subsec:eff-chi2-specpi}

Data-to-MC corrections to the $\chi^2$ efficiency for $\pi\pi\gamma(\gamma)$ derived above from $\mu\mu\gamma(\gamma)$ and simulation account for the well-simulated common sources of inefficiencies in pions and muons. In contrast, simulation of pion-specific inefficiencies may be unreliable, necessitating specific data/MC corrections.

\subsubsection{Pion-induced fake photons}

\begin{figure}[tbp]
  \centering
\includegraphics[width=\linewidth]{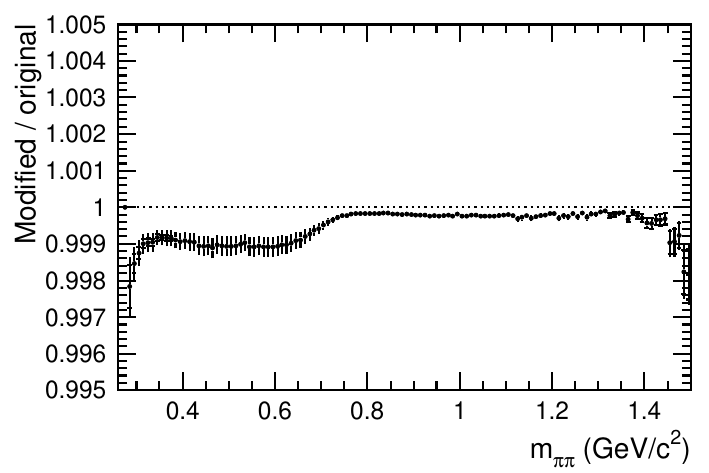}
  \caption{\small \justifying The effect of fake photon corrections on the $\chi^2$ selection efficiency as a function of mass. 
  At low and high masses, statistical uncertainties (inner error bar) dominate while in the central region the correlated uncertainties across the mass bins from the data-to-MC correction factors are dominant.
 }
  \label{fig:fake-photon-mass}
\end{figure}

A detailed investigation of spurious photons created by pion interactions in the EM calorimeter in data and simulation has been 
conducted for the study of additional FSR in $\pi\pi$ events ~\cite{BaBar:2023xiy}. The spurious photon contributions are studied with three discriminant variables, the energy deposit in the electromagnetic calorimeter, $E_\mathrm{cal}$, the moment of the energy deposit of the EMC cluster, Zernike20, which characterizes the lateral shape of the shower, and the $\chi^2_\textrm{LA}$ value of the LA kinematic fit. It is found that the region with $E_\mathrm{cal}>0.4$\gev and $\text{Zernike20}<0.85$ is dominated by the fake photon contributions. The fake photon fraction is determined from a template fit to the $\chi^2_\textrm{LA}$ distribution to be $0.377(4)(7)$ in simulation, 
and $0.458(4)(10)$ in data, where the first uncertainty is statistical and the second one systematic. These values are used to derive a correction to the simulated pion $\chi^2$-selection  efficiency shown in Fig.~\ref{fig:fake-photon-mass}.

\subsubsection{Effect of pion secondary interactions}
\label{sec-int-study}

\begin{figure}[tbp]
  \centering
\includegraphics[width=\linewidth]{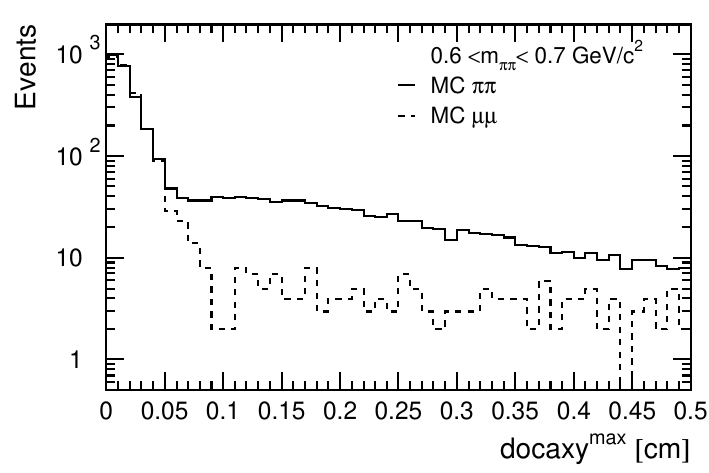}
\includegraphics[width=\linewidth]{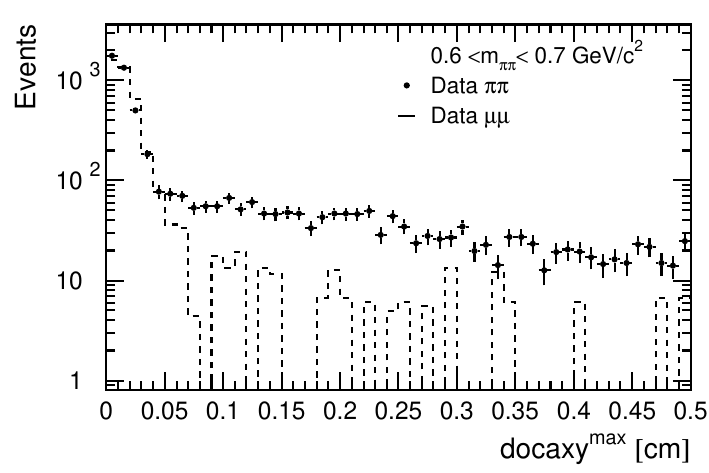}

  \caption{\small \justifying Distributions of $\mathrm{doca}_{xy}^\textrm{max}$ comparing events satisfying pion and muon ID in the interaction-enriched zone in the 2d-$\chi^2$ plane for MC (top) and data (bottom)  in the mass interval 0.6--0.7\gevcc.} 
  \label{fig:docamax-zone}
\end{figure}

The effects of secondary interactions are mostly
seen in the tracking efficiency because of the tight selections imposed on the 
track pointing to the interaction region. Some residual effects in the 
2d-$\chi^2$ and $V_{xy}$ selection efficiencies are also expected. 
The study of the effect of secondary interactions in data and simulation is based on the distributions of the $\mathrm{doca}_{xy}^\textrm{max}$ variable, which is the largest distance of closest approach to the beam axis in the $x$-$y$ plane of the two good tracks, a larger deviation indicating a kink suffered by the interacting track. Comparing the magnitudes of the $\mathrm{doca}_{xy}^\textrm{max}$ tail above 0.1\cm
allows one to measure the relative rate of secondary interactions in data and simulation in a restricted region in the 2d-$\chi^2$ plane near the boundary to enhance the effect. The method is illustrated in Fig.~\ref{fig:docamax-zone}. The data/MC ratio is determined in 100\,MeV/c$^2$-wide mass intervals, reduced to 50\,MeV/c$^2$ below 0.5\gevcc. On the $\rho$ resonance the ratio is 1.64(13)(4) with statistical and systematic uncertainties indicated.

\begin{figure}[htbp]
  \centering
\includegraphics[width=\linewidth]{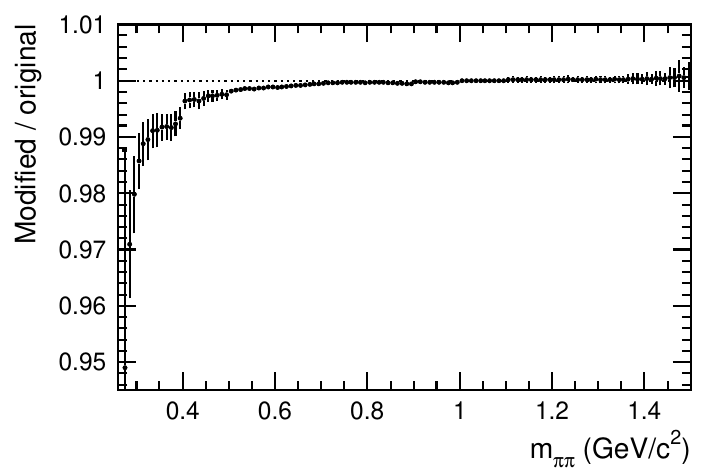}
  \caption{\small \justifying The effect of pion interaction corrections on the $\chi^2$ selection efficiency as a function of mass. The first bin is dominated by the statistical uncertainty while the remaining bins are  dominated by the correlated uncertainties across the mass bins from the data-to-MC correction factors.
 }
  \label{fig:modified-mass}
\end{figure}

 Using those data/MC ratios, the effect of pion interaction corrections on the $\chi^2$ selection efficiency is shown in Fig.~\ref{fig:modified-mass}. 

\section{Backgrounds}
\label{sec:bkg}

\subsection{Summary on backgrounds}

Signal events are from the ISR processes $e^+e^-\rightarrow \mu^+\mu^-\gamma, \pi^+\pi^-\gamma, K^+K^-\gamma$ whose different components are separated through fits of the angular distributions. Other processes are considered as background and subtracted. They are estimated with MC generators corrected with input from measured cross sections and checked with direct tests in data. A special case is the $e^+e^-\gamma$ final state, which is
strongly reduced at different levels of the event reconstruction and whose contribution is accounted for at the fit level as discussed later (Sections~\ref{sec:eeg} and \ref{sec:spectra}).

\begin{figure}[hbp] 
\centering
    \includegraphics[width=\linewidth]{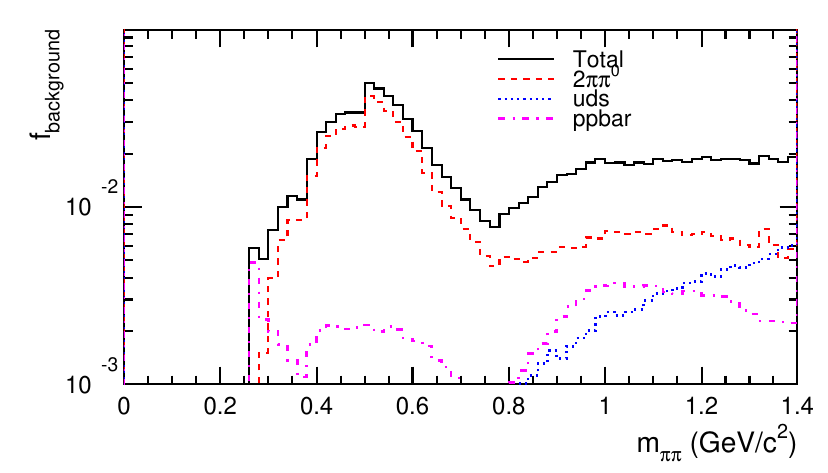}
    \includegraphics[width=\linewidth]{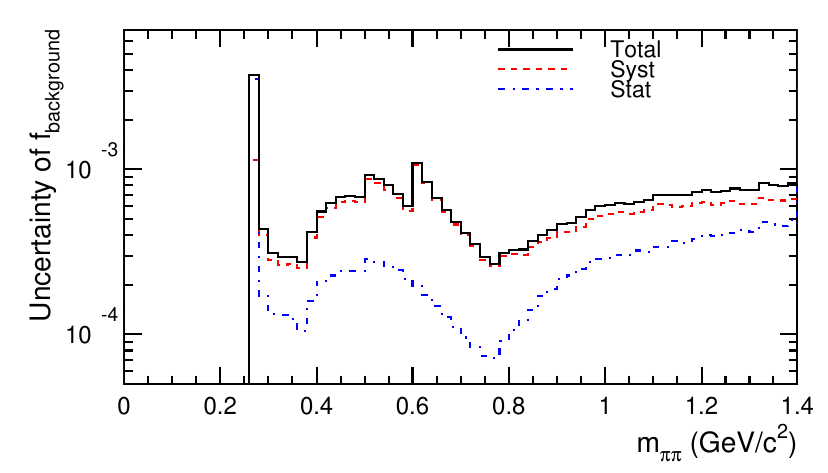}
    \caption{\label{fig:bkg}   \small \justifying
 Fractional contribution of the background to the total number of events, $f_\textrm{background}$
 (top) and its uncertainty (bottom) after the 2d-$\chi^2$ selection.}
\end{figure}

The background contribution over the total selected events and its uncertainty are shown in Fig.~\ref{fig:bkg}. The dominant background processes are $p\overline{p}$ at threshold (but this background is very small at higher mass), $\pi^+\pi^-\pi^0\gamma$ over the full mass range, and \qqbar at the top edge of the mass distribution. Minor contributions are from other ISR multihadronic and $\tau^+\tau^-$ processes. The total uncertainty includes the uncertainties in the measured cross sections used to normalize the different processes, and uncertainties obtained from direct tests of the \qqbar simulation detailed in the next section, and is limited to around and below 0.1\%. 

\subsection{Multihadrons from the \qqbar process}
\label{bkg-had}
Hadronic processes  
are a significant source of background for masses above 0.9\gevcc.  
Their contribution is estimated using simulated samples generated with \textsc{JETSET}. 
However, since the prediction for low-multiplicity final states may be unreliable, the rate is normalized using data.

In a \qqbar event mimicking an ISR process, the ISR photon
candidate most often originates from the decay of an energetic
$\pi^0$. Such a signature is searched for, both in data
and in MC. Resolved $\pi^0$ are identified by pairing the ISR photon candidate with all
detected additional photons having an energy larger than 50\mev. 
The pair with $\gamma\gamma$
mass closest to the nominal $\pi^0$ mass is retained.
However, for energetic $\pi^0$'s, the two photon showers are likely to be merged into a single calorimeter cluster, elongated in the transverse distribution. Such a topology is searched for using 
the Zernike20 value of the $\gamma_{\rm{ISR}}$ candidate in the EMC crystals. 
In this case, the $m_{\gamma\gamma}$ calculation involves a second photon either from an extra $\pi^0$ or from beam background. Thus, the \qqbar
background has two main populations corresponding to resolved and unresolved $\pi^0$, which are clearly separated in the ($m_{\pi\pi}$, Zernike20) plane.

To select the hadronic processes in the accepted $\chi^2$ region, a BDT-based multivariate technique is exploited in three separate $\pi\pi$ mass windows: the low mass window between threshold and 0.6\gevcc, the intermediate one between 0.6 and 0.9\gevcc, and the high mass window between 0.9 and 1.4\gevcc. For each $\pi\pi$ mass window, two separate BDT-based selections are performed for the resolved and unresolved $\pi^0$ samples, providing a huge reduction of the non-\qqbar events. Typically the obtained \qqbar/non-\qqbar ratio is about 2.5 for a \qqbar efficiency of roughly 50\%. Detailed fits to $\gamma\gamma$ mass distributions are provided for the resolved and unresolved $\pi^0$ in different mass intervals. It is found that \textsc{JETSET} overestimates the rate of selected low-multiplicity events: in the mass range 0.9--1.4\gevcc the factors are $2.2(0.2)$ and $1.6(0.1)$ for resolved and unresolved $\pi^0$, respectively. The  factors obtained are used to scale down the MC predictions.

\subsection{Multihadronic ISR processes}
\label{bkg-isr}
The multihadronic background is estimated using $e^+e^-\to X\gamma_\mathrm{ISR}$ processes simulated by \textsc{AfkQed} with 14 final states $X$ considered, of which the most important is $\pi^+\pi^-\pi^0$, with $\pi^+\pi^-2\pi^0$, $2\pi^+2\pi^-$, $\eta\pi^+\pi^-$, $K_SK_L$ sub-dominant. The mass spectra of these processes are scaled to the measured cross sections based on the combination using HVPTools~\cite{Davier:2010fmf}. 

A BDT approach 
similar to that applied to the \qqbar background study is performed to determine the normalization factor of the $3\pi$ MC samples in two $m_{\pi\pi}$ mass regions: below 0.6\gevcc and between 0.6 and 0.9\gevcc. The $3\pi$ process is dominated by the production of the $\omega$ and $\phi$ narrow resonances, which are fitted both in data and MC.
Each sample is subdivided into two cases depending on the number of reconstructed photons (including $\gamma_\mathrm{ISR}$), $n_\gamma>2$ and $n_\gamma=2$. A BDT is optimized for each case, the separation between $3\pi$ and non-$3\pi$ being better for the first one.

The normalization factor is obtained from the ratio of the total BDT-selected candidates in data in the above mass ranges after non-$3\pi$ background subtraction over the event yield of the $3\pi$ MC sample.
The normalization factors to be applied to the MC prediction are given in Table~\ref{tab:nf-3pi}. They are consistent with one but their uncertainties are taken as systematic uncertainties for the dominant $3\pi$ background subtraction.

\begin{table}[tbp]
\centering
\caption{\small \justifying Normalization factors for the $3\pi$ background in the two $\pi\pi$ mass windows (2nd and 3rd columns) and for $n_\gamma>2$ (2nd row) and $n_\gamma=2$ (3rd row). The first uncertainty is statistical and the second (when shown) systematic.}
\label{tab:nf-3pi}
\vspace{2mm}
\begin{tabular}{c|ll} \hline
  & $m_{\pi\pi}<0.6$\gevcc & $0.6<m_{\pi\pi}<0.9$\gevcc \\\hline
$n_\gamma>2$ & $1.003(10)(11)$ & $1.013(21)(14)$ \\
$n_\gamma=2$ & $1.009(37)          $ & $1.097(118)         $ \\\hline
\end{tabular}
\end{table}

\subsection{Background from $\tau\tau$ process}

The 
$\tau\tau$ background is small (about 0.2 per mil) in the accepted 2d-$\chi^2$ region. It dominates, however, by far in the rejected 2d-region, particularly in the upper corner region. This provides the opportunity to check the normalization of the $\tau\tau$ sample, after a selection on the measured total energy $E_\mathrm{tot}<9$\gev.
The integrated data and MC event yields give a data/MC ratio of $1.12\pm 0.04$. Given the $\tau\tau$ process is well modeled, we keep the normalization of the process unchanged but quote a systematic uncertainty of 12\% for the $\tau\tau$ background subtraction. 

\subsection{Remaining $ee\gamma$ background}\label{sec:eeg}
 
Radiative Bhabha events are strongly reduced by the electron veto applied in the track selection  
and by an event veto rejecting two-track events with
$[(E_\mathrm{cal}/p)_1-1]^2 + [(E_\mathrm{cal}/p)_2-1]^2 < (0.35)^2$. Remaining events
have variables outside of the selections or one of them not defined. This is the case in particular 
with tracks below $p_\mathrm{T}\sim 0.3$\gevc, which do not reach the EM calorimeter. Since this
configuration is essential for muons at large $|\cos\theta^*_\pi|$ values, it is
important to understand the behavior of the leftover Bhabha events.
Although largely removed by the $V_{xy}$ selection, another source of $ee\gamma$ background is from the process 
$e^+e^- \rightarrow \gamma \gamma$ where one of the photons converts into an
$e^+e^-$ pair in the beam pipe or material at larger radii. 
 
 The small $ee\gamma$ background results from several levels of suppression that are difficult to accurately model with MC simulation. Furthermore this background is strongly dependent on $|\cos\theta^*_\pi|$. As a consequence, it is not subtracted from the mass spectrum, but taken into account in the angular distribution fit of the different components. 
In the following, the $|\cos\theta^*_\pi|$ distribution  obtained from data by selecting a proper $ee\gamma$ sample is described.

\subsubsection{Cut-based \eeg\ selection}
\label{subsubsec:cutb}

An initial cut-based selection is achieved with the following requirements, labeling 1 (2) the track with the higher (lower) momentum:
\begin{itemize}
  \item 
 $\left(E_\mathrm{cal}/p\right)_1 > 0.3$,
  \item 
 $\left(\dedx\right)_1 > 600$,
  \item 
 $\left(\dedx\right)_2 > 500 +60\cdot p_2 +8.9\cdot p_2^2$,
  \item 
 $\mathrm{isElectron}_1 < 10$ or $\mathrm{isElectron}_2 < 10$, 
\end{itemize}
 where 
 $\mathrm{isElectron} =\left(\frac{E_\mathrm{cal}/p-1}{0.15}\right)^2+\left(\frac{\dedx-690}{150}\right)^2$.

\subsubsection{BDT-based \eeg\ selection}
\label{subsubsec:BDT}

In order to improve the signal purity of the selected \eeg\ sample, a selection is devised based on a BDT, taking as training features ten variables with discriminating power: $(E_\mathrm{cal}/p)_1$, $(\dedx)_1$, $(\dedx)_2$, $p_1$, $p_2$, $V_{xy}$, $\mathrm{isElectron}_1$, $\mathrm{isElectron}_2$, $m_{\pi\pi}$, and $m_{\mu\mu}$. The training is performed on all events that pass the cut-based selection, the signal being defined as data events, and the background as MC simulated  \mumug, \pipig, and $KK\gamma$ events with proper normalization.

Parameters of the BDT, such as the maximum depth of a tree, and the learning rate, are optimized in order to improve its accuracy. In order to avoid the bias that would come by applying a BDT to a sample on which it was trained, the sample taken as input is split for the purpose of training and validation.
Data events that have a 60\% or higher probability of being \eeg\ signal are selected to determine the $|\cos\theta^*_\pi|$ distribution of the \eeg\ process. 

\subsubsection{$\pi\pi\gamma$ background in the $ee\gamma$ sample}
\label{subsubsec:pibkg}

Some residual background remains in the selected $ee\gamma$ sample, mostly from pions, but also from muons and kaons. It is not  
expected that the simulation describes well the behavior of pions in the $E_\textrm{cal}$ and $dE/dx$ distribution tails. Therefore a data/MC comparison is performed using an electron-free and muon-free sample. 
This is realized by the set of selections:
$(E_\mathrm{cal})_1>0.5$\gev,
$(E_\mathrm{cal})_2>0.3$\gev,
$0.6<m_{\pi\pi}<0.9$\gevcc,
and $|\cos\theta^*_\pi|<0.9$. 
All the selections are effective at removing $\mu\mu\gamma$ events while the last one is designed to veto $ee\gamma$ background.

Applying the cut-based $ee\gamma$ selection to this $\pi\pi\gamma$ sample, it is found that the simulation underestimates the rate of $\pi\pi\gamma$ that pass the set of selections. The discrepancy comes mainly from the $(E_\mathrm{cal}/p)_1>0.5$ requirement, while the $(\dedx)_{1,2}$ selections are better reproduced, leading to a scale factor of about 3.2 applied to the MC.
Due to the inherent uncertainty in this background scaling method, the actual \eeg\ distributions are defined as the average of the background-subtracted distributions with and without scaling.

\subsubsection{Distributions of $ee\gamma$ background}
\label{subsubsec:eetemp}
 
\begin{figure}[tbp]
    \centering
    \includegraphics[width=\linewidth]{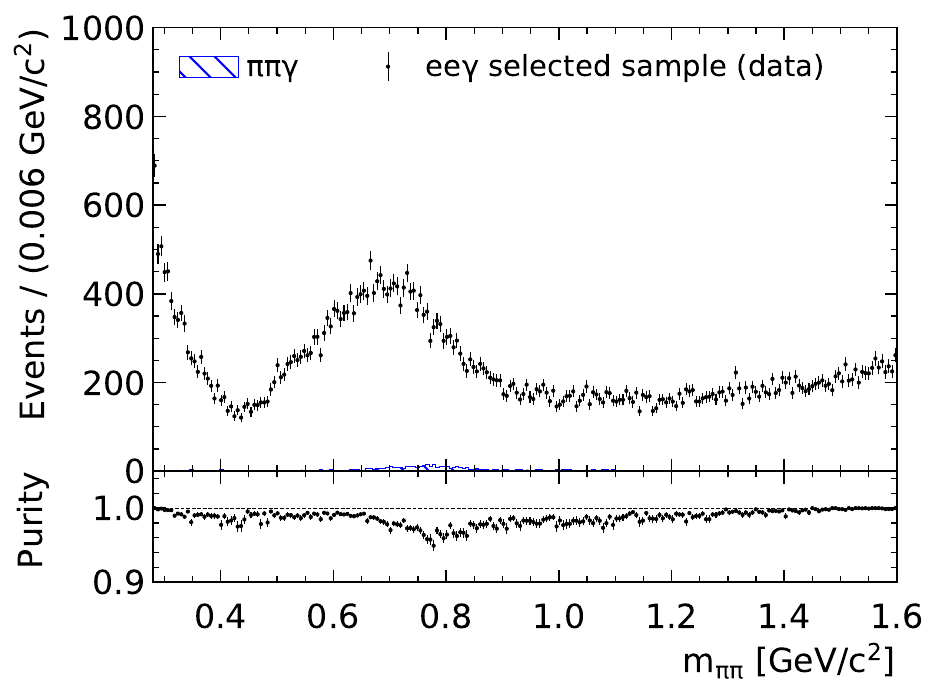}
\caption{\label{esel-mass} \small \justifying
The $m_{\pi\pi}$ (top panel) 
mass distribution of the selected $ee\gamma$ sample in data (black points) and data-corrected MC \pipig\ background (hatched histograms). 
The purity is the proportion of data events after background subtraction.
}
\end{figure}

The mass distribution of the $ee\gamma$ selected sample is shown in Fig.~\ref{esel-mass} together with the data-corrected background. 
The \pipig\ MC background is normalized according to the scale factor determined in Section~\ref{subsubsec:pibkg}. The \mumug\ and \KKg\ backgrounds are negligible. 
The mass spectrum of the selected $ee\gamma$ events presents a large peak at threshold due to symmetric photon conversions from diphoton events. The $m_{ee}$ mass of these events is below $0.05\gevcc$. The wide bump centered around $m_{\pi\pi}=0.675\gevcc$ is due to asymmetric photon conversions, where the low momentum electron does not reach the EMC and is not identified. The bump corresponds to the tail of the $m_{ee}$ distribution due to angular resolution of the electron tracks, scattering in the beam pipe or SVT, or intrinsic photon virtuality.  The combined effect of the decreasing $m_{ee}$ distribution and the increasing
asymmetry, hence increasing electron misidentification, creates the dip around $0.45\gevcc$ in the $m_{\pi\pi}$ spectrum seen in Fig.~\ref{esel-mass}. As expected for these events, the cosine of the angle between the two tracks is strongly peaked at 1, especially at threshold. In addition to photon conversions, radiative Bhabha events span the full range of the $m_{\pi\pi}$ spectrum.

\begin{figure}[!htp] \centering
    \includegraphics[width=\linewidth]{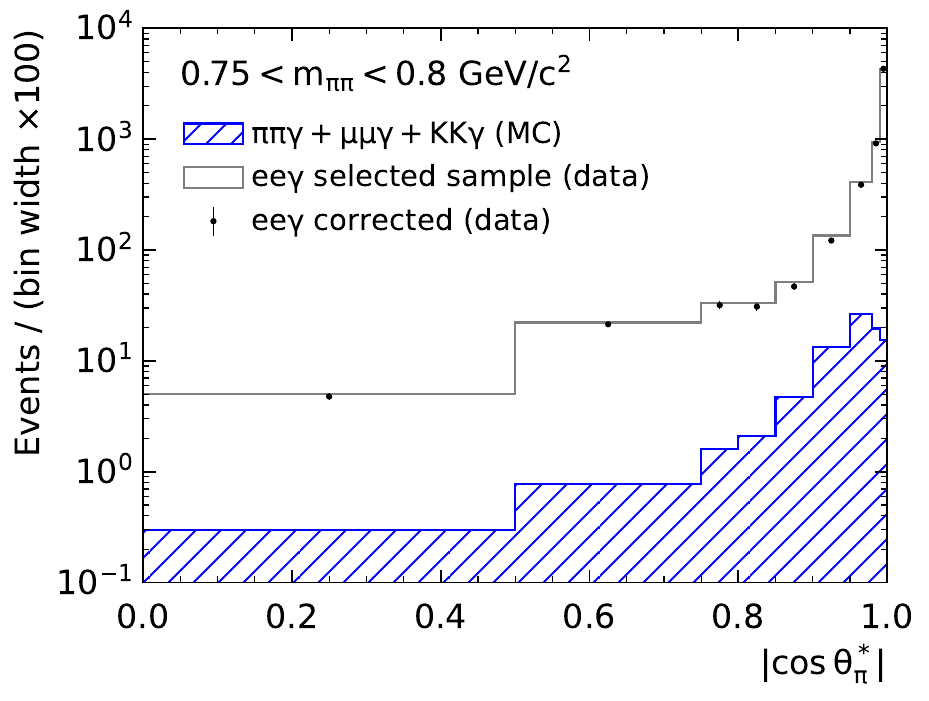}
  \caption{\label{eeg-cosths} \small \justifying 
The $|\cos\theta^*_\pi|$ distributions of the MC simulation background (blue hatched histogram), $ee\gamma$ selected sample in data (gray histogram) and the $ee\gamma$ corrected sample in data after background subtraction (black points) around the $\rho$ resonance. Bin edges are 0, 0.50, 0.75, 0.80, 0.85, 0.90, 0.95, 0.98, 0.99 and 1.
} 
\end{figure}

The $|\cos\theta^*_\pi|$ distributions of the $ee\gamma$ background in each mass interval are obtained in a similar way as the mass distribution. The distribution at the $\rho$ resonance mass is shown in Fig.~\ref{eeg-cosths} 
before and after background subtraction.
The background-subtracted $ee\gamma$ distributions are included in the angular fits to data distributions, as explained in Section~\ref{sec:spectra}.

\section{Trigger efficiency}
\label{sec:trigger}
\subsection{Method}

The response of each of the three levels of the \babar\ trigger (hardware L1, online software L3, and prompt reconstruction background filters BGFilter)~\cite{BaBar:2001yhh,BaBar:2013byz} is built from the individual responses of conditions (lines) defining the trigger.
In L1 and BGFilter, line definitions mix information from different detectors. In L3, there are two physics lines only, one relying on DCH quantities, the other on EMC quantities. The efficiencies at each level are computed by cross-checking the responses of lines (or groups of lines) that are as orthogonal as possible to minimize common inefficiencies. 

Trigger efficiency measurements are performed using control samples selected with the same criteria in data and MC. Since the analysis relies on mass spectra and templates of $|\cos\theta^*_\pi|$ distributions of $\mu\mu\gamma$ and $\pi\pi\gamma$ events, trigger efficiencies are measured in the two signal channels separately, using PID to select a pure sample of each. For both signals, trigger efficiencies and data/MC corrections are determined under the pion mass hypothesis for the charged tracks. In the initial stages of the analysis, the efficiencies in MC are blinded with a constant multiplicative offset. The efficiencies and data/MC corrections shown here are the unblinded ones.

To be as uncorrelated as possible with the tracking efficiency measurement, the trigger efficiency is determined with samples selected with minimal requirements on the number of measured tracks in the event. Data and MC events are required to pass a one-constraint (1C) fit to $e^+e^-\to \pi\pi\gamma_{\rm{ISR}}$ that uses only the 4-momenta of one track assumed to have the pion mass, and the ISR photon. The only requirement put on the expected second track is for it to lie within the detector acceptance ($0.40<\theta<2.45\rad$ in the laboratory frame). The primary track is required to be a 
track (so-called `standard track') that meets relaxed criteria with respect to the 
`good' tracks defined in Section~\ref{evt-sel}: $\mathrm{doca}_{xy}<2.5\cm$ and no requirement on the number of hits in the DCH. 

The biases induced by the selection of control samples are studied, as well as the bias due to the non-orthogonality of the crossed lines of the trigger.  

\subsection{Trigger efficiency results for muons}

The L1 efficiency is larger than 0.9998 in data and MC and no data/MC correction is needed. This is also the case for BGFilter, in which dedicated di-muon lines are very efficient. On the contrary, sizable inefficiencies are measured for the online software trigger L3. 

The DCH-based L3 line b1 requires either two reconstructed tracks with minimal transverse momentum $p_\mathrm{T}>0.25\gevc$ or one single track with $p_\mathrm{T}>0.6\gevc$. When tracks are close to each other in azimuthal angle, the online tracking algorithm may not disentangle the overlapping hits in the drift chamber, and misses one track or even the two tracks altogether. Track overlap in the DCH leads to the dominant loss for the b1 trigger.
The EMC-based L3 line b2 relies solely on cluster finding. It requires either two high energy clusters ($E^\ast>0.35\gev$ in CM) or at least four clusters of $E>0.1\gev$. This is hardly fulfilled by $\mu\mu\gamma$ events with mostly minimum ionizing particle deposits.  The global L3 efficiency is thus driven by the DCH-based line, while the statistical precision of its determination is limited by the large inefficiency of the EMC-based line.

 Identification of the primary track as a muon is required, using the standard \babar\ BDT-based muon selector (\hbox{BDTMuonSelector})~\cite{BaBarBelle2014}.
 Even after muon ID, some background coming from $\pi\pi\gamma$ events remains and is subtracted out, resulting in an uncertainty of $7 \times 10^{-4}$ in the data/MC correction at the $\rho$ peak but is negligible elsewhere. 

\begin{figure}[tbp] 
    \centering
    \includegraphics[width=\linewidth]{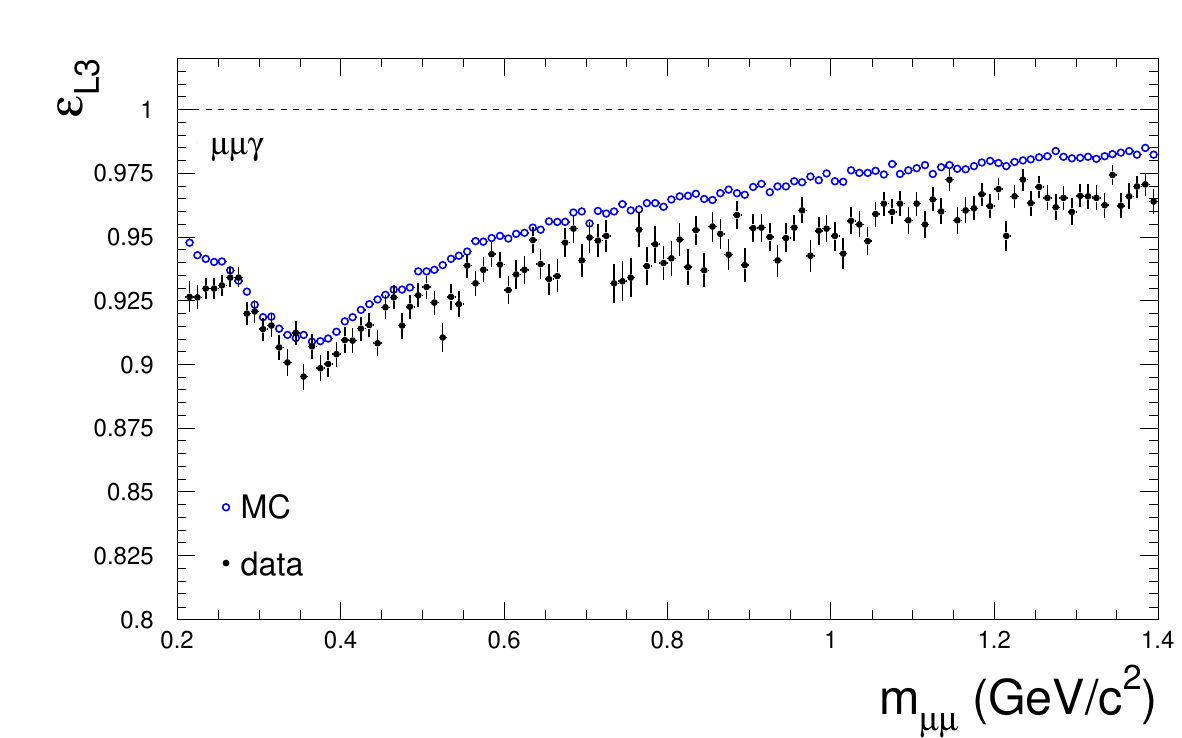}    
    \includegraphics[width=\linewidth]{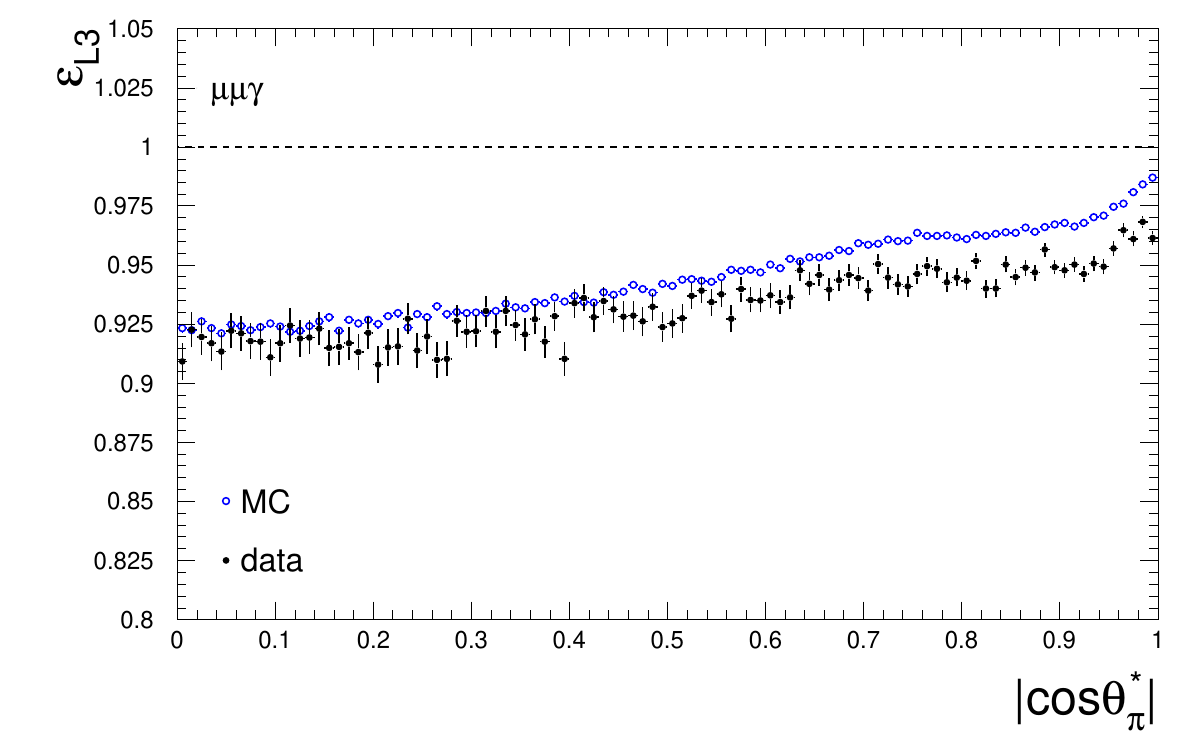}
    \caption{\label{fig:MuTrigEff_mass-cstst}   \small \justifying
    The L3 efficiency measured in $\mu\mu\gamma$ events in MC and data (after background subtraction) as a function of (top) $m_{\mu\mu}$ mass and
    (bottom) \acosthpi.
}
\end{figure}
\begin{figure}[htbp] 
    \centering
   \includegraphics[width=\linewidth]{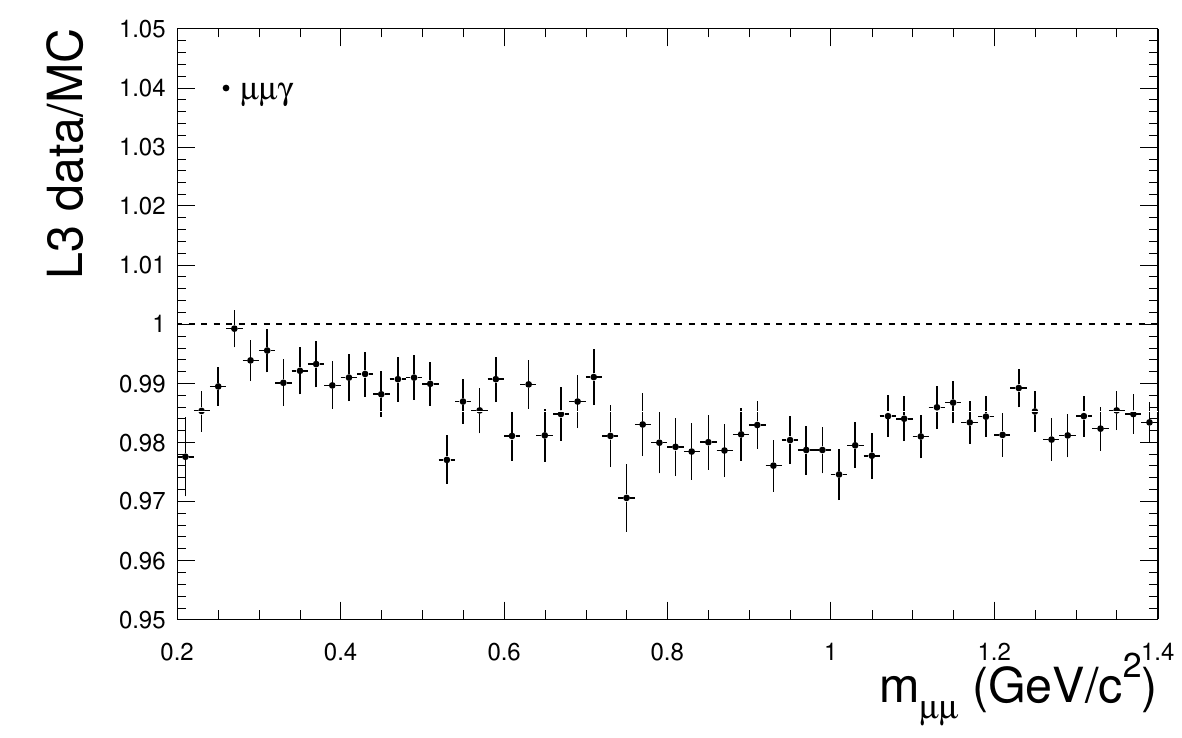}
   \includegraphics[width=\linewidth]{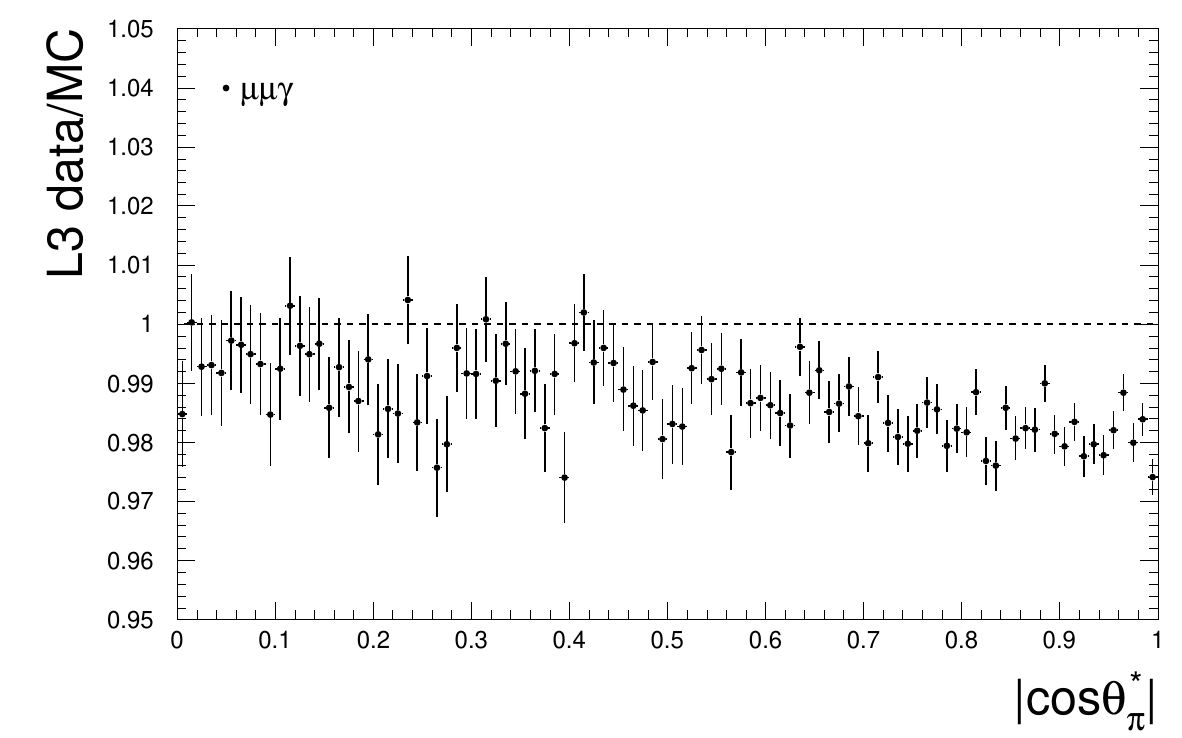}
   \vspace{-5mm}
     \caption{\label{fig:MuTrigEff_dataMCcorr}   \small \justifying
    The data/MC L3 efficiency correction in $\mu\mu\gamma$ events: (top) as a function of $m_{\mu\mu}$, and
    (bottom) \acosthpi.
}
\end{figure}
The global efficiency determined by cross-checking the two L3 lines b1 and b2 is shown in Fig.~\ref{fig:MuTrigEff_mass-cstst} (top) as a function of mass, and (bottom) as a function of \acosthpi\ integrated over the full mass range. As expected, the largest inefficiency is reached around 0.35\gevcc, where the effect of track overlap in the DCH is maximal. The efficiency measured in data departs from that in MC mostly at large \acosthpi, where one of the two muons, the primary one or the expected one, is of very low momentum. Data/MC corrections to the MC templates are applied bin by bin, while systematic uncertainties are expressed as changes of curvature of the \acosthpi\ distribution.  

The difference between the L3 efficiency on the full sample with zero or more tracks and the efficiency using the selected sample with at least one standard
track is determined in the simulation. That bias is controlled with data, using the efficiencies measured separately with events with one or two standard tracks. 
The bias induced by requiring muon ID is studied using the \hbox{BDTMuonSelector} with different identification levels, which can be compared in data and simulation. 
The data/MC correction is taken from the ratio measured at the very loose level, with an uncertainty given by the difference with the ratio at the loose level.  
The bias on the global efficiency due to the measurement method is estimated with MC as the difference between the rate of events passing the trigger (true efficiency) and the measured efficiency calculated by cross-checking the trigger lines. It affects the data/MC efficiency correction by less than 1 per mil.   

The data/MC efficiency correction 
is given in Fig.~\ref{fig:MuTrigEff_dataMCcorr}. The total systematic uncertainties summarized in Table~\ref{tab:syst-trigger} amount to $1.1\times10^{-3}$ in the absolute value and an uncertainty in the \acosthpi\ template shape of $5.\times 10^{-3}\times\cos^2\theta^\ast_\pi$.  This results in an uncertainty in the cross section that is much smaller than the statistical uncertainty in each bin (see Section~\ref{unfolding}).

\subsection{Trigger efficiency results for pions}

The trigger efficiency for $\pi\pi\gamma_\mathrm{ISR}$ events is computed in the same way as for $\mu\mu\gamma_\mathrm{ISR}$, with additional selections to reduce the large muon and multihadron background that dominates in the low and high mass ranges. To this end, a positive PID, based on a standard \babar\ selector~\cite{BaBarBelle2014}, is required for the primary track, and events with identified $\pi^0$ (resolved and unresolved photon pairs) are removed. 
The remaining backgrounds from muons and electrons, which are not removed by the standard \babar\ PID selector, are reduced by a muon veto 
and an electron veto 
applied to any track pointing to the event vertex. 

In addition, 
events are subjected to a four-constraint (4C) fit to $e^+e^-\to\pi\pi\gamma_\mathrm{ISR}$, which makes use of 
any reconstructed \babar\ track (with at least 1 hit in the SVT) in addition to the required primary standard track and ISR photon. The second track is required to be of opposite sign to the primary. In case several second track candidates are available, the (primary, second) track pair leading to the smallest $\chi^2$ of the 4C fit is chosen. The more constrained 4C fit retains 65\% of the pion signal in MC, while reducing the multihadronic background substantially. 
The L3 and BGFfilter efficiencies are measured on the pion sample satisfying the 4C fit. As for muons, the hardware trigger L1 has a negligible inefficiency and no data/MC correction is needed.

\subsubsection{L3 efficiency results for pions}

\begin{figure}[tbp] 
    \centering
    \includegraphics[width=\linewidth]{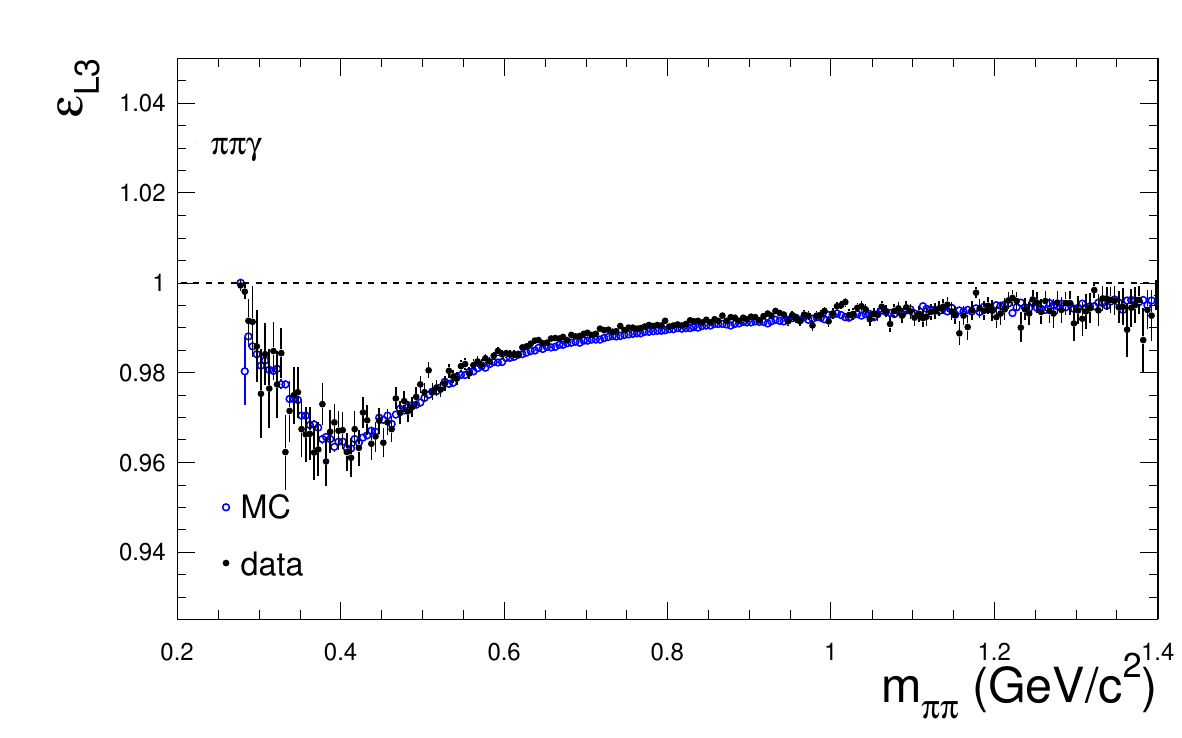}
    \includegraphics[width=\linewidth]{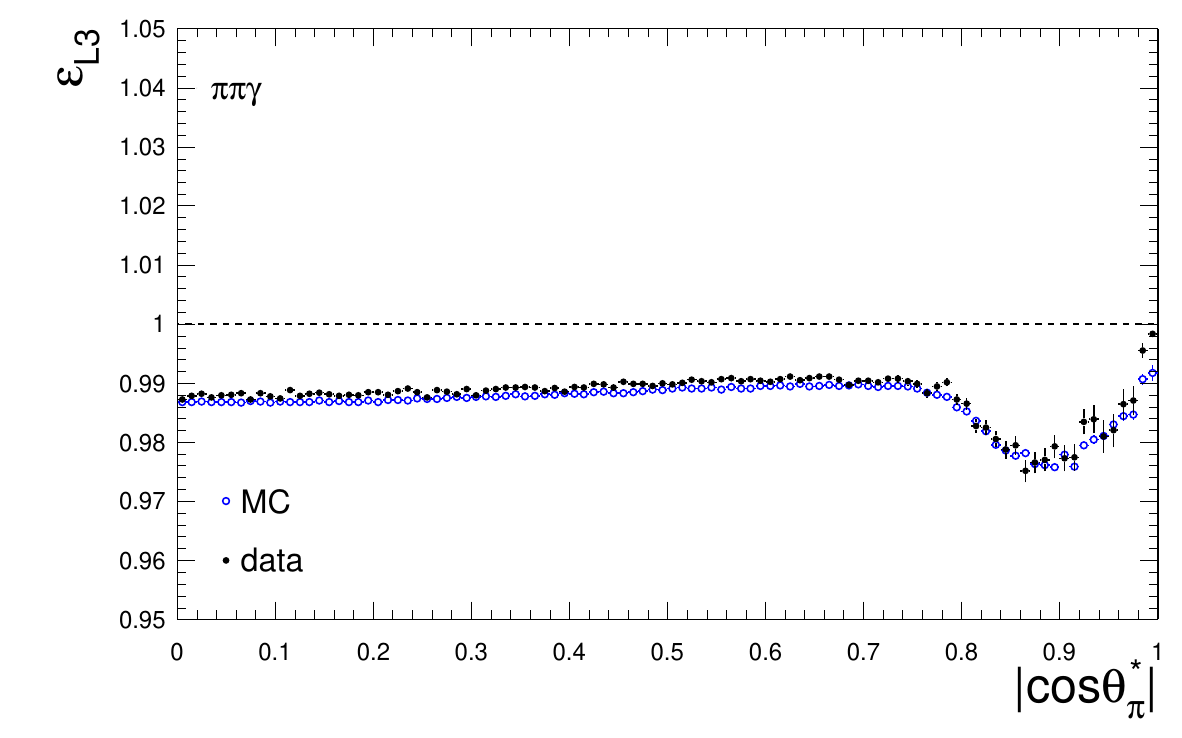}
    \caption{\label{fig:PiTrigEff_mass-cstst}   \small \justifying
    The L3 efficiency measured in $\pi\pi\gamma$ events in MC and data (after background subtraction) as a function of (top) $m_{\pi\pi}$ mass and
    (bottom) \acosthpi\ integrated over mass.
}
\end{figure}

 The efficiency of the DCH-based line b1 for pions is similar to that observed for muons, which is expected since the track overlap effect is dominant. As the EMC-based line b2 is very efficient on pions, the global L3 efficiency is much higher than for muons, as shown in Fig.~\ref{fig:PiTrigEff_mass-cstst}. 
\begin{figure}[htbp] 
    \centering
    \includegraphics[width=\linewidth]{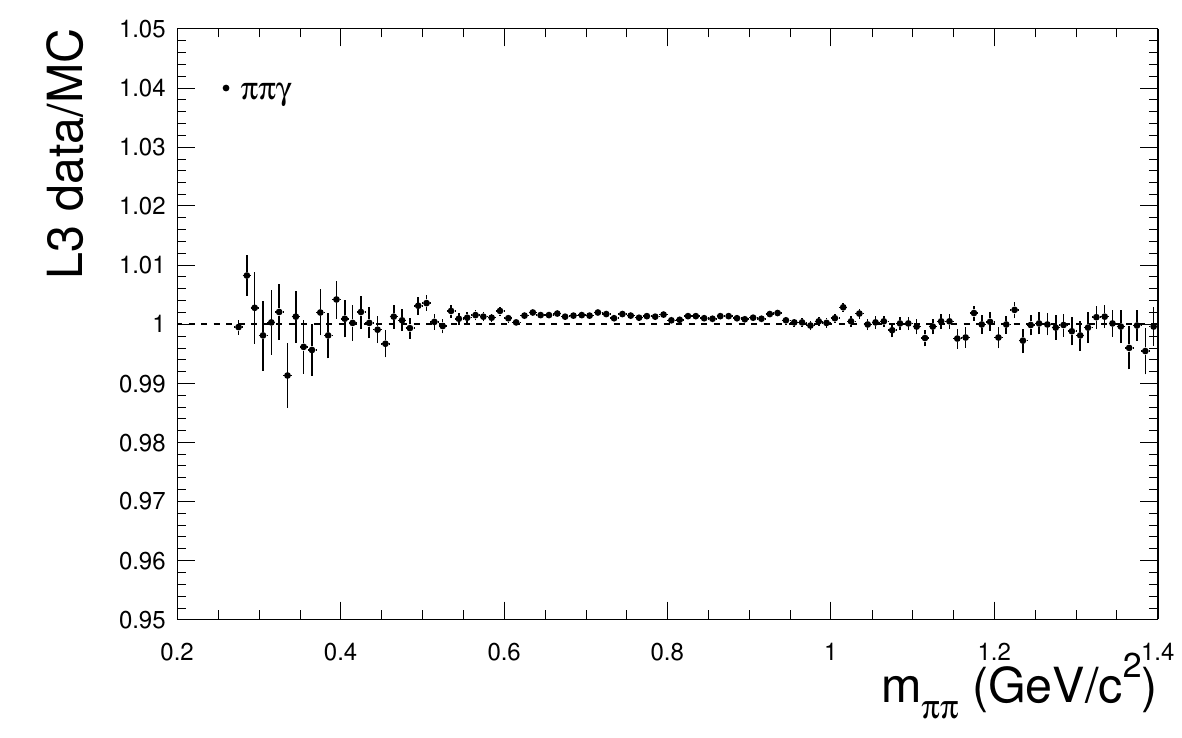}
    \includegraphics[width=\linewidth]{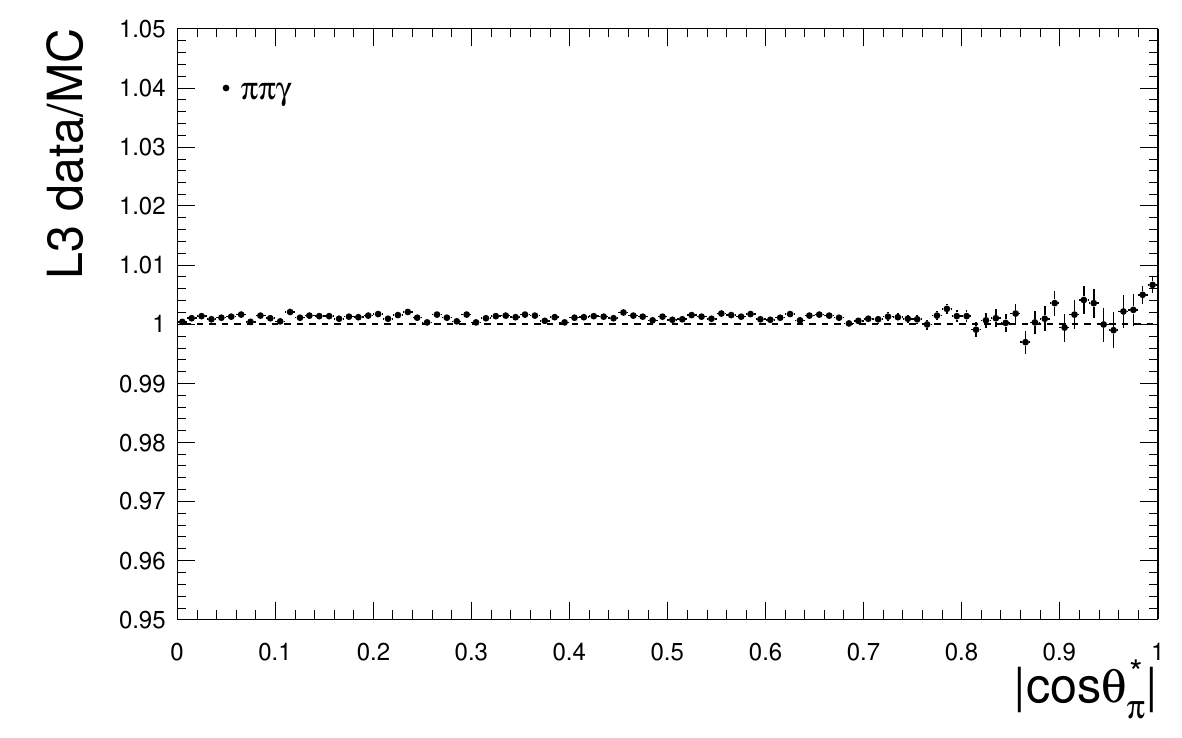}
    \caption{\label{fig:PiTrigEff_dataMCcorr}   \small \justifying
    The data/MC L3 efficiency correction in $\pi\pi\gamma$ events as a function of (top) $m_{\pi\pi}$, and (bottom) \acosthpi\ integrated over mass.
}
\end{figure}

Since the 4C fit requires two reconstructed tracks, the potential bias on the measured trigger efficiency due to the control sample selection is validated by the 1C fit, which requires only one track. As the backgrounds prevent the 1C fit to be used in low and high mass regions, the validation of the 4C fit is restricted to the $\rho$ peak region. In the mass range 0.65--0.85\gevcc, the absolute data/MC corrections using the 4C fit or the 1C fit differ by $8\times10^{-4}$, which is taken as the uncertainty in the correction estimated with the 4C fit.
The difference between the L3 efficiency on events with zero or more tracks and with at least one standard track is obtained from simulation. As in the muon channel, this bias is controlled on data by the difference between the efficiencies measured on events with one standard track or more and strictly two standard tracks. The uncertainty induced in the data/MC correction is estimated in this way to be $2\times10^{-4}$. 

The measured L3 efficiency is rather insensitive to the PID required on the primary track of the event, with a change of $4\times10^{-4}$ in the data/MC correction when applying a very tight or a very loose PID selection.
The uncertainty in the correction induced by electron backgrounds is found to be $3\times10^{-4}$ in the mass range 0.65--0.85\gevcc, with a null effect on the $|\cos\theta^\ast_\pi|$ template curvature, as the electrons affect only the high end of the distribution ($|\cos\theta^\ast_\pi|>0.95$). The muon background subtraction induces an uncertainty in the data/MC correction of $5\times10^{-4}$ in the $\rho$ mass region and $4\times10^{-4}$ in curvature of the $|\cos\theta^\ast_\pi|$ template. 
 
The data/MC correction of the L3 efficiency for pions is given in Fig.~\ref{fig:PiTrigEff_dataMCcorr} and their systematic uncertainties in Table~\ref{tab:syst-trigger}. 

\subsubsection{BGFilter efficiency results for pions}
\label{subsubsec:BGF_pions} 

\begin{figure}[t!] 
    \centering
    \includegraphics[width=\linewidth]{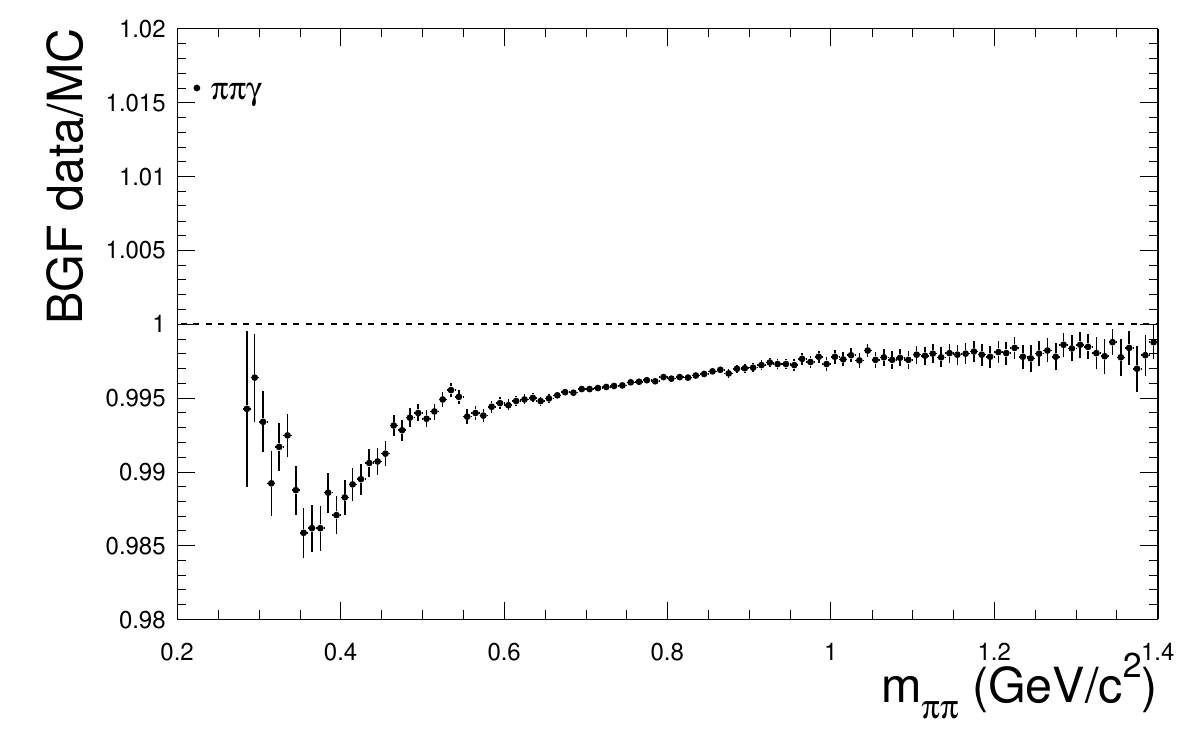}
    \includegraphics[width=\linewidth]{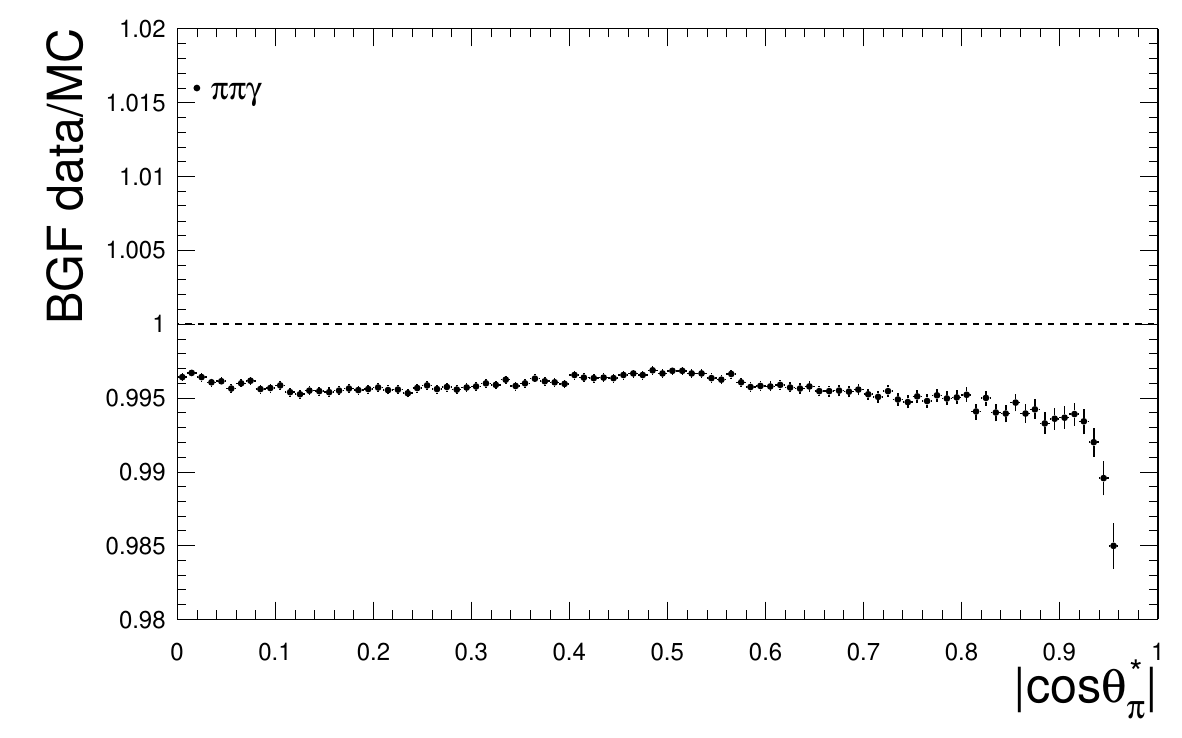}
    \caption{\label{fig:PiBGEff_dataMCcorr}   \small \justifying
    The data/MC BGFilter efficiency correction in $\pi\pi\gamma$ events as a function of (top) $m_{\pi\pi}$, and (bottom) \acosthpi\ integrated over mass.
}
\end{figure}

 While BGFilter di-muon dedicated lines lead to a negligible inefficiency in $\mu\mu\gamma$, no filter applied at the prompt reconstruction level selects $\pi\pi\gamma$ specifically. Two-pion events are, however, triggered by standard multihadronic filters or filters designed for other channels. 
 Since L3 trigger and BGFilter are applied sequentially, the BGFilter efficiency is estimated on events that passed L3. To compute the BGFilter efficiency, lines are grouped in two sets: ISR-dedicated lines and other lines, which are not fully orthogonal.

A similar validation of the efficiencies and data/MC corrections obtained with the 4C fit is performed for BGFilter as for L3 in the restricted mass range 0.65--0.85\gevcc. The data/MC corrections using the 4C fit or the 1C fit differ by $8\times10^{-4}$ in that range, which is taken as a systematic uncertainty. The differences determined on simulation between the BGFilter efficiencies on the full sample, including zero-track events, and on events with at least one standard track, is negligibly small ($<10^{-4}$), and so is the bias on the data/MC correction.
PID has little effect on the BGFilter efficiency and affects the low mass only by less than $5\times10^{-4}$.

Because all BGFilter lines mix information from tracking and from clusters in the EMC calorimeter, a large correlated loss is expected, which escapes the efficiency measurement based on cross-checking the responses of lines\footnote{The correlated loss was overlooked for the preliminary results in 2025. This is one of the changes summarized in Appendix~\ref{app:changes}.}. This is determined from simulation as the difference between the rate of events that pass BGFilter (true efficiency) and the efficiency measured by cross-checking lines. The correlated loss is dominated again by the overlap of tracks in the DCH, around $\Delta\phi=0.1\rad$ and mass around $0.35\gevcc$, where $\Delta\phi$ is the azimuthal difference between the positive and negative tracks~\cite{BaBar:2012bdw}. The effect of the pion shower overlap in the EMC is also observed in a broad range around $\Delta\phi=0.5\rad$ and peaks at about $0.55\gevcc$. Since uncorrelated and correlated dominant inefficiencies result from the same causes, they are indeed proportional to each other in the overlap regions. Their relative sizes observed in simulation allow an estimate of the correlated loss in data. This estimate is controlled on events that by-pass the \hbox{BGFilter}, on which the true inefficiency can be measured, although with limited statistical accuracy. By including the correlated loss, the data/MC efficiency ratio is changed by $3\times10^{-3}$ in the $\rho$ peak region. Its uncertainty is accounted for in the bin-by-bin corrections. 

The final data/MC correction of the BGFilter efficiency on pions, including the correlated loss, is shown in Fig.~\ref{fig:PiBGEff_dataMCcorr} and the systematic uncertainties in Table~\ref{tab:syst-trigger}. 

\begin{table}[tbp]
\centering
\caption{\small \justifying Systematic uncertainties for the data/MC trigger corrections.}
  \label{tab:syst-trigger}
\begin{tabular}{c|cc}\hline
Source   & Mass   & $\cos\theta^\ast_\pi$ curvature \\
         & (per mil)& ($\times{{\cos^2\theta^\ast_\pi}}$, per mil) \\\hline
\multicolumn{3}{c}{L3 muons}         \\\hline 
0-track events        &0.2   &4     \\
Primary $\mu$ ID      &0.4   &3.5   \\
Correlated loss       &1.0   &--    \\\hline
Sum muons             &1.1   &5.0   \\\hline
\multicolumn{3}{c}{L3 pions}        \\\hline 
4C-1C fit consistency &0.8   &0.3   \\
0-track events        &0.2   &0.2   \\
Primary $\pi$ ID      &0.4   &0.5   \\
Electron veto         &0.3   &--    \\
Muon veto             &0.5   &0.4    \\\hline
Sum L3 pions          &1.0   &0.7     \\\hline
\multicolumn{3}{c}{BGFilter pions}              \\\hline 
4C-1C fit consistency &0.8   &0.2   \\
Primary $\pi$ ID      &0.5   &--     \\\hline
Sum BGFilter pions    &1.0   &0.2     \\\hline
\end{tabular}
\end{table}
\section{Tracking efficiency}
\label{sec:tracking}
\subsection{Method}

The cross-section measurements in this analysis require two reconstructed `good' tracks 
with enhanced quality, as described in Section~\ref{evt-sel}.
A larger set, called `standard tracks', have relaxed
selections, $\mathrm{doca}_{xy}<2.5$\,cm and no requirement on $N_\textrm{DCH hits}$.
Because the `good' track selection includes tracks of very low $p_\mathrm{T}$, the reconstruction efficiency of low momentum tracks is specially studied as reported below.

The tracking control samples of $\mu\mu\gamma_{\rm{ISR}}$ ($\pi\pi\gamma_{\rm{ISR}}$) events 
are selected through the efficiency-dedicated 1C fit described above for the trigger studies, and the tracking efficiency is measured using a tag\&probe method. 
In order to determine separately the muon and pion efficiencies, PID is required for the tag track.

The events are required to pass the triggers and the online filter and are 
selected without specific requirements on the second reconstructed track, 
if any. Biases, affecting the tracking efficiency
measurement as a result of the selection of the primary track and of the
event-level background selections, are identified and studied with the simulation,
and evaluated on the data. The predicted track is required to lie within the angle and momentum 
acceptance.

\sloppy
The method therefore determines the efficiency to
reconstruct a given track with a charge opposite to that of the primary track in the
geometrical acceptance, no matter how close or distant this track is with
respect to the expected one, to account for decays or secondary interactions. 
The possible mismatch in momentum and/or 
angles affects the full kinematic reconstruction of the event, and its 
effect is included in the efficiency of the corresponding 2d-$\chi^2$ 
selections used for the physics sample.

The tag\&probe method works well for tracks with large enough momenta. In the low-momentum range and near the edges of the 
angular acceptance it is no longer reliable 
because the resolution on the predicted parameters of the probe track becomes too poor
and does not allow the angular acceptance selections to be applied safely. 
Therefore, the method is extended to proceed in two steps: (1) standard tag\&probe 
(now called method 1) when the predicted track is in a safe region with respect to the
resolution issue, {\it i.e.},\ $p_\textrm{T}^\textrm{predicted}>0.6$\gevc and 
$0.5<\theta^\textrm{predicted}<2.45$\,rad, and (2) method 2 for the complementary conditions.

In method 2, 
the predicted probe track parameters 
are ignored and standard tracks are considered as final good 
track candidates. All standard reconstructed tracks, except the
tag one, are considered in turn, and the standard-tag pair with the ISR photon
candidate is chosen on the basis of the best 4C kinematic fit. 
The efficiency is measured as the rate of these selected standard tracks that fulfill the good track conditions.
This procedure does not yield the full tracking efficiency as it is conditioned by
the prior existence of a standard track. However it retains the dominant
contributions to the inefficiency from the stricter requirements on $\mathrm{doca}_{xy}$  
and $N_\textrm{DCH hits}$. 
Therefore, the respective behavior of the missing part of
the inefficiency in data and simulation 
is investigated by comparing the 
correction factors obtained in methods 1 and 2 for the sample where method 1 is safe, 
representing a large majority of the events. 
Overall, the impact of the small bias introduced by the
incomplete track efficiency determination in method 2 is 0.13 per mil  and reaches 1.5 per
mil for $|\cos{\theta_\pi^*}|>0.8$. 
The bias on the tracking correction to the mass spectrum is 0.25 per mil. 

The individual track efficiency is determined assuming
that the efficiencies of the two tracks are uncorrelated. However,
a very sharp increase of the tracking inefficiency is observed when
tracks overlap in the DCH. Not only the individual track efficiency is 
locally reduced, but a correlated loss of the two tracks occurs.
In addition, as the final physics sample is required to have two and only two good tracks 
with opposite charges, the understanding of the tracking involves not only track 
losses, but also the probability to reconstruct extra tracks as a result of 
secondary interactions with the detector material or the presence of 
beam-background tracks.

The full tracking efficiency is then the product of the square of the 
single-track efficiencies, the probability for not losing the two tracks in a 
correlated way (loss probability $=f_0$), and the probability for not having 
an extra reconstructed good track (loss probability $=f_3$). 
The mass-dependent quantities $f_0$
and $f_3$ are in the 0.5--2.5 per mil range. The event tracking correction $C_{\rm tracking}$ 
to be applied to the MC efficiency, corresponding to the product of the data/MC ratios of each term, is then expressed as:
\beq
\label{Ctracking}
 C_{\rm tracking} = \frac {[(1-f_0-f_3)\epsilon^2]_{\rm data}}
                         {[(1-f_0-f_3)\epsilon^2]_{\rm MC}},
\eeq
where the average track efficiency $\epsilon$, $f_0$, and $f_3$ depend on the
values of $m_{\mu\mu}$ and $m_{\pi\pi}$ for the muon and pion samples, respectively, and of $|\cos{\theta_\pi^*}|$ for both samples.

The main sources of track loss are well identified: the track overlap in the DCH 
is important for both muons and pions as well as a small intrinsic reconstruction loss, while secondary interactions and decays affect mainly pions. 
The two effects are separated by using the distribution of the tracking inefficiency 
as a function of the two-track opening angle in the plane transverse to the beams $\Delta \phi=\phi^+-\phi^-$. 

Throughout the analysis, MC efficiency values are blinded with a constant multiplicative offset. The offset does not play a role in the angular distribution fits, as the correction only affects the normalized template shape. The efficiencies and data/MC ratio are given below after unblinding. 

\subsection{Tracking efficiency results for muons}

Final results on the track efficiency in data and MC are given for muons using the 
combination of methods 1 and 2. Reference values correspond to the following tagging 
conditions: good track with tight BDT muon ID~\cite{BaBarBelle2014}.  

The tag\&probe method gives access to the tracking efficiency in biased conditions. 
It is important to identify and assess the different sources of bias and to determine 
their impact on the final results. For each source, the level of impact is studied 
first with the simulation, which allows one to look for correlations with observables 
accessible in data. Then tests are made in data and simulation for these observables to
finally estimate the actual bias level and its associated uncertainty. For the tracking 
efficiency, there are two sources of global bias due to the event selection: the LO 
assumption for the 1C tag fit and the non-$\mu\mu$ background subtraction. 

In the first 
case, the bias originates from hard additional radiation which modifies the predicted 
track parameters. We know from the specific \babar\ analysis~\cite{BaBar:2023xiy} that the \textsc{Phokhara}
generator at NLO level does not account correctly for the hard additional ISR photons due to 
the neglect of NNLO radiation. However, there is some
compensation between the NLO and NNLO effects, which is precisely evaluated, resulting in a final negligible bias 
less than 0.1 per mil in the absolute value for the mass distribution, and less than
0.3 per mil in the $|\cos{\theta_\pi^*}|$ distribution relevant for the muon template
in the angular fit. 

The second source of global bias is from background subtraction, dominated by 
$\pi\pi\gamma$ peaking at the $\rho$ mass, 
quantified from the mass 
distribution both in data and simulation. The background level in data estimated in 
this way takes into account data/MC differences in the tag particle ID, and the 
statistical uncertainty is used as a basis for the systematic uncertainty in the 
data/MC ratio of track efficiencies, which turns out to be 0.25 per mil at the $\rho$ peak. 

Two sources of bias occur because the tag track selection 
has a small level
of correlation with the reconstruction of the predicted track. The first is from the
particle ID definition and the second from the track quality. In both cases, more 
demanding cuts increase the chance of correlation. These effects are 
compared in data and simulation by varying the quality cuts both for the track 
particle ID and track reconstruction.

The two-track loss probability is given by $(1-\epsilon)^2$ for the non-correlated part, directly obtained in data and simulation,
whereas the correlated part $f_0$ is only accessible in MC. Studies with the simulated sample show that the correlated loss occurs when
the two tracks overlap in the detector. 
The $f_0$ contribution broadly peaks around a mass of 0.4\gevcc, at a level of a few per mil, small but much larger than the non-correlated two-track loss at the 0.1 per mil level. 
Because the tracking loss is dominated by 
two-track overlap, a correlation is expected and found between the tracking
inefficiency $(1-\epsilon)$ and the two-track correlated loss $f_0$, allowing to estimate in data the small $f_0$ distribution in mass.

Finally the probability $f_3$ that a muon-pair event is lost because of more than two reconstructed good tracks is found to be about 2 per mil in simulation without significant variation over the full mass range. The determination in data is more difficult as pion background contributes significantly. A clean test is performed by requiring one of the tracks to be identified as a muon (in addition to the primary one), showing good consistency with the simulation within 0.1 per mil. 

\begin{figure}[tbp] \centering 
  \includegraphics[width=\linewidth]{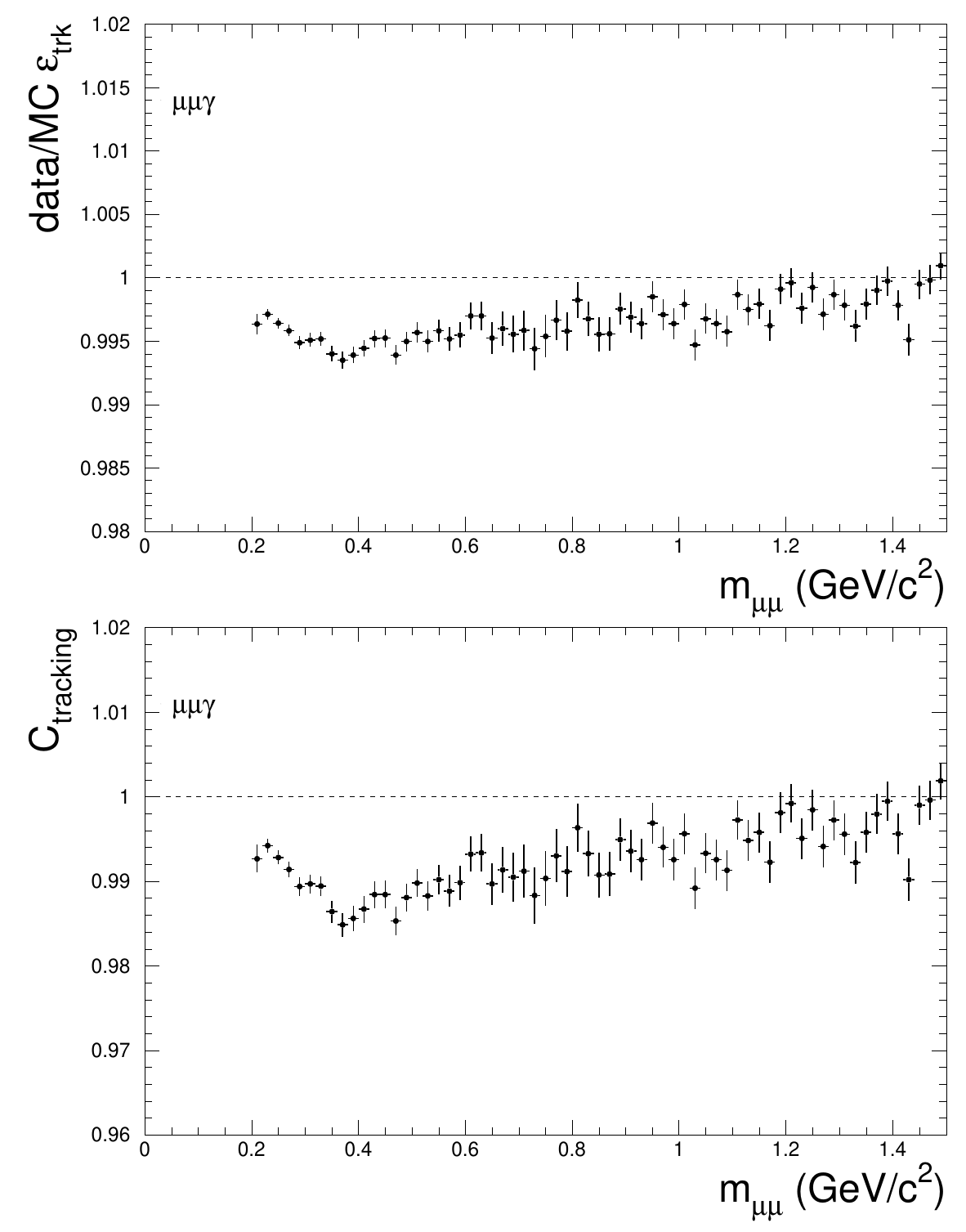}
  \caption{\label{TrackCorr-muon-massmu} \small \justifying (Top) the data/MC track efficiency ratio for muons as a function of $m_{\mu\mu}$, (bottom) the full data/MC event tracking correction $C_{\rm tracking}$.
  }
  \end{figure}

\begin{figure}[tbp] \centering
  \includegraphics[width=\linewidth]{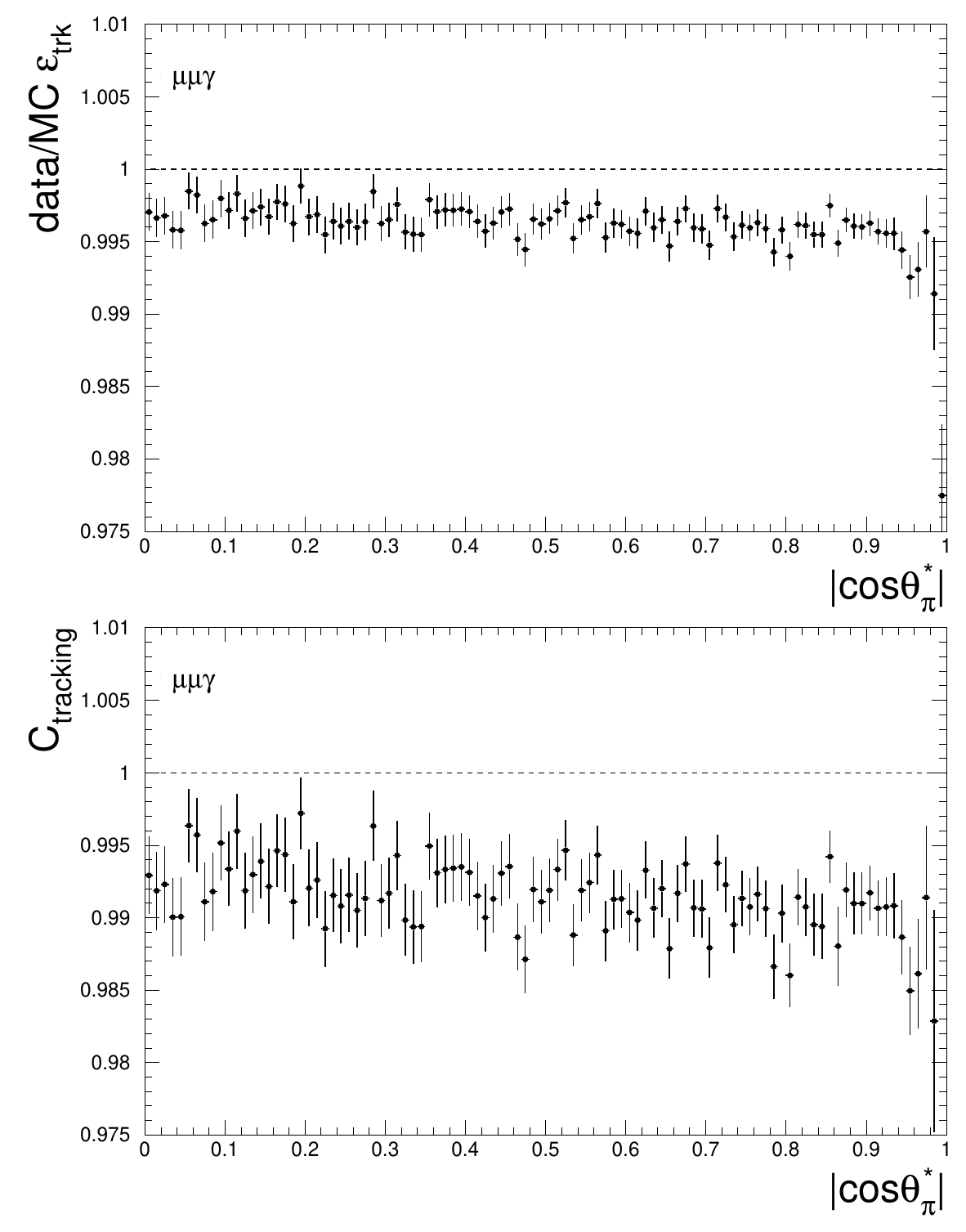}
  \caption{\label{TrackCorr-muon-costhpi} \small \justifying  (Top) the data/MC track efficiency ratio for muons as a function of $|\cos{\theta^*_\pi}|$, (bottom) the full data/MC event tracking correction $C_{\rm tracking}$. 
  }
  \end{figure}

The complete data/MC event tracking corrections for muons assembles all the components according to Eq.~(\ref{Ctracking}). The results for muons as a function of $m_{\mu\mu}$ and 
$|\cos{\theta_\pi^*}|$ are given 
in Figs.~\ref{TrackCorr-muon-massmu} and \ref{TrackCorr-muon-costhpi}, respectively. The latter one is applied to the muon
templates when performing the fits in bins of 
$m_{\pi\pi}$, 
while the first one is the tracking correction to be applied to the muon derived spectrum in bins of 
$m_{\mu\mu}$. Statistical uncertainties are given at each point and are uncorrelated. They also include the systematic uncertainties from the $f_0$ and $f_3$ corrections, given point-by-point.

  Finally, uncertainties to be applied to the muon mass spectrum after the angular fits are detailed in Table~\ref{tab:syst-tracking-muon}. They sum quadratically to 1.0 per mil.

\begin{table}[tbp]
\centering
\caption{\small \justifying Systematic uncertainties for the data/MC muon tracking correction.}
  \label{tab:syst-tracking-muon}
\begin{tabular}{c|cc}\hline
Source  & per track& per event \\
              & (per mil)& (per mil) \\\hline

Consistency methods 1/2& 0.25&0.5  \\
1C LO fit (NLO/NNLO)& 0.1&0.2 \\
Pion background subtraction&0.25&0.5 \\
Tag $\mu$ ID& 0.2&0.4 \\
Tag track quality&0.2&0.4 \\
two-track correlated loss& &point-by-point \\
Extra tracks& &0.2 \\\hline
Sum& & 1.0 \\\hline
\end{tabular}
\end{table}

\subsection{Tracking efficiency for pions}
\label{subsec:tracking_pions}
The study of the tracking for pions follows the same path as for muons using a sample of selected events with one leading good track identified to be a pion using 
a standard \babar\ selector~\cite{BaBarBelle2014}. The background is relatively large at low masses from mis-identified $\mu\mu$ events and multihadronic ISR processes, and at large masses from $q\overline{q}$. A veto on the extra neutral energy (not including the ISR photon) is applied when the sum of energies of additional EM clusters exceeds 0.5\gev.  Even with this selection, the background level achieved when using the same method as for muons (methods 1 and 2 combined) is not satisfactory at small and large masses for a reliable tracking efficiency determination, leading to a change of strategy, using only method 2, which provides more handles. A veto from $\mu$-ID on the standard candidate track and a loose selection on the 4C fit $\chi^2<40$ reduce the background to a satisfactory level\footnote{The selection was more strict for the preliminary results in 2025. There was in addition a bookkeeping error for handling the histograms discovered for the final analysis. This change is the most important one summarized in Appendix~\ref{app:changes}.}.
Systematic uncertainties in the background subtraction are assigned in the different mass regions (Table~\ref{tab:syst-tracking-pion}).

  The consistency check of the more constrained method 2 is performed by comparing to the results obtained with method 1, unbiased with respect to the standard candidate track. Since the background level severely limits the applicability of method 1 outside of the $\rho$ region, the test is restricted to a narrow region 0.7--0.8\gevcc on the $\rho$ peak. A small observed difference leads to a correction of $-1.7$ per mil applied to the data/MC correction obtained by method 2 and a systematic uncertainty of 0.6 per mil is assigned.

  Unlike muons with method 1, the determination of the pion track efficiency does not rely on the 1C fit and is therefore immune to the LO assumption in that fit. It is not biased either by the 4C fit, also assuming LO, since the candidate track is measured independently of the fitted predicted track. The selection $\chi^2<40$ could introduce a small bias, but varying the selection value does not show any significant effect. For the same reason, no bias is expected from the tag track selection (pion ID or quality of track reconstruction), which only plays a role in the 4C fit. Again here no effect is found within statistical variations less than 0.1 per mil. The bias from using method 2 with incomplete coverage of all track inefficiencies is tested
  against method 1 measuring the full efficiency in the $\rho$ resonance region, as discussed above.  Finally, the bias caused by the pion ID and muon veto on the standard track candidate is investigated in a way similar to the muon case for the tagging track using four different ID conditions. 

  The determination of the data/MC corrections for the correlated two-track loss ($f_0$) and the probability to have more than two  
  `good' tracks ($f_3$) follows the same procedure as for muons. 

\begin{figure}[tbp] \centering 
  \includegraphics[width=\linewidth]{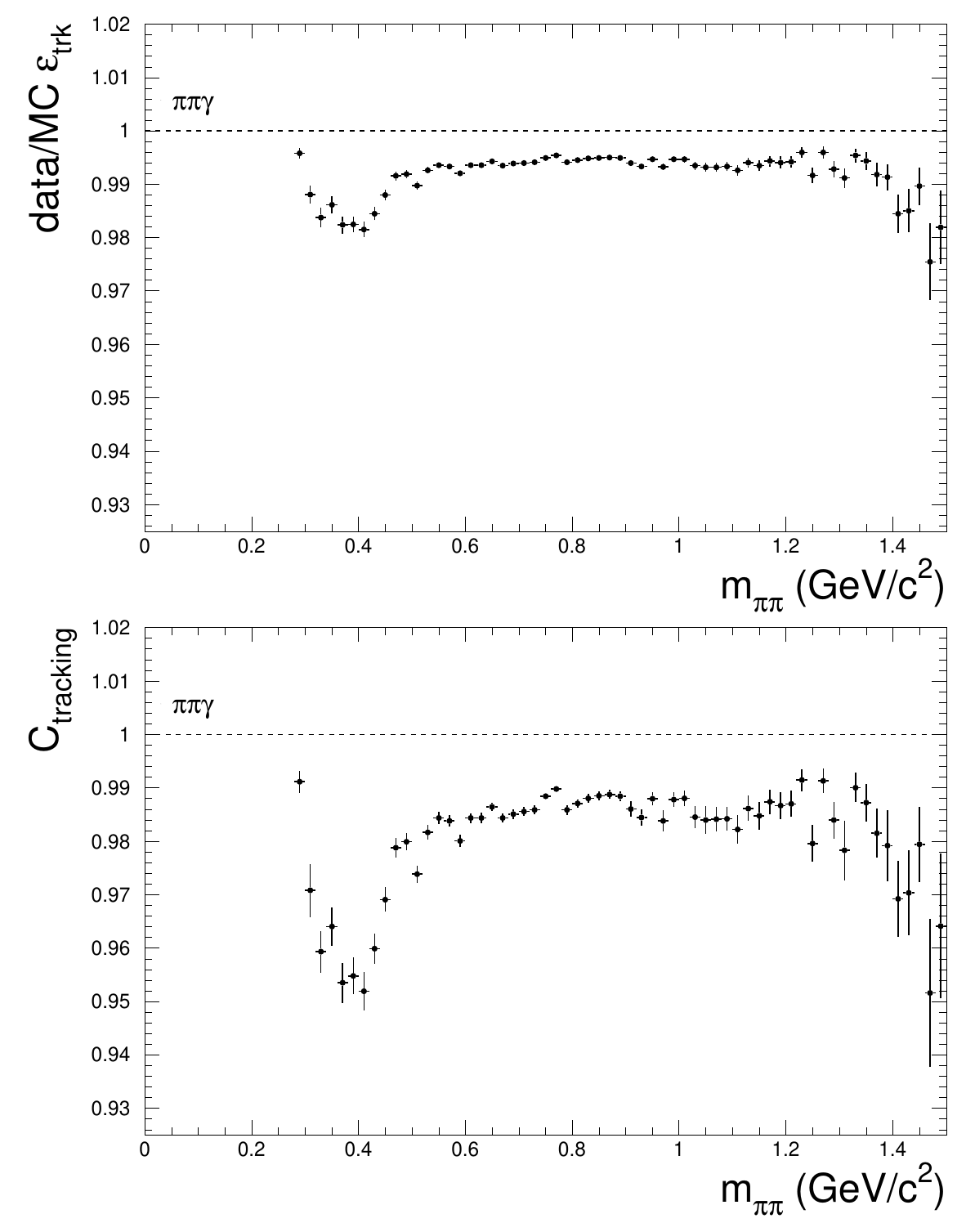}
  \caption{\label{TrackCorr-pion-masspi} \small \justifying (Top) the data/MC track efficiency ratio for pions as a function of $m_{\pi\pi}$, (bottom) the full data/MC event tracking correction $C_{\rm tracking}$.
  }
  \end{figure}

\begin{figure}[tbp] \centering
  \includegraphics[width=\linewidth]{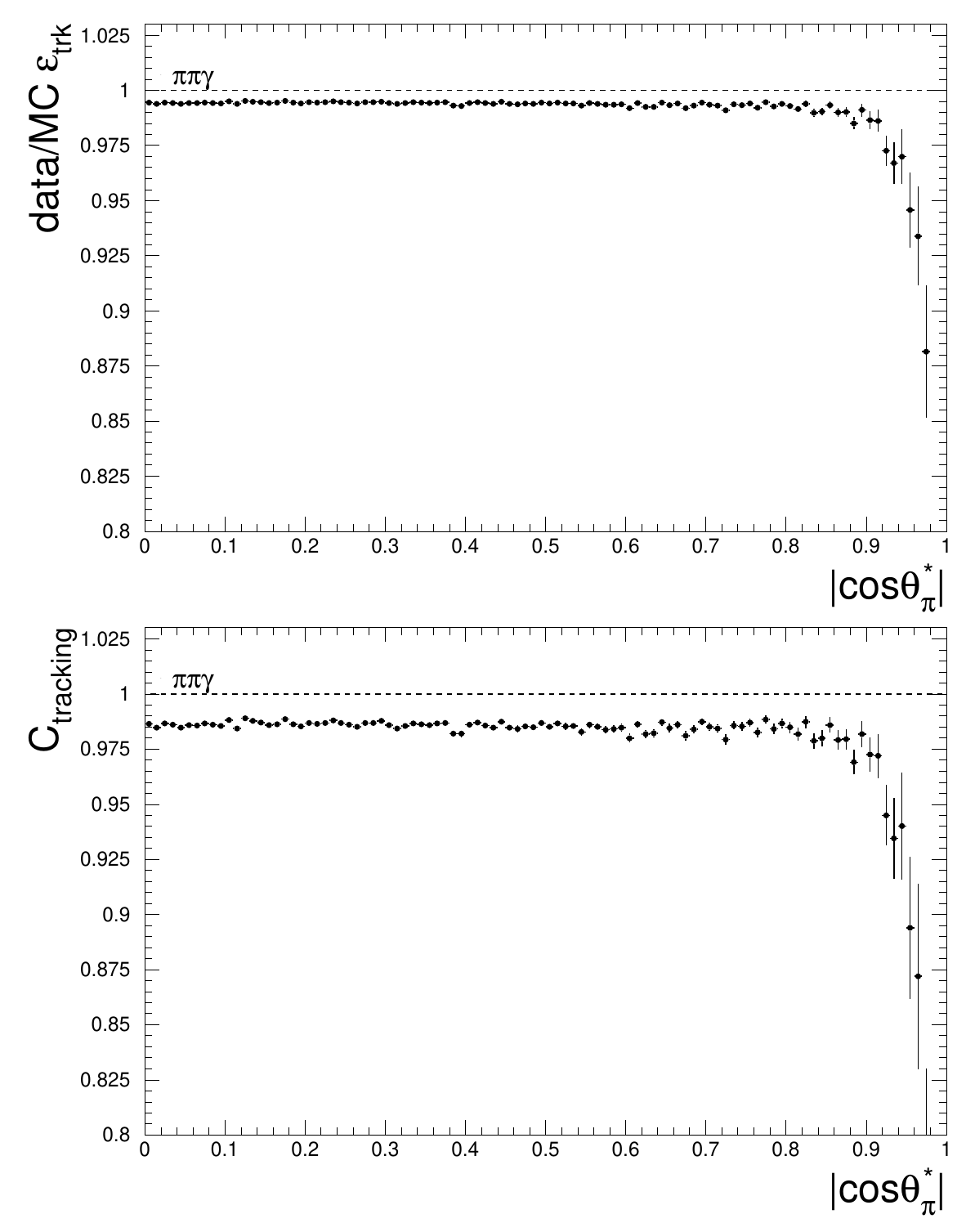}
  \caption{\label{TrackCorr-pion-costhpi} \small \justifying  (Top) the data/MC track efficiency ratio for pions as a function of $|\cos{\theta^*_\pi}|$, (bottom) the full data/MC event tracking correction $C_{\rm tracking}$.
  }
  \end{figure}

  The complete data/MC event tracking corrections for pions collect all the components according to Eq.~(\ref{Ctracking}). The results as a function of $m_{\pi\pi}$ and $|\cos{\theta_\pi^*}|$ 
  are given in Figs.~\ref{TrackCorr-pion-masspi} and \ref{TrackCorr-pion-costhpi}, respectively. The latter one is applied to the pion
templates when performing the angular fits in bins of $m_{\pi\pi}$, 
while the former one is the tracking correction to be applied to the pion derived mass spectrum. 
Statistical uncertainties are given at each point and are uncorrelated. They also include the systematic uncertainties from the $f_0$ correction, given point-by-point.

  The values for the global systematic uncertainties are
  detailed in Table~\ref{tab:syst-tracking-pion} to be applied to the pion mass spectrum after the fits, summing up quadratically to 1.4 per mil in the 0.5--1.0\gevcc mass region and larger outside.

\begin{table}[tbp]
\centering
\caption{\small \justifying Systematic uncertainties for the data/MC pion tracking correction.}
  \label{tab:syst-tracking-pion}
\begin{tabular}{c|cc}\hline
Source  & per track& per event \\
              & (per mil)& (per mil) \\\hline
Consistency methods 2/1&  0.6&  1.2  \\
Background subtraction&10.  $(<0.3)$&20.   \\
(mass range in \gevcc) &3.  (0.3--0.4)&6.   \\
                      &1.  (0.4--0.5)&2.   \\
                      &0.2  (0.5--1.0)&0.4   \\
                      &1.  $(>1.0)$&2.   \\
Candidate $\pi$ ID& 0.2&0.4 \\
Two-track correlated loss& &point-by-point \\
Extra tracks& &0.3 \\\hline
Sum& & 20.   $(<0.3)$ \\
                   &   &6.1  (0.3--0.4)   \\
                   &   &2.4  (0.4--0.5)   \\
                   &   &1.4  (0.5--1.0)   \\
                   &   &2.4  $(>1.0)$   \\\hline
\end{tabular}
\end{table}
              
\subsection{Tracking efficiency for pions and muons at low momentum}

\begin{figure}[tbp] \centering
  \includegraphics[width=\linewidth]{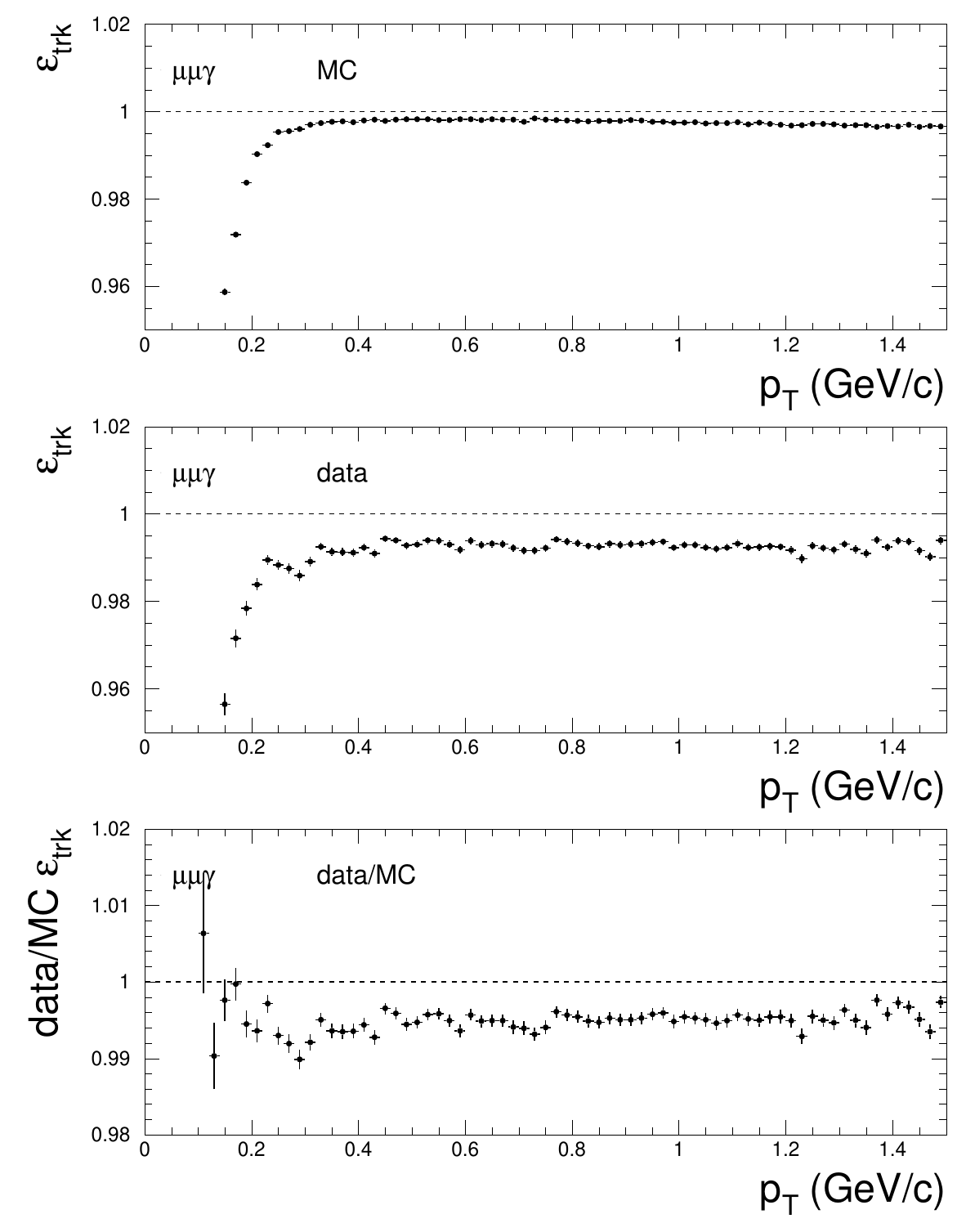}
  \caption{\label{muon-lowpt} \small \justifying  The track efficiency for muons determined in MC (top), data (middle) and their ratio (bottom) as a function of $p_\mathrm{T}$.
  }
  \end{figure}
  \begin{figure}[tbp] \centering
  \includegraphics[width=\linewidth]{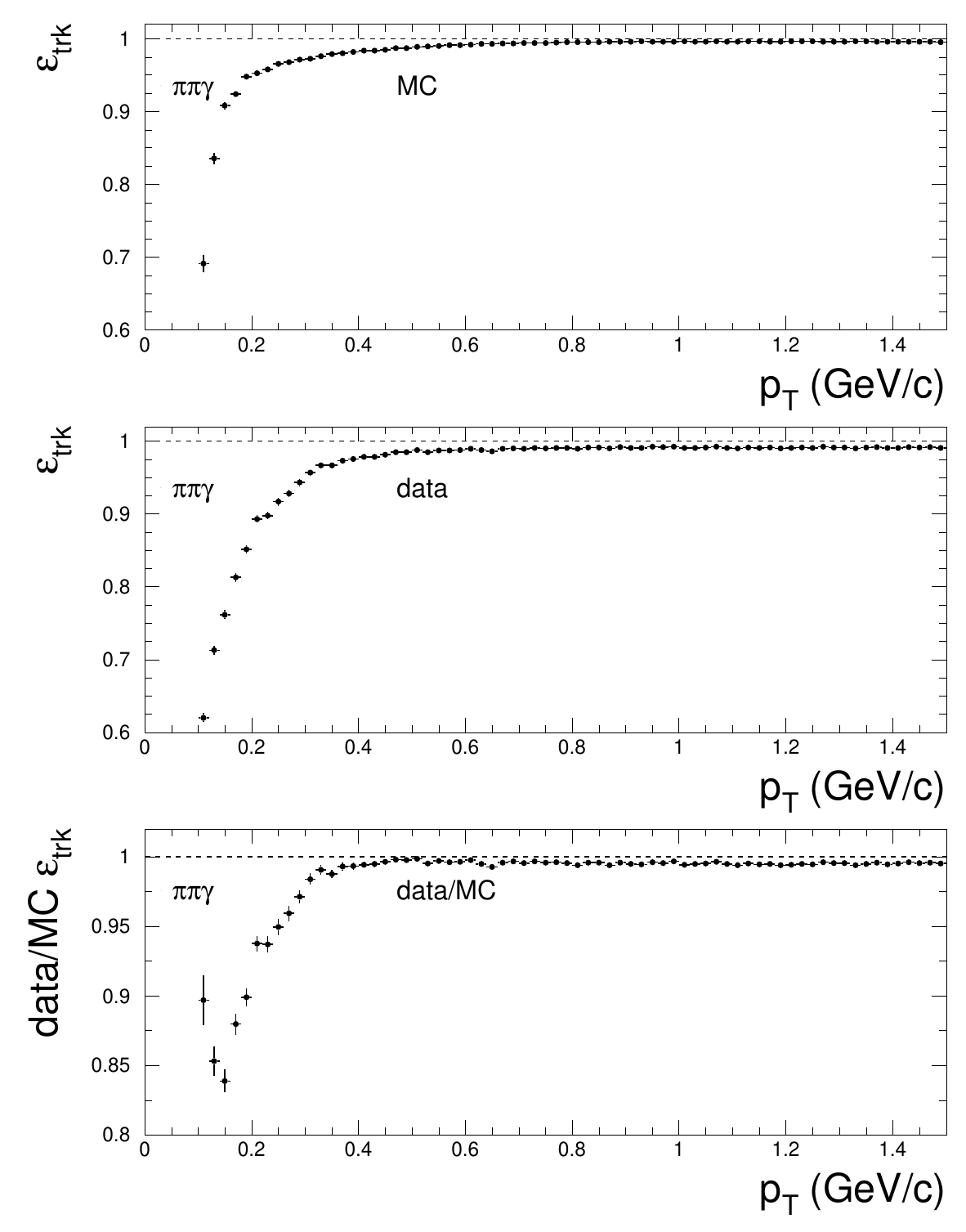}  
  \caption{\label{pion-lowpt} \small \justifying  The track efficiency for pions determined in MC (top), data (middle) and their ratio (bottom) as a function of $p_\mathrm{T}$.
  }
  \end{figure}

The statistical separation of $\pi\pi$ and $\mu\mu$ by the angular distribution fit depends on being able to use asymmetric events where one track has low transverse momentum down to 100\,MeV/c. So far, the results on the tracking efficiency have been expressed as a function of $m_{\pi\pi,\mu\mu}$ or $|\cos{\theta^*_{\pi}}|$. Thus, the behavior of the efficiency at low momenta is not explicitly visible. Here, we concentrate on this aspect to check if these configurations are well taken into account in the data/MC corrections.

Pions and muons share the same track reconstruction problems. Additional losses are present for pions through decays and secondary interactions, both enhanced at low momentum. Low-$p_\mathrm{T}$ tracks are only accessible with method 2. Data and MC track efficiencies are compared as a function of $p_\mathrm{T}$ in Fig.~\ref{muon-lowpt}.  
For muons, although still above 95\% at 0.15\gevc, the efficiency is seen to plunge below 0.4\gevc as a result of strong track curvature, including looping tracks. 
The efficiency drop is more gradual for pions, from 0.9\gevc downward (Fig.~\ref{pion-lowpt}). In this region, the pion statistical precision is lower compared to the muons. 
The efficiency behavior at low $p_\mathrm{T}$ is rather well reproduced by simulation for muons: the ratio of efficiencies data/MC is flat from 0.1 to 1.5\gevc within statistical uncertainties, with a negligible impact on systematic uncertainties. However, for pions, the efficiency drop at low-$p$ in data beyond the standard reconstruction loss seen for muons is enhanced by decays and secondary interactions -- an effect not well reproduced in the simulation. Nevertheless this data/MC difference is well-measured and accounted for. It occurs 
for all $\pi\pi$ masses and it is responsible for the small bias at large $|\cos{\theta^*_{\pi}}|$ values involving very asymmetric pion pairs. It is also the place where a significant electron background remains, sharply peaking at very low momenta, causing the apparent efficiency to rise again. Both effects occur in a region where the pion yield is negligibly small.
\section{Efficiency of the $V_{xy}$ selection}
\label{sec:vxy}

\begin{figure}[hbtp]
    \centering
    \includegraphics[width=\linewidth]{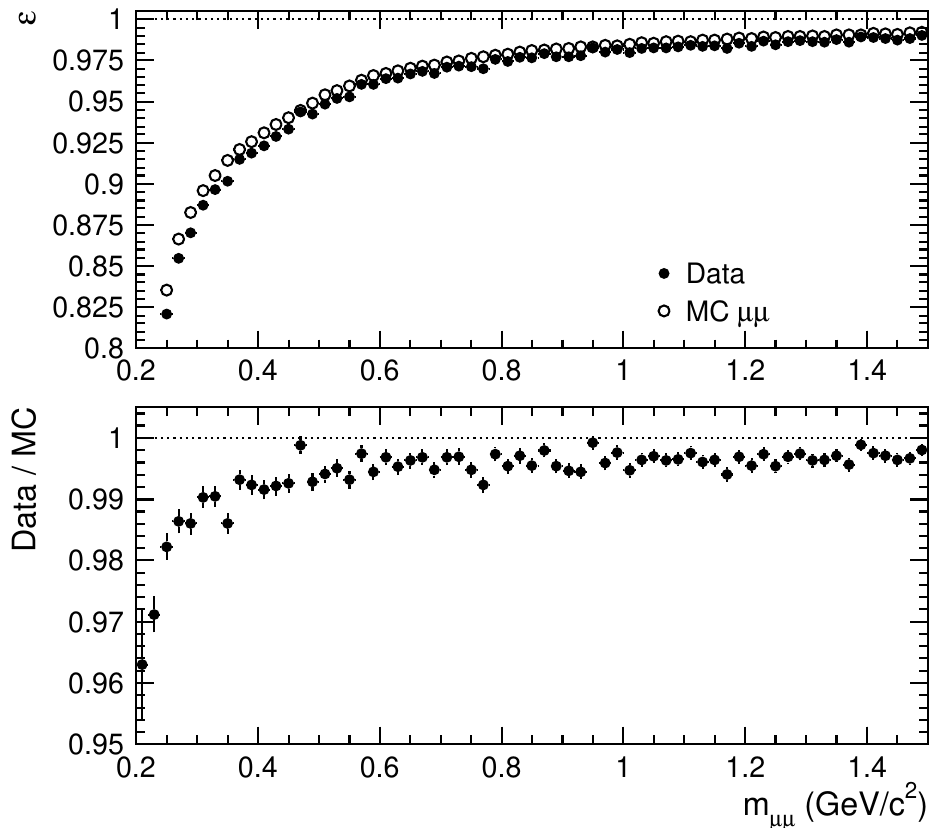}
\caption{\small \justifying The \vxy\ efficiency for $\mu\mu\gamma$ data (after background subtraction) and $\mu\mu\gamma$ MC simulation and their ratio as a function of mass.}
\label{fig:vxy_eff_mass_muon}
\end{figure}

Electron background arises from photon conversion from the $e^+e^-\to \gamma\gamma$ process at low mass and from Bhabha scattering at high mass where one of the final electrons (positrons) undergoes bremsstrahlung in the beam pipe, 
leading to displaced two-track vertices. Therefore, a selection  
$V_{xy}<0.5$\,cm is applied. 
For the electron-background-free $\mu\mu\gamma$ samples, with both muons identified using the tight BDT muon selector~\cite{BaBarBelle2014}, the $V_{xy}$ selection efficiency as a function of $m_{\mu\mu}$ is presented in Fig.~\ref{fig:vxy_eff_mass_muon}. The efficiency is found to be insensitive to the stringency level of muon ID, both in data and simulation. 

\begin{figure}[bp]
    \centering
    \includegraphics[width=\linewidth]{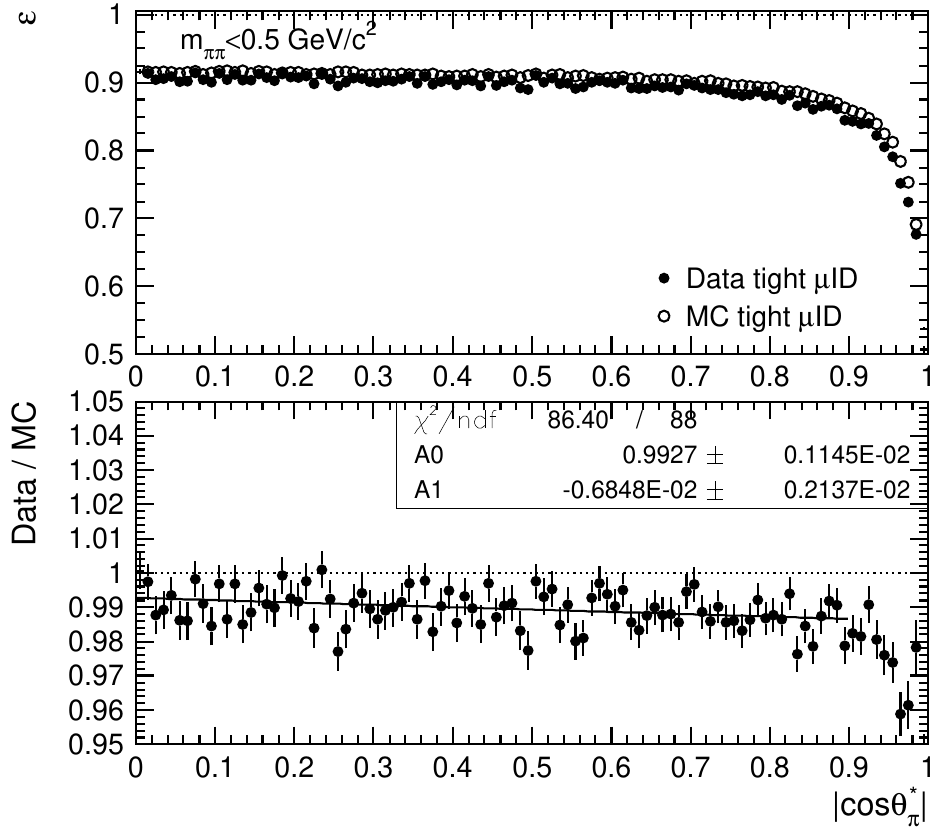}
    \includegraphics[width=\linewidth]{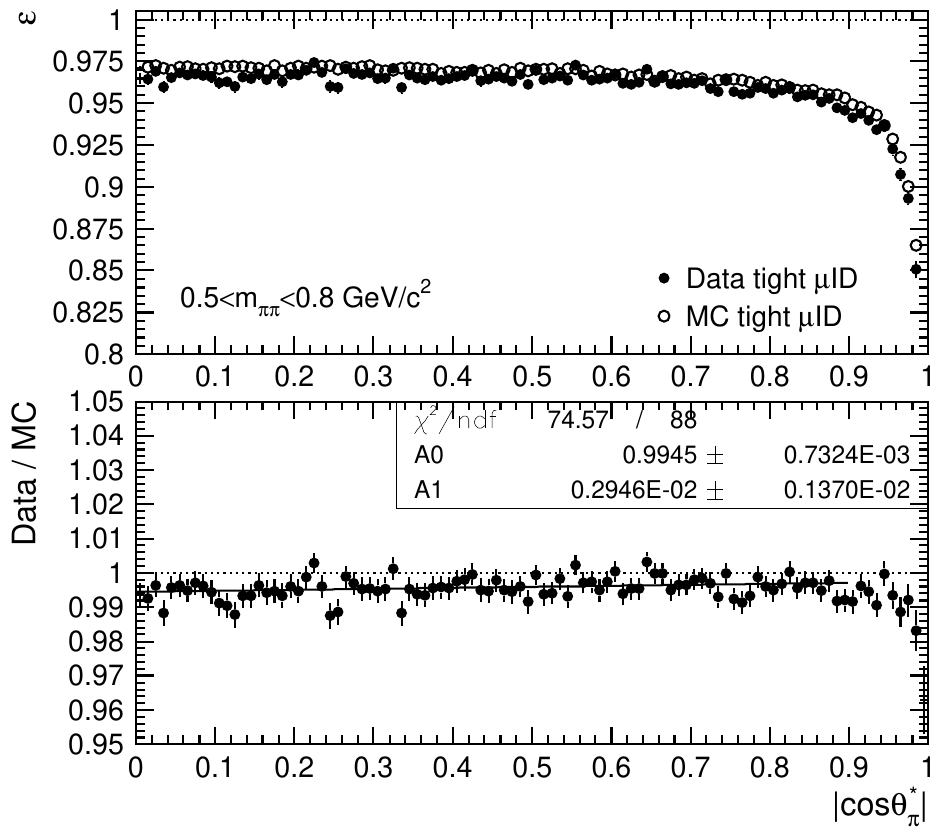}\\
    \includegraphics[width=\linewidth]{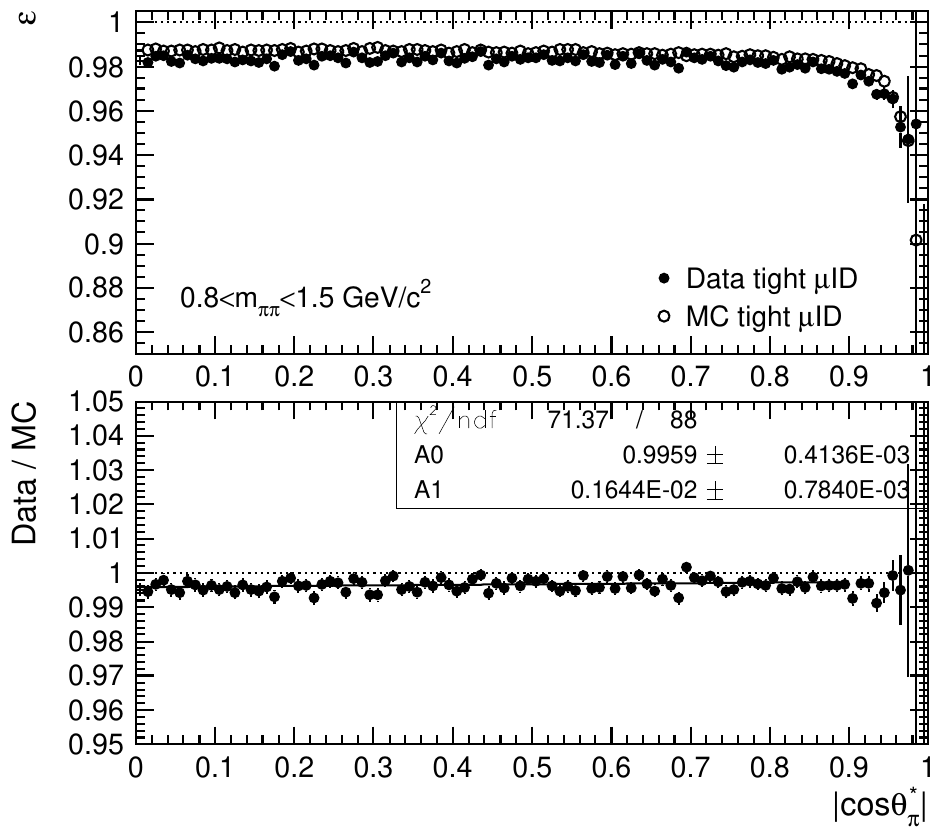}
\caption{\small \justifying The \vxy\ efficiency for $\mu\mu\gamma$ data (after background subtraction) and $\mu\mu\gamma$ MC simulation and their ratio as a function of $|\cos\theta^\ast_{\pi}|$ in three different mass regions defined with the pion mass hypothesis.}
\label{fig:vxy_eff_costh_muon}
\end{figure}

The \vxy\ efficiencies as a function of $|\cos\theta^\ast_\pi|$ are also determined in three mass intervals, between threshold and $0.5\gevcc$, 0.5--$0.8\gevcc$, and 0.8--$1.5\gevcc$, and shown in Fig.~\ref{fig:vxy_eff_costh_muon}. The fit is illustrative and the non-fitted corrections are directly applied to the templates obtained from MC simulation in the corresponding mass regions.

\begin{figure}[tbp]
    \centering
    \includegraphics[width=\linewidth]{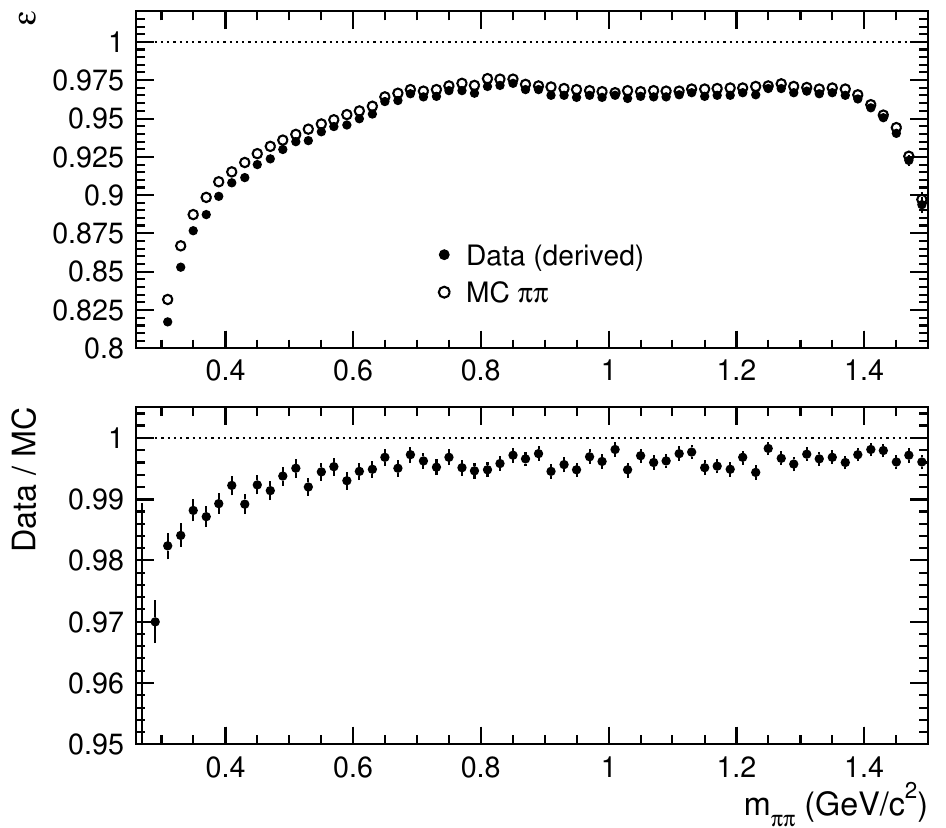}
\caption{\small \justifying The \vxy\ efficiency for $\pi\pi\gamma$ data and $\pi\pi\gamma$ MC simulation and their ratio as a function of mass.}
\label{fig:vxy_eff_mass_pion}
\end{figure}

\begin{figure}[tbp]
    \centering
    \includegraphics[width=\linewidth]{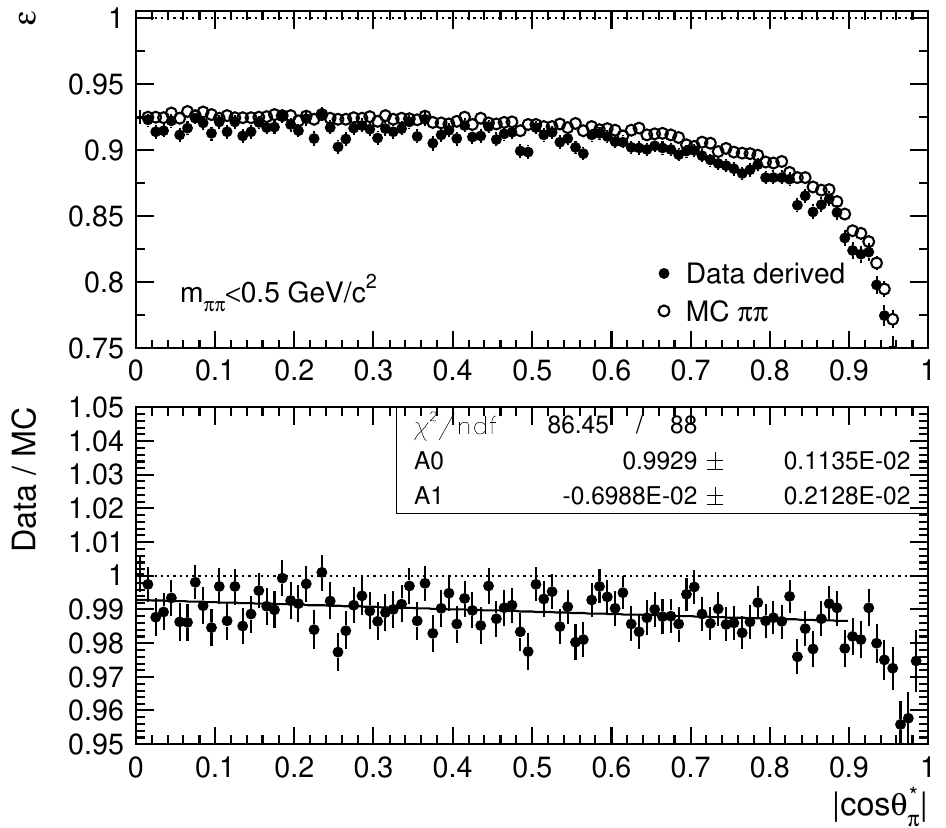}
    \includegraphics[width=\linewidth]{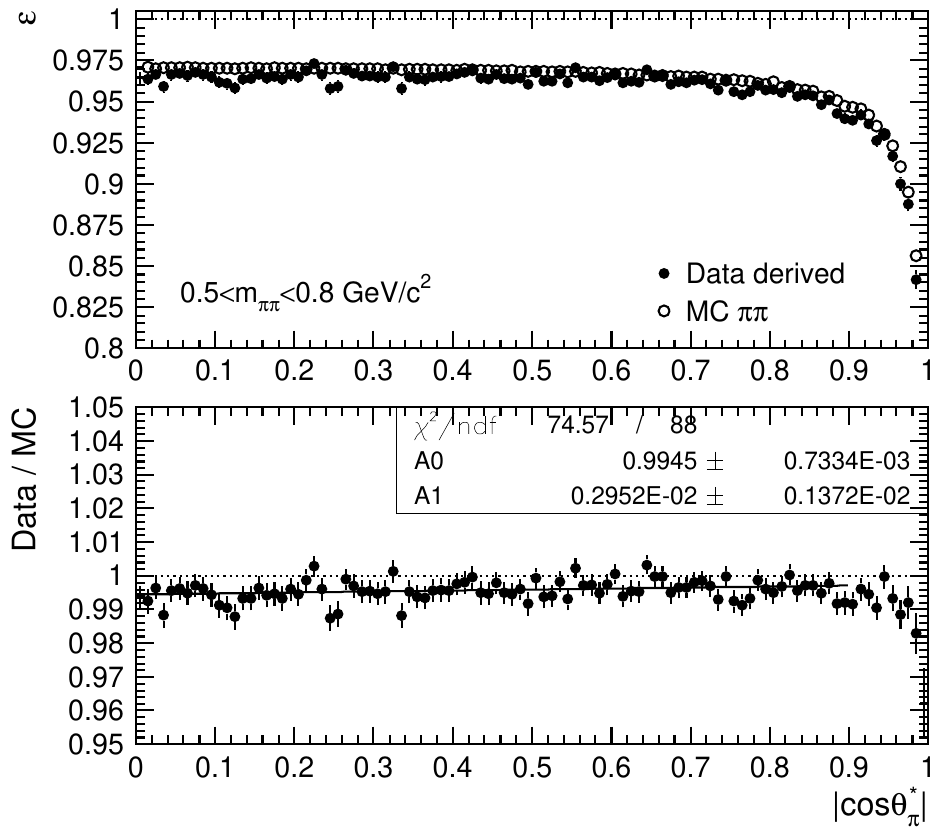}\\
    \includegraphics[width=\linewidth]{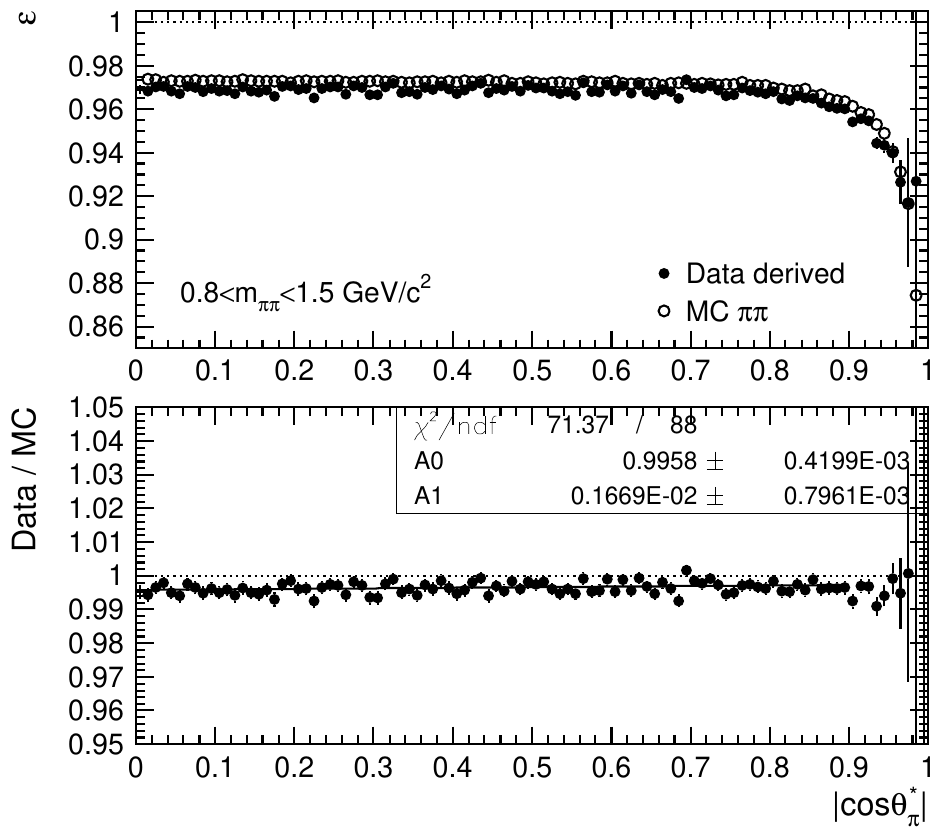}
\caption{\small \justifying The \vxy\ efficiency for $\pi\pi\gamma$ data and $\pi\pi\gamma$ MC simulation and their ratio as a function of $|\cos\theta^\ast_{\pi}|$ in three different mass regions defined with the pion mass hypothesis.}
\label{fig:vxy_eff_costh_pion}
\end{figure}

For $\pi\pi\gamma$ data samples, the background contribution is much more important, so a direct determination of the efficiency in data is not feasible. Since the main contribution comes from the geometrical nature of the vertex reconstruction, common to muon and pion track pairs, we estimate the efficiency for pions in data starting with the muon result also in data, corrected for explicit differences taken from the simulation:
\begin{equation}
\epsilon^{\pi\pi\gamma, \textrm{data}}_{V_{xy}}=\epsilon^{\mu\mu\gamma, \textrm{data}}_{V_{xy}}+\epsilon^{\pi\pi\gamma, \textrm{MC}}_{V_{xy}}-\epsilon^{\mu\mu\gamma, \textrm{MC}}_{V_{xy}},
\end{equation}
where the three terms are the corresponding efficiencies in the $\mu\mu\gamma$ data, $\pi\pi\gamma$ MC, and $\mu\mu\gamma$ MC samples, respectively. The results as a function of $m_{\pi\pi}$ and $|\cos\theta^\ast_{\pi}|$ are shown in Figs.~\ref{fig:vxy_eff_mass_pion} and \ref{fig:vxy_eff_costh_pion}. While the absolute efficiency as a function of $m_{\pi\pi}$ differs slightly from that of the $\mu\mu\gamma$ sample, the data-to-MC ratio is comparable in both samples.

\section{Charged particle momentum calibration}
\label{sec:calib}

\subsection{Calibration method with control samples}

The absolute momentum calibration  
in the data affects the calculation of $|\cos\theta^*_\pi|$ and hence
the $\pi\pi$/$\mu\mu$ separation and the mass scale of the derived spectra. Possible sources 
of mis-calibration are: (1) the value of 
the magnetic field used in the reconstruction, producing a $\Delta p_\textrm{T}/p_\textrm{T}=\Delta p/p$ shift in data, (2) the energy loss, which is primarily in the beam pipe and the SVT, and
(3) effects affecting the momentum and angle reconstruction. To investigate these issues, 
control samples with ISR-produced narrow resonances are considered: 
$J/\psi \to \mu^+\mu^-$ and $K_S \to \pi^+\pi^-$ from $\phi \to K_S K_L$.

For each sample, the available phase space ($p_\mathrm{low},p_\mathrm{high}$) matches well the region of interest. The features of potential biases
are best assessed for low momentum tracks, so fits in $p_\mathrm{low}$ intervals of 
the difference between the reconstructed mass and the PDG value of the resonance 
mass are performed for data and MC in order to extract the mass shift 
$\Delta m=m_\mathrm{reco} - m_\mathrm{true}$, while integrating over all the corresponding 
$p_\mathrm{high}$ values. Then any bias in $\Delta p$,
$\Delta p/p$, or $\Delta\alpha$, where $\alpha$ is the opening angle of the track pair, can be expressed as a shift in $\langle\Delta m\rangle$, assuming each
one to be solely responsible for the shift. 

Corrections for a bias in the assumed energy loss at the data reconstruction level are
applied prior to this procedure, producing 
very small momentum shifts for our sample, of order  
0.1\mevc. 

\subsection{Muons from $J/\psi$} 

The muon sample from the ISR-produced $J/\psi$ decays is obtained according to the physics 
analysis selections, particle ID being unnecessary. The spectra of the mass
difference between the reconstructed mass and the PDG value are fitted
in the 3.0--3.2\gevcc range to a resonance lineshape using two Gaussians and taking into account interference with QED and a smooth background in $p_\mathrm{low}$ bins up to 3\gevc.

The mass shifts are determined to be $\sim$2\mevcc\ independent of $p_\mathrm{low}$, within the statistical
accuracy of 0.4\mevcc\ per point. There is no evidence for a $\Delta p$ effect, nor a significant bias on the opening angle of the muon pair. In fact, the mass shift is fully explained by a
$\Delta p/p$ bias of $-(5.63 \pm 0.46)\times 10^{-4}$.

A consistency check is provided by correcting the momentum of every track by the 
above value and refitting 
the mass distributions. 
The resulting mass shifts are then all consistent with zero. 

The same exercise is repeated for the simulation. But since our \textsc{Phokhara}-v9 production did not include the generation of $J/\psi$, we proceed instead by computing directly $\Delta m=m_\mathrm{reco} - m_\mathrm{true}$ for
QED events in the same 3.0--3.2\gevcc mass range. The mass shifts in MC are one order of magnitude lower than in data, typically $\sim$0.2\mevcc. 
As in data, the preferred bias model is a shift in $\Delta p/p$, however 10 times smaller and with the opposite sign, 
$+(0.58 \pm 0.10)\times 10^{-4}$. Since the bias cannot originate from the value of the magnetic field, it is likely that it is caused at the track reconstruction level. Whereas data and MC are calibrated separately,
it is the relative effect which matters for the $|\cos\theta^*_\pi|$ fit, 
corresponding to a correction of $\Delta p/p = -(6.21 \pm 0.47)\times 10^{-4}$. 

\subsection{Pions from $K_S$}

The $K_S$ calibration sample is obtained from ISR-produced $\phi$ decays to
$K_SK_L$ selected in the standard analysis acceptance, but without $\chi^2$
selections, as most of these events are in the rejected region because of the 
missing $K_L$. The $K_S$ candidates are 
taken from the \babar\ reconstruction of displaced vertices, the $\pi^+\pi^-$ 
mass being evaluated at the decay point. Loose selections are applied on the accepted
mass range (0.472--0.522\gevcc), the $\chi^2$ of the vertex fit ($<7$), the 
transverse decay length ($>3$\,mm), and the pointing to the interaction point (cosine of the angle
between the line of flight and the reconstructed $K_S$ direction $>0.95$). 
A selection is applied on the missing mass squared ($>0.2\gev^2/c^4$) and the 
$K_SK_L$ mass is assumed to be equal to the PDG value of the $\phi$ meson mass. The selected $K_S$ sample is clean, with a flat background
of 4.8\% in a 20\mevcc window encompassing the peak. A similar analysis as for $J/\psi$ yields a consistent correction of $\Delta p/p = -(6.87 \pm 0.89)\times 10^{-4}$.

\subsection{Summary of calibration results and systematic uncertainties}\label{sec:calib-syst}

Calibration studies with muons ($J/\psi$) and pions ($K_S$ from $\phi$)
yield a consistent picture of a data/MC $\Delta p/p $ bias
with average 
$-(6.35 \pm 0.42)\times 10^{-4}$.

Systematic uncertainties arise from
the parametrization of the $J/\psi$ lineshape, differences between pions
and muons, and the level of overall consistency
between the different results. A very conservative limit is taken from
the respective measurements above, yielding an
uncertainty of $\pm 1.0\times 10^{-4}$.
The tails from QED-$J/\psi$ interference and FSR are small and computed
at lowest order. However, they displace the mean fitted mass by $-0.4$\mevcc  
due to their asymmetric shape. The conservative assumption of an uncertainty of 
$\pm$10\% on their contribution reflects into a systematic uncertainty of
$5\times 10^{-5}$ on the $\Delta p/p$ correction.

Since the small bias observed in the MC in the $J/\psi$ region was not expected
and could possibly be indicative of an effect not considered in the analysis, it
is taken in full as a systematic uncertainty. 

Thus, the total estimate of the $\Delta p/p$ data-MC correction amounts to $\Delta p/p = -(6.4 \pm 1.2)\times 10^{-4}$, which is applied to track momenta in the MC.

The effect of the $\Delta p/p$ correction on the $\mu\mu$ and
$\pi\pi$ MC templates used in the $|\cos\theta_\pi^*|$ fit is studied using a toy MC to mimic data. The 
deviation found for the derived $\mu\mu$ spectrum reaches a maximum value of  
0.74 per mil at a mass of 0.75\gevcc, leading to a similar effect for the $\pi\pi$/$\mu\mu$ ratio. Larger effects, up to 1\%, are observed near threshold. Given the accuracy reached in the determination of the $\Delta p/p$ bias, the systematic uncertainty in the $\mu\mu$ and $\pi\pi$ results is negligible. 
\section{Pion form factor description in \pipig\ simulation}
\label{sec:pionFF}

\subsection{Vacuum polarization correction} 
\label{subsec:vp-correction}
\begin{figure}[tbp]
    \centering
    \includegraphics[width=1\linewidth]{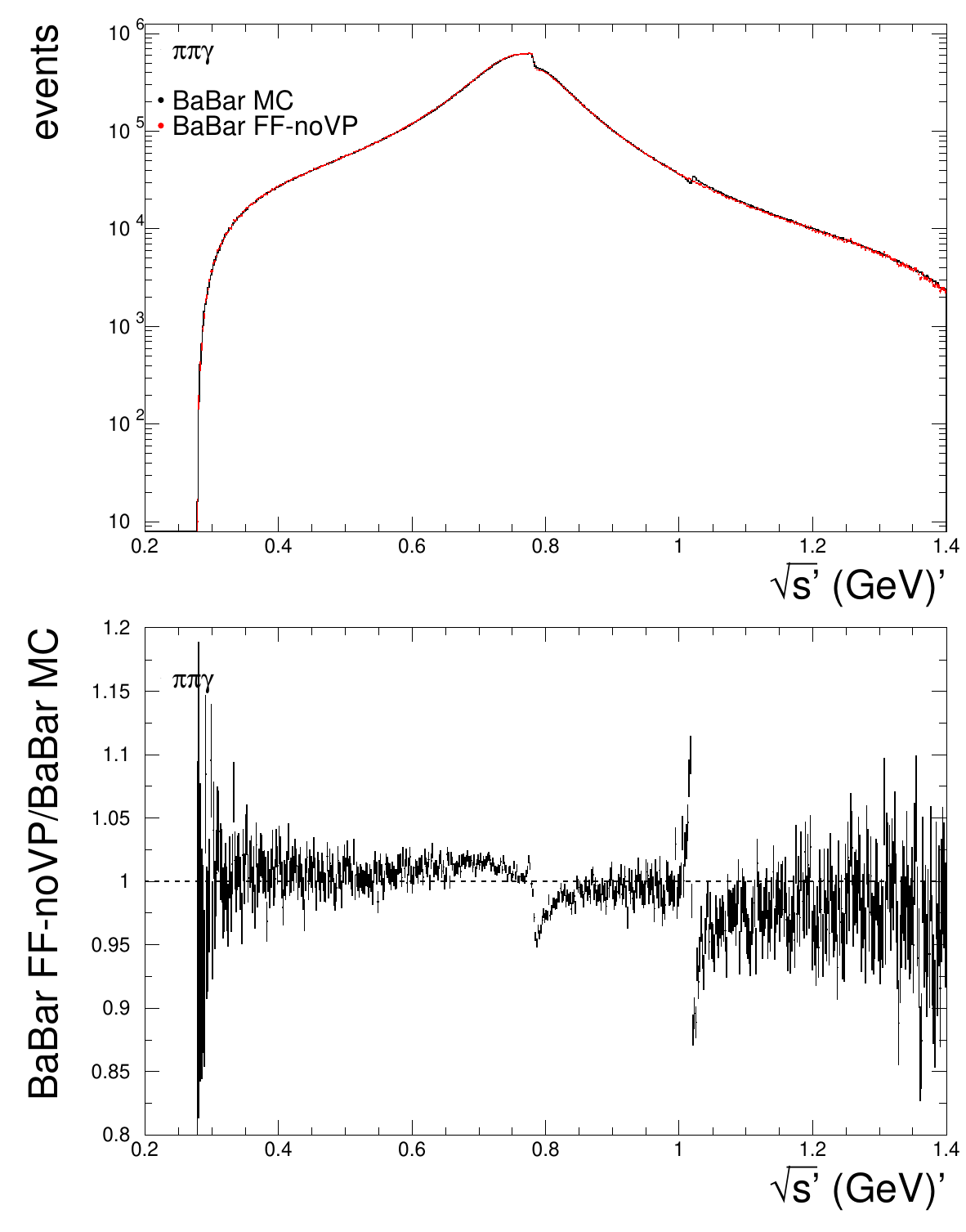}
\caption{\label{fig:BbrnoVP-Bbr_spectrum} \small \justifying 
(top) Spectrum of \pipig\ events in \babar\ MC and in \textsc{Phokhara} simulation using the \babar\ pion form factor without extra VP correction. (bottom) Ratio of the above spectra.}
\end{figure}

\begin{figure}[tbp]
    \centering
    \includegraphics[width=1\linewidth]{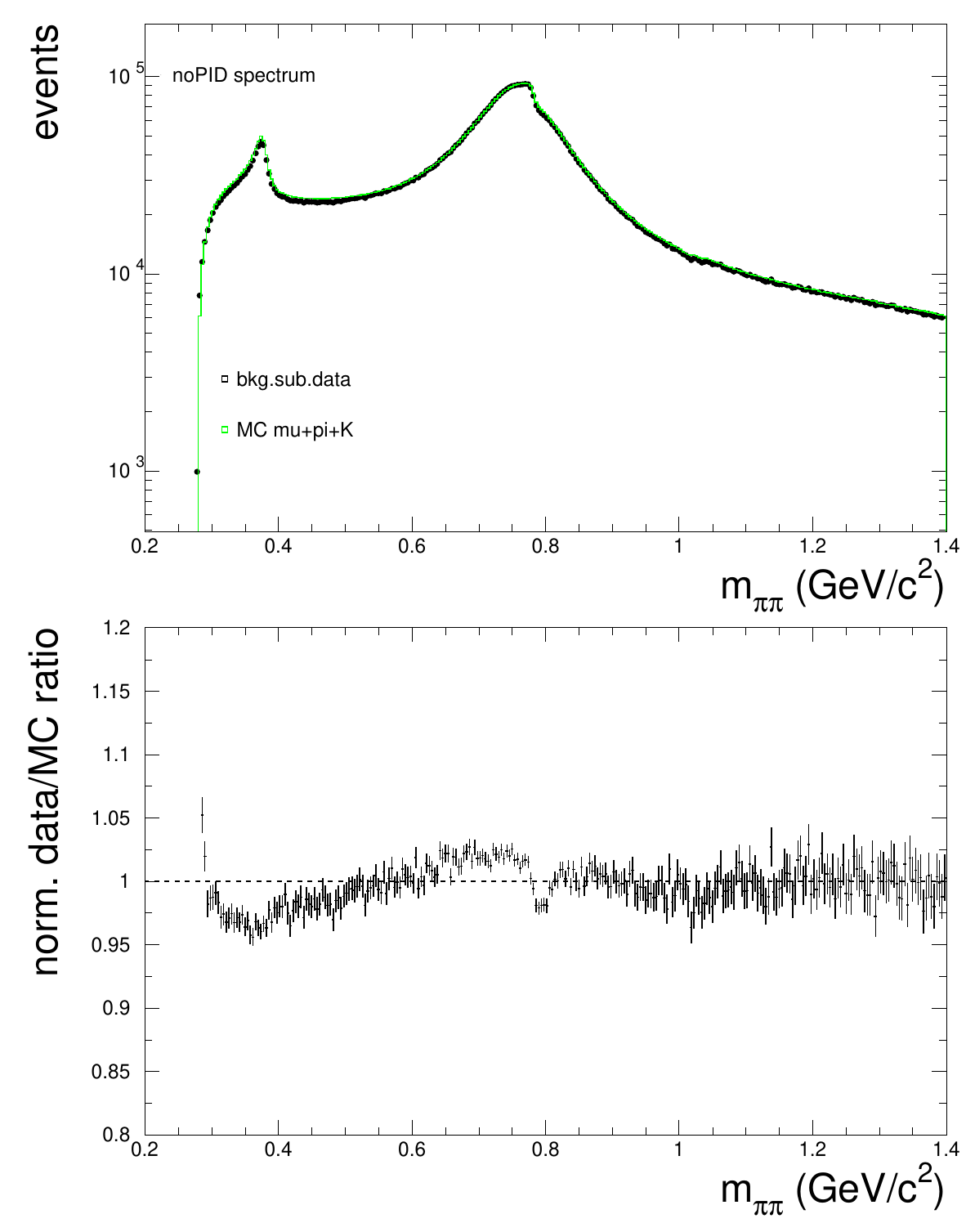}
\caption{\label{fig:data-MC_noPID} \small \justifying 
(top) Spectra of data events with no PID selection and \babar\ simulation of \mumug, \pipig\ and \KKg\ events. (bottom) Normalized ratio of the above spectra.}
\end{figure}

The signal \pipig\ events were produced with \textsc{Phokhara} using the pion form factor measured by \babar~\cite{BaBar:2009wpw} together with a vacuum polarization factor (VP) from F. Jegerlehner~\cite{Jegerlehner:2017gek}. However, the published \babar\ form factor is dressed, i.e., includes a VP correction, while \textsc{Phokhara} expects a bare one~\cite{Czy__2010}. This mismatch induces an incorrect $\sqrt{s'}$ spectrum of \pipig\ MC as seen in Fig.~\ref{fig:BbrnoVP-Bbr_spectrum}, which shows the comparison of the \babar\ MC to a \textsc{Phokhara} sample produced without the additional VP factor. 
The discrepancy can be observed as well in the ratio of the data and MC spectra of all $\mu/\pi/K$ events as shown in Fig.~\ref{fig:data-MC_noPID} as a function of $m_{\pi\pi}$. The impact of the improperly applied VP factor is clearly seen in the mass range dominated by the \pipig\ channel. 

The double VP correction has no impact on the $\sqrt{s'}$ distribution of \pipig\ events measured in data since discrepancies between data and simulated $m_{\pi\pi}$ or $\sqrt{s'}$ spectra are canceled through the iterative unfolding method (see Section\ref{sec:unfolding}).
However, the double counting of vacuum polarization affects the pion templates. 
While templates for LO and ISR NLO events are insensitive to vacuum polarization, this is not the case for FSR NLO events. 
Since the $|\cos\theta^*_\pi|$ quantity is computed from tracks only, it is sensitive to the additional FSR photon momentum and FSR rate, hence to the $\sqrt{s'}$ distribution. Although FSR NLO events are only a small fraction of all \pipig\ events (2.4\% with an additional FSR photon $E_{\mathrm{FSR}}>50$\mev), the impact of the double VP correction on the \pipig\ templates used in the analysis is small but not negligible.
To address that issue, a weight is assigned to the \babar\ MC events depending on the $\sqrt{s'}$ value, 
where this quantity is computed with the true values of track and photon momenta in the event. To avoid large fluctuations due to the limited statistics of the special \textsc{Phokhara} production with no double VP correction, the assigned weight is taken from the analytical formula of vacuum polarization from F. Jegerlehner.  
The pion templates in bins of $m_{\pi\pi}$ used in the $|\cos\theta^*_\pi|$ fits of the data are deduced from these weighted events\footnote{VP weighting was not applied for the preliminary results in 2025 as reported in Appendix~\ref{app:changes}}. The resulting changes in the $m_{\pi\pi}$ and $m_{\mu\mu}$ spectra amount to an oscillation of $(-1,+2)$ per mil and $(+2,-6)$ per mil, respectively, at the $\rho$-$\omega$ interference mass.

\subsection{$\rho$-$\omega$ interference description}
\label{sec:CMD3-BaBar}

The pion form factor parametrization included in our MC simulation is taken from the \babar~\cite{BaBar:2009wpw} measurement. Since then, several experiments have published similar results, the most precise of which is from CMD-3. 
A fit of their published data with the \babar\ parametrization gives slightly different values of the resonance parameters, $\rho$ and $\omega$ mass and width, and interference phase, compared to \babar. The ratio of CMD-3 to \babar\ fitted form factors serves as an event-by-event reweighting factor applied to the \pipig\ simulation to assess the uncertainty associated with the modeling of the $\rho$-$\omega$ interference.
The uncertainty induced on the \pipig\ spectrum after unfolding is negligible. An oscillation of $\pm 7$ per mil is observed at the $\rho$-$\omega$ interference on the \mumug\ spectrum with respect to the one obtained from the angular fit with the \babar\ nominal pion templates. The impact on the  
effective luminosity and \pipig\ cross section is negligible.
\clearpage
\section{Mass spectra determination}
\label{sec:spectra}

After all selections, including the mass interval restriction $0.28<$\mpipi$<1.5$\gevcc, the total data sample entering the angular fits amounts to 
7\,617\,066 events, of which 135\,249 events are subtracted as backgrounds before entering the fits. 
\subsection{Fits of \acosthpi\ distributions in \mpipi\ mass intervals}
\label{subsubsec:fit_strat}

With the pion mass hypothesis for the charged tracks, the fits to the \acosthpi\ angular distributions determine the fraction of each $X=\pi/\mu/K/e$ component in data in \mpipi\ bins. 

The mass binning consists of intervals of 10\mevcc from 0.28 
to 0.5\gevcc and from 1 to 1.5\gevcc. Between 0.5 and 1\gevcc, the fits are performed in finer intervals of 2\mevcc.
The data \acosthpi\ distribution in each mass bin, from which non-\eeg\ backgrounds are subtracted, is adjusted by a linear combination $N$ of the normalized templates $x_i \pm \Delta x_i$ ($i=1,...,k$, with $k=4$ 
for $\pi/\mu/K/e$) that contributes to the $\chi^2$ of the fit:
\begin{widetext}
\begin{eqnarray}
\label{eq:template_sum}
    N &=& M\left[ \sum_{i=1}^{k-1} f_ix_i+\left(1-\sum_{i=1}^{k-1} f_i\right)x_k \right] \pm 
    M\sqrt{\sum_{i=1}^{k-1}\left(f_i\Delta x_i\right)^2 + \left[ \left(1-\sum_{i=1}^{k-1} f_i\right)\Delta x_k\right]^2}.\nonumber
\end{eqnarray}
\end{widetext}
Here, $M$ is the integral of the fitted data distribution and $f_i \in [0,1]$ are the parameters of the fit, i.e., the scale factors of the first $k-1$ templates. The fit returns these scale factors, 
with statistical uncertainties, by minimizing the $\chi^2$.
From threshold to 0.4\gevcc the fit takes into account four contributions from \pipig, \mumug, \KKg, and \eeg\ processes. Beyond 0.4\gevcc, the \KKg\ \acosthpi\ distribution, scaled down by 0.937 to account for efficiencies, is subtracted from data and the corresponding template is removed from the fit, as this process becomes negligible at higher masses and deteriorates the statistical uncertainties in the fitted \pipig\ scale factors.

For \pipig, \mumug, and \KKg, the templates are taken from the respective \textsc{Phokhara} simulations. Distributions for \pipig\ and \mumug\ are further corrected for the data and MC differences described in the previous sections. The \eeg\ distributions are obtained from data as described in Section~\ref{sec:eeg}. To compensate for the low number of \eeg\ events beyond the initial peak at threshold and thereby reduce the statistical uncertainty, \eeg\ templates are defined everywhere in bins of 10\mevcc, with sliding windows of 30\mevcc\ for masses above 0.4\gevcc.

With a single fit of all the contributions along the full \acosthpi\ range, the \eeg\ component affects the uncertainties in the \pipig\ and \mumug\ mass spectra in a way that is not negligible, adding a systematic uncertainty in the latter that would reach the percent level in the $\rho$ peak mass region. Consequently, a three-step strategy, dependent on various \acosthpi\ ranges, is devised to mitigate the impact of the \eeg\ background on the final results:
\begin{itemize}
    \item a first fit for \acosthpi\ between 0 and 1,
    \item a second fit between 0.9 and 1, using the results obtained in the first step as constraints (initial value and range) on the fitted scale factors to help it converge,
    \item a third fit between 0 and 0.9, where the \eeg\ distribution normalized according to the second step is subtracted from data.
\end{itemize}
At the end of the third step, the resulting normalization of the \pipig\ and \mumug\ distributions between 0 and 0.9 is extrapolated to the full \acosthpi\ range, which finally gives the pion and muon yields in the fitted mass bin. 

The \acosthpi\ distributions are fitted according to a combined Neyman-Pearson (CNP) chi-square test statistic proposed in Ref.~\cite{Ji:2019yca} as an alternative to both Neyman's and Pearson's chi-squares, in order to reduce their inherent biases. 
In a closure test that consists of replacing data with a simulation sample of comparable size, the CNP chi-square method showed the best compatibility (within $1\sigma$) between the fitted and true \pipig\ or \mumug\ fractions for masses larger than 0.5\gevcc.

The statistical uncertainty in the scale factors, as estimated by the fit algorithm, has been compared to the corresponding standard deviations obtained from toy replicas generated using the bootstrap method\footnote{In the bootstrap method the weight of each event is fluctuated according to a Poisson(1) distribution~\cite{Efron1992,HANLEY200657,Bohm:2005bu,ATLAS:2021kho}. 
The seed of the random number generator for every event is deterministically and uniquely set, on the basis of the event number.
Doing so provides a coherent~(i.e.\ statistically correlated) series of pseudo-experiments when a given event is employed at multiple analysis steps, while these random number series are independent among different events.}.
The fit tends to underestimate the uncertainty,  
down to around $-5$\% in the central mass region.
Statistical uncertainties provided by the bootstrap replicas, featuring good coverage properties, are therefore used\footnote{It was tested that in the current study even asymptotic approaches would match reasonably well the expected scaling properties discussed in Ref.~\cite{Alexe:2024smm}. }.

\begin{figure*}[tbp]
    \centering
    \includegraphics[width=1\linewidth]{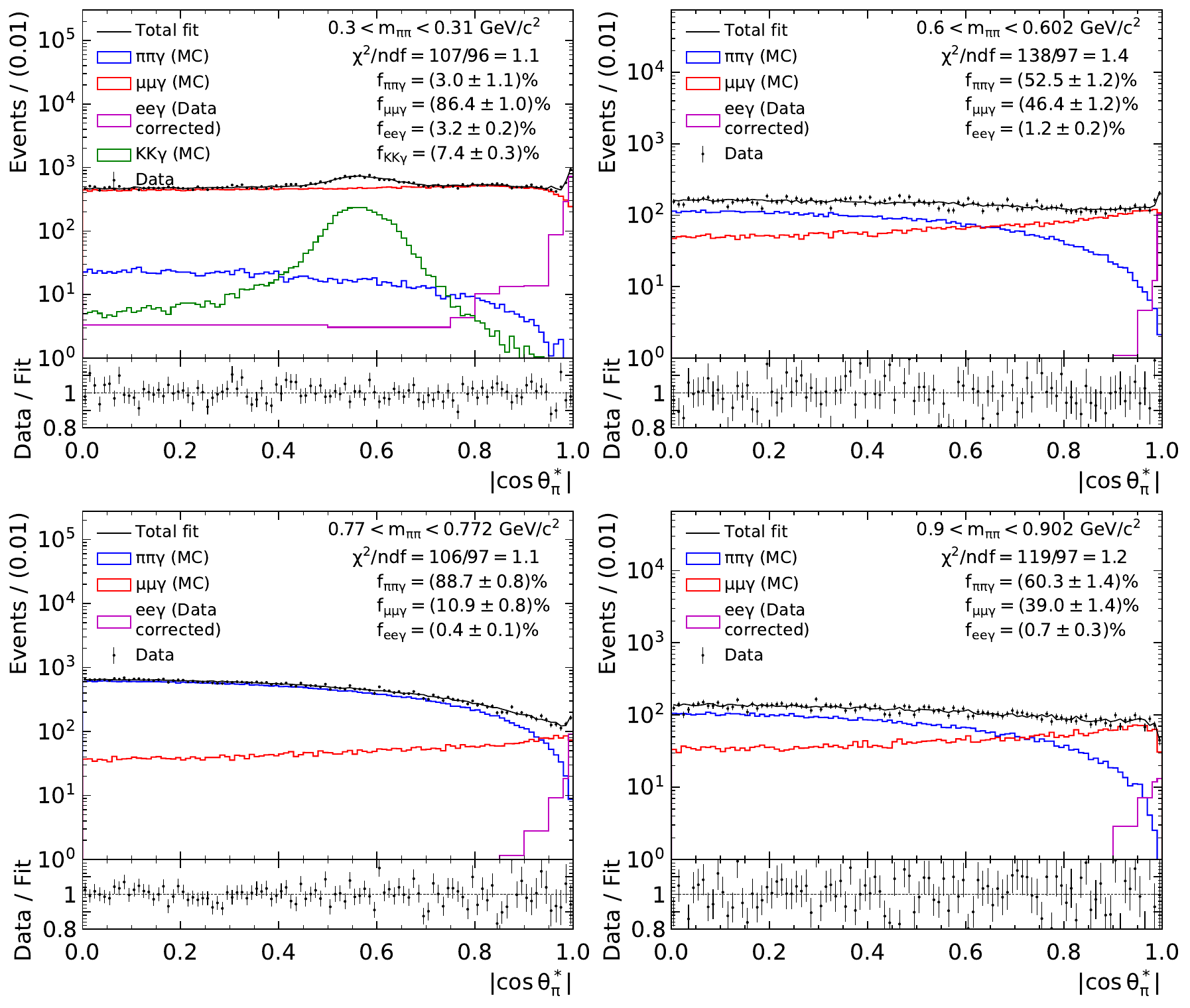}
\caption{\small \justifying Combination of the results from the $\mathrm{2^{nd}}$ (\eeg) and $\mathrm{3^{rd}}$ (\pipig, \mumug, \KKg) step fits of \acosthpi\ data distributions (black points) in four representative $m_{\pi\pi}$ intervals and $0<$\acosthpi$<1$. The result of a fit is the sum of the \pipig\ (blue histogram), \mumug\ (red histogram), \KKg\ (green histogram), and \eeg\ (magenta histogram) contributions given by the solid line. The proportion of each component is indicated as $f_{XX\gamma}$ ($X=\pi,\mu,K,e$).}
\label{fig:fits_costh_tot}
\end{figure*}

\begin{figure}[tbp]
\begin{center}
  \makebox[0pt]{
  \centering
      \includegraphics[width=\linewidth]{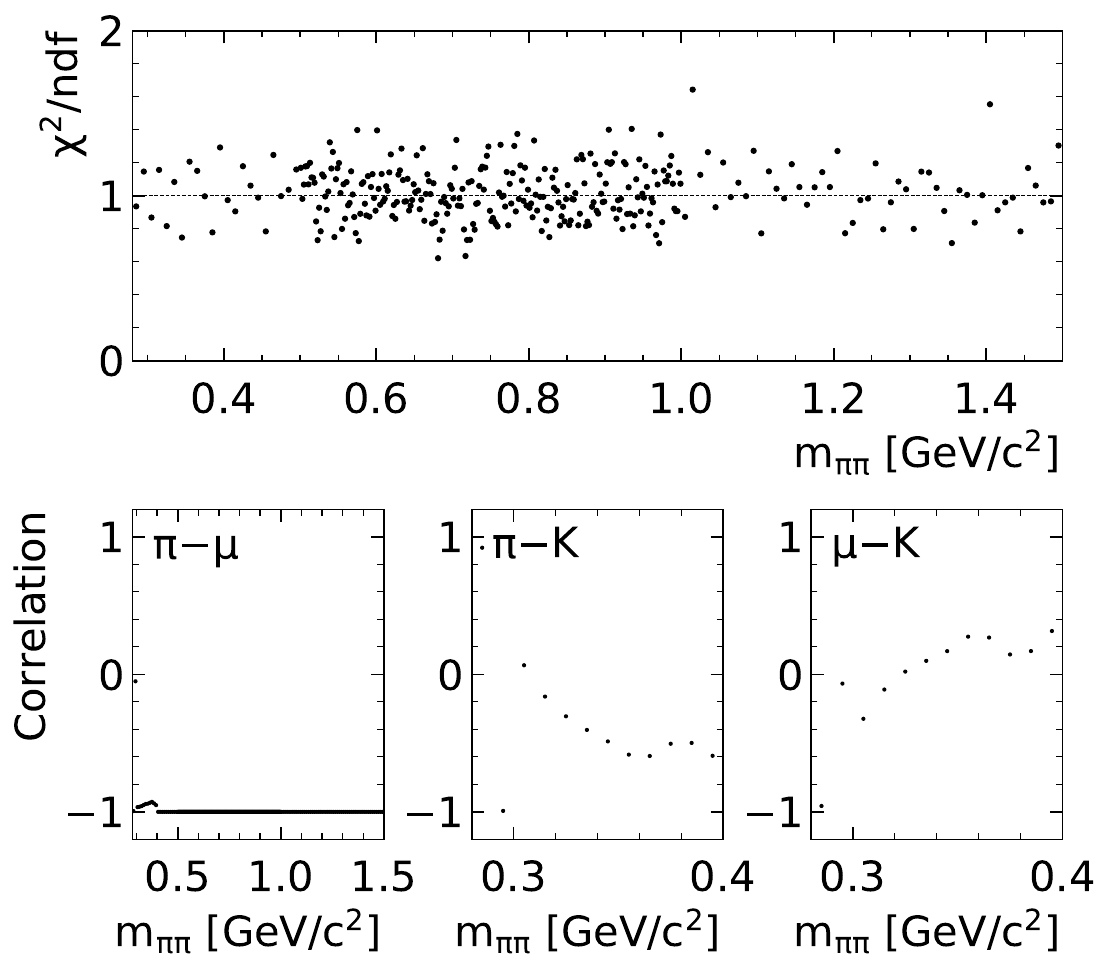}
  }
  \end{center}
\caption{\small \justifying \acosthpi\ fit $\chi^2$ divided by the number of degrees of freedom (top) and off-diagonal correlation matrix elements of the fitted scale factors (bottom), obtained with the $\mathrm{3^{rd}}$ fit.}
\label{fig:chi2_corr_pi}
\end{figure}

Four representative fits 
are illustrated in Fig.~\ref{fig:fits_costh_tot}, after the results of the third-step fit are extrapolated to the full angular range. The quality of the fits is good as seen from the $\chi^2$ over the number of degrees of freedom (ndf) in \Cref{fig:chi2_corr_pi}. 
Below 0.4\gevcc\ when all components are fitted, the correlation matrices show a strong anti-correlation between $\pi - \mu$, and between $\pi - K$ for masses above 0.3\gevcc. Above 0.4\gevcc, $\pi$ and $\mu$ are fully anti-correlated.

\subsection{Mass spectra derived from the fits}
\label{subsec:mumu_strat}

The \pipig\ mass spectrum is readily obtained 
as the distribution of pion yields determined in the angular fits.
For \mumug, as the analysis requires the mass spectrum to be determined along the \mmumu\ variable, it is necessary to translate the \mumug\ spectrum from pion mass hypothesis to muon mass. An alternate option has first been considered, namely to carry out a second angular fit of the data using the $|\cos\theta^*_\mu|$ and $m_{\mu\mu}$ variables under the muon mass hypothesis. This procedure, used for the preliminary results in 2025, was however found to be very sensitive to the detailed description of the pion form factor and was consequently dropped. Two other approaches have been assessed with closure tests: the first one consists in weighting the data according to the fitted fractions of \mumug\ events in two-dimensional \mpipi\ vs \acosthpi\ bins, then drawing the \mmumu\ distribution. This statistical procedure however fails because the shift from the pion to muon mass hypotheses depends on the true nature $X=\mu,\pi$, or $K$ of the $XX\gamma$ events.

\begin{figure}[htp]
  \centering
  \includegraphics[width=\linewidth]{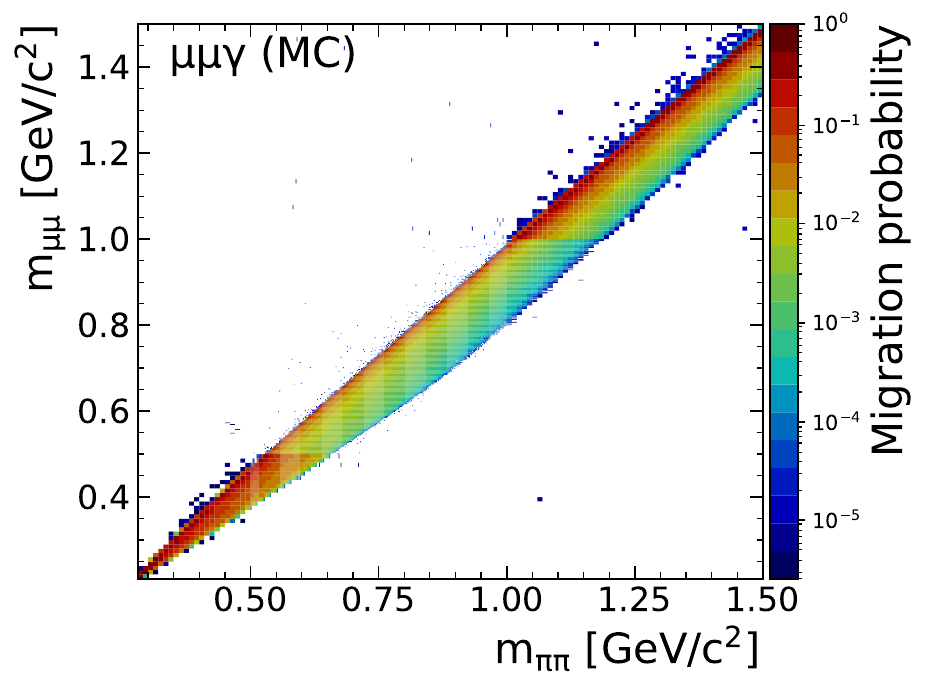}
\caption{\small \justifying Transfer matrix from \mpipi\ to \mmumu, based on simulated \mumug\ events and defined as the two-dimensional event yield divided by the integral over \mpipi.}
\label{fig:migrationMatrix}
\end{figure}

The second option is to recover the \mmumu\ spectrum using a transformation matrix from \mpipi\ to \mmumu, based on \mumug\ simulation, that gives the migration probability as shown in Fig.~\ref{fig:migrationMatrix}. A closure test that replaces data with a simulation sample of equivalent size, repeated for ten uncorrelated samples of this sort, shows that this procedure does not induce a bias in the \mmumu\ spectrum obtained through transformation, compared to the reconstructed spectrum. In addition, this procedure proved to be robust with respect to the uncertainties in the dominant \pipig\ channel. It is thus applied throughout the present analysis. 
Because the transfer matrix is bounded at 1.5\gevcc on \mpipi, the shifted dimuon spectrum is depleted in the 1.4 to 1.5\gevcc range. This region does not contribute to the rest of the analysis, however the last few bins below 1.4\gevcc are corrected for the missing migration of \mumug\ events from beyond 1.5\gevcc. This effect is small, of the order of 8 per mil at most, and affects only a mass window of 50\mevcc\ at the high end of the spectrum.

\begin{figure}[tbp]
\centering
\includegraphics[width=\linewidth]{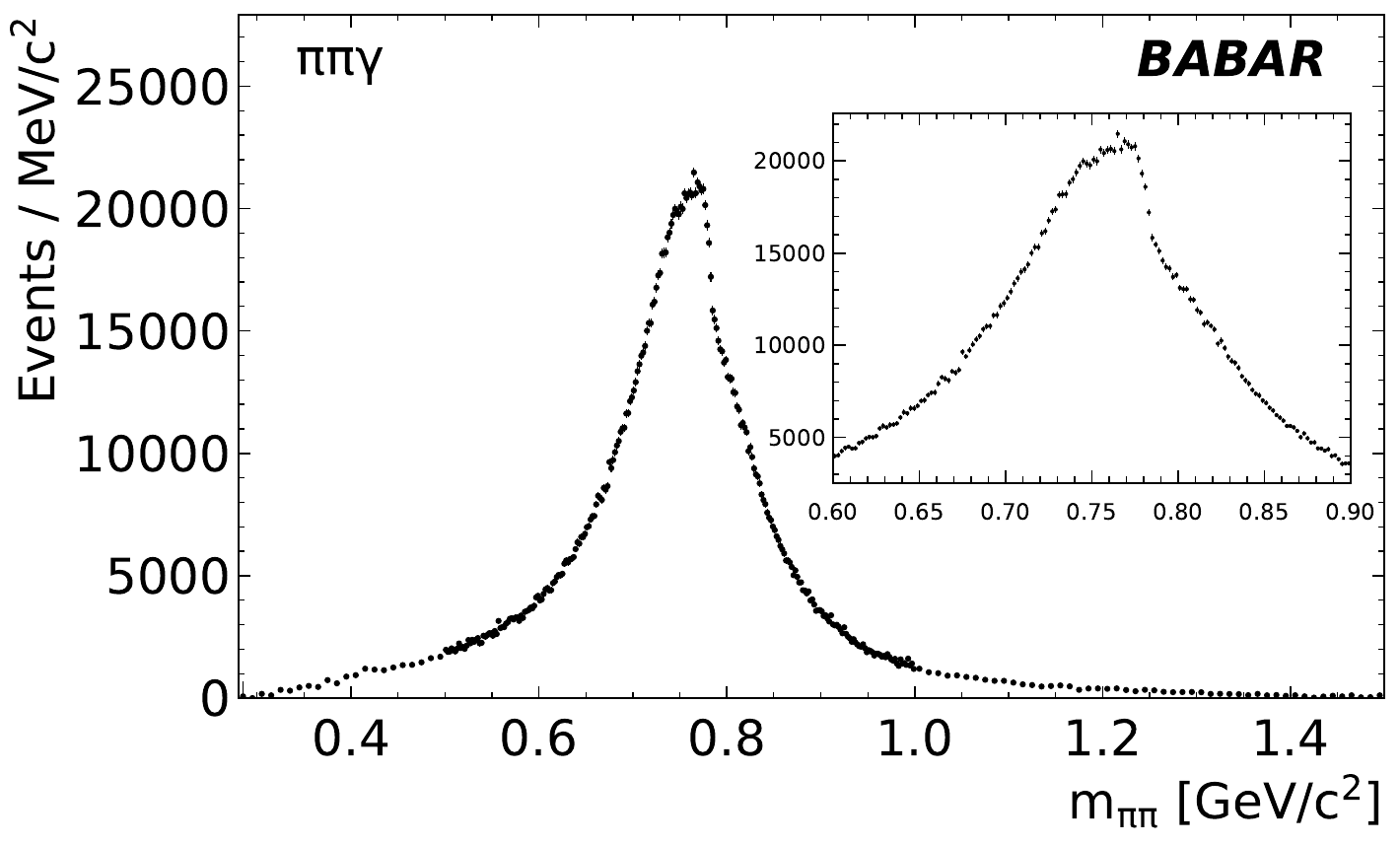}\\
\includegraphics[width=\linewidth]{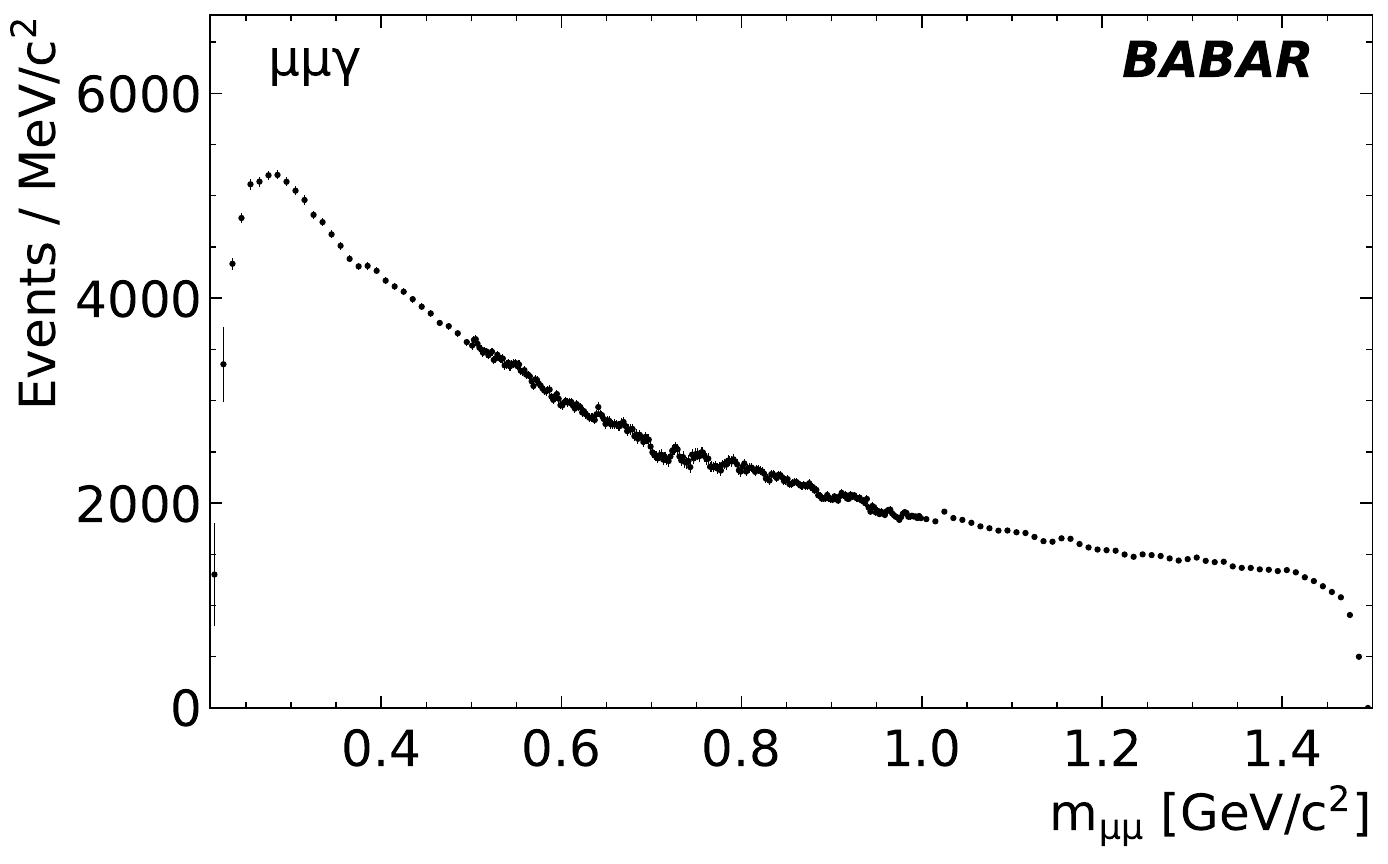}
\caption{\small \justifying Mass spectra of the \pipig\ (top) and \mumug\ (bottom) processes, obtained from fits to the \acosthpi\ data distributions in \mpipi\ bins. The \mumug\ spectrum is transformed to the \mmumu\ basis with a transfer matrix, as explained in the text. Template and mass-dependent corrections are applied. The error bars include statistical uncertainties and systematic uncertainties  
coming from data/MC corrections. The number of events is divided by the bin width in\mevcc.}
\label{fig:sig_spectra}
\end{figure}

Figure~\ref{fig:sig_spectra} presents the \pipig, and \mumug\ mass spectra obtained from the fits, in bins of 2\mevcc between 0.5 and 1\gevcc and bins of 10\mevcc outside this range. The \pipig\ spectrum is directly derived from the angular fits while the \mumug\ mass spectrum is obtained through the transfer matrix procedure as explained above. Both spectra are corrected for the mass-dependent data/MC absolute ratio of efficiencies. 

\subsection{Systematic uncertainties in the mass spectra}

\subsubsection{Impact of systematic uncertainties of the \eeg\ templates}
\label{subsec:eeg_syst}

\begin{figure}[tbp]
  \centering
  \includegraphics[width=\linewidth]{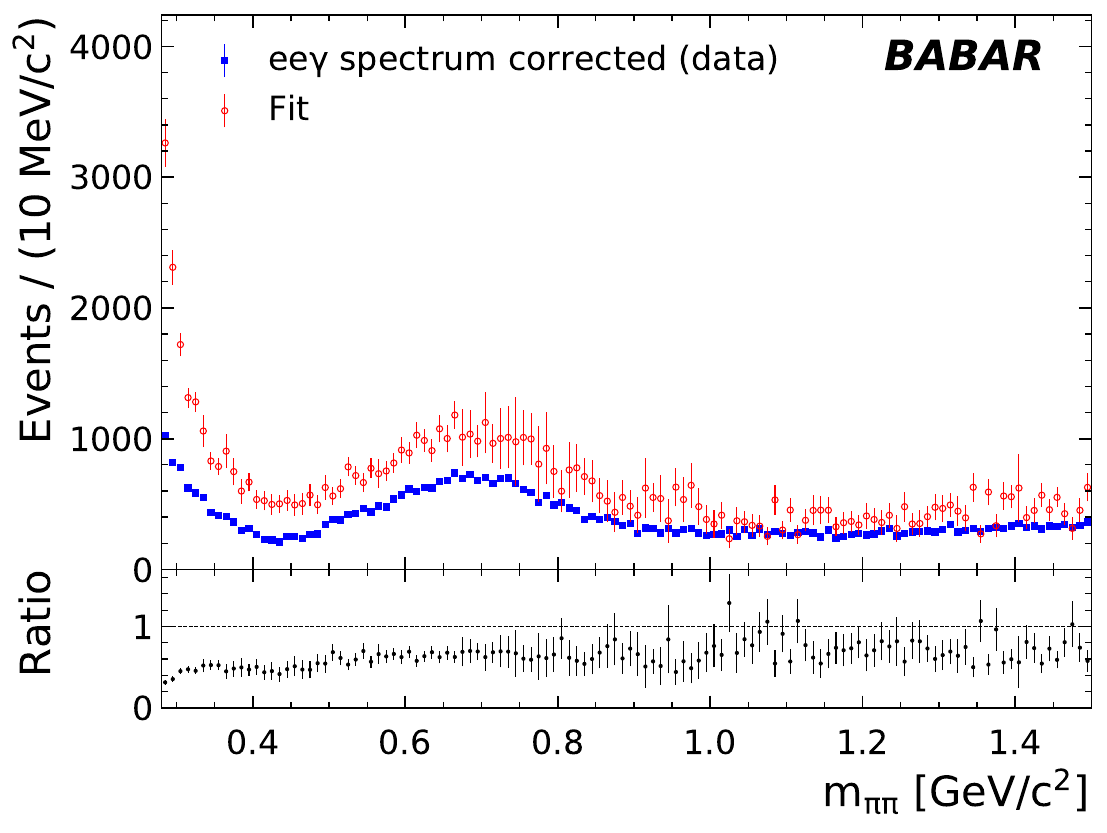}
\caption{\small \justifying \mpipi\ mass spectrum of the \eeg\ process derived from fits on \acosthpi\ data distributions (open red circles), compared to the spectrum of the \eeg\ data control sample  (blue squares, see Section~\ref{sec:eeg}).
}
\label{fig:eeg_spectra}
\end{figure}

Figure~\ref{fig:eeg_spectra} compares the \mpipi\ mass spectrum of \eeg\ derived from the angular fits to the spectrum observed for the data control sample in Section~\ref{sec:eeg}. The shapes of the fitted and expected spectra are consistent while the efficiency of the 
control sample selection appears to be constant for most of the range except at threshold.
This validates the data-driven procedure to determine the \eeg\ templates (Section~\ref{subsubsec:eetemp}). 

Two known sources of systematic uncertainty affect the \eeg\ templates and propagate to the fitted spectra:
\begin{enumerate}
    \item the choice of the selection value on the BDT output (Section~\ref{subsubsec:BDT}). The value is fluctuated by $\pm 0.1$, compared to the nominal selection set at 0.60.
    \item the scaling of the \pipig\ background that contaminates the \eeg\ sample before subtraction from data. The background is either not scaled, or scaled according to the method presented in Section~\ref{subsubsec:pibkg}.
\end{enumerate}

\begin{figure}[tbp]
\centering
   \includegraphics[width=0.95\linewidth]{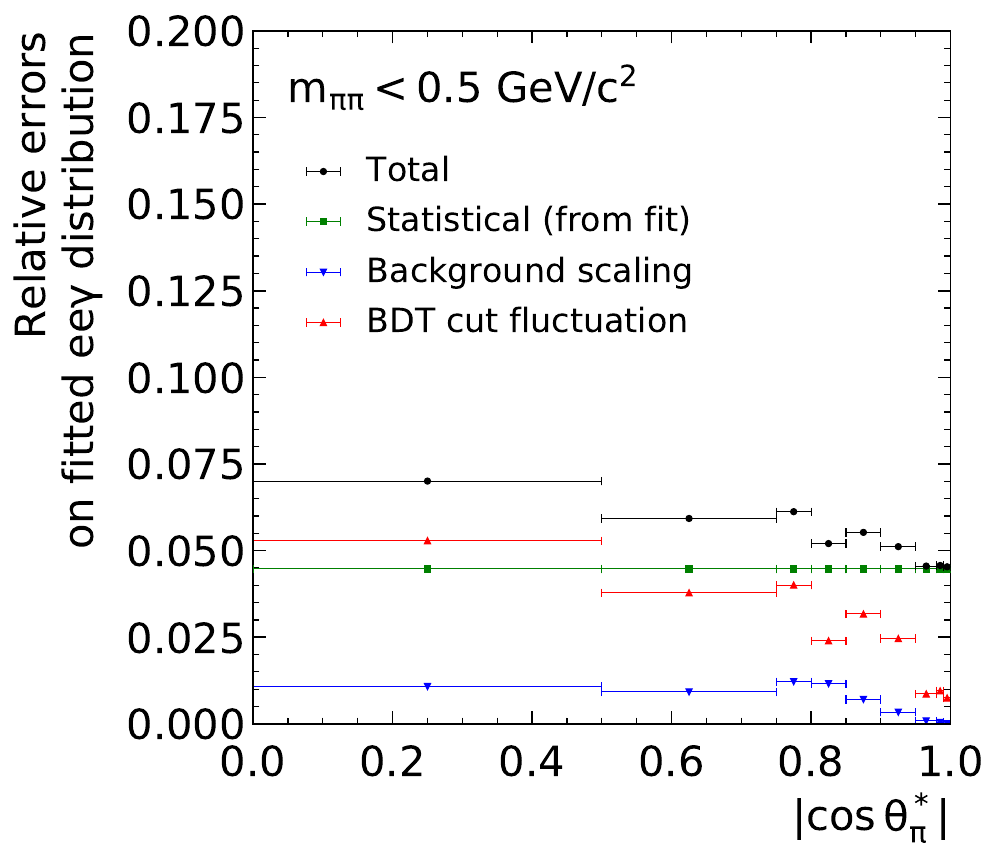}\\
   \includegraphics[width=0.95\linewidth]{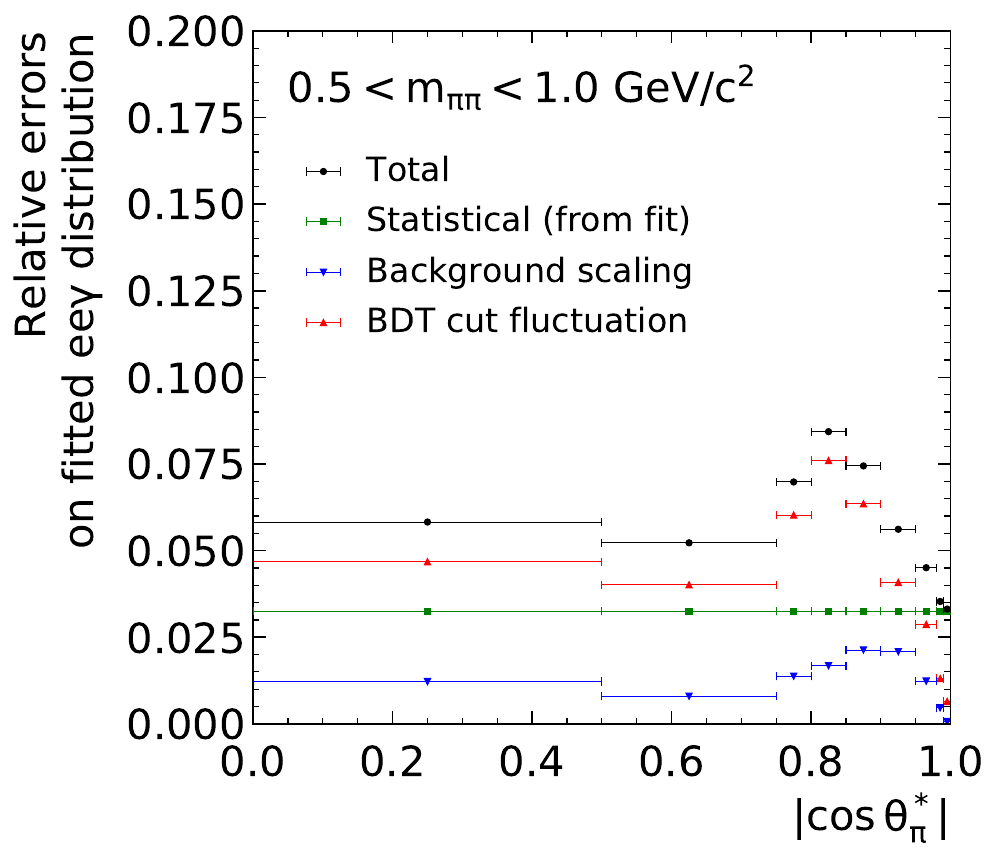}\\
   \includegraphics[width=0.95\linewidth]{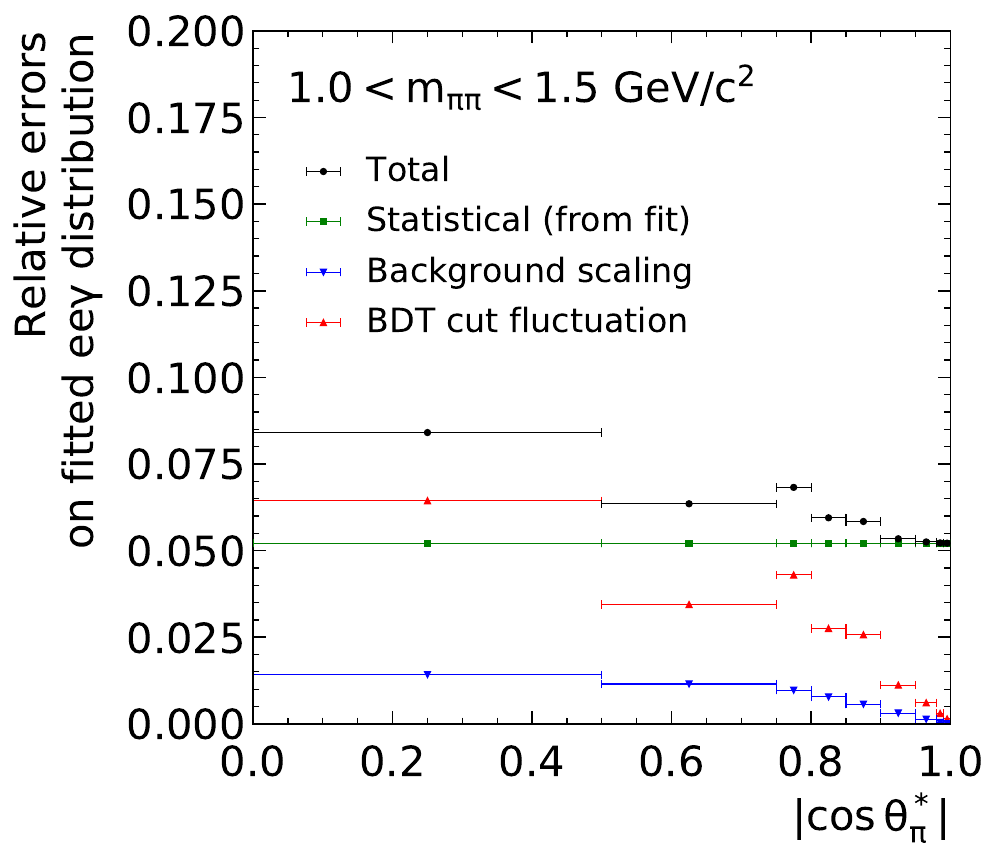}\\   
\caption{\small \justifying Relative uncertainties in the fitted \eeg\ contribution to be subtracted from data before the $\mathrm{3^{rd}}$ angular fit. The uncertainties are given in three large mass bins and are composed of the statistical uncertainty determined by the $\mathrm{2^{nd}}$ fit (green squares), the systematic uncertainty originating from the scaling applied to background events that contaminate the \eeg\ sample (downward blue triangles), and the systematic uncertainty inherent to the nominal BDT selection by fluctuation of the selection value (upward red triangles), while the total is represented with black points.}
\label{fig:relErrors}
\end{figure}

These contributions, along with the statistical uncertainty that comes from the fit itself, are evaluated with the bootstrap method, reproducing the fit procedure described in Section~\ref{subsubsec:fit_strat}. Changes are introduced independently to measure the individual systematic effects.
The relative uncertainties in the \eeg\ fitted contributions are given in Fig.~\ref{fig:relErrors} as a function of \acosthpi, in three large mass bins. 
In the central mass region, the BDT selection is the dominant source of systematic uncertainty and reaches a few percent at \acosthpi\ $>0.9$, where \eeg\ events are not negligible.

\begin{figure*}[tbp]
    \centering
\includegraphics[width=\linewidth]{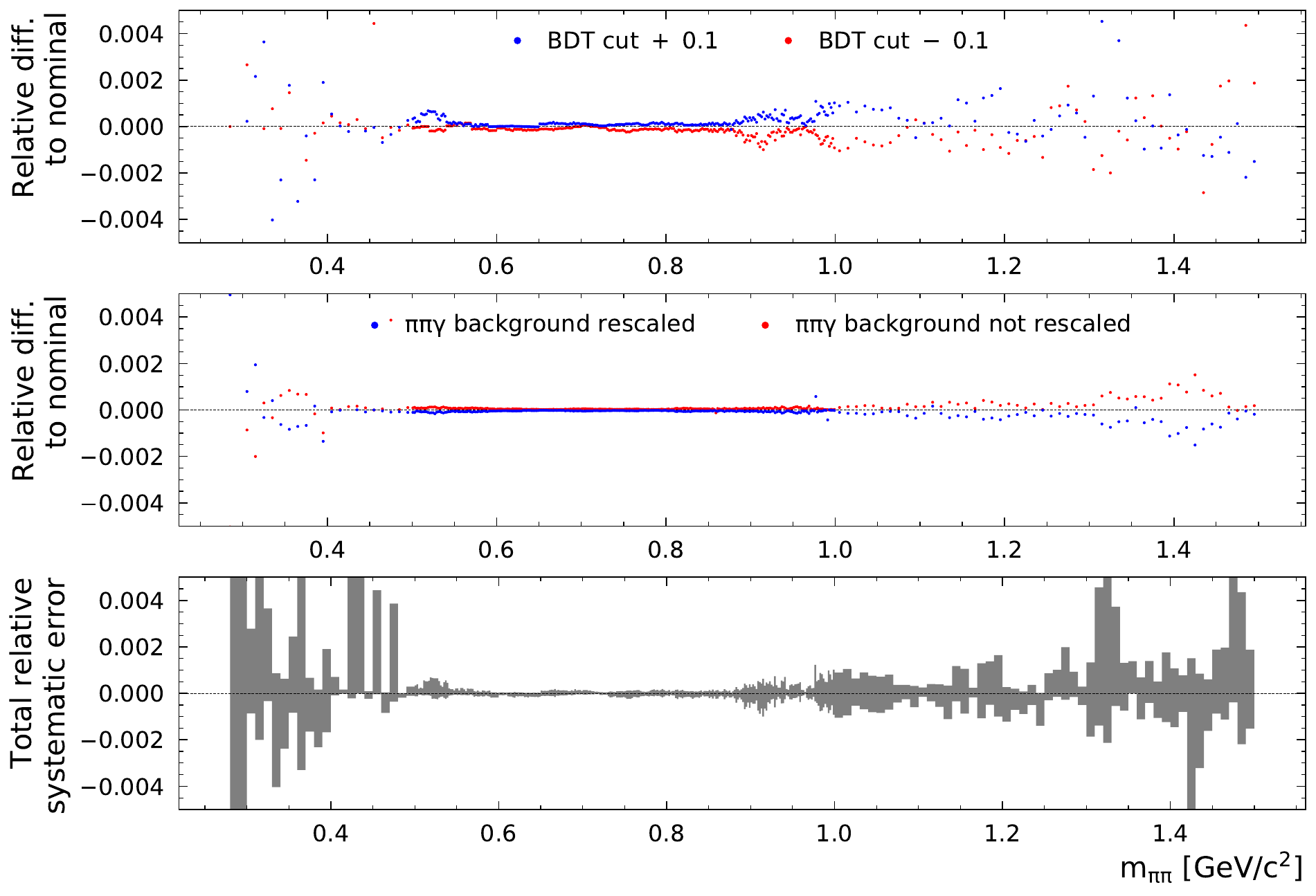}
\includegraphics[width=\linewidth]{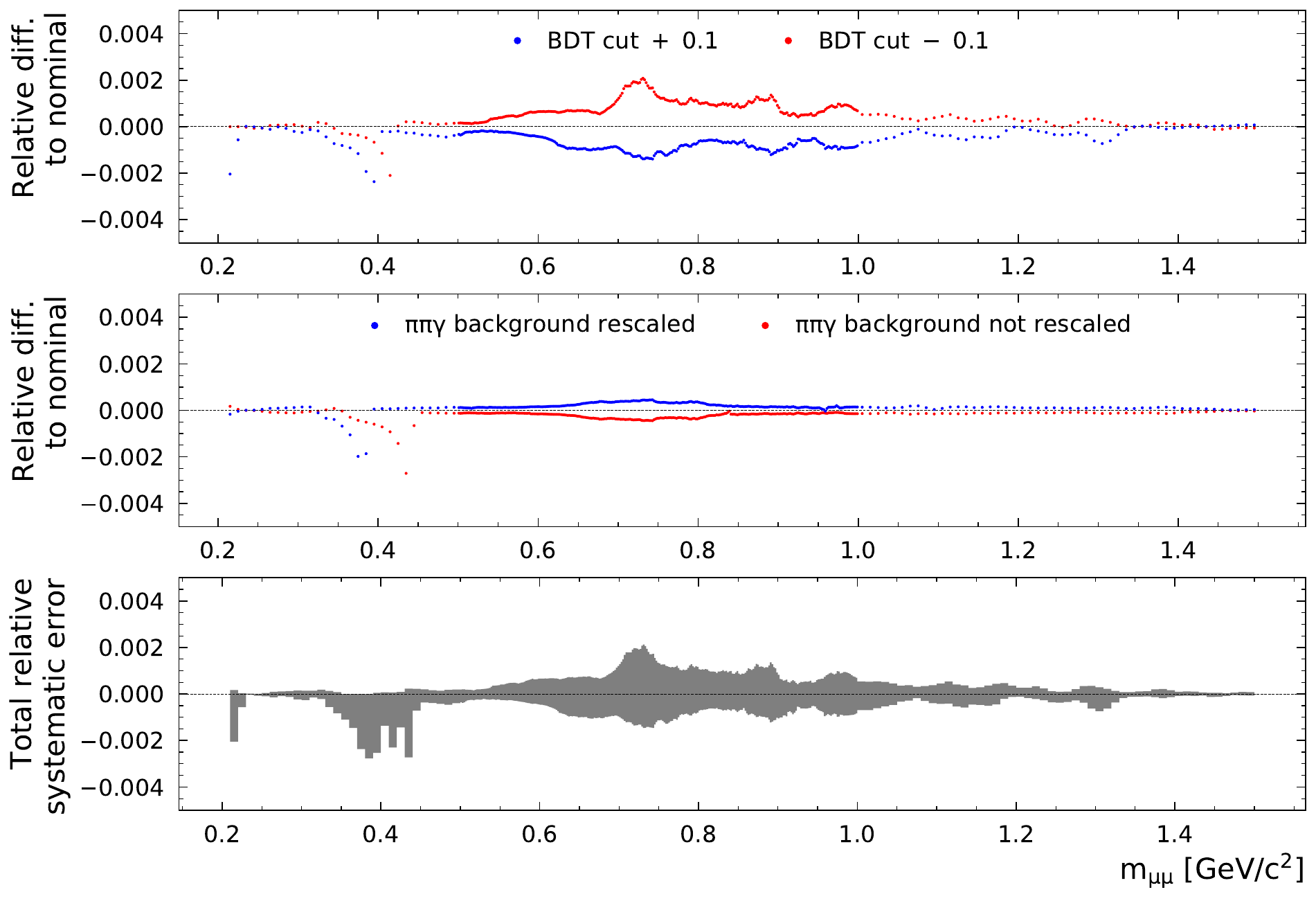}
\caption{\label{fig:relErrorsMass} \small \justifying 
Relative systematic uncertainties due to \eeg\ in the fitted \pipig\ (top three panels) and \mumug\ (bottom three panels) mass spectra. In both the top and bottom panel groups, the upper two plots show the relative difference of the fitted spectra due to each variation of the \eeg\ determination method, while the lower plot shows the total related systematic uncertainties.
}
\end{figure*}

Figure~\ref{fig:relErrorsMass} shows the relative systematic uncertainties in the fitted \pipig\ and \mumug\ mass spectra due to the \eeg\ contribution. They are negligible (below 1 per mil) around the $\rho$ peak in the \pipig\ channel and around $\pm 2$ per mil for \mumug. 

\subsubsection{Impact of selection efficiency corrections on the fitted mass spectra}
\label{subsec:corr_impact}

\begin{figure*}[!htp]
  \centering
  \includegraphics[width=\linewidth]{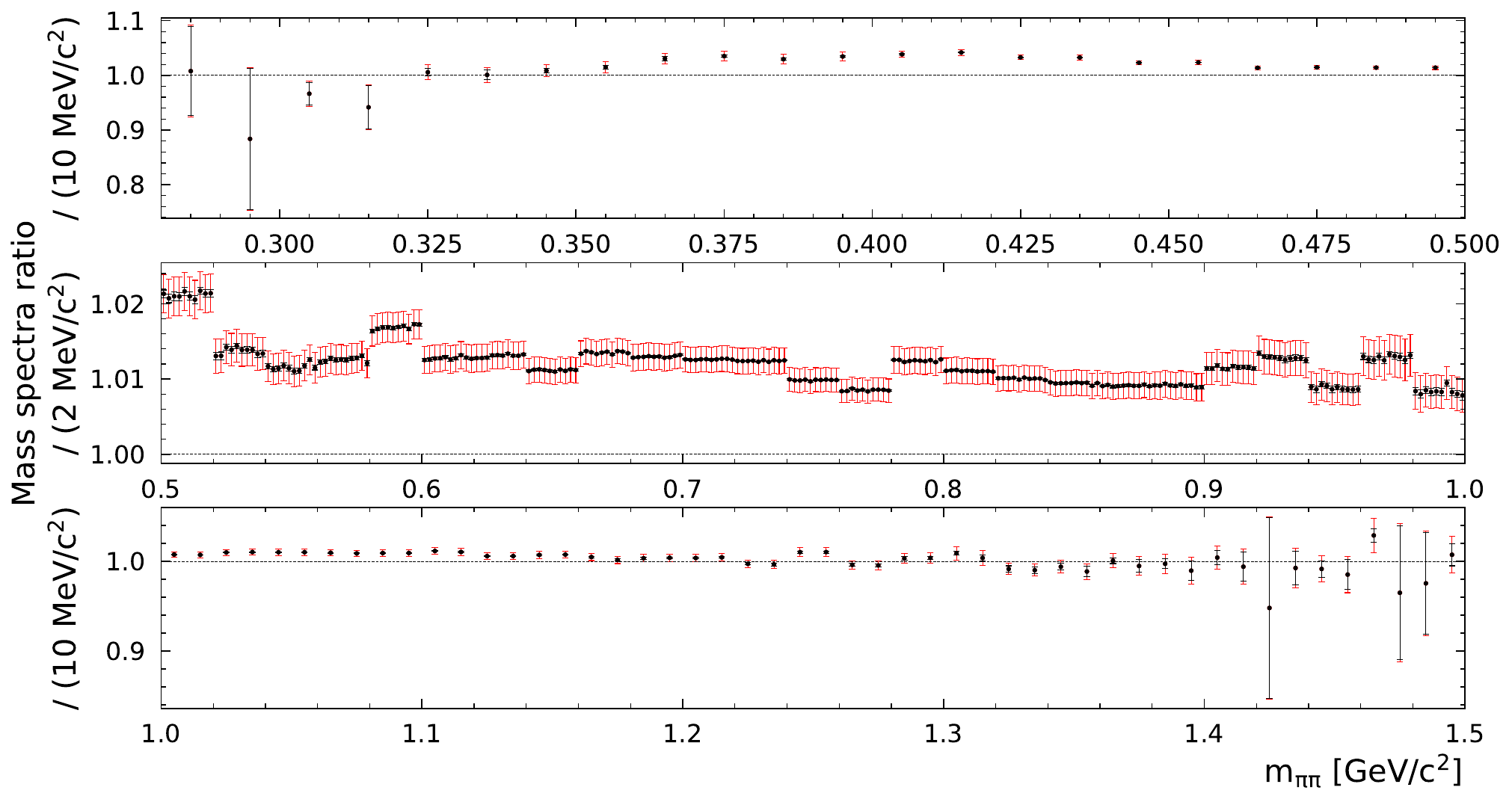}
  \centering
  \includegraphics[width=\linewidth]{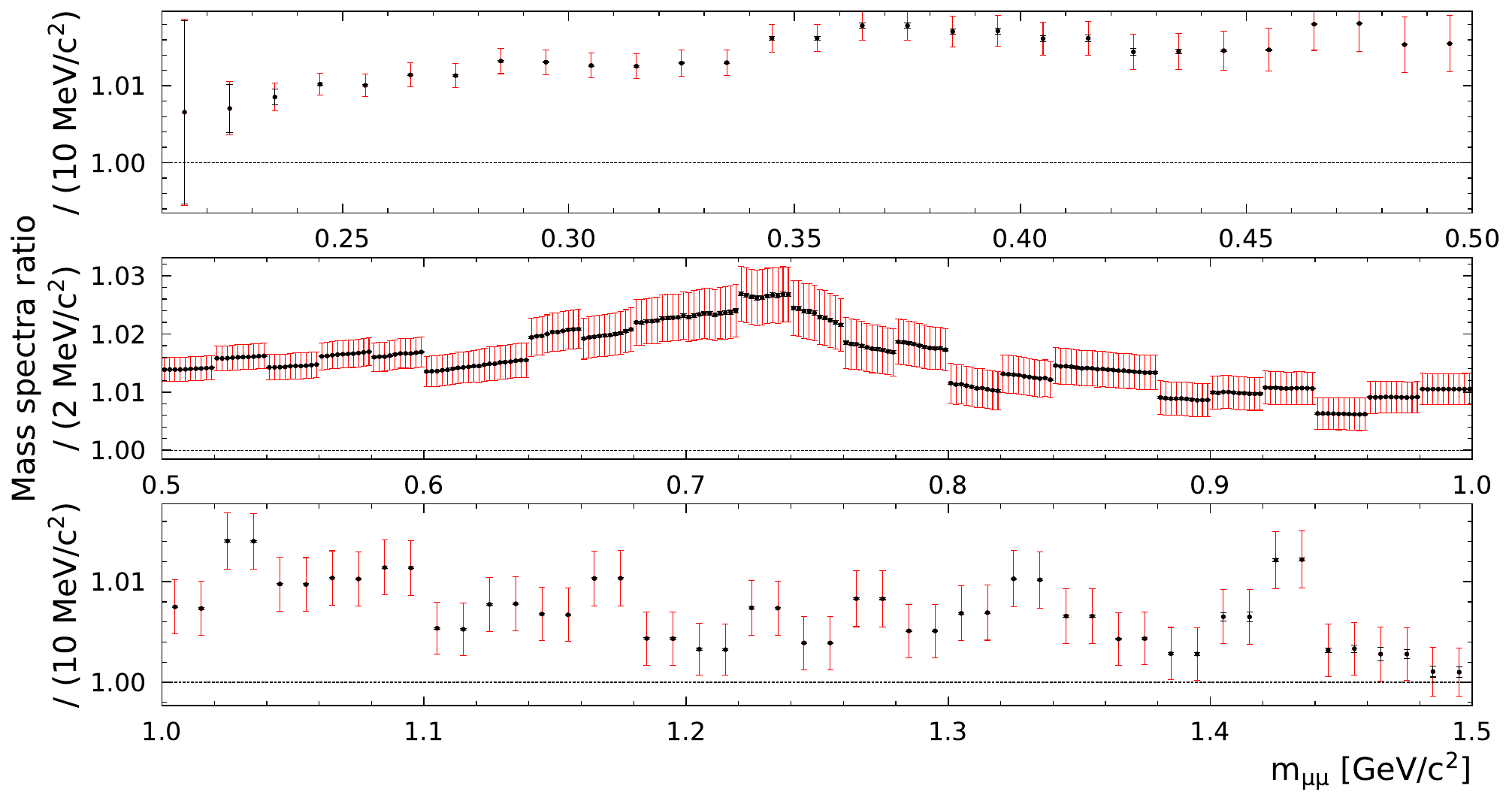}
  \caption{\label{fig:trk_m_ratio} \small \justifying 
Ratio of fitted \pipig\ (top three panels) and \mumug\ (bottom three panels) mass spectra obtained after/before applying the 
tracking efficiency corrections on templates and mass. The first error bars (in black) represent the statistical uncertainties computed with 1000 bootstrap replicas and taking correlations into account, while the second (in red) reflect the contribution of uncertainties in the corrections.}
\end{figure*}

\begin{figure}[tbp]
\centering
    \subfloat{\includegraphics[width=\linewidth]{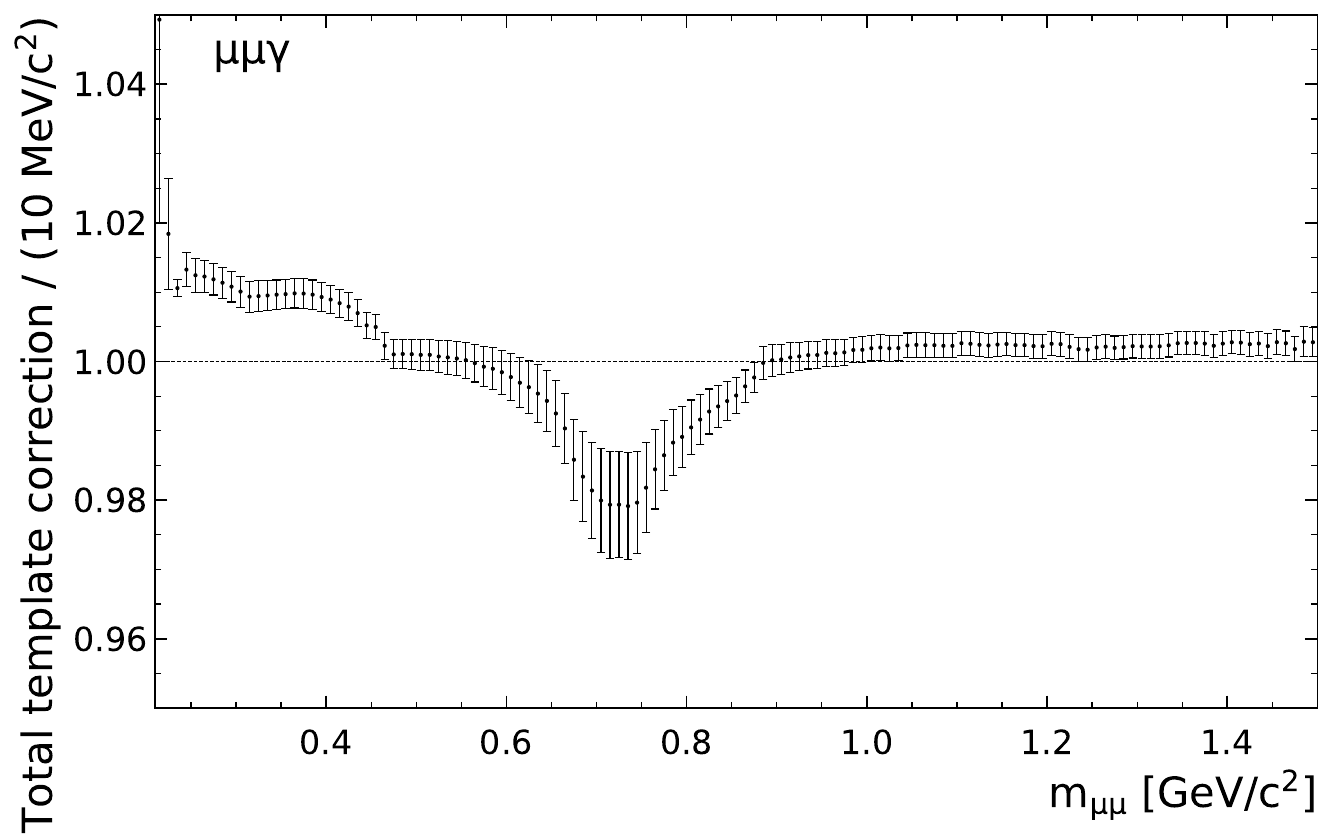}}\\
    \subfloat{\includegraphics[width=\linewidth]{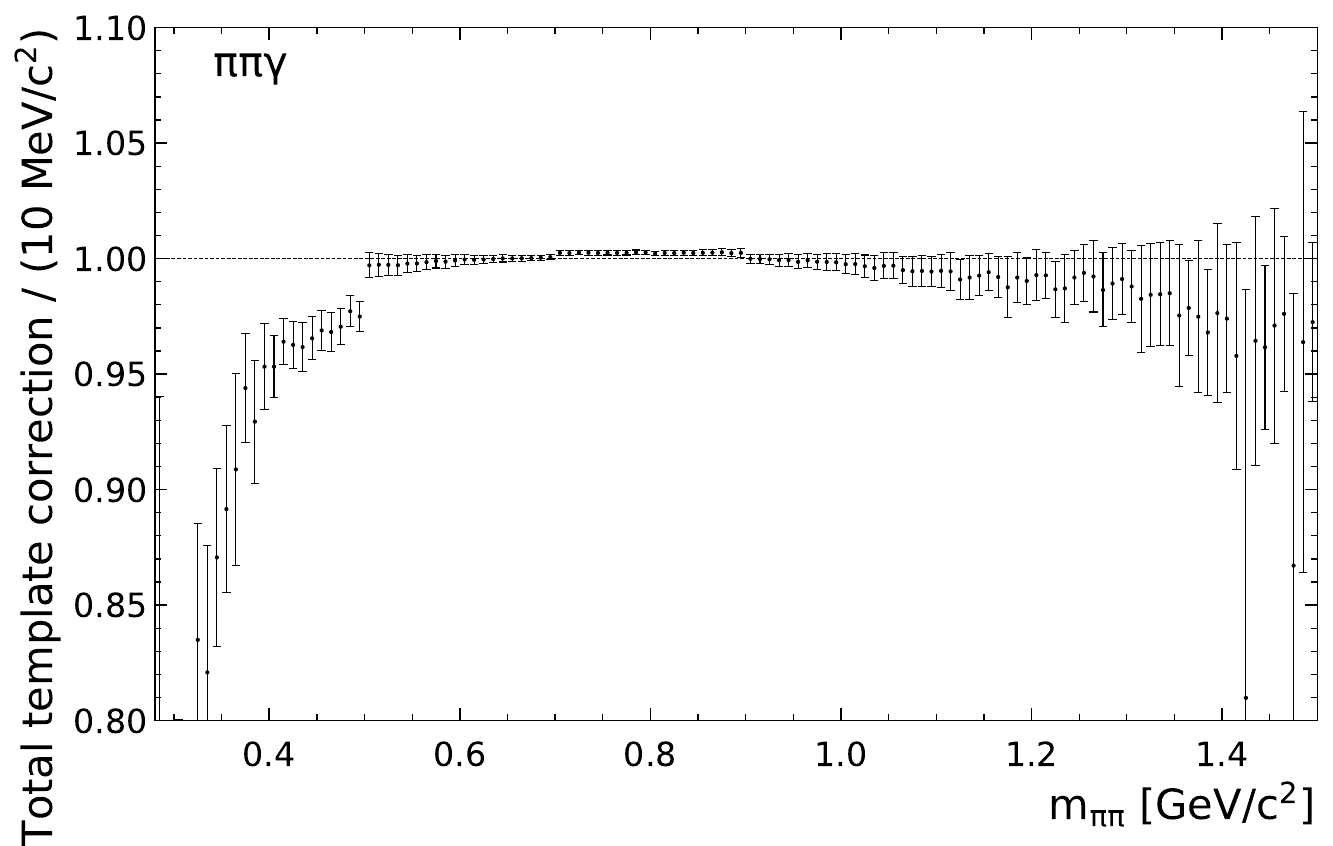}}\\
    \subfloat{\includegraphics[width=\linewidth]{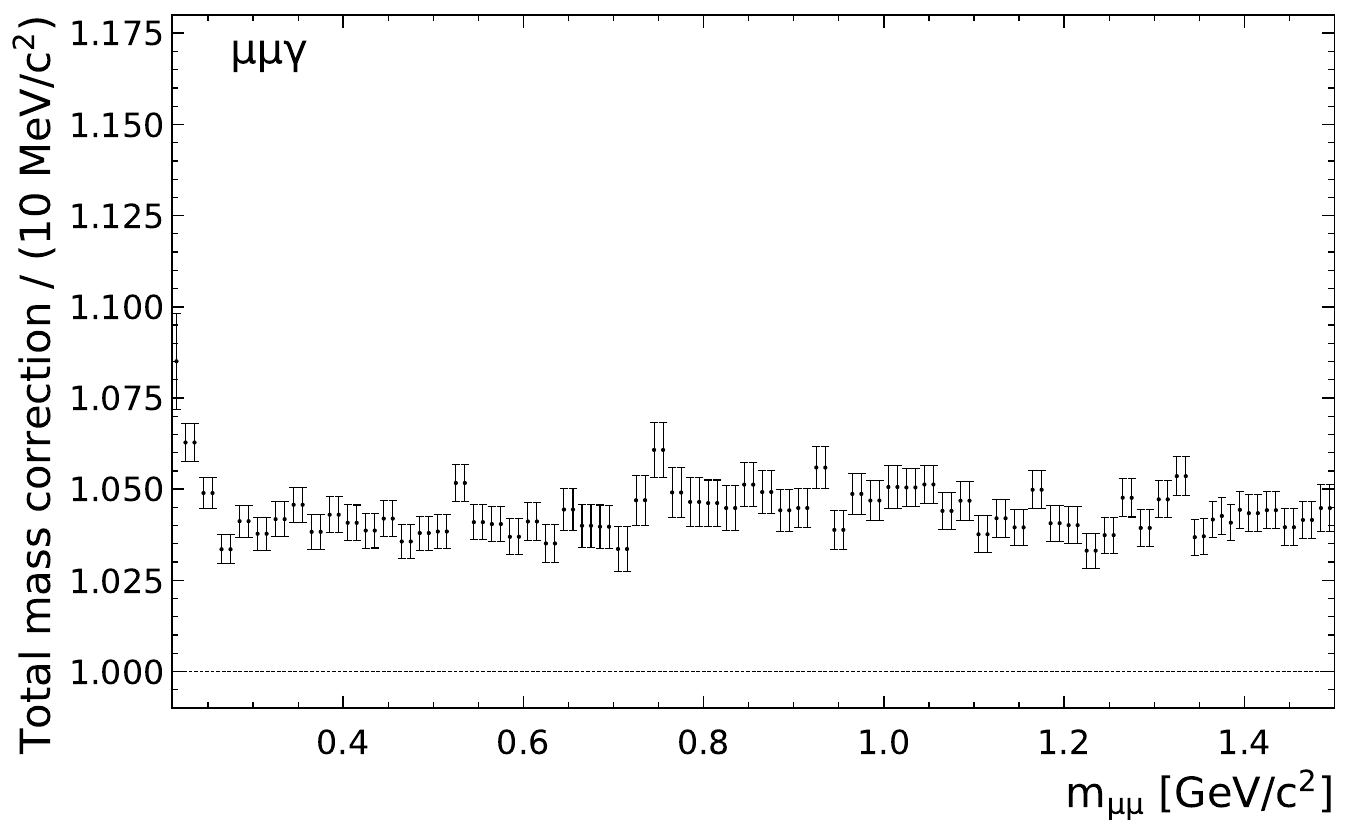}}\\
    \subfloat{\includegraphics[width=\linewidth]{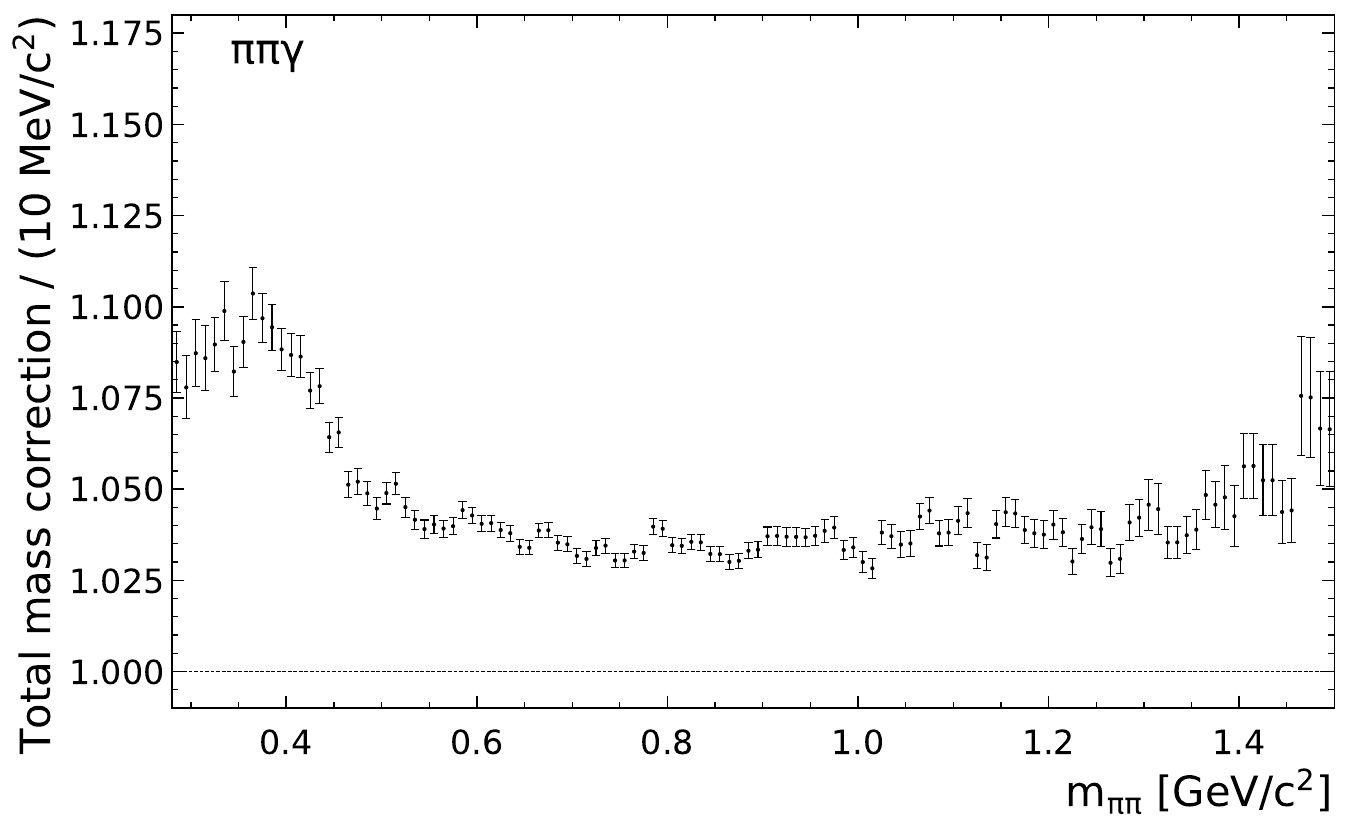}}
\caption{\small \justifying Total impact of data/MC template (top two panels) and mass  corrections (bottom two panels) on the data $\mu\mu$ and $\pi\pi$ mass spectra.}
\label{fig:total_corr}
\end{figure}

The impact of selection efficiency corrections on the pion and muon mass spectra has been studied for the L3 trigger line and BGFilter efficiencies described in Section~\ref{sec:trigger}, the tracking (Section~\ref{sec:tracking}), the $V_{xy}$ selection (Section~\ref{sec:vxy}), the 2d-$\chi^2$ selection, and pion secondary interactions and fake photons (Section~\ref{sec:kin-fit}). An example is presented in Fig.~\ref{fig:trk_m_ratio} for the tracking corrections, which have one of the largest impacts. Overall, the \mumug\ spectrum is the most sensitive to the effect of the corrections at the $\rho$ peak, mainly from the tracking and the 2d-$\chi^2$ selection with opposite effects, and to a smaller extent from the other corrections. Each correction
is well determined experimentally and since they have different signs their overall effect is small. Indeed, the total impact of template and mass corrections on the muon and pion data spectra from all the measured efficiencies present a slow and smooth variation with mass as shown in 
Fig.~\ref{fig:total_corr}. 

\section{Unfolding of $\pi\pi$ and $\mu\mu$ spectra}
\label{sec:unfolding}

The background-subtracted and efficiency-corrected spectra for $\pi\pi(\gamma)\gamma_{\mathrm{ISR}}$ and $\mu\mu(\gamma)\gamma_{\mathrm{ISR}}$ events are obtained as described in the previous section as a function of the invariant mass \mpipi\ and \mmumu, respectively (Fig.~\ref{fig:sig_spectra}). 
Unfolding is then applied to the mass spectra to recover the true underlying distribution of produced events as a function of $\sqrt{s'}$, statistically accounting for resolution effects and additional FSR.

\subsection{Unfolding}

The mass spectra are unfolded to the effective center-of-mass energy after ISR emission, \(\sqrt{s'}\), defined as the center-of-mass energy of the \(XX(\gamma)\) system (\(X=\pi,\mu\)) including possible FSR from the charged tracks. 

The determination of $\sqrt{s’}$ requires distinguishing FSR photons from large-angle ISR photons. This distinction is not directly provided by the \textsc{Phokhara} generator, since ISR--FSR interference is included at NLO. In Ref.~\cite{BaBar:2023xiy}, the separation was investigated using the \textsc{AfkQed} generator as an alternative reference sample, as it does not include large-angle ISR emission at NLO. The study used the distribution of the minimum angle between an additional large-angle photon, $\gamma_{\mathrm{LA}}$, and the charged tracks, $\theta_{\min}(\mathrm{track},\gamma_{\mathrm{LA}})$. This distribution was fitted with FSR templates obtained from \textsc{AfkQed}, together with large-angle ISR templates derived from the difference between the \textsc{Phokhara} and \textsc{AfkQed} predictions. Following the outcome of that study, the present analysis adopts the criterion $\theta_{\min}(\mathrm{track},\gamma_{\mathrm{LA}})<20^\circ$ to classify photons as FSR.

In simulation, \(\sqrt{s'}\) is then computed using the true charged-track information by adding the four-momentum of any additional photon within a cone of \(\theta_{\min}^{\mathrm{true}}<20^\circ\) around a charged track. Then \(\sqrt{s'}\) and the invariant mass \(m_{XX}\) are used to build a two-dimensional response matrix for the unfolding.

An iterative unfolding method, as described in Refs.~\cite{Malaescu:2009dm,Malaescu:2011yg}, is implemented. The response matrix $A_{ij}$, constructed from simulated events, reflects the number of events generated in true ($\sqrt{s^{\prime}}$) bin $j$ and reconstructed in ($m_{XX}$) bin $i$, thus accounting for event migration across bins. The folding probability, i.e., the probability for an event generated in bin $j$ to be reconstructed in bin $i$, is given by $P_{ij} = A_{ij} / \sum_k A_{kj}$. Conversely, the unfolding probability representing the likelihood that an event reconstructed in bin $i$ originated from true bin $j$ is defined as $P^{'}_{ij} = A_{ij} / \sum_k A_{ik}$.

Unfolding relies on the assumption that the MC simulation provides an accurate description of the true spectrum and of the resolution smearing of the detector response. If significant discrepancies between the data and the reconstructed MC spectrum remain after an initial unfolding, an iterative procedure is applied. In each iteration, the transfer matrix is refined by reweighting the true MC spectrum to better match the unfolded data, while the folding probabilities are held fixed. The updated spectrum is then folded back and compared to the data, with the differences attributed to deficiencies in the prior MC model. To mitigate instabilities caused by statistical fluctuations, a (weak)~regularization is applied to limit the unfolding of the corresponding data-MC deviations. This ensures convergence toward a stable solution without introducing artificial bias.
The choices of the number of iterations and of the binning of the spectrum provide another implicit regularization of the unfolding.

\subsection{Procedure and data-driven test}
\label{unfolding}

\begin{figure}
    \centering
    \includegraphics[width=\linewidth]{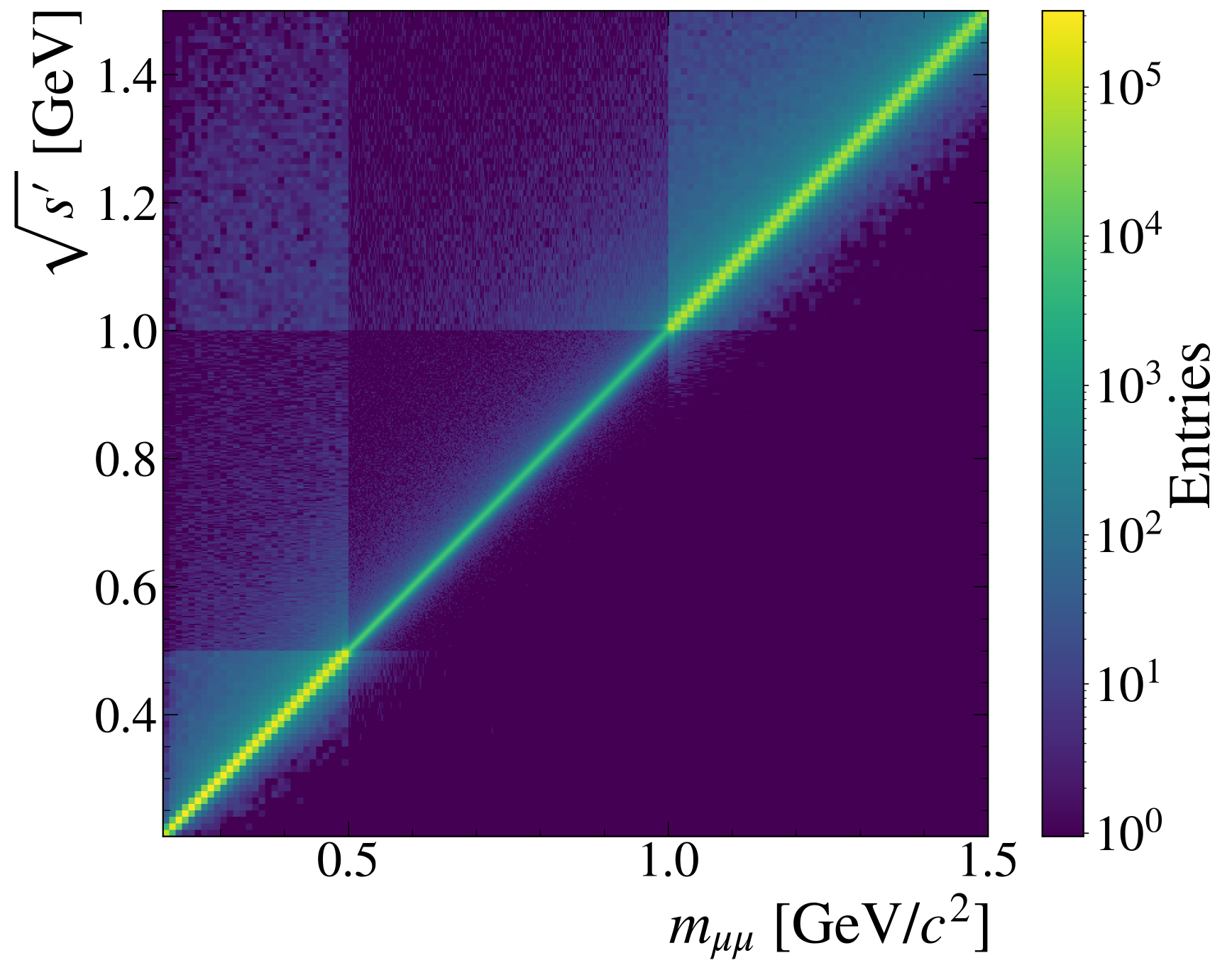}
    \includegraphics[width=\linewidth]{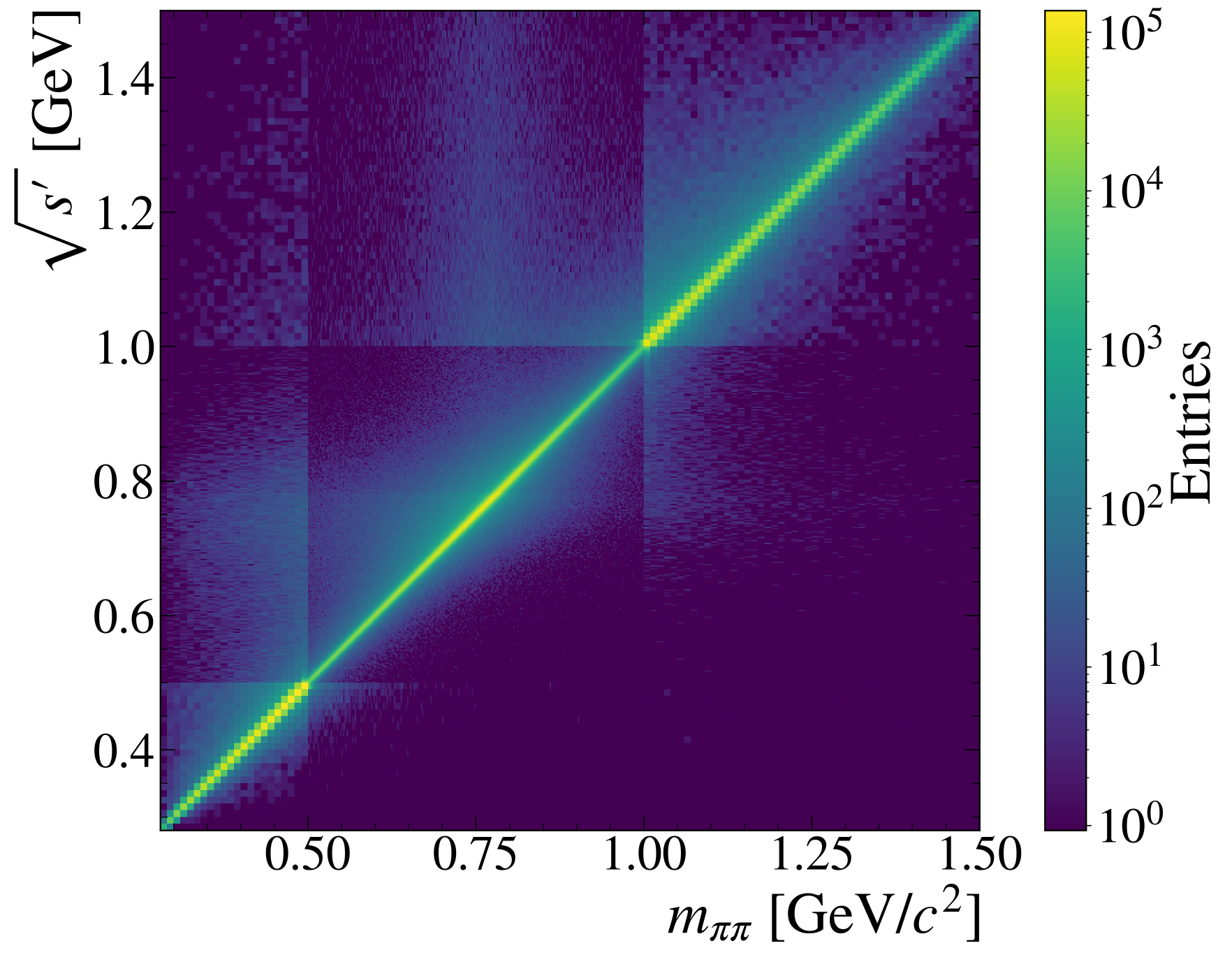}    
    \caption{\small \justifying Response matrices $A_{ij}$ from simulation. The $z$ axis corresponds to the number of events generated with a true $\sqrt{s^{\prime}}$ in a bin $j$ and a reconstructed mass $m_{XX}$ in a bin $i$ with $\theta_\mathrm{min}^\mathrm{true}<20^\circ$ for (top) $\mu\mu\gamma$ and (bottom) $\pi\pi\gamma$.}
    \label{fig:unfolding_transfer_matrices}
\end{figure}

The response matrices for $\mu\mu(\gamma)\gamma_{\mathrm{ISR}}$ and $\pi\pi(\gamma)\gamma_{\mathrm{ISR}}$ are constructed using the binning and mass ranges defined in Section~\ref{subsubsec:fit_strat} (10\mevcc over the tails and  2\mevcc for the central $\rho$ region), resulting in matrix dimensions of 329$\times$329 and 322$\times$322, respectively. Resolution effects are minimal in the $\mu\mu$ sample, where bin migration is mainly due to FSR. The initial matrices are shown in Fig.~\ref{fig:unfolding_transfer_matrices}.

Once the response matrix is constructed, the unfolding procedure is performed in three steps to account separately for migration effects and unmatched detector- and particle-~(i.e., truth-)~level events.

First, the data distribution, $D$, is corrected for events not represented in the response matrix via the correction factor $C^{\mathrm{det}}_i$ as
\begin{equation}
D^{\mathrm{corr}}_i=C^{\mathrm{det}}_i D_i, \hspace{0.5cm} C^{\mathrm{det}}_i=
\frac{N^{\mathrm{matched}}_{i,\mathrm{det}}}
{N^{\mathrm{total}}_{i,\mathrm{det}}},
\end{equation}
where $N^{\mathrm{matched}}_{i,\mathrm{det}}=\sum_j A_{ij}$ is the detector-level projection of the matrix and $N^{\mathrm{total}}_{i,\mathrm{det}}$ is the total detector-level distribution in simulation.

For the $\mu\mu(\gamma)$ and $\pi\pi(\gamma)$ spectra these efficiency corrections are close to unity over most of the mass range, showing that the fraction of detector-level events not represented
in the response matrix is small. In the $\mu\mu(\gamma)$ channel the deviation from unity is below about $0.2\%$ for $m_{\mu\mu}<1\gevcc$, including the $\rho$-resonance region, and increases at high mass, reaching approximately $2.5\%$ above $1.4\gevcc$. In the $\pi\pi(\gamma)$ channel the deviation from unity 
is smaller in the central mass region, remaining within about $0.1\%$ 
below $1\gevcc$, and only becomes sizable in the last mass bins, where it reaches about $2.5\%$.

The corrected distribution is then unfolded using the response matrix, $\mathrm{UR}_j=\mathrm{Unfold}(D^{\mathrm{corr}}_i,A_{ij})$. Finally, the unfolded spectrum, $\mathrm{UR}$, is corrected for particle-level events failing the reconstruction or matching requirements via the factor, $C^{\mathrm{part}}_j$ as
\begin{equation}
\mathrm{UR}^{\mathrm{final}}_j=
C^{\mathrm{part}}_j \,\mathrm{UR}_j,
\hspace{0.5cm}
C^{\mathrm{part}}_j=
\frac{N^{\mathrm{total}}_{j,\mathrm{part}}}
{N^{\mathrm{matched}}_{j,\mathrm{part}}},
\end{equation}
where $N^{\mathrm{matched}}_{j,\mathrm{part}}=\sum_i A_{ij}$ is the particle-level projection of the matrix and $N^{\mathrm{total}}_{j,\mathrm{part}}$ is the total particle-level distribution in simulation.

The corresponding particle-level efficiency corrections for both the $\mu\mu(\gamma)$ and $\pi\pi(\gamma)$ channels, $C^{\mathrm{part}}$, remains very close to unity over most of the $\sqrt{s'}$ range, with deviations at the per mil level below approximately $1.3\gev$ and negligible corrections in the $\rho$-resonance region. The correction increases only near the upper
edge of the spectrum. In the $\mu\mu(\gamma)$ channel it reaches about $1\%$ around $\sqrt{s'}\simeq 1.47\gev$ and approximately $3\%$ in the last bin. The $\pi\pi(\gamma)$ channel shows a similar
trend, with values around $1.3\gev$, reaching the percent level in the last few bins and about $3\%$ at the highest energies.

To evaluate potential biases from the unfolding procedure under realistic, data-driven conditions, a closure test is performed in both the $\mu\mu$ and $\pi\pi$ channels. The test checks whether the unfolding procedure recovers a known input truth distribution from pseudo-data obtained after applying the detector response, with agreement within uncertainties indicating the absence of significant bias. The input spectra are modified (``pseudo-true'') distributions designed to mimic data-MC discrepancies, obtained by reweighting the MC truth to reproduce the differences observed between data and reconstructed MC.

This pseudo-truth is convoluted with the response matrix to produce pseudo-data\footnote{The pseudo-data are validated by a direct comparison with the actual data.}, which is then unfolded. Comparing the unfolded result to the input pseudo-truth allows direct assessment of the unfolding method's robustness, isolating effects from regularization, prior dependence, and bin migrations.

\begin{figure}
    \centering
    \includegraphics[width=\linewidth]{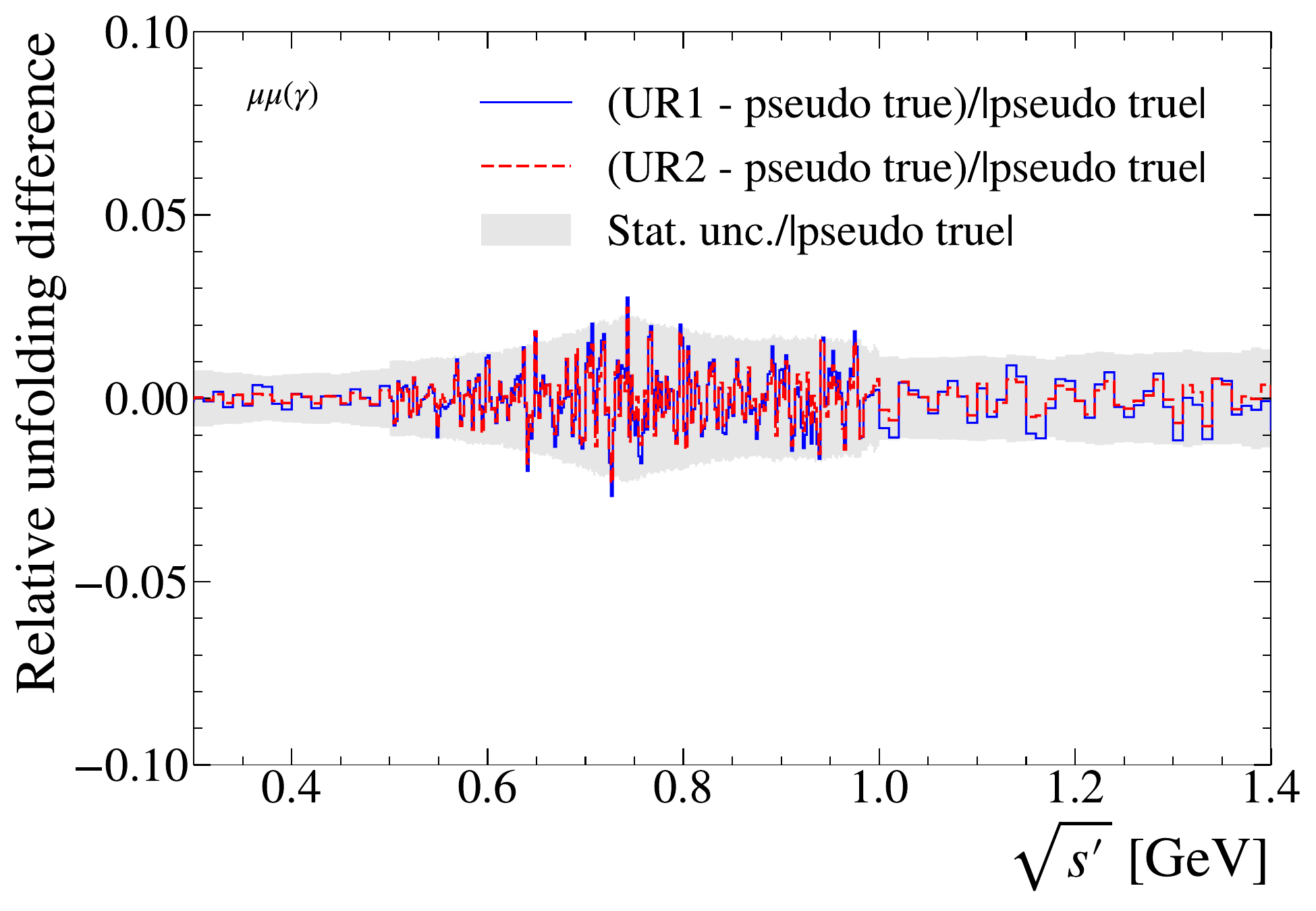}
    \includegraphics[width=\linewidth]{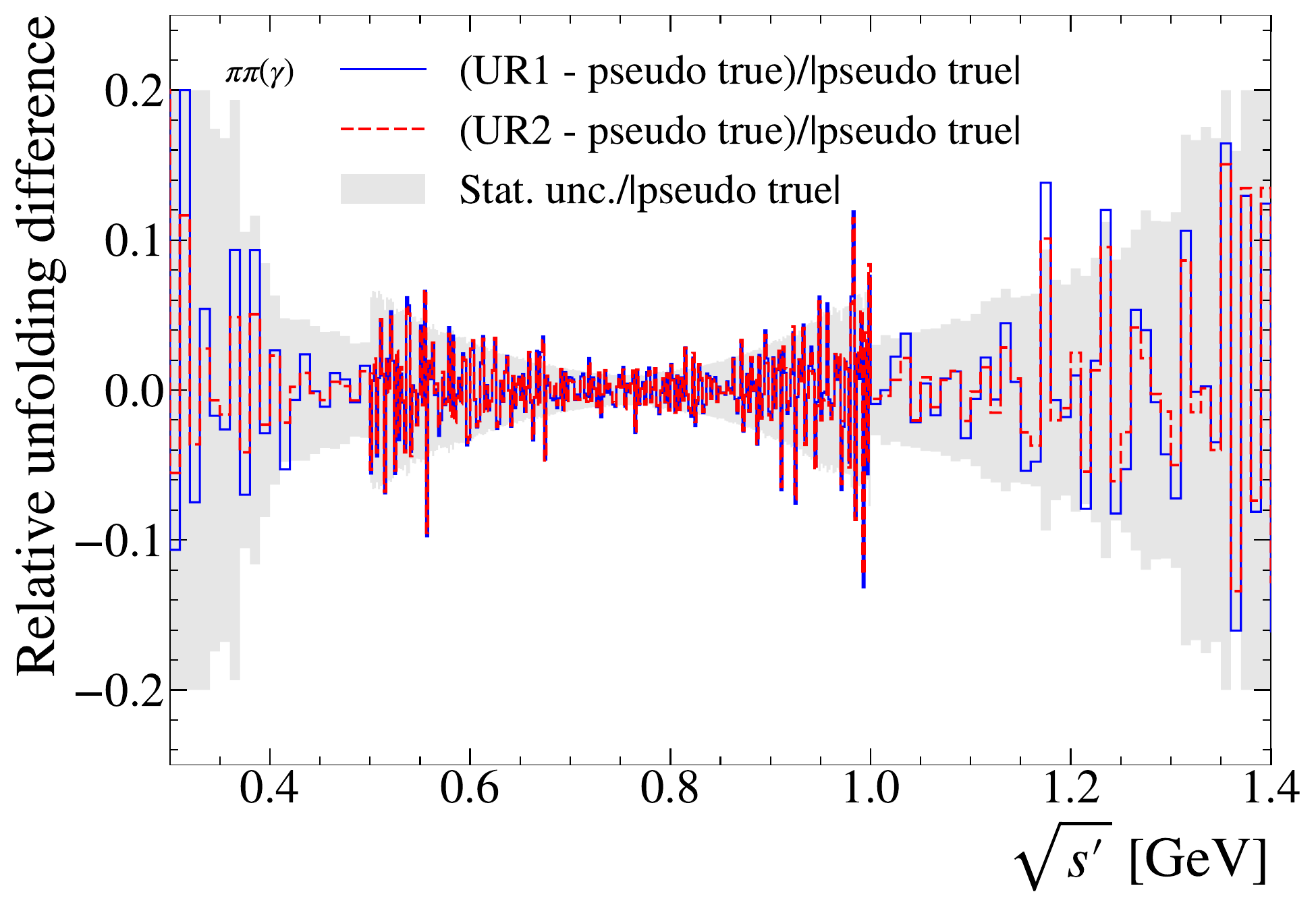} 
    \caption{\small \justifying Data driven closure test: relative residual of the unfolding with respect to the pseudo-truth for (top) $\mu\mu$ and (bottom) $\pi\pi$.}
    \label{fig:data_driven}
\end{figure}

In the $\mu\mu$ channel (Fig.~\ref{fig:data_driven} top), residuals remain within uncertainties, with fast convergence between the first and second iterations, confirming minimal systematic unfolding bias. 
For the $\pi\pi$ channel (Fig.~\ref{fig:data_driven} bottom), where larger structures are present in the spectrum and in the initial data-reco MC differences, the method still recovers the pseudo-truth within statistical uncertainties. The agreement across iterations further validates the stability of the procedure. 
Overall, these closure tests demonstrate that the unfolding introduces negligible methodological bias and reliably recovers distorted spectra under realistic conditions.

\begin{figure*} 
  \begin{minipage}[t!]{0.475\textwidth}
    \centering
    \includegraphics[width=\linewidth]{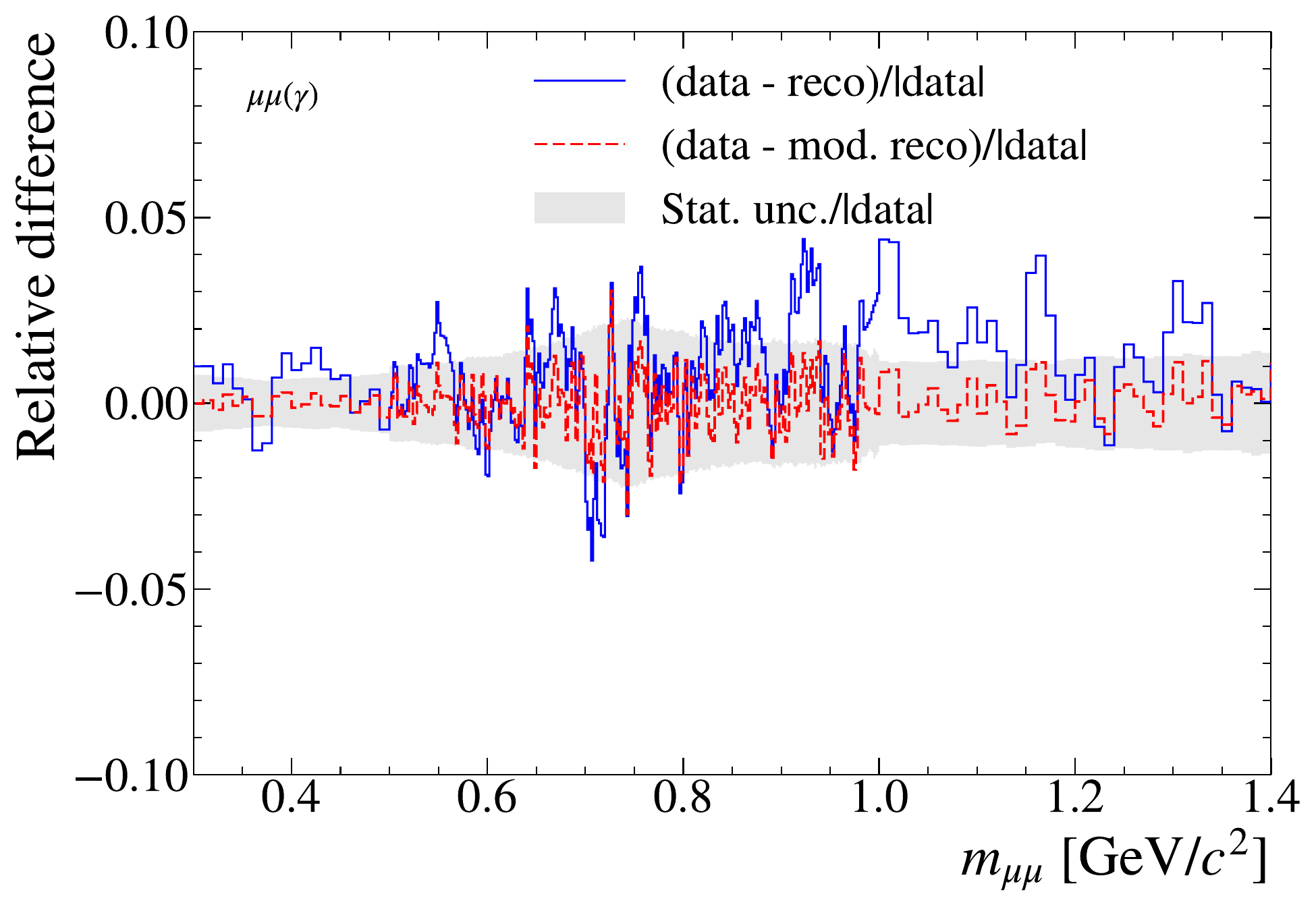}
    \caption{\small \justifying Relative differences between the $\mu\mu$ mass distributions of data and reconstructed MC (reco) before unfolding and the modified MC (mod.\ reco) after one iteration.}
    \label{fig:unfolding_data_reco_mu}
  \end{minipage}\hfill
  \begin{minipage}[t!]{0.475\textwidth}
     \centering
    \includegraphics[width=\linewidth]{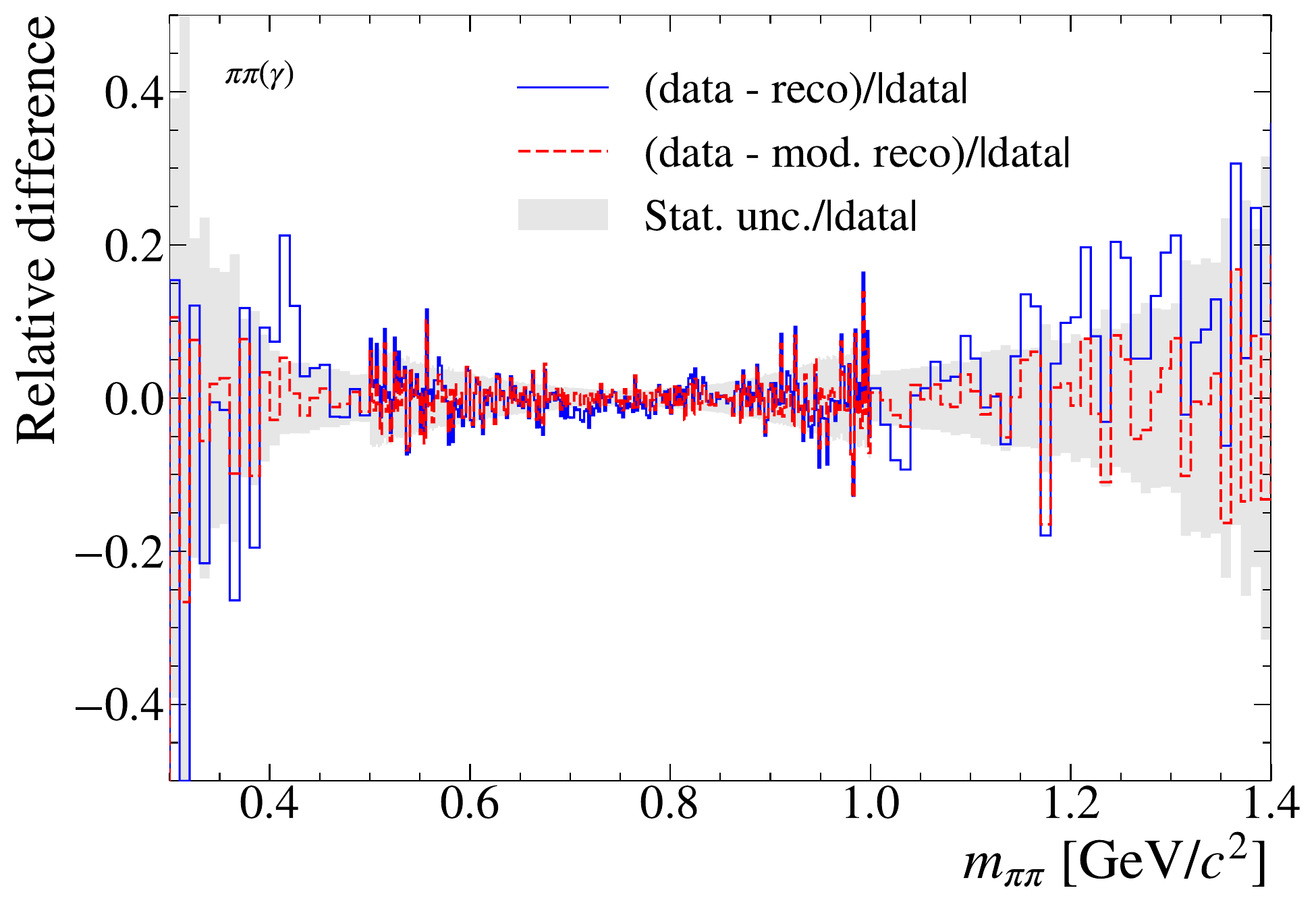}
    \caption{\small \justifying Relative differences between the $\pi\pi$ mass distributions of data and reconstructed MC (reco) before unfolding and the modified MC (mod.\ reco) after one iteration.}
    \label{fig:unfolding_data_reco_pi}
  \end{minipage}
\end{figure*}

Figures~\ref{fig:unfolding_data_reco_mu} and~\ref{fig:unfolding_data_reco_pi} show the relative variation between data, reconstructed MC (reco), and the reco after iterations (mod.\ reco) for $\mu\mu$ and $\pi\pi$, respectively. Small deviations within statistical uncertainties are visible in the $\rho$ lineshape, particularly in the tails, the peak region, and the $\rho$-$\omega$ interference region. These discrepancies are attributed to differences in the generated mass distribution in the MC, while resolution effects between data and simulation are handled separately. After the first iteration, agreement improves considerably: for muons, residuals fall well within uncertainties across the spectrum, and for pions, differences are similarly reduced including in the $\rho$-$\omega$ region.

\begin{figure}
    \centering
    \includegraphics[width=\linewidth]{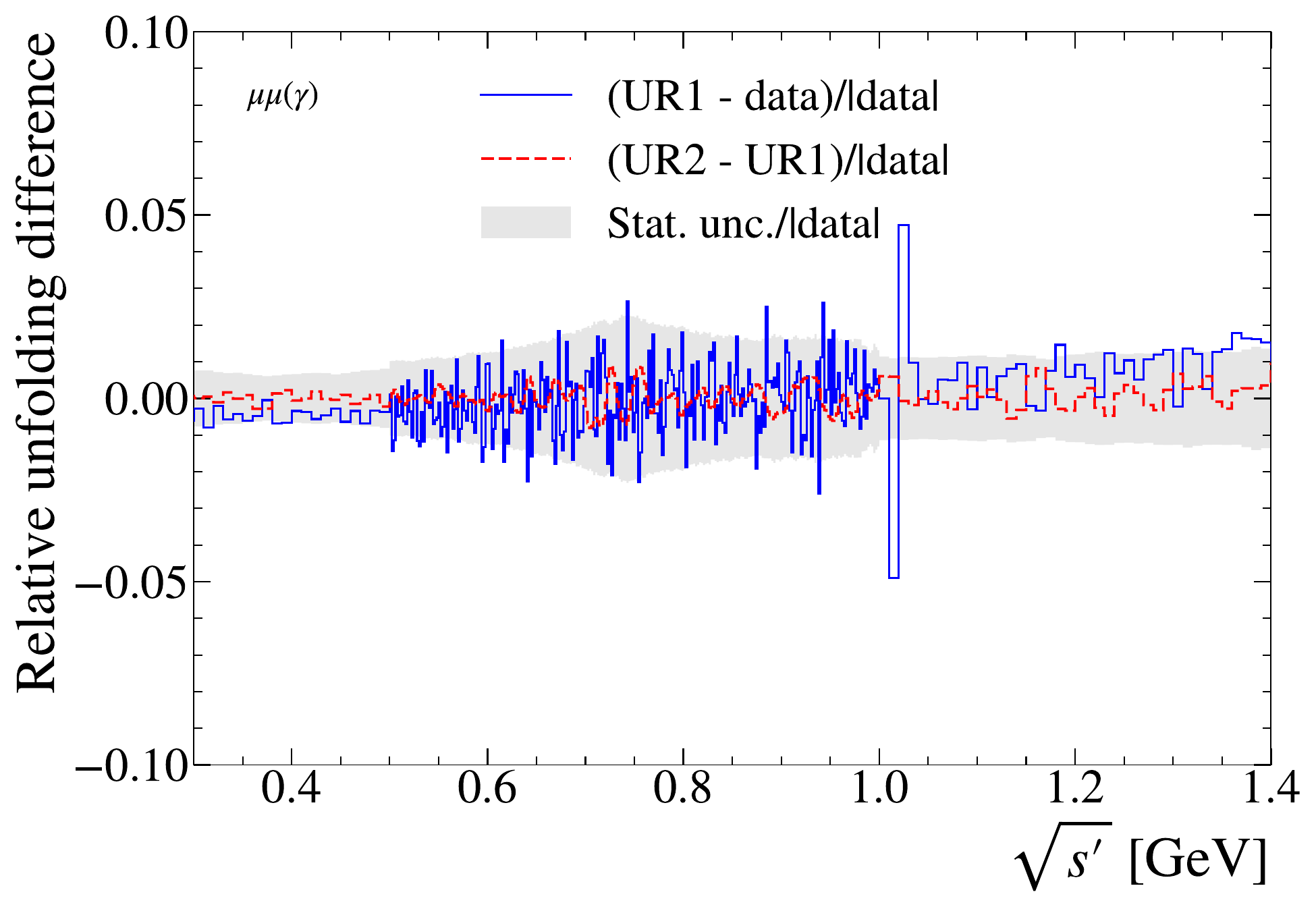}
    \includegraphics[width=\linewidth]{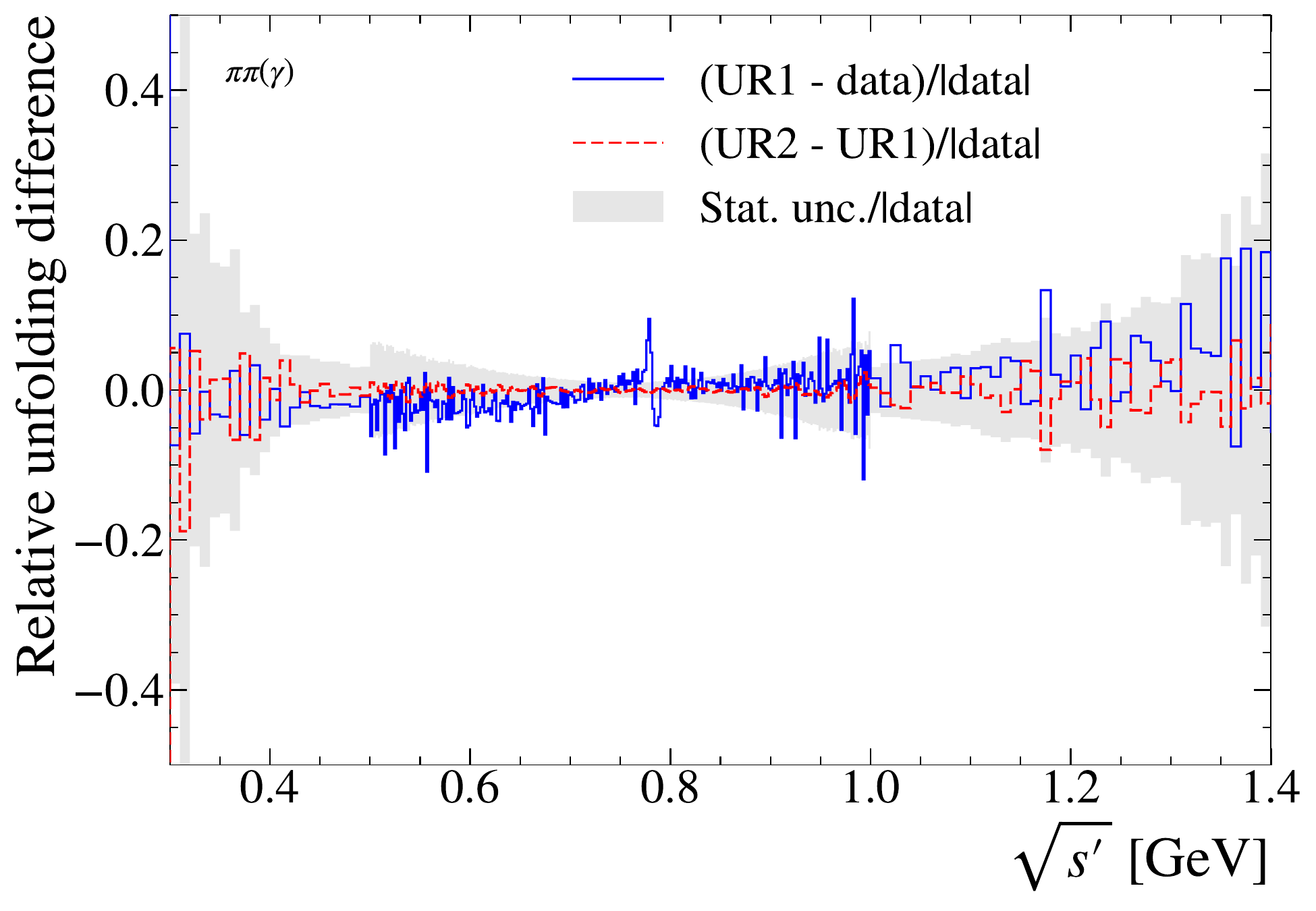}
    \caption{\small \justifying Relative differences between the first unfolding result (UR1) and the iterative unfolding result (UR2) with respect to the (top) $\mu\mu$ and (bottom) $\pi\pi$ data.}
   \label{fig:unfolding_UR_data_mupi}
\end{figure}

The residual data-reco MC systematic differences have a small effect on the result of the unfolding, as shown in Fig.~\ref{fig:unfolding_UR_data_mupi} (top) and (bottom) for $\mu\mu$ and $\pi\pi$, respectively. 
In the muon case, the first unfolding iteration result is very close to the initial data within the statistical uncertainties. Further iterations have a negligible effect, and the unfolded spectrum remains stable, indicating a minimal systematic impact from mismodeling the generated mass distribution. 
In the case of the pions, the first unfolding step induces a correction typically smaller than the data statistical uncertainties, except in the $\rho$-$\omega$ interference region, with no improvement from additional iterations. The unfolding result after the first iteration (UR2) is taken as the nominal result.

\subsection{Uncertainties in the unfolded spectra}
\label{sec:unfold-error}

\subsubsection{Statistical uncertainty}
\label{sec:stat-error}

Statistical uncertainties and their bin-to-bin correlations from the unfolding are estimated using pseudo-experiments. Thousands of toy datasets~(bootstrap replicas) are generated by fluctuating the data and response matrix coherently, as described in Section~\ref{subsubsec:fit_strat}, and each is unfolded independently using the iterative method.

\begin{figure}
    \centering
    \includegraphics[width=\linewidth]{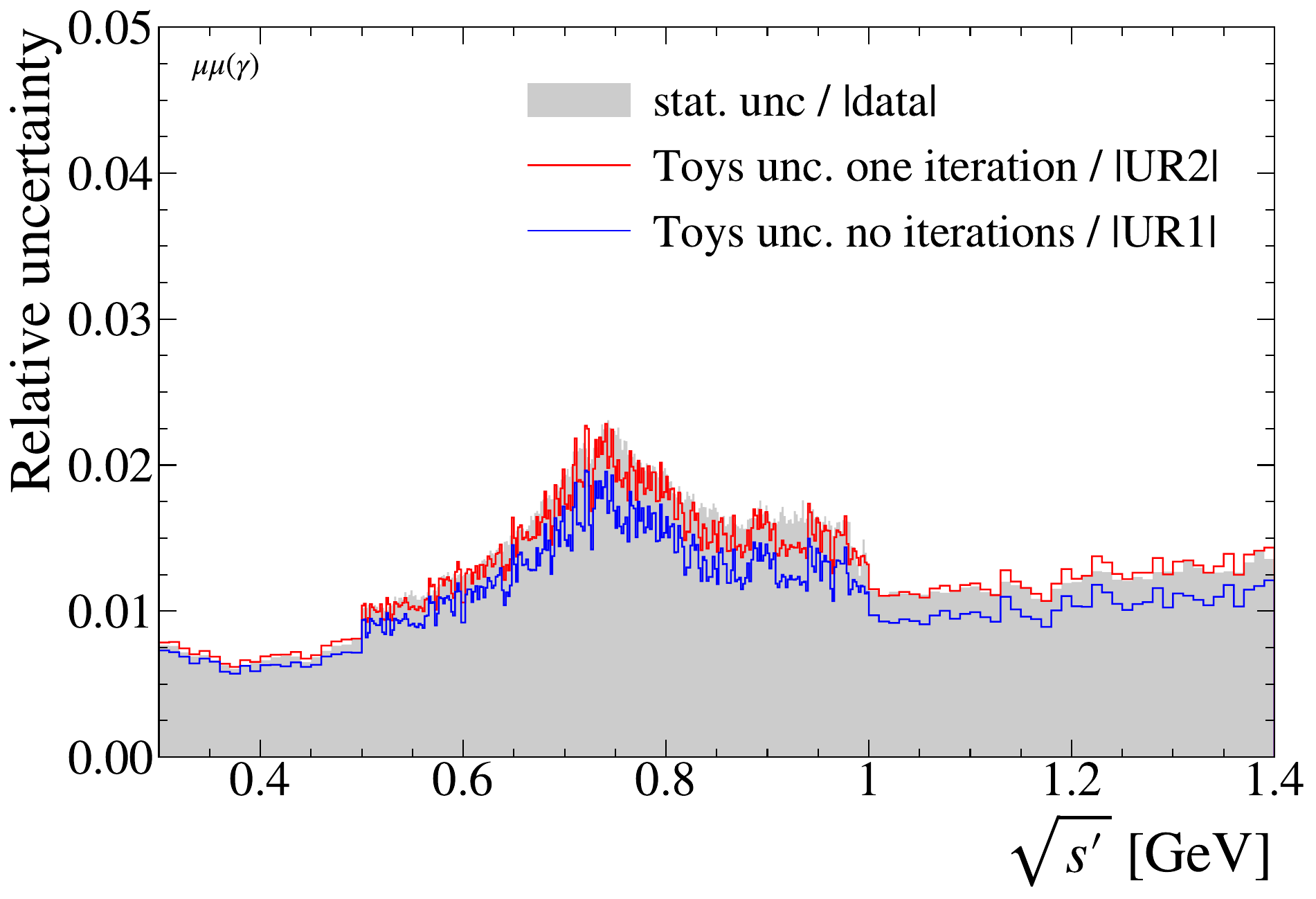}
    \includegraphics[width=\linewidth]{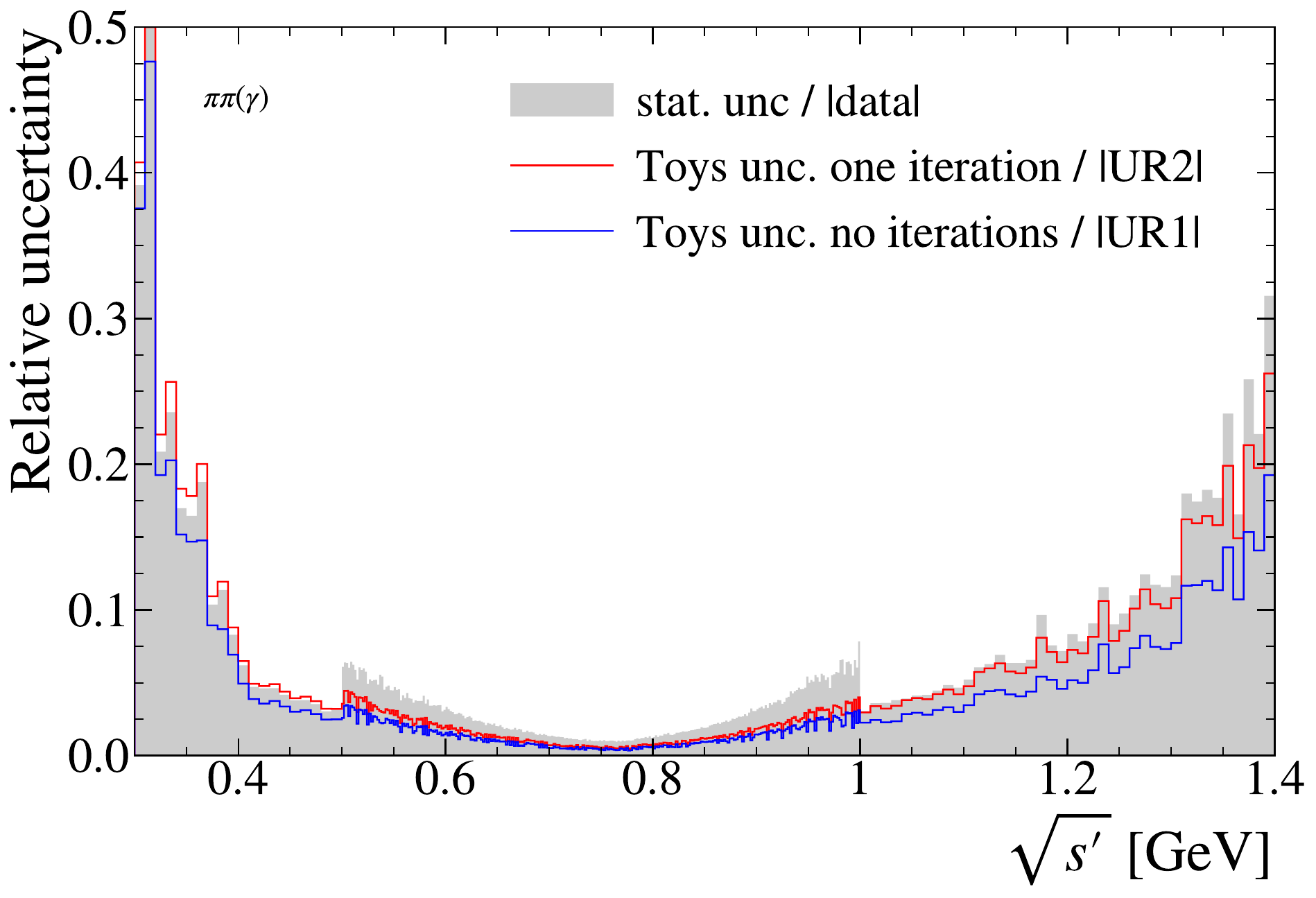}
    \caption{\small \justifying
    Total relative statistical uncertainty estimated from the bootstrap replicas for data and for the unfolding results after each iteration.}
    \label{fig:unfolding_toys}
\end{figure}

The statistical covariance matrix is computed from the ensemble of results, with statistical uncertainties per bin derived from the diagonal elements. 
Relative diagonal uncertainties with respect to the unfolding result and their comparison with the relative uncertainties of the input data are presented in Fig.~\ref{fig:unfolding_toys}. 

\subsubsection{Systematic uncertainties from imperfect response matrix}

The systematic uncertainty from the imperfect modeling of the response matrix is studied by varying the angular selection on $\theta_{\text{min}}^{\text{true}}$ used in defining $\sqrt{s'}$. 
Response matrices are generated for selections between $10^\circ$ and $30^\circ$ (in $5^\circ$ steps), and data are unfolded accordingly. Based on the studies described in Ref.~\cite{BaBar:2023xiy}, selections above \(20^\circ\) are excluded to avoid contamination from events with multiple ISR photons, while a cut of \(10^\circ\) is found to reject an excessively large fraction of FSR photons.
The final uncertainty is evaluated by comparing the unfolded results with selections at $15^\circ$ and $20^\circ$, yielding a conservative estimate.

To assess the systematic uncertainty associated with possible mismodeling of the detector response and event composition, an alternative response matrix for $\mu$ and $\pi$ is constructed using a reweighting procedure based on the two-dimensional 2d-$\chi^2$ selection plane. This plane is divided into six regions, each corresponding to different signal and background compositions. For each region, the fractions of $\mu^+\mu^-$ and $\pi^+\pi^-$ events are determined separately in data and simulation, and the ratio of these fractions is used to define a reweighting factor. A response matrix is built for each region and reweighted according to the corresponding factor. The final reweighted response matrix is then obtained as the sum of the six region-specific reweighted matrices. The unfolding is performed both with the nominal (unreweighted) and the reweighted response matrices. The difference between the resulting unfolded spectra is taken as a systematic uncertainty associated with the modeling of the response matrix and the event composition in the 2d-$\chi^2$ plane.

To cover potential mismodeling of event migration effects above the upper bound of the response matrix at $1.4\gev$, an additional systematic uncertainty associated with the unfolding of the reconstructed $m_{\mu\mu}$ spectrum to the true $\sqrt{s'}$ distribution is assigned ``Unfolding muons HM (High Mass)''.

An additional systematic uncertainty is assigned in the $\phi$ resonance region to account for imperfections in the modeling of the resonance line shape for the muon channel. This uncertainty covers possible discrepancies between data and simulation affecting the description of the $\phi$ peak, ``$\phi$-resonance unfolding (VP)''.

\subsubsection{Systematic uncertainty propagation}
\label{Sec:syst-uncertainties}
Systematic uncertainties in the unfolded spectrum are evaluated through coherent bin-wise variations of the input data. 
For each source of uncertainty, the measured spectrum is shifted 
by its corresponding $1\sigma$ bin uncertainty, and each variation is unfolded using the \textit{nominal} response matrix.

The resulting unfolded spectra are compared to the nominal result, and the 
deviation per bin is assigned as the systematic uncertainty for that source. This procedure is repeated independently for all sources, after which each is smoothed using a Gaussian kernel applied to the binned distribution. 
For long-range uncertainties~(typically impacting the full mass range), the smoothed bin content is computed as a weighted average of neighboring bins within a sliding window of $60\mev$, reflecting the expected scale over which the systematic uncertainties can vary. 
A narrower window of 5\mev is employed for uncertainty sources impacting a restricted mass range.
This choice provides a balance between suppressing statistical fluctuations and preserving the shape of the distribution.

The total systematic uncertainty is obtained by adding all contributions in quadrature. The final uncertainty per bin includes both statistical and systematic components combined in quadrature.

The systematics sources described in Sections~\ref{subsec:eeg_syst} and~\ref{subsec:corr_impact}, are propagated accordingly for both templates and mass corrections. The post-unfolding systematic components are calculated with respect to the nominal unfolding result of the first iteration (UR2). 

\begin{figure}
    \centering
    \includegraphics[width=\linewidth]{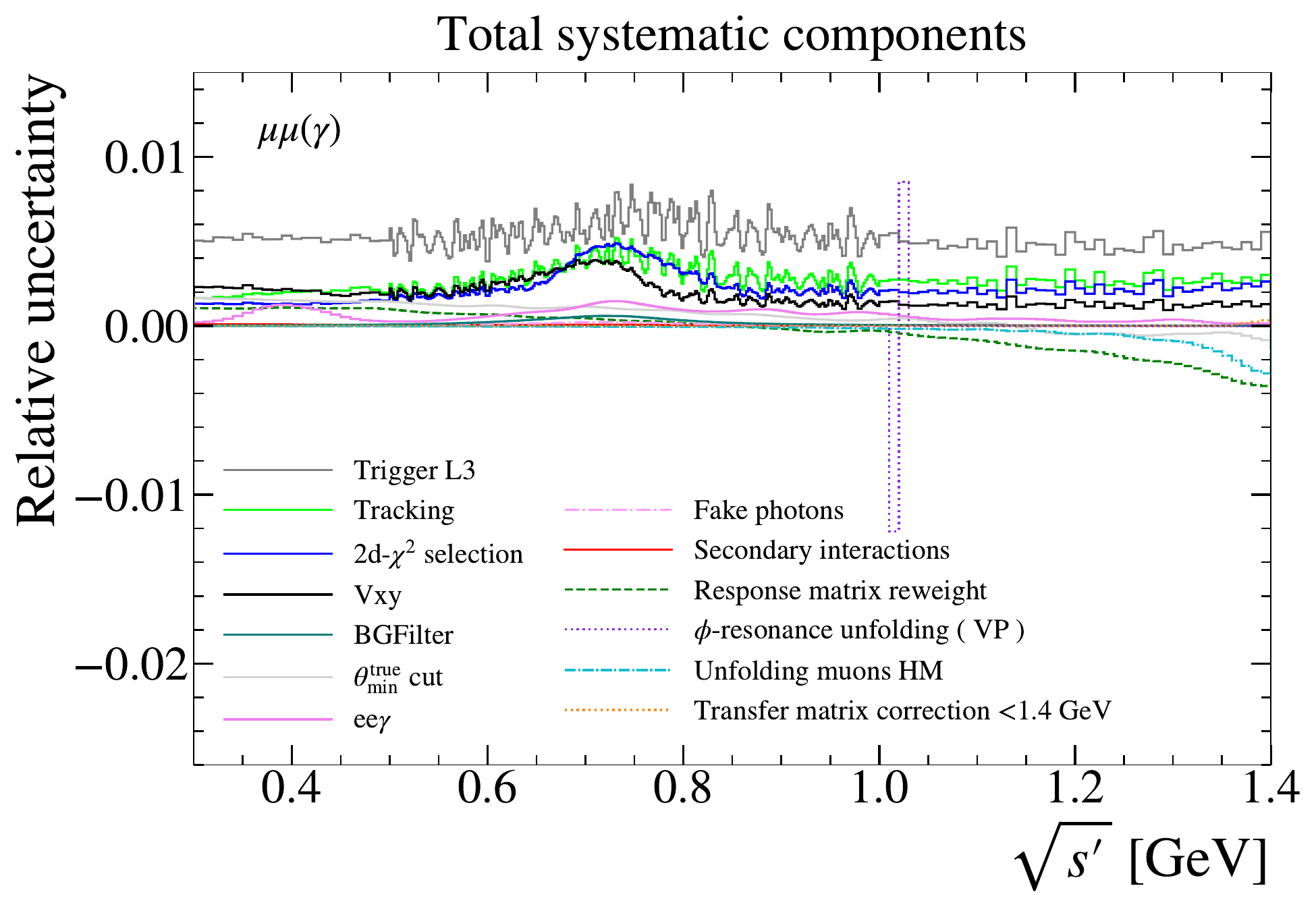}
    \includegraphics[width=\linewidth]{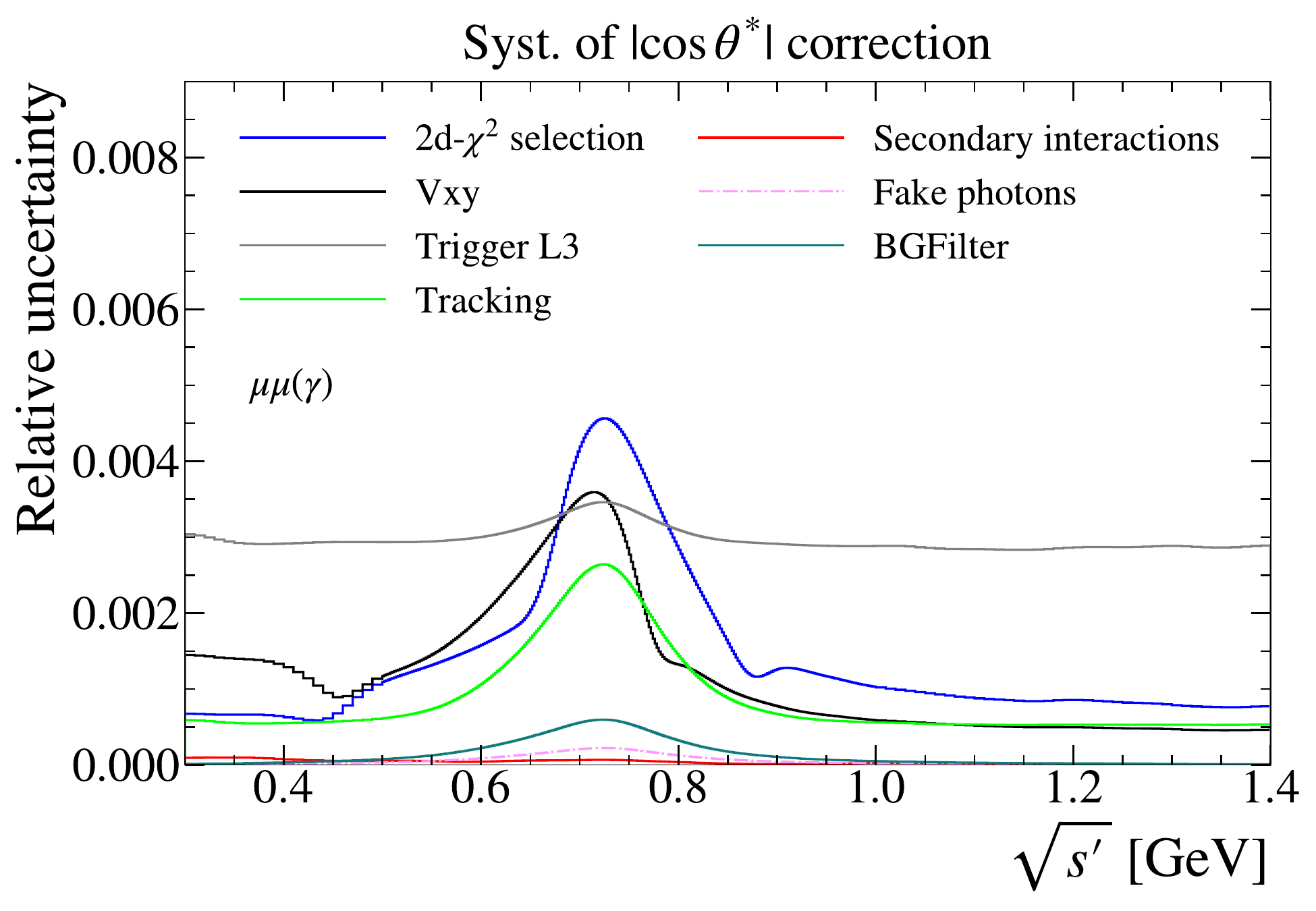}
    \includegraphics[width=\linewidth]{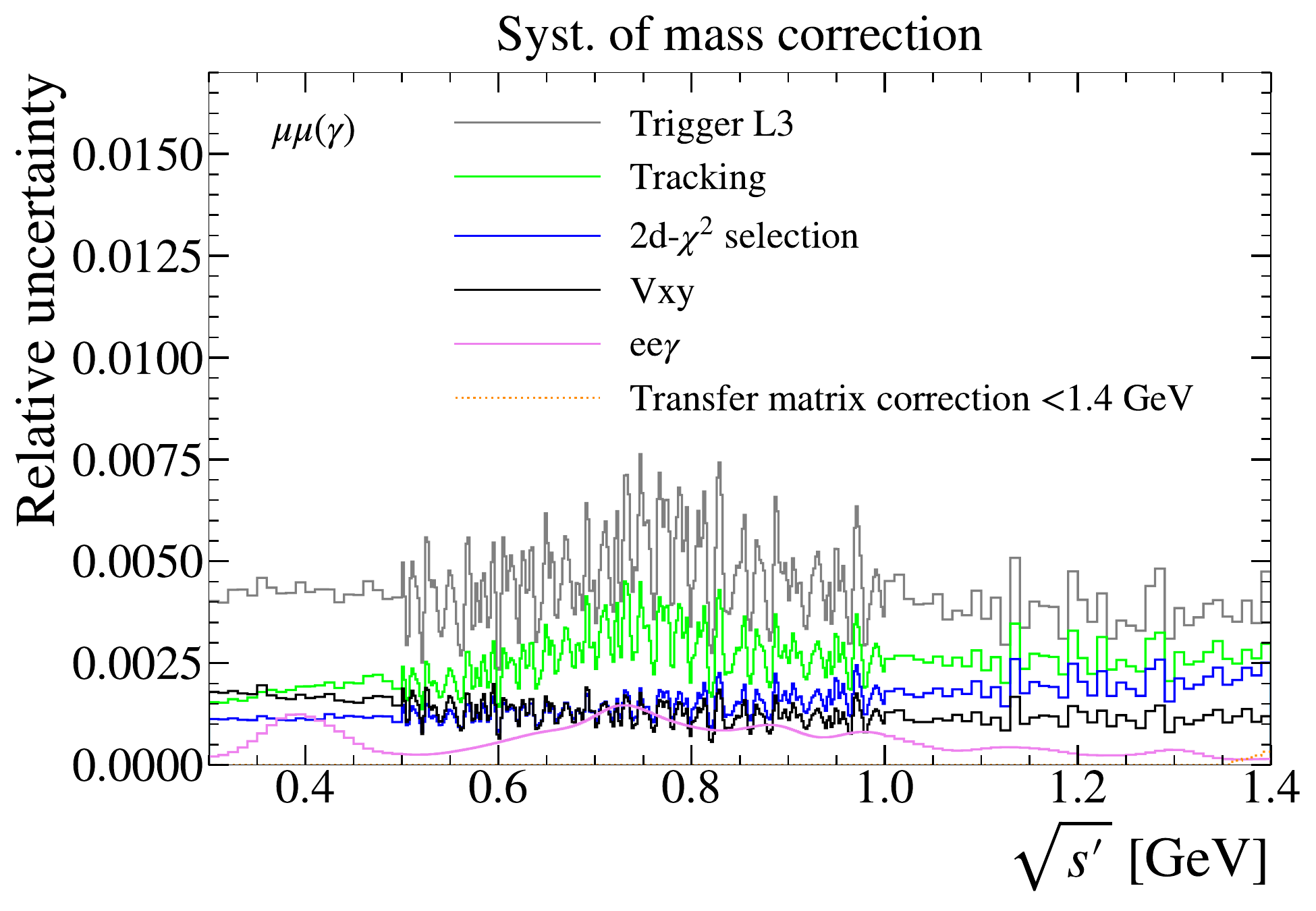}    
    \caption{\small \justifying (top) Systematic uncertainty components after the $\mu\mu\gamma$ unfolding. 
    Systematic contributions from \acosthpi\ template corrections~(middle) and mass correction~(bottom). The local statistical noise impacting some uncertainty components at this level is subject to further smoothing in the propagation towards the cross section.}
    \label{fig:syst_total_mu}
\end{figure}

Figure~\ref{fig:syst_total_mu} (top) shows the systematic uncertainties 
in the process $e^+e^- \to \mu^+\mu^-(\gamma)$. The dominant contributions arise from the $\phi$-resonance unfolding (VP), 2d-$\chi^2$ selection, trigger L3, vertex $V_{xy}$, and tracking performance, particularly in the region $\sqrt{s'} \approx 0.6\text{--}0.9\gev$ where 2d-$\chi^2$ selection uncertainties reach up to 0.5\%. 
The L3 trigger also significantly affects the event selection efficiency. Additional sources such as background filter, fake photon, and secondary interactions, are included but remain subdominant. The systematic uncertainties associated with the \(\theta_{\min}^{\mathrm{true}}\) cut and the response-matrix reweighting are positive at low mass and become negative at higher mass, reflecting an (anti-)correlated shape variation induced through the bin-to-bin migrations encoded in the unfolding matrix.

Figure~\ref{fig:syst_total_mu} (middle) shows the uncertainties related to the template corrections in \acosthpi, while Fig.~\ref{fig:syst_total_mu} (bottom) shows the systematics related to mass corrections. Across all corrections (total, angular, and mass), 2d-$\chi^2$ selection, trigger L3, vertex $V_{xy}$, and tracking are consistently the dominant sources of systematic uncertainty in the measurement of the $e^+e^- \to \mu^+\mu^-(\gamma)$ process.

\begin{figure}
    \centering
    \includegraphics[width=\linewidth]{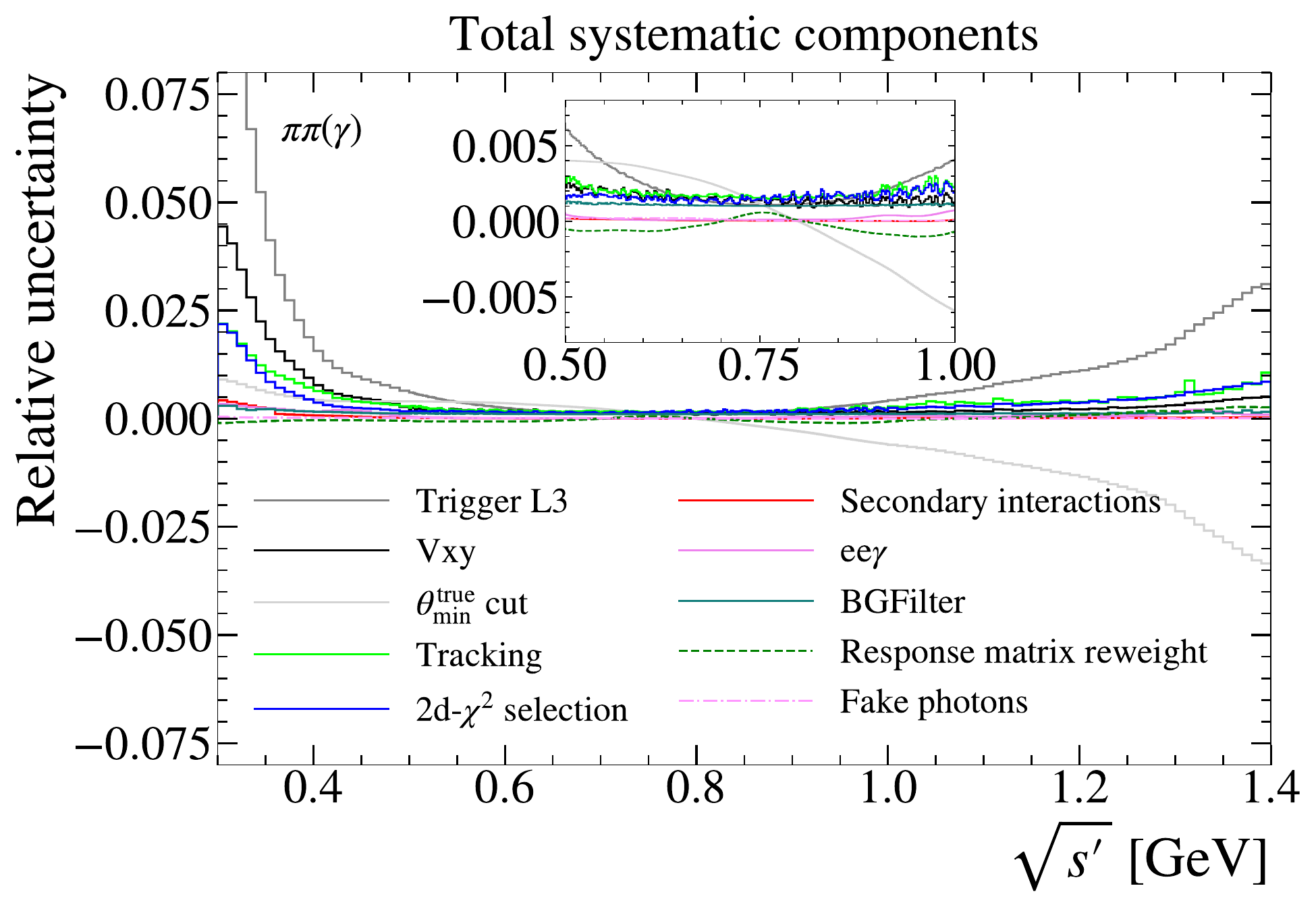}
    \includegraphics[width=\linewidth]{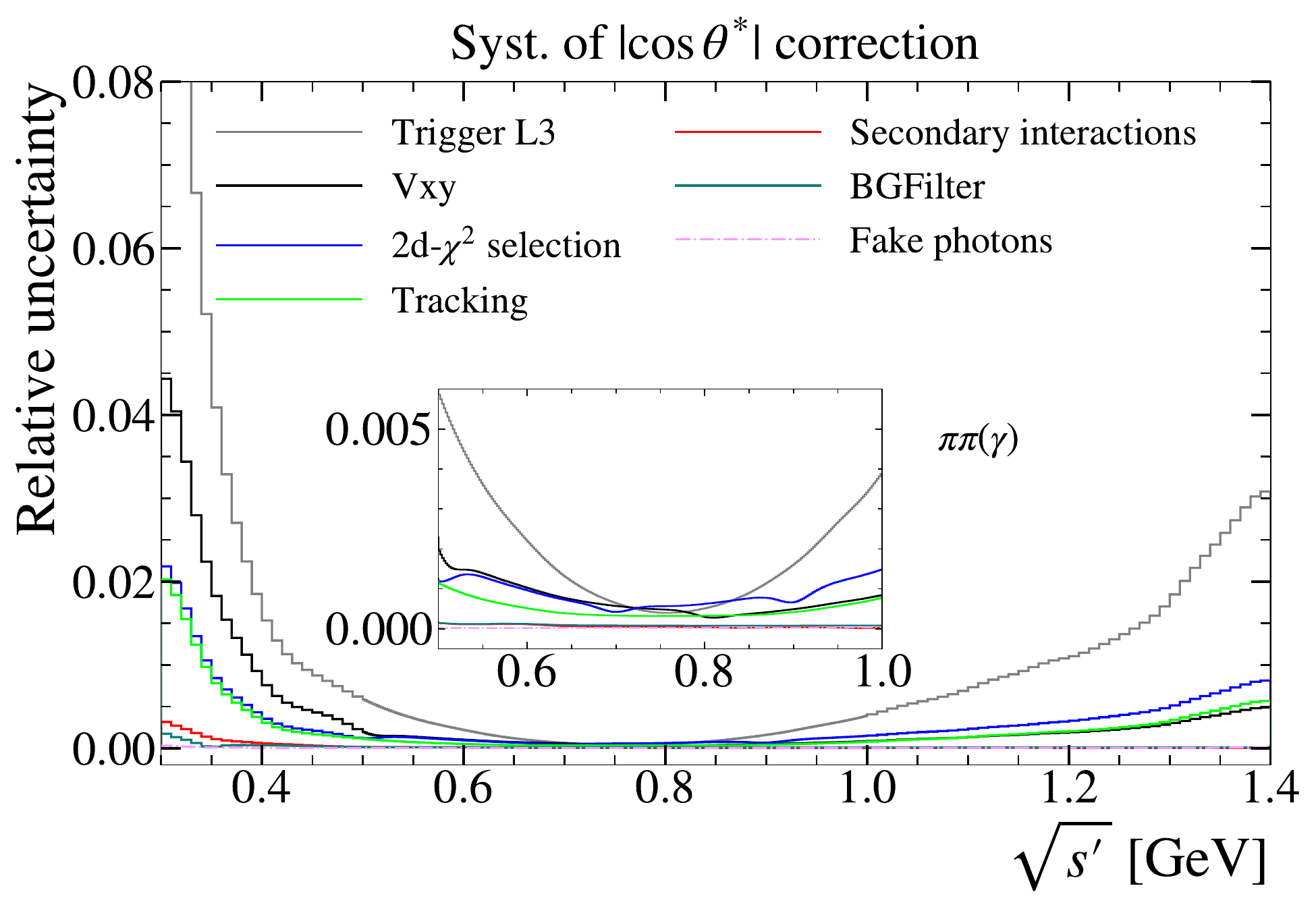}
    \includegraphics[width=\linewidth]{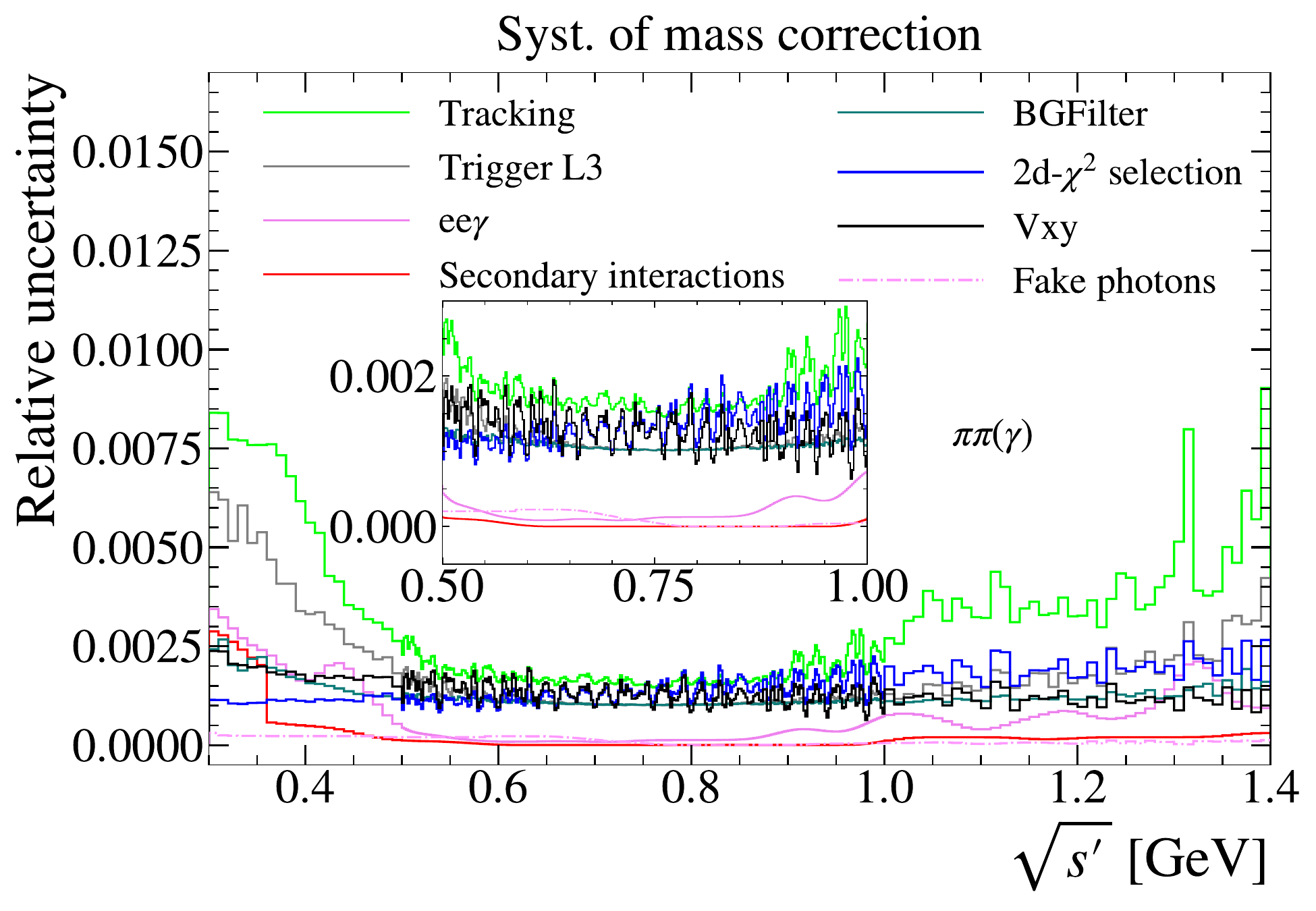}    
    \caption{\small \justifying Total systematic uncertainty components after the $\pi\pi\gamma$ unfolding~(top). Systematic contributions from \acosthpi\ template corrections~(middle) and mass correction~(bottom). The local statistical noise impacting some uncertainty components at this level is subject to further smoothing in the propagation towards the cross section.} 
    \label{fig:syst_total_pi}
\end{figure}

Figure~\ref{fig:syst_total_pi} shows the systematic uncertainties after unfolding for the process $e^+e^- \to \pi^+\pi^-(\gamma)$. Figure~\ref{fig:syst_total_pi} (middle) presents the uncertainties related to template corrections in 
\acosthpi\ and Fig.~\ref{fig:syst_total_pi} (bottom) shows the uncertainties from mass corrections.
The trigger L3, vertex \(V_{xy}\), 2d-\(\chi^2\) selection, and tracking contributions are dominant at the spectrum edges while the $\theta_\text{min}^\text{true}$ and L3 trigger dominate above $\sqrt{s'} > 1.2\gev$, where the total uncertainty remains below 4\%.

Overall, the $\theta_\text{min}^\text{true}$, tracking, trigger L3, and 2d-\(\chi^2\) selection are the leading sources of systematic uncertainty, as shown in Fig.~\ref{fig:syst_total_pi} (top), especially at low energies, due to intrinsic challenges in pion reconstruction because of secondary interactions. While mass corrections are subdominant, angular and selection-related effects contribute non-negligibly and are carefully assessed. At low and high $\sqrt{s'}$, these sources can reach 2--3\%, mainly due to reduced pion yields and detector resolution. However, in the central region $\sqrt{s'} \in [0.5,1.0]\gev$, where the $\rho-\omega$ resonance dominates, the total systematic uncertainty from all \textit{experimental} sources remains well below 0.3\%. 

\subsubsection{Total uncertainty}

\begin{figure}
    \centering
    \includegraphics[width=\linewidth]{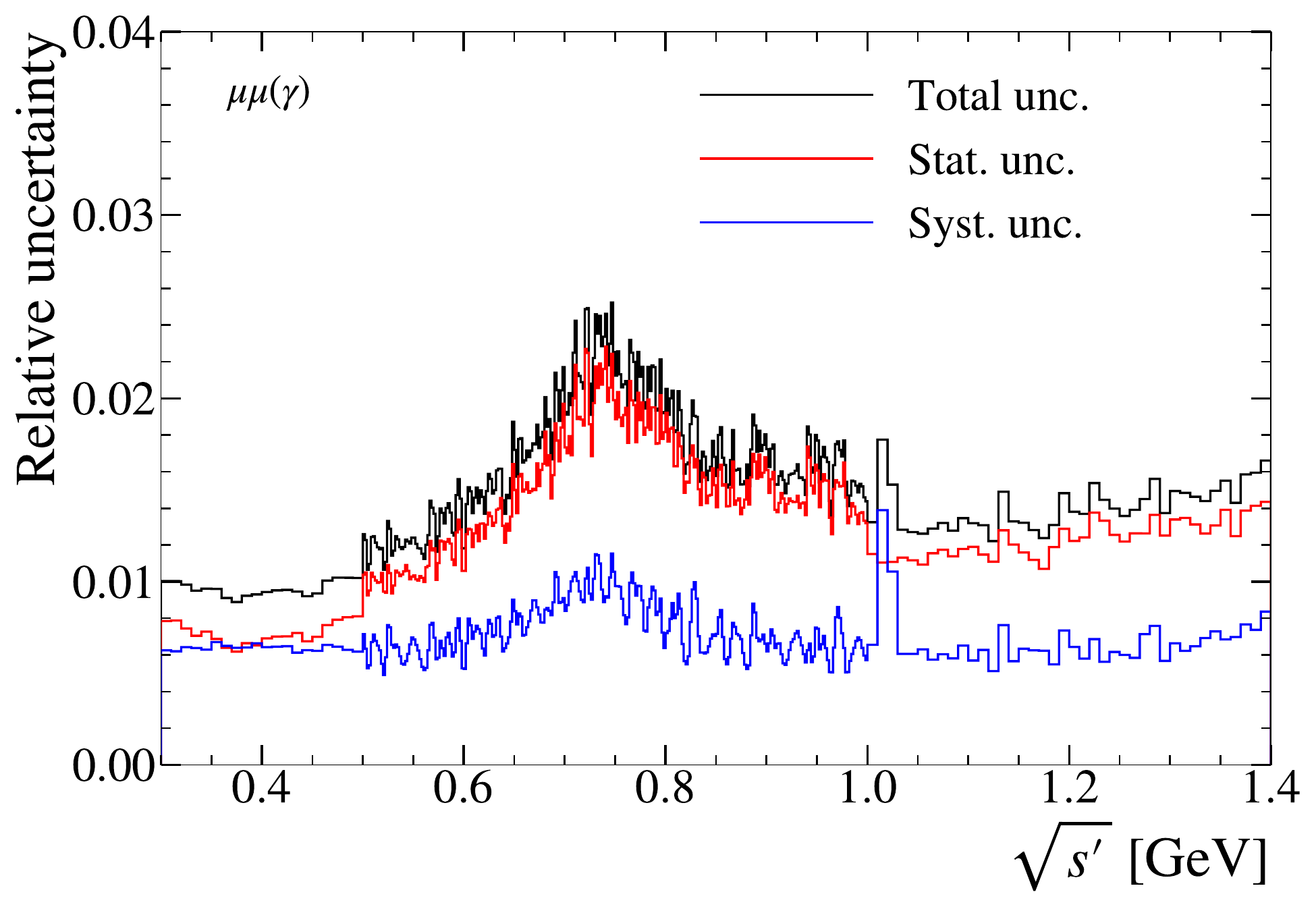}
    \includegraphics[width=\linewidth]{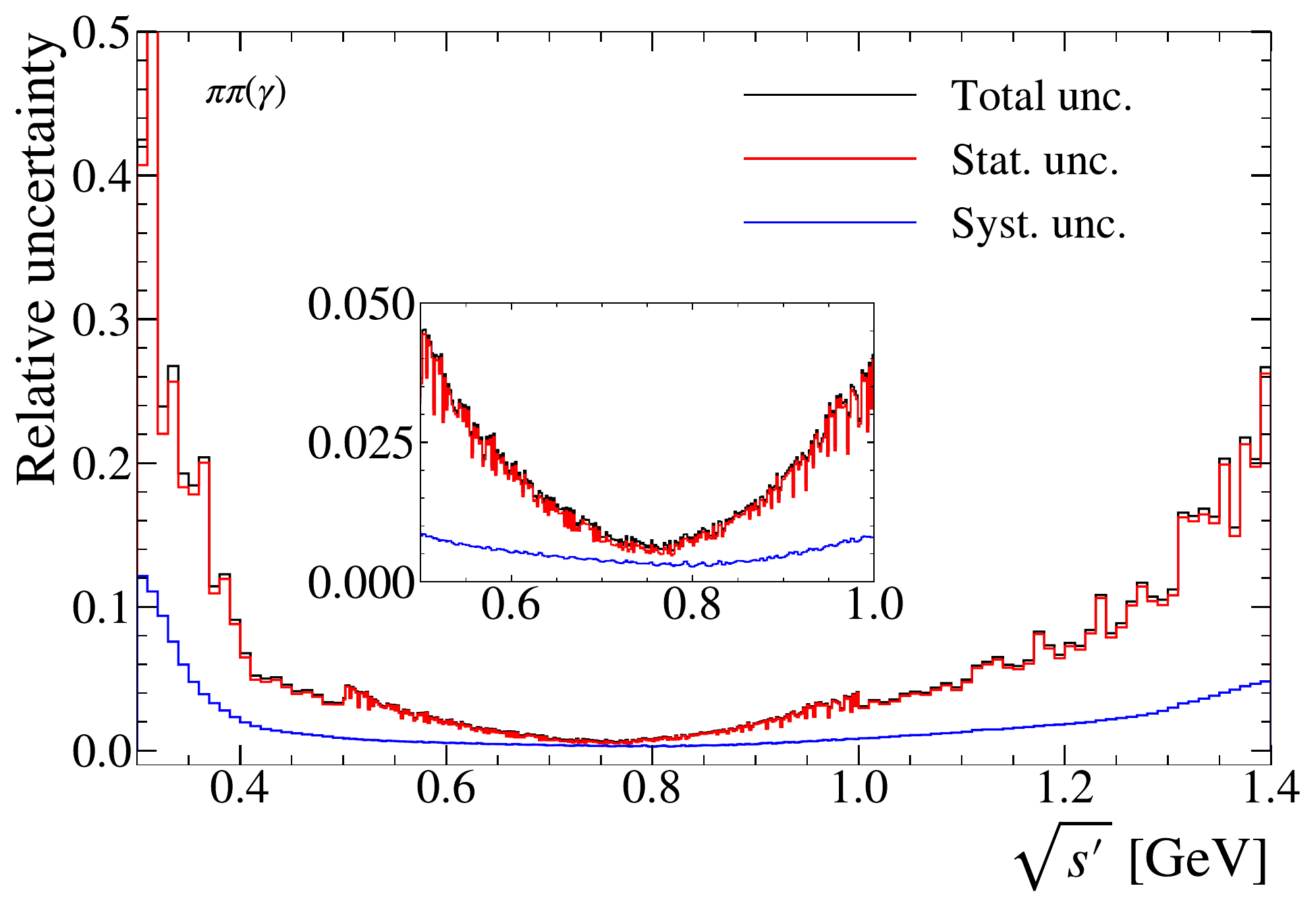}
    \caption{\small \justifying Total, statistical and systematic uncertainty components after the unfolding for the (top) muon and (bottom) pion. The local statistical noise impacting some uncertainty components at this level is subject to further smoothing in the propagation towards the cross section. }
    \label{fig:total_unc_unfold}
\end{figure}

The total uncertainty consists of two main components: statistical, estimated via toy experiments~(bootstrap replicas) as shown in Fig.~\ref{fig:unfolding_toys}, and systematic, obtained by summing individual sources in quadrature (Figs.~\ref{fig:syst_total_mu} and \ref{fig:syst_total_pi}).
Figure~\ref{fig:total_unc_unfold} decomposes the total uncertainty into its statistical and systematic contributions for the muon and pion channels, respectively. 

In the muon channel (Fig.~\ref{fig:total_unc_unfold} top), the total uncertainty is dominated by the statistical component across most of the spectrum with the systematic uncertainty becoming comparable in the $\phi$-resonance region. Around the $\rho$ resonance ($\sqrt{s'} \approx 0.75\gev$), statistical fluctuations peak around 3\%, while systematic uncertainties remain relatively flat and below 1\%. Outside the peak region, the total uncertainty stays below 2\%.

In the pion channel (Fig.~\ref{fig:total_unc_unfold} bottom), the total systematic uncertainty is dominated by the statistical component, which at low ($<$0.5\gev) and high ($>$1\gev) energies reaches values 
approaching \(50\%\) in regions where detector resolution degrades or background levels rise. 
The pion channel reaches a precision better than \(5\%\) in the central region and below 1\% in the $\rho-\omega$ resonance.

\section{Results on $e^+e^-\to\mu^+\mu^-(\gamma)\gamma_\mathrm{ISR}$ and ISR luminosity}
\label{sec:qed-test}

As explained in Section~\ref{pimuratio}, the $\pi\pi(\gamma)$ cross section is measured from the ratio of the $\pi^+\pi^-$ to $\mu^+\mu^-$ spectra as a function of $\sqrt{s'}$, in which several effects cancel out. Prior to the $\pi\pi(\gamma)$ cross section determination using the effective ISR luminosity from muons, the comparison of the $\mu\mu(\gamma)$ cross section to QED provides a validation of the full procedure. This absolute comparison to QED as implemented in the \textsc{Phokhara} NLO generator necessitates corrections that cancel in the $\pi^+\pi^-$ to $\mu^+\mu^-$ ratio, as detailed below. 

\subsection{ISR photon efficiency correction}
\label{isr-photon-QED-test}

Loss of the ISR photon can occur because of inefficient EMC crystals, cracks between crystals, and conversions in the early detector material. These loss sources are well described in the detector simulation. It is, however, necessary to crosscheck and establish a mass-dependent data/MC correction. This study is based on a sample of muon-pair events fitted to the $\mu\mu\gamma$ hypothesis without using any measured information on the radiated photon (1C fit). Because of the very loose constraints on the ISR photon energy, the only significant loss occurs as a function of the polar angle, when integrating over the azimuthal distribution. From the detailed study of Refs.~\cite{Lees_2012,hafner-thesis}, the loss is a bit larger in data and the ratio data/MC is quantified in broad angular regions.  By reweighting the mass distribution in the simulation with these ratios, one gets the ISR photon efficiency correction as a function of $m_{\mu\mu}$. The result, given in Fig.~\ref{ISR-photon-corr} (top), shows a very weak mass dependence with an average inefficiency in data compared to MC of 1.21\%.

Systematic uncertainties have been evaluated in detail in Ref.~\cite{hafner-thesis}. They include the differences induced by the photon selection criteria between data and simulation and are evaluated to be 0.33\%.

\subsection{NLO-NNLO \textsc{Phokhara} correction}
\label{NLO-NNLO-corr-QED-test}

\begin{figure}[tbp] \centering
\includegraphics[width=\linewidth]{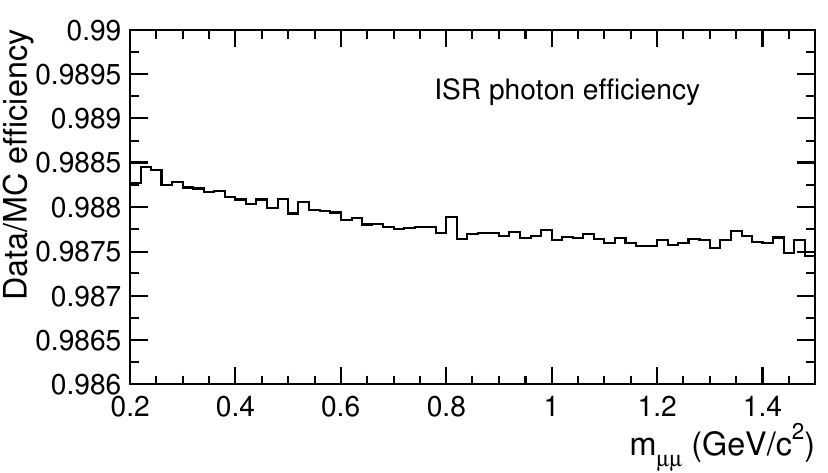}
\includegraphics[width=\linewidth]{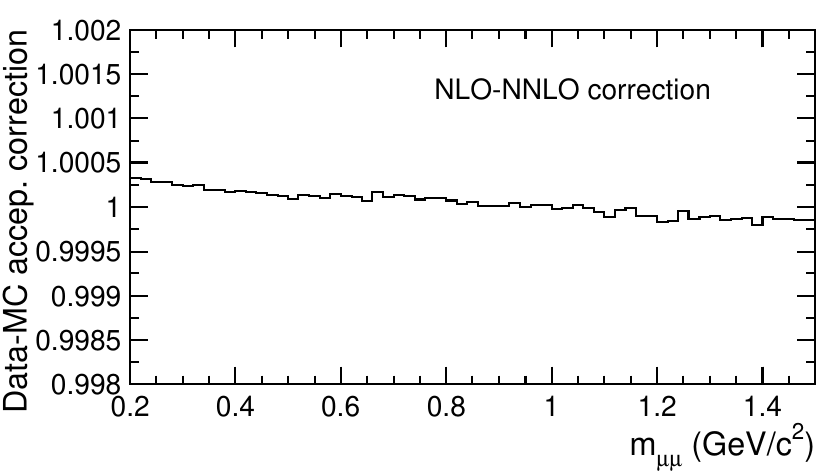}
    \caption{\label{ISR-photon-corr}   \small \justifying
(top): The ratio of ISR photon efficiencies in data and simulation for $\mu\mu\gamma$ events as a function of $m_{\mu\mu}$. (bottom):
The acceptance correction for NLO-NNLO data/\textsc{Phokhara} mismatch for $\mu\mu\gamma$ events as a function of $m_{\mu\mu}$.
  } 
\end{figure}

Since it was established in the \babar\ analysis of additional radiation in the ISR $\mu\mu$ and $\pi\pi$ samples~\cite{BaBar:2023xiy} that compared to data, the rate of ISR NLO events predicted by \textsc{Phokhara} was overestimated and NNLO contributions are not provided, it is necessary to establish a correction procedure in the muon sample acceptance. The mismatch between the predicted NLO and the measured NLO+NNLO fractions leads to a small excess on the data side which induces a slight bias when computing the acceptance given the selections used on the muon tracks and the ISR photon. In the absence of a full NNLO generator, the effect is estimated using the NLO \textsc{Phokhara} assuming that NNLO and NLO configurations generate similar deviations in the acceptance. This assumption is justified by the very small effect of the estimated bias. The data/MC radiative mismatch is propagated to the MC samples of LO-like events (defined as true LO + NLO with additional photon energy in the CM $E_\gamma^*$ less than 50\mev) and hard-NLO events (defined with $E_\gamma^*>50\mev$). The acceptance correction is shown in Fig.~\ref{ISR-photon-corr} (bottom): the bias is indeed tiny leading to a variation of only 0.47 per mil from threshold to 1.5\gevcc. As the NLO+NNLO data/MC mismatch is known to 10\% accuracy~\cite{BaBar:2023xiy} the systematic uncertainty in the acceptance correction is below 0.1 per mil and neglected.

\subsection{Vacuum polarization correction}
\label{VP-corr-QED-test}
\begin{figure}[tbp] \centering
\includegraphics[width=\linewidth]{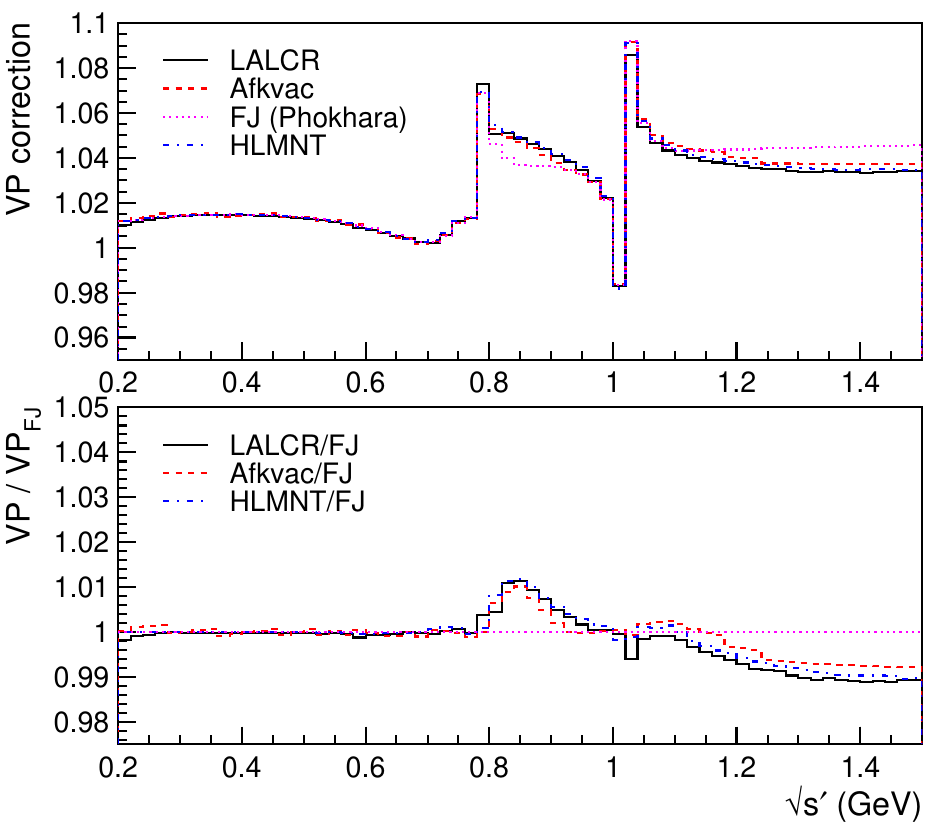}
    \caption{\label{fig:comp_vp}   \small \justifying
A comparison of various vacuum polarization corrections as a function of $\sqrt{s'}$. 
  } 
\end{figure}

The vacuum polarization correction is obtained by a dispersion integral over $e^+e^-$ data. The signal \textsc{Phokhara} MC samples were produced using a parametrization from Jegerlehner (FJ)~\cite{Jegerlehner:2017gek}. Its dependence as a function of $m_{\mu\mu}$ is shown and compared with a few other results in Fig.~\ref{fig:comp_vp}, where LALCR is a private routine used in some of our earlier studies, Afkvac is the routine used in \textsc{AfkQed}, and HLMNT~\cite{Hagiwara_2011} is an alternative choice  available in \textsc{Phokhara}. The comparison shows that the correction from FJ differs from the others in the mass region beyond 0.8\gevcc by about $\pm 1\%$. This difference is clearly related to shortcomings in FJ's parametrization. Therefore, a correction taking into account the more recent determinations cited above is applied to our \textsc{Phokhara} MC samples. In practice, the ratio HLMNT/FJ is used as a correction and the difference between Afkvac and HLMNT results is taken as an uncertainty in the correction.

\subsection{Test before unblinding}

As a reminder, 
three blinding offsets have been introduced in the analysis, which affect the comparison between the absolute $\mu\mu$ mass spectrum derived from the $|\cos{\theta^*_\pi|}$ fits: one offset each for the trigger and tracking data/MC corrections and a third one in the fitted $\mu\mu$ fraction. These offsets are constant with $m_{\mu\mu}$ and each one is known only by a single person responsible for the corresponding analysis task. In this way the overall resulting blinding offset is unknown. The result of the comparison of the $\mu\mu$ mass spectra in data and NLO \textsc{Phokhara} duly corrected for NLO/NNLO shortcomings as quantified in Section~\ref{NLO-NNLO-corr-QED-test} amounts to a direct comparison between the derived $\mu\mu$ mass spectrum and QED. With the assumption that the QED prediction indeed holds true, the result of the test is a powerful check of the $\mu\mu$-$\pi\pi$ separation using the angular fits. Thus the importance of performing this test using a rigorous blinding procedure is stressed.

Unblinding has only been performed once the analysis has been thoroughly checked to the best of our knowledge as was done for all the data/MC corrections to the efficiencies. All the mass-dependent corrections are applied, including the three discussed in Sections~\ref{isr-photon-QED-test}--\ref{VP-corr-QED-test}. In the end, the crucial test before unblinding is to check the constancy of the ratio of $\mu\mu$ mass spectra in data and MC. At this level, any significant deviation from flatness would readily invalidate the QED test and necessitate further investigation. The outcome of the pre-unblinding test is satisfactory as the ratio value remains constant within statistical uncertainties throughout the full investigated mass range from $\mu\mu$ threshold to 1.4\gevcc. This quantitative comparison is performed for the total spectrum and in restricted mass regions down to 100\mevcc intervals. The results confirm the compatibility of the ratio with a flat distribution over the full spectrum. They are presented in Table~\ref{tab:qed_test_100MeV} after unblinding since removing the offsets does not affect the shape
but only the absolute normalization.

\subsection{Test of the procedure: $\mu\mu$ mass spectrum data/QED ratio}
\label{QED-test}

\begin{figure}[tbp]
  \centering
  \includegraphics[width=\linewidth]{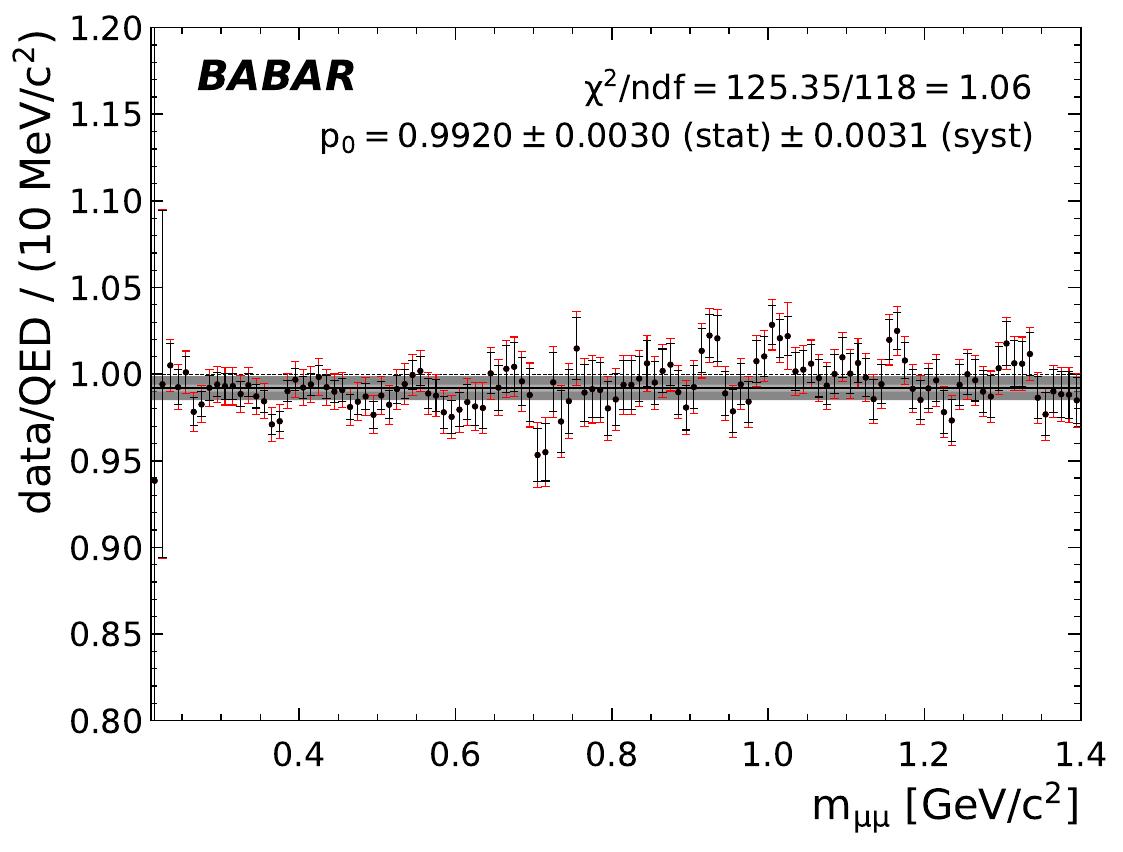}
\caption{\small \justifying The ratio of the $\mu\mu$ mass spectrum over the corrected \textsc{Phokhara} (QED) prediction. The full horizontal line represents a constant fit to the points. The gray band drawn around shows the total systematic uncertainty (including contributions from the ISR photon data/MC efficiency and the $e^+e^-$ luminosity).}
\label{fig:qed_test}
\end{figure}

Having checked the entire procedure and observed a flat ratio of the data/MC $\mu\mu$ mass spectra, unblinding of the three offsets is performed allowing a determination of whether the true observed ratio is consistent with unity. Figure~\ref{fig:qed_test} shows the result of the test with unblinded data as a function of the $\mu\mu$ mass in 10\mevcc bins from threshold to 1.4\gevcc. There is good agreement within the point-to-point uncertainties dominated by statistical fluctuations in the fits of the \acosthpi\ distributions. The difficulty in statistically separating the $\pi\pi$ and $\mu\mu$ components is reflected in the size of the statistical uncertainties, which affects particularly the $\mu\mu$ fraction in the 0.6--0.8\gevcc region strongly dominated by the large $\rho$ resonance. 

\begin{table}
    \centering
    \caption{\small \justifying The deviation from unity (expressed in per mil) of the unblinded ratio of the $\mu\mu$ mass spectrum over the corrected \textsc{Phokhara} (QED) prediction fitted to a constant in 100\mevcc mass intervals. The quoted uncertainties are statistical and systematic in that order.}
    \label{tab:qed_test_100MeV}
     \begin{tabular}{c|r|c}
\hline
Mass range & Fit to a constant & \multirow{2}{*}{$\chi^2/$ndof} \\
(\gevcc) & (shift from unity in per mil) & \\
\hline
0.21 -- 1.40 & $-8.0 \pm 3.0 \pm 1.7$ & $1.06$ \\
\hline 
0.21 -- 0.35 & $-9.3 \pm 3.6 \pm 1.6$ & $1.22$ \\
0.35 -- 0.45 & $-12.4 \pm 4.2 \pm 1.9$ & $1.23$ \\
0.45 -- 0.55 & $-12.9 \pm 3.5 \pm 1.7$ & $0.91$ \\
0.55 -- 0.65 & $-14.0 \pm 4.7 \pm 1.9$ & $0.77$ \\
0.65 -- 0.75 & $-13.6 \pm 8.3 \pm 2.7$ & $1.37$ \\
0.75 -- 0.85 & $-5.5 \pm 5.6 \pm 2.2$ & $0.41$ \\
0.85 -- 0.95 & $1.1 \pm 6.1 \pm 1.9$ & $0.99$ \\
0.95 -- 1.05 & $5.7 \pm 6.5 \pm 1.8$ & $1.40$ \\
1.05 -- 1.15 & $-0.8 \pm 4.0 \pm 1.7$ & $0.57$ \\
1.15 -- 1.25 & $-2.7 \pm 6.4 \pm 1.6$ & $1.32$ \\
1.25 -- 1.35 & $0.1 \pm 4.6 \pm 1.6$ & $0.98$ \\
1.35 -- 1.40 & $-14.5 \pm 4.7 \pm 1.6$ & $0.39$ \\
\hline
\end{tabular}
\end{table}

A global comparison test is performed over the full mass range. The ratio is overall consistent with unity within the statistical and systematic uncertainties, discussed in detail in Sections~\ref{sec:stat-error} and \ref{Sec:syst-uncertainties}. Fitting to a single constant value $R_{\mu\mu}$ yields:
\begin{eqnarray}
R_{\mu\mu}=0.9920&\pm& 0.0030_\mathrm{stat}\pm0.0031_\mathrm{syst}\nonumber\\
 &\pm& 0.0033_{\gamma\,\mathrm{ISR}}\pm 0.0043_{\mathrm{lumi}\,ee},
\end{eqnarray}
where the first uncertainty is statistical (data and measured data/MC efficiency corrections) and the last three are systematic (respectively from the templates in the \acosthpi\ fits, the ISR photon data/MC efficiency, and the $e^+e^-$ luminosity~\cite{BaBar:2013agn}).
Taking into account both the statistical and systematic uncertainties, together with their bin-to-bin correlations, included in a total covariance matrix, we obtain for this fit a value of $\chi^{2}/{\rm ndf} = 125 / 118$.
This test validates our $\pi\pi$/$\mu\mu$ separation procedure against NNLO QED as given by our data-corrected \textsc{Phokhara} prediction with a precision of 0.69\%. 
This precision of the QED test is dominated by the uncertainties from the $e^+e^-$ luminosity and the ISR photon efficiency, both of which cancel when taking the $\pi\pi$/$\mu\mu$ ratio to determine the $\pi\pi$ cross section.

\subsection{Determination of ISR luminosity}

\subsubsection{Lowest-order FSR correction}

\begin{figure}[tbp]
  \centering
  \includegraphics[width=\linewidth]{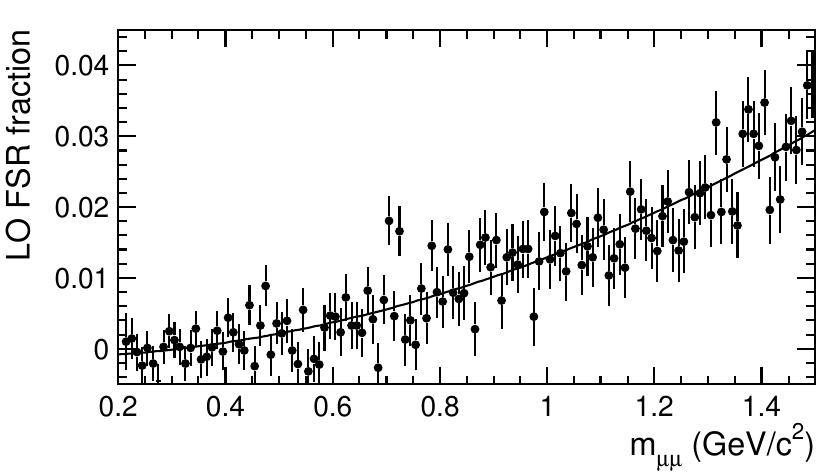}
\caption{\small \justifying The 
LO FSR event fraction $f_\text{LO FSR}$ in \textsc{Phokhara}.}
\label{fig:LO-FSR-mumu}
\end{figure}

In this analysis the photon with the highest energy in the CM is assumed to be emitted by the initial state (the so-called ISR photon). However, in principle, both ISR and FSR origins are possible, corresponding to the diagrams shown in Fig.~\ref{diagrams}, at LO and NLO levels. The ISR assumption is an excellent approximation for the $\pi\pi$ final state, since the pion structure strongly depresses the hard LO FSR process at 10.58\gev. This issue was investigated and the expectation confirmed in a dedicated \babar\ analysis of the ISR-FSR interference~\cite{BaBar:2015onb}, particularly sensitive to the small FSR amplitude. However for the point-like muons the LO FSR contribution is significant and must be subtracted out when evaluating the ISR luminosity necessary to compute the $\pi\pi$ cross section. The LO FSR fraction is obtained by running \textsc{Phokhara} at the 4-vector level in two modes, one with $|\text{LO ISR}|^2$ only and the other with $|\text{LO ISR} + \text{LO FSR}|^2$, which includes the interference, although the interference vanishes in a charge-symmetric situation. The result for the 
LO FSR contribution $f_\text{LO FSR}$ to the $\mu\mu(\gamma)\gamma_\mathrm{ISR}$ cross section is given in Fig.~\ref{fig:LO-FSR-mumu} where the expected smooth distribution is fitted to a second-order polynomial.

\subsubsection{The effective ISR luminosity for the $\pi\pi(\gamma)\gamma_\mathrm{ISR}$ cross section}

The effective ISR luminosity is determined from the unfolded $\mu\mu$ spectrum $dN_{\mu\mu}^\mathrm{ISR}/d\sqrt{s^\prime}$, corrected for the data/MC differences of all the relevant efficiencies and the LO FSR fraction, as
\begin{equation}
\frac{dL^\mathrm{eff}_\mathrm{ISR}}{d\sqrt{s^\prime}}=\frac{dN_{\mu\mu}^\mathrm{ISR}/d\sqrt{s^\prime}}{\epsilon_{\mu\mu}(\sqrt{s^\prime})\ \sigma^0_{\mu\mu}(\sqrt{s^\prime})},
\end{equation}
where $\epsilon_{\mu\mu}(\sqrt{s^\prime})$ is the acceptance of the selection obtained with \textsc{Phokhara} and $\sigma^0_{\mu\mu}(\sqrt{s^\prime})$ is the bare cross-section for the $\mu\mu$ process with the vacuum polarization contribution removed, but including the FSR contribution. The latter small component is calculated at NLO taking into account loop diagrams and the emission of soft and hard photons,  with large cancellations between the different components leading to a finite result (see for example Ref.~\cite{Bystritskiy:2005ib}).

\begin{figure}[tbp]
  \centering
  \includegraphics[width=\linewidth]{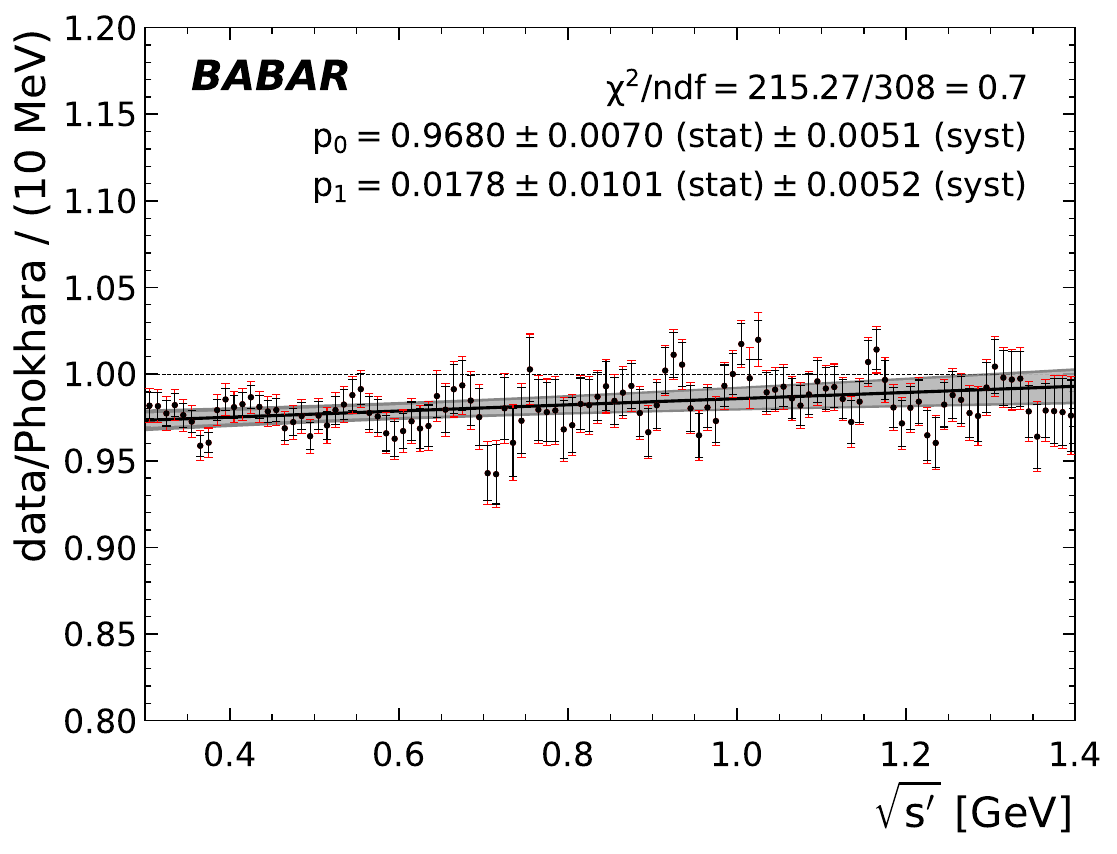}
\caption{\small \justifying The ratio of the unblinded $\mu\mu$ mass spectrum in $\sqrt{s^\prime}$ over the \textsc{Phokhara} spectrum with the vacuum polarization correction of Section~\ref{VP-corr-QED-test} applied. The full line represents a linear fit to the points. The gray band drawn around shows the total systematic uncertainty. The fit is performed on bins of 2\MeVM\ in the central mass region and 10\MeVM\ outside, while the ratio points are displayed in 10\MeVM\ bins for better readability. }
\label{fig:lumi_p1_fit}
\end{figure}

As the measured spectrum $dN_{\mu\mu}^\mathrm{ISR}/d\sqrt{s^\prime}$
suffers from large statistical fluctuations in the \acosthpi\ fits, especially in the range corresponding to the dominating $\rho$ resonance, it is advantageous to smooth it. 
This is done by replacing the ratio $(dN_{\mu\mu}^\mathrm{ISR}/d\sqrt{s^\prime})/\epsilon_{\mu\mu}(\sqrt{s^\prime})$ by:
\begin{equation}
\frac{dN^\text{MC gen}_{\mu\mu}}{d\sqrt{s^\prime}}\times (1-f_\text{LO FSR}) \times f_{\mu\mu}(\sqrt{s^\prime}),
\end{equation}
where the small LO-FSR contribution is explicitly removed from the generated MC spectrum $dN^\text{MC gen}_{\mu\mu}/{d\sqrt{s^\prime}}$, and $f_{\mu\mu}(\sqrt{s^\prime})$ corresponds to the fitted ratio of the unfolded data and MC $\mu\mu$ spectra to a polynomial function. This procedure is justified by (1) the test of the measured $\mu\mu$, consistent with QED within uncertainties throughout the mass spectrum and (2) the slow and smooth variation with mass of the global data/simulation correction from all the measured efficiencies at the levels of the fit templates and the mass distribution (Fig.~\ref{fig:total_corr})\footnote{The only fast varying (but small) correction on the HVP contribution used in our \textsc{Phokhara} production as discussed in Section~\ref{VP-corr-QED-test} is applied separately.}. Whereas the test with QED implied correcting for all known effects affecting our \textsc{Phokhara} sample (NLO-NNLO and HVP), here with  
$f_{\mu\mu}(\sqrt{s^\prime})$ we are dealing with the difference between the measured spectrum and the simulation based on \textsc{Phokhara}. Thus it differs from unity by (1) small differences not taken into account by the measured corrections for efficiencies, (2) differences between the true ISR radiator function and the \textsc{Phokhara} prediction, and (3) especially the data/simulation correction on the ISR photon efficiency. It turns out that a fit to a constant is already acceptable, however a 
linear fit yields a small improvement. Since adding a quadratic term does not improve the $\chi^2$, the linear fit is retained 
for the nominal result, as shown in Fig.~\ref{fig:lumi_p1_fit}. 
Thus the fit for $f_{\mu\mu}(\sqrt{s^\prime})$ incorporates all the features of the measured $\mu\mu$ spectrum needed for the determination of the effective ISR luminosity, preserving the positive features of using a $\pi\pi$/$\mu\mu$ ratio in order to obtain the $\pi\pi$ cross section. 
A systematic uncertainty is associated with the choice of the functional form of the $f_{\mu\mu}(\sqrt{s^\prime})$ fit.
It is evaluated by comparing the nominal linear fit with a constant one.

\begin{figure}[tbp]
  \centering
  \includegraphics[width=\linewidth]{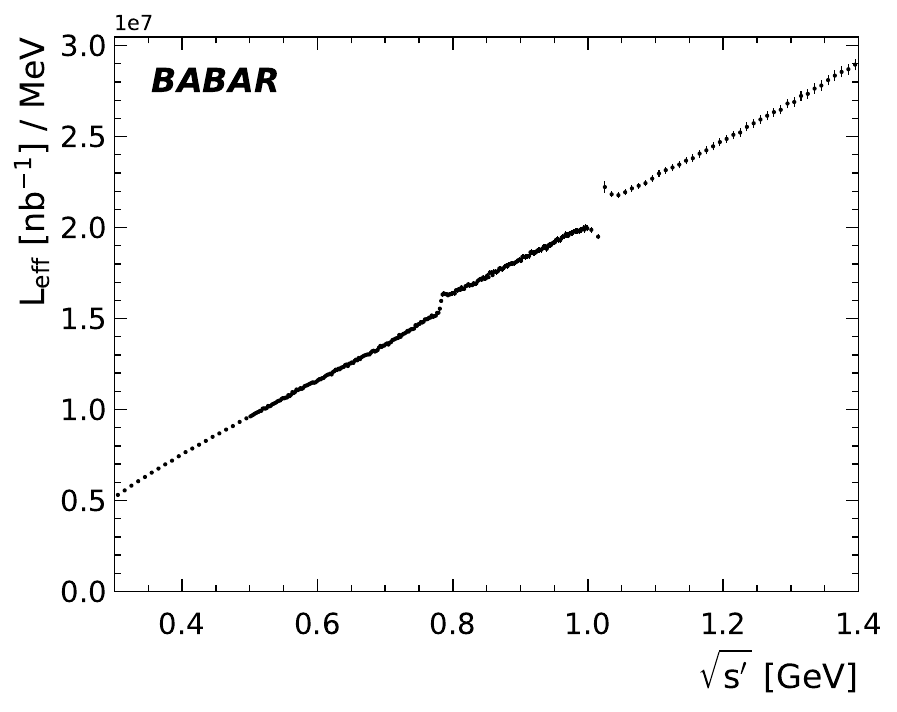}
\caption{\small \justifying The effective ISR luminosity.}
\label{fig:ISR-lumi}
\end{figure}

The derived effective ISR luminosity is displayed in Fig.~\ref{fig:ISR-lumi}. The two discontinuities are due to the vacuum polarization contributions that are included in the generated \textsc{Phokhara} energy spectrum but not in the bare cross section. The effective ISR luminosity is small at low energy resulting in a larger statistical uncertainty there for processes with low cross section. It is larger at larger energy where, however, the $\pi\pi$ cross section also drops.

\section{Measurement of the $e^+e^-\to\pi^+\pi^-(\gamma)\gamma_\mathrm{ISR}$ cross section}
\label{sec:pionXsect}

\begin{figure}[tbp]
  \centering
  \includegraphics[width=\linewidth]{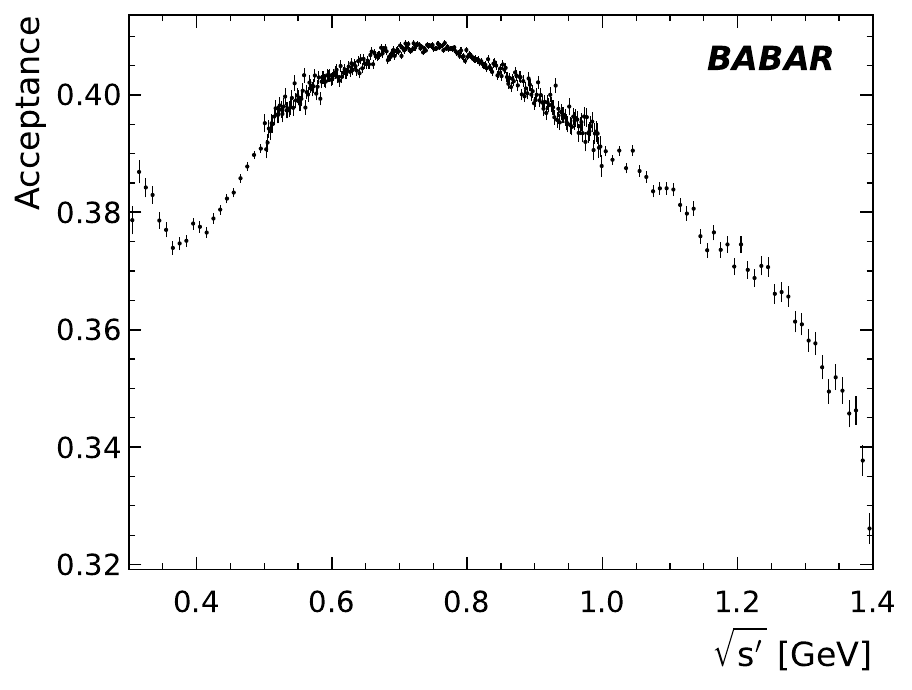}
\caption{\small \justifying The acceptance of \pipig\ events computed with \textsc{Phokhara}. All data/simulation corrections are included.
}
\label{fig:acceptance-pi}
\end{figure}

\begin{figure}[tbp]
  \centering
  \includegraphics[width=1.\linewidth]{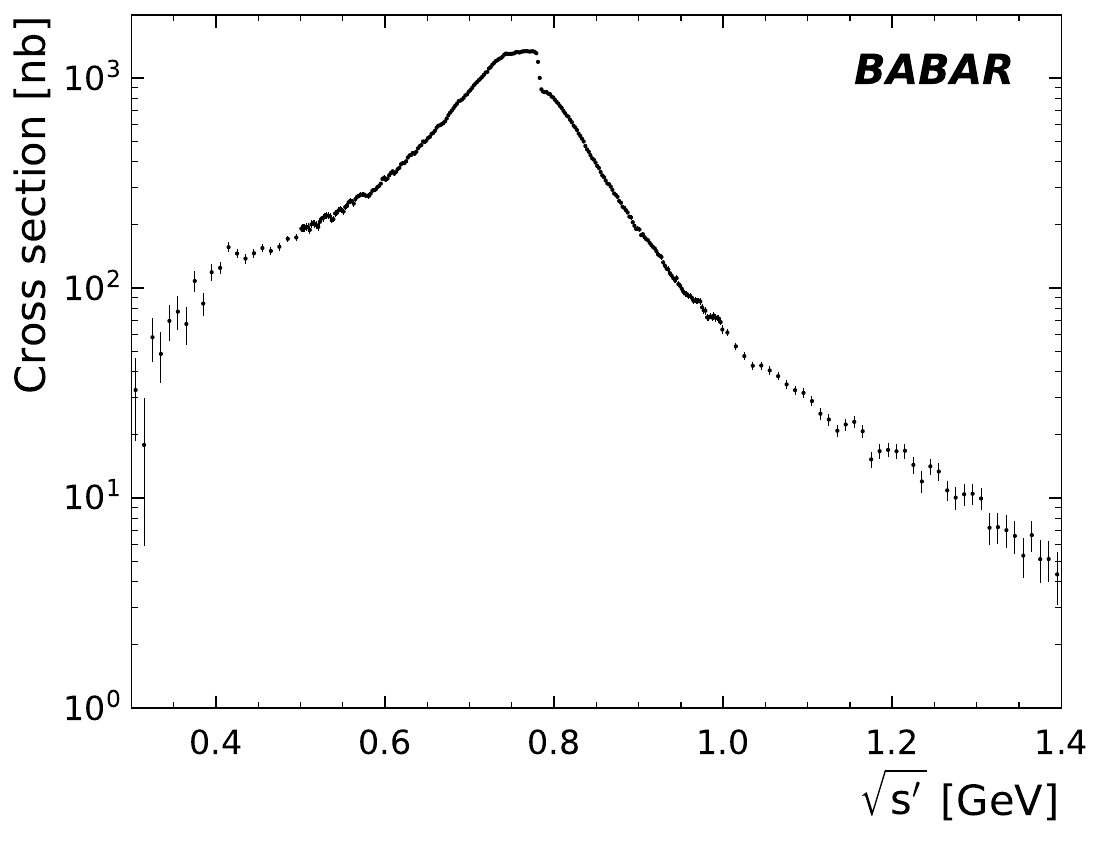}
\caption{\small \justifying The measured cross section for $e^+e^- \to \pi^+\pi^-(\gamma)$ over the full energy range.}
\label{fig:pipi-cross_section}
\end{figure}

\begin{figure}[tbp]
  \centering
  \includegraphics[width=1.\linewidth]{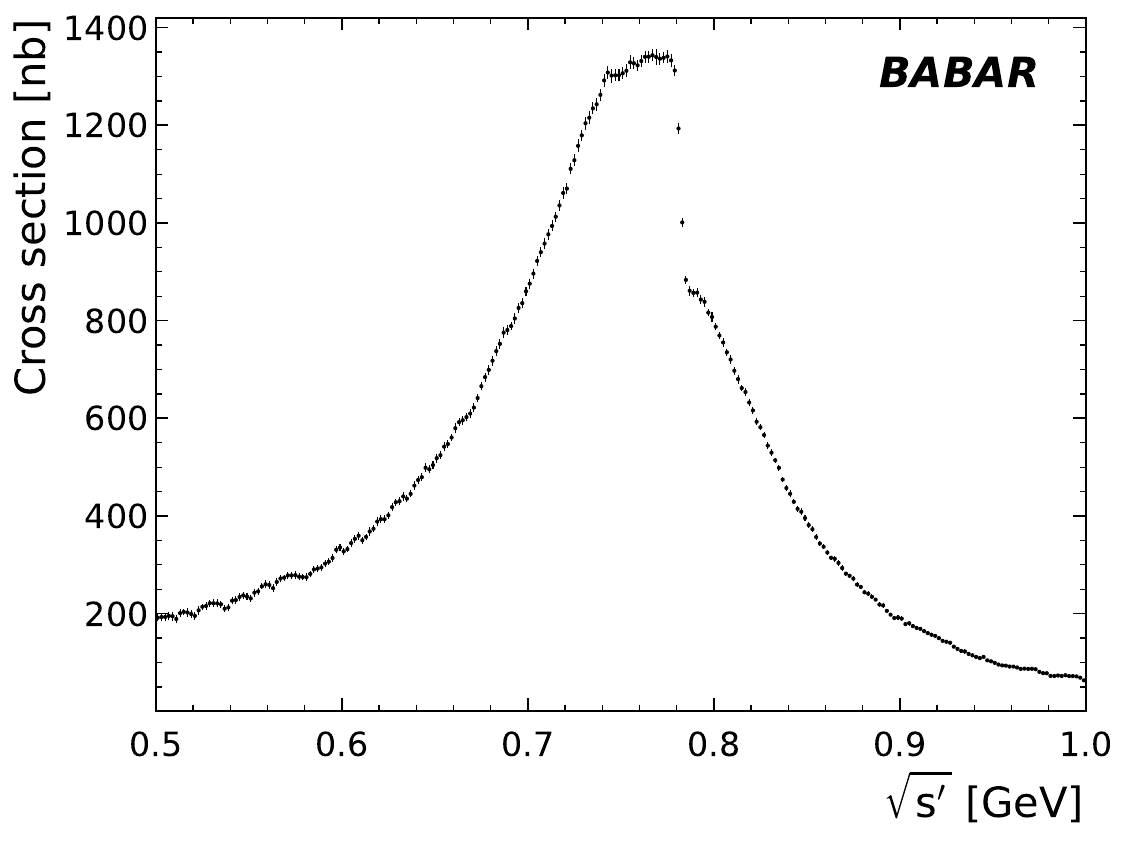}
\caption{\small \justifying The measured cross section for $e^+e^- \to \pi^+\pi^-(\gamma)$ in the central energy region.}
\label{fig:pipi-cross_section-central}
\end{figure}

The bare cross section for the $\pi\pi$ process is obtained from the unfolded $\pi\pi$ spectrum using
\begin{equation}
\sigma^0_{\pi\pi}(\sqrt{s^\prime})=\frac{dN_{\pi\pi}/d\sqrt{s^\prime}}{\epsilon_{\pi\pi}(\sqrt{s^\prime})\ dL^{\rm eff}_{\rm ISR}/d\sqrt{s^\prime}},
\end{equation}
where $\epsilon_{\pi\pi}(\sqrt{s^\prime})$ is the acceptance of the selection for the $\pi\pi$ process (Fig.~\ref{fig:acceptance-pi}).
Replacing the effective ISR luminosity using the formula from the previous subsection, one gets
\begin{eqnarray}
\sigma^0_{\pi\pi}(\sqrt{s^\prime})&=&\sigma^0_{\mu\mu}(\sqrt{s^\prime})\nonumber\\
&\times&\frac{dN_{\pi\pi}/d\sqrt{s^\prime}}{dN^\text{MC gen}_{\mu\mu}/d\sqrt{s^\prime}\ (1-f_\text{LO FSR})\ f_{\mu\mu}(\sqrt{s^\prime})}\nonumber\\
&\times&\frac{1}{\epsilon_{\pi\pi}(\sqrt{s^\prime})}\,.
\end{eqnarray}
Therefore, effectively, the bare cross section of the $\pi\pi$ process is the bare cross section of the $\mu\mu$ process weighted by the ratio of the unfolded data spectra of the $\pi\pi$ and  $\mu\mu$ processes, resulting in reduced systematic uncertainties due to the cancellation of the common uncertainty sources or corrections. 
The statistical and systematic uncertainties, discussed in detail in 
Section~\ref{sec:unfold-error}, are fully propagated with their correlations to the $\pi\pi$ cross section measurement.
The unblinded results are shown 
in Figs.~\ref{fig:pipi-cross_section} and \ref{fig:pipi-cross_section-central}.

Preliminary results of this analysis have been presented in 2025~\cite{Zhang:2026jvb,Polat:2026ysh}. The present paper incorporates substantial improvements in the methodology and many additional studies, as well as correcting several shortcomings. Details on the main changes are presented in Appendix~\ref{app:changes}.

The results of this analysis are compared to those obtained previously by \babar ~\cite{BaBar:2009wpw,BaBar:2012bdw}, referred to as 2009 in the following. It is worth pointing out here the main differences between the two analyses:
\begin{itemize}
    \item The $\pi\pi$/$\mu\mu$ separation method is completely different: while in 2009 the separation was made using particle identification for both particles, in the present 2026 analysis a statistical separation is used as described above using the different angular distributions of the two processes in their center-of-mass frame. 
    \item In 2009 only about one half of the final \babar\ data sample was available while the full sample is now used with the new method.
    Furthermore, since no PID of the charged particles is now used, the lower momentum requirement applied in 2009 is relaxed, allowing a large statistical increase.
    However, because of the statistical separation in 2026, the final statistical uncertainties are comparable.
    \item The dominant systematic uncertainties are uncorrelated between the two analyses. The main contributions were from particle identification in 2009 which was complicated because of the 
    degrading performance of the muon detection system. As a consequence, detailed run-by-run control of efficiency maps over the detector's active area had to be implemented, including cuts to exclude low-efficiency regions. In the 2026 analysis, particle ID is used only for particle tagging and the determination of efficiencies for trigger, tracking, and kinematic fits. It does not play a role in separating
    the muon and pion channels. A large source of uncertainty in 2026 is related to the data/simulation corrections, which affect the shape of the templates used to perform the angular fits.
    \item The new method is less competitive than the particle-ID-based approach in the threshold region and at energies larger than about 1\gev where the $\pi\pi$ signal is overwhelmed by the large $\mu\mu$ cross section.
\end{itemize}
Two important features are shared by both methods: (1) the use of the $\pi\pi$/$\mu\mu$ ratio to determine the $\pi\pi$ cross section with cancellation of the effects of the ISR photon efficiency, $ee$ luminosity, and vacuum polarization, and (2) the inclusion in the measurement of 
one photon in the final state in addition to the main ISR photon in order to be largely insensitive to the MC modeling of radiative corrections.

\begin{figure}[tbp]
  \centering
  \includegraphics[width=1.\linewidth]{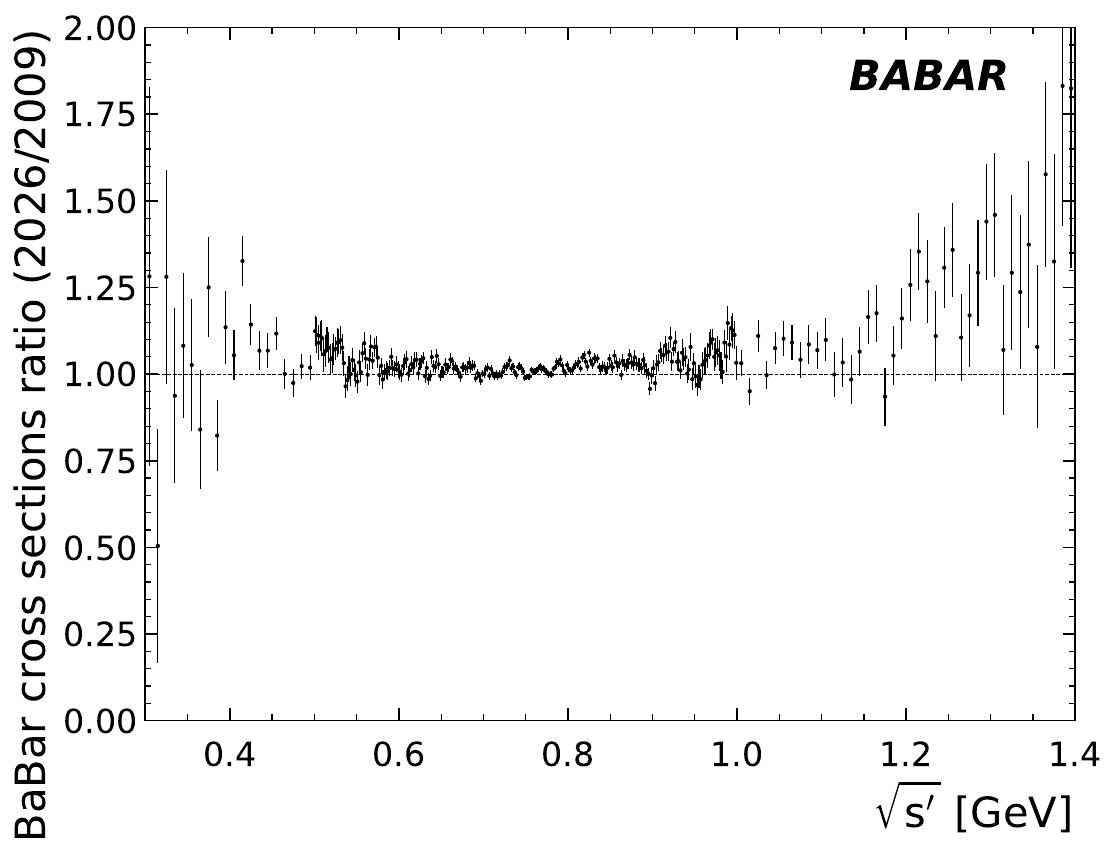}
\caption{\small \justifying Ratio of the measured cross section for $e^+e^- \to \pi^+\pi^-(\gamma)$
  between the current 2026 analysis and the previous 2029 \babar\ analysis, from 0.3 to 1.4\gev.}
\label{fig:pipi-cross_section-ratio}
\end{figure}

The comparison of the $\pi\pi$ 
measured cross section with the previous result~\cite{BaBar:2009wpw,BaBar:2012bdw} is shown in Fig.~\ref{fig:pipi-cross_section-ratio}.
The 2026 measurement is found to be about 2\% higher in the $\rho$-resonance region than the 2009 one with larger differences in the lower and higher energy  regions though the statistical precision there is limited. A more quantitative comparison in different energy regions is given in Section~\ref{sec:amu}. 

\subsection{Cross section uncertainties}

\begin{figure}
    \centering
    \includegraphics[width=\linewidth]{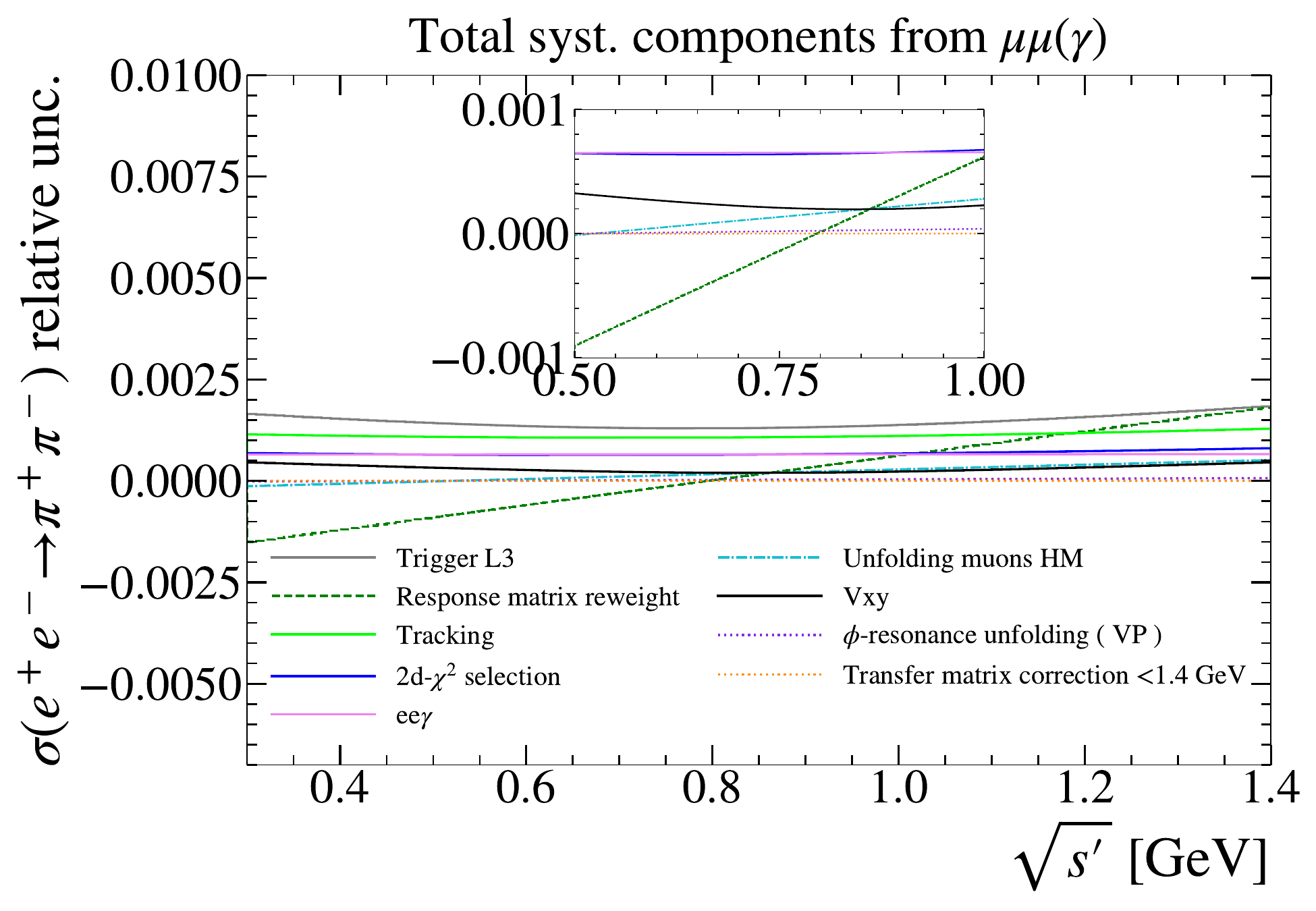}
    \caption{\small \justifying Systematic uncertainty components in $\sigma(e^+e^- \to \pi^+\pi^-)$ from muons. }
    \label{fig:xsec_syst_total_mu}
\end{figure}

\begin{figure}
    \centering
    \includegraphics[width=\linewidth]{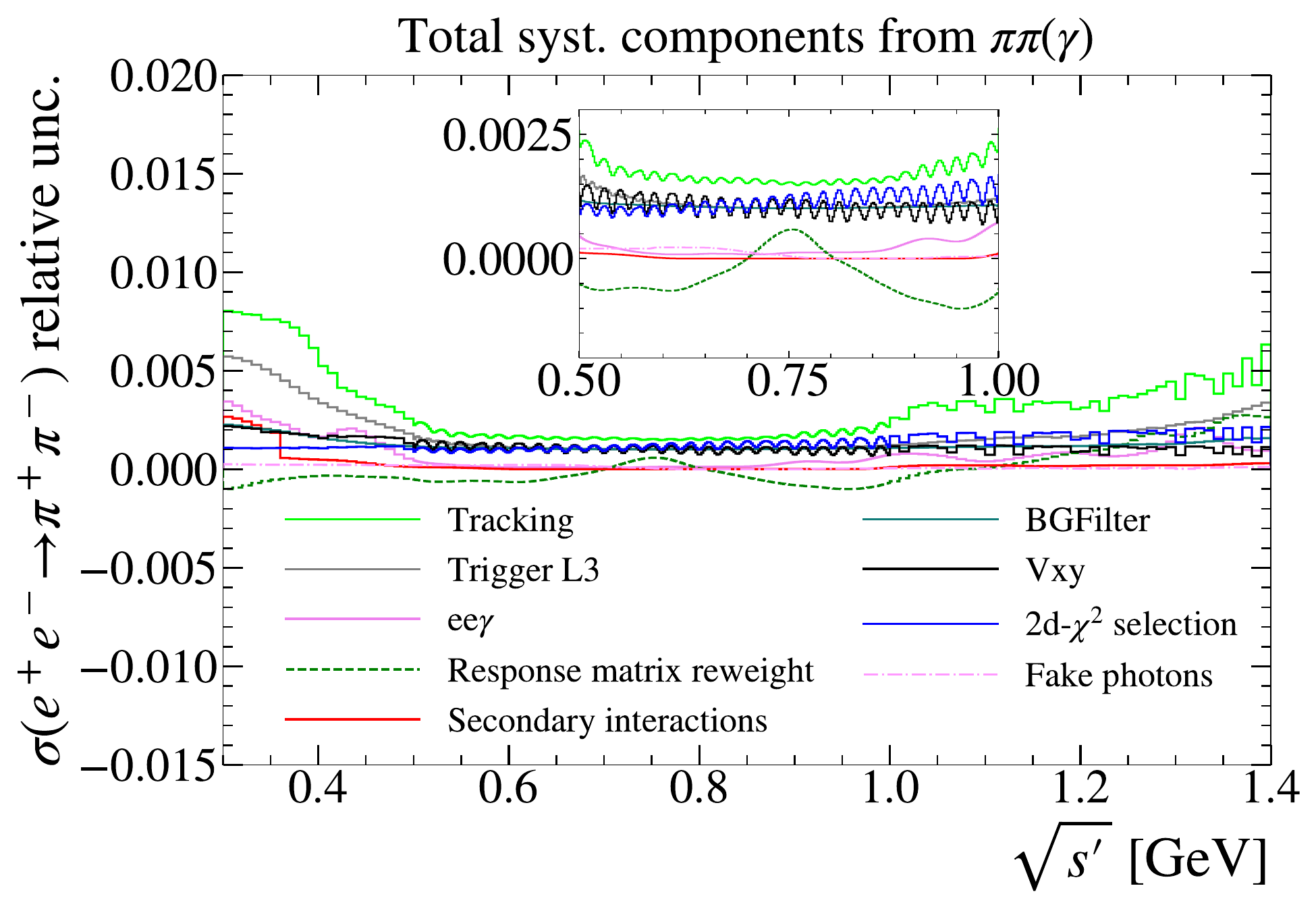}    
    \caption{\small \justifying Systematic uncertainty components in $\sigma(e^+e^- \to \pi^+\pi^-)$ from pions. }
    \label{fig:xsec_syst_total_pi}
    
\end{figure}

\begin{figure}
    \centering
    \includegraphics[width=\linewidth]{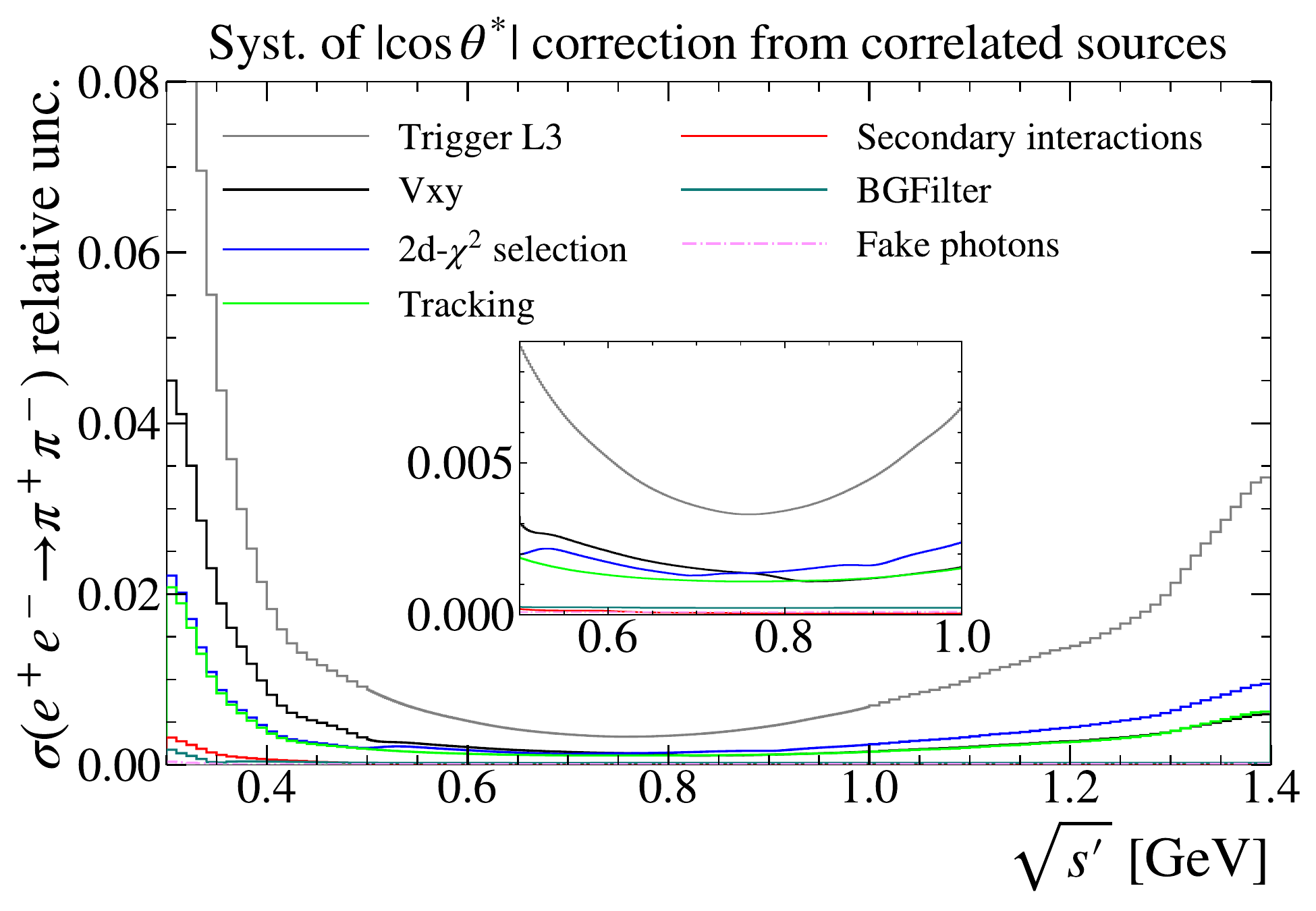}
    \includegraphics[width=\linewidth]{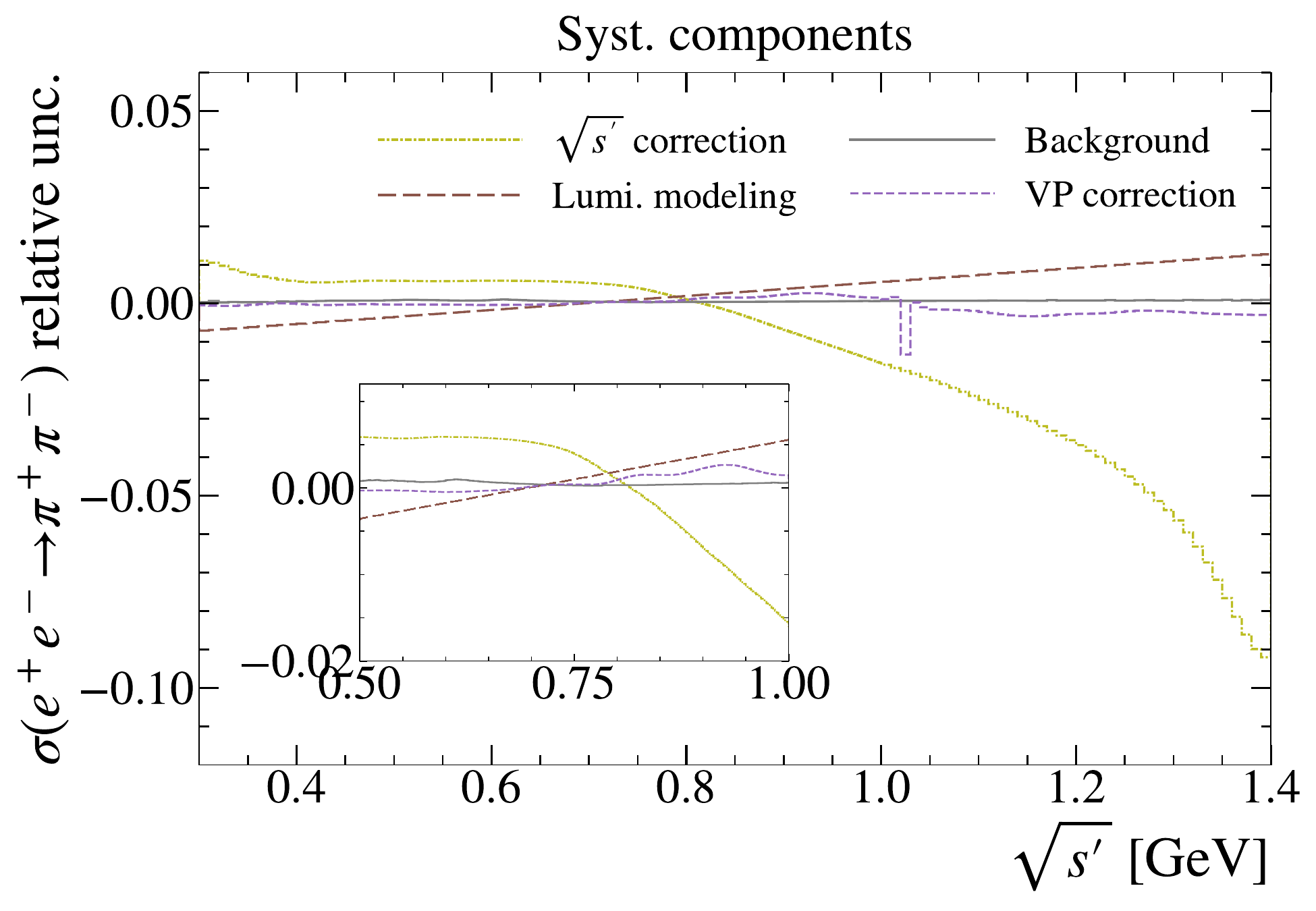}    
    \caption{\small \justifying (top) Systematic uncertainty components in $\sigma(e^+e^- \to \pi^+\pi^-)$ from correlated sources among the $\pi\pi(\gamma)$ and $\mu\mu(\gamma)$ channels and (bottom)~systematic uncertainties from the impact of FSR on the energy definition, as well as uncertainties impacting the luminosity determination. }
    \label{fig:xsec_syst_total_correlated}
\end{figure}

\begin{figure}
    \centering
    \includegraphics[width=\linewidth]{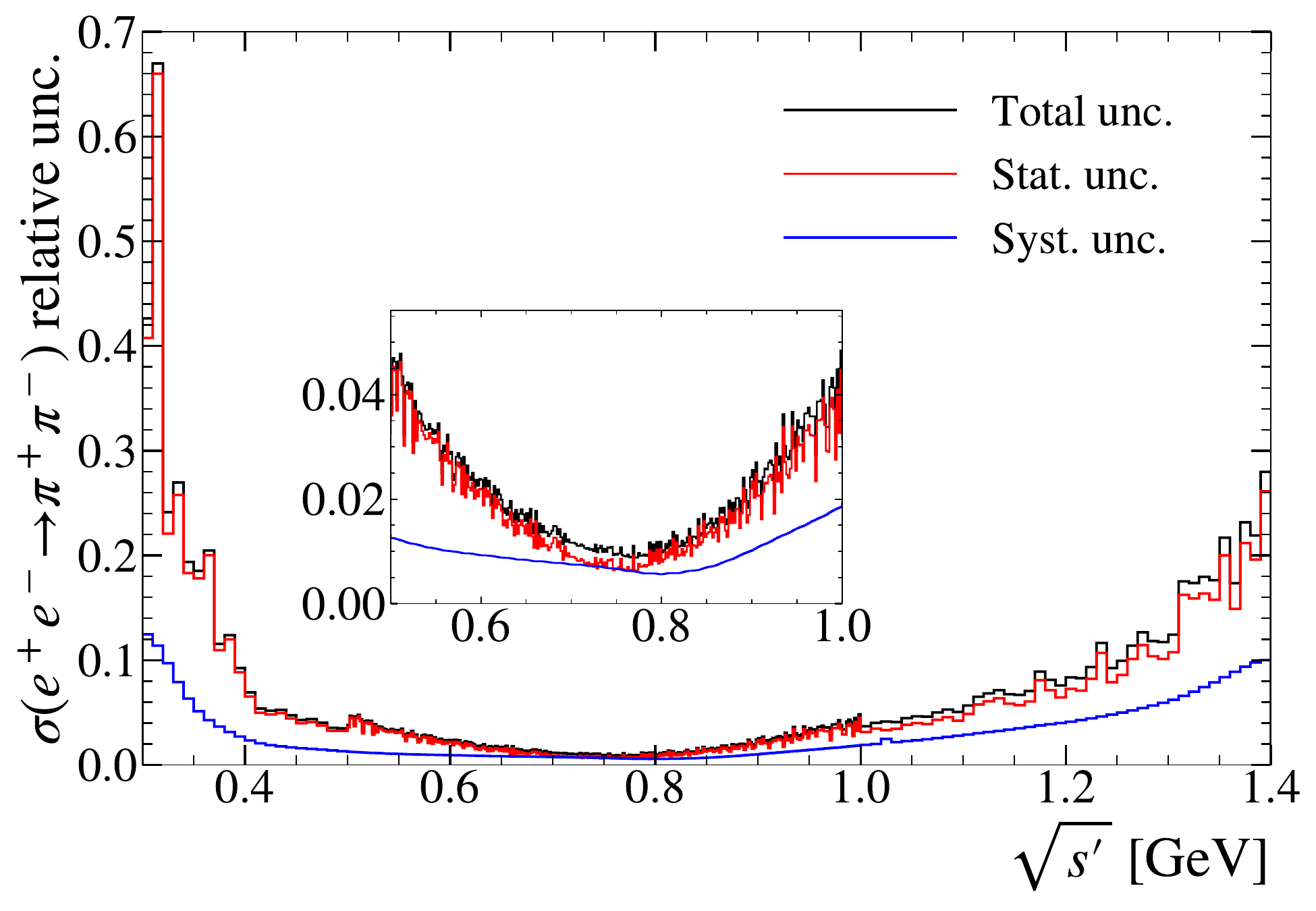}
    \caption{\small \justifying Total, statistical and systematic uncertainty components in $\sigma(e^+e^-\to\pi^+\pi^-)$.}
    \label{fig:total_unc_Xsec}
\end{figure}

The individual uncertainty components from the $\pi\pi(\gamma)$ and $\mu\mu(\gamma)$ spectra discussed above are propagated to the $\sigma(e^+e^- \to \pi^+\pi^-)$ cross section, taking into account the (anti-)correlations among the two.  
They are displayed in Figs.~\ref{fig:xsec_syst_total_mu}--\ref{fig:xsec_syst_total_correlated}.
Figure~\ref{fig:xsec_syst_total_correlated} also shows the VP and modeling uncertainties impacting the determination of the luminosity, while Fig.~\ref{fig:total_unc_Xsec} displays the statistical, systematic, and total uncertainties impacting the cross-section measurements.
Therein, the statistical uncertainties include the contributions from the unfolded $\pi\pi(\gamma)$ and $\mu\mu(\gamma)$ spectra~(the correlations of which are determined through a bootstrap procedure), as well as the uncertainties of statistical nature impacting the corrections applied at the various analysis steps.
In the derivation of the corresponding covariance matrices, each source of systematic uncertainty is treated as fully (anti-)correlated between different phase-space~(i.e.\ energy and \costhpi) regions, while different sources are treated as independent between each other.

\subsection{Quantitative comparison with the 2009 \babar\ measurement}
\label{sec:comp09}

\begin{figure}
    \centering
    \includegraphics[width=\linewidth]{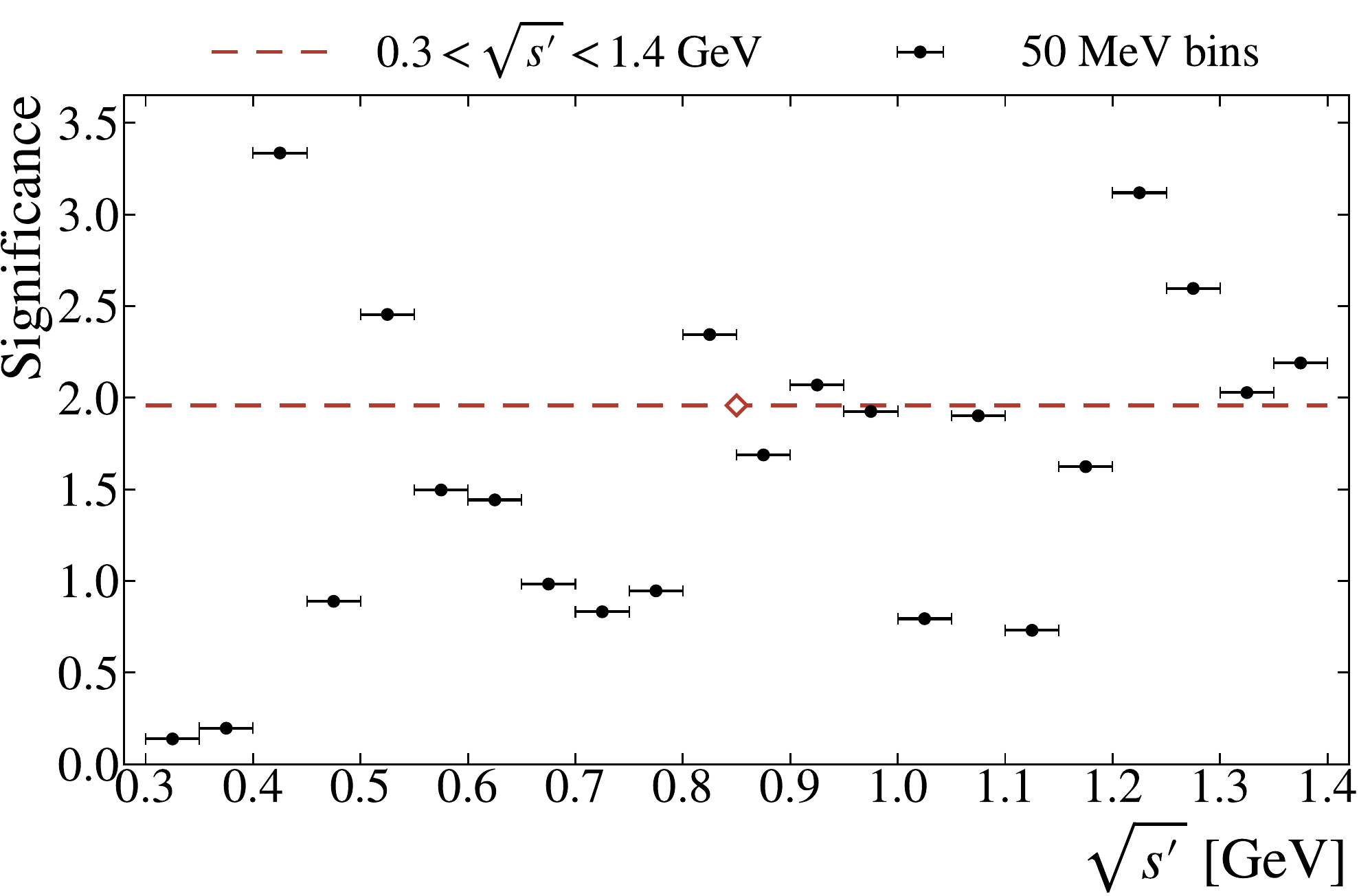}
    \caption{\small \justifying Statistical significance~(expressed in terms of standard deviations) for the difference between dispersion integrals $\mathrm{a}_{\mu}^{\pi\pi}$ computed with the current measurement of the $\sigma(e^+e^-\to\pi^+\pi^-)$ and the 2009 one, taking into account their full set of statistical and systematic uncertainties, with the corresponding correlations between different energy regions. These uncertainties are treated here as independent between the two measurements. }
    \label{fig:Significance_amu_09_26}
\end{figure}

Given the larger data sample employed in the current analysis, the looser track-momentum selection, and different methodology compared to the 2009 \babar\ measurement, one expects the statistical and systematic correlations between the two measurements to be relatively small.
Treating the two measurements as independent between each other, while taking into account the correlations of their corresponding uncertainties between different energy regions, we determine the statistical significance of the difference between the dispersion integrals 
$\mathrm{a}_{\mu}^{\pi\pi}$ (see Section~\ref{sec:amu}) computed either on the full energy range of overlap, or for a scan of $50$~MeV-wide bins~(Fig.~\ref{fig:Significance_amu_09_26}).
Besides a few exceptions, these significances are typically below $2.5$ standard deviations.
Larger fluctuations are observed for a comparison performed through a local $\chi^2$ per number of degrees of freedom, evaluated within the original bins of the measurements~(see Fig.~\ref{fig:LocalWeightsChi2}).

The bootstrap method discussed in Section~\ref{subsubsec:fit_strat} is employed in order to evaluate the statistical correlations between the current measurement and the 2009 \babar\ result~\cite{BaBar:2009wpw,BaBar:2012bdw}.
Bootstrap replicas are propagated for the current analysis through the full analysis chain, from the event-level, to the angular fits and the resulting reconstructed mass spectra, the unfolding, the luminosity evaluation, and the extraction of the cross section.
Coherent bootstrap replicas are also propagated for the 2009 measurement, from the event-level to the reconstructed mass spectra of selected events.
These coherent sets of replicas allow the determination of the correlations among, e.g., the 2009 reconstructed mass spectra and either the similar spectra from the current analysis and/or the resulting unfolded cross section.
The former category of correlations~(among reconstructed mass spectra) are slightly larger than~(although very similar to) the correlations among the 2009 energy spectra and the current cross section.
A relatively weak level of statistical correlation between the two measurements is expected, given the doubled integrated luminosity of the new measurements and its looser track-momentum selection, as well as the very different analysis methodology.
The impact of these correlations at the level of the statistical~(total) covariance matrices of the two cross-section measurements, yields correlation coefficients averaged in bins of $50$~MeV amounting to about $3-5\%$~($2-3\%$) on the $\rho$ peak, increasing to about $6-9\%$~($4-7\%$) in the $0.4-0.5$\gev or $0.85-1.3$\gev ranges, while decreasing fast at lower and higher energies.
On this basis, it was checked that these statistical correlations among the cross sections induce a relative impact on the significance of the difference among corresponding dispersion integrals performed in $50$\mev bins~(Fig.~\ref{fig:Significance_amu_09_26}) of about $1\%$ on the $\rho$ peak, up to $3-4\%$ at intermediate energies, and much less at lower and higher energies.

At the level of the systematic uncertainties, several sources are clearly independent between the two analyses, being either very small or inexistent in one of the two analyses, mainly due to the change of methodology.
For example, the particle-identification uncertainties that were among the largest in the 2009 analysis are not relevant for the current analysis, while the relatively small unfolding uncertainty component evaluated for the 2009 measurement has a negligible impact for the new measurement.
For the luminosity, the fit-related uncertainty introduced in the new measurement is not relevant for the 2009 analysis, where a different approach was employed.
Furthermore, the uncertainty sources related to the effective $\sqrt{s'}$ definition are specific to the new analysis.

Several systematic uncertainty sources, related to the trigger, tracking, background, and the event selection based on kinematic fits are nevertheless potentially correlated between the two analyses.
A challenge in the treatment of these correlations originates from the fact that~(due to the different analysis approach) for the 2009 analysis some sources originating from either the pion or muon channels were merged, while a more detailed split into sub-components has to be propagated for the new analysis.
In cases where this difference in the uncertainty treatment induces some ambiguity~(i.e., an uncertainty in the correlations), a conservative approach is adopted.
Several correlation scenarios are studied for such sources.
The conservative approach retained for the evaluation of the significance of the differences typically consists in correlating one uncertainty source from the 2009 analysis with the largest corresponding sub-component from the new analysis.
In this sense, the trigger efficiency uncertainty from 2009 is correlated with the L3 component of the muon channel impacting the $|{\rm cos}\theta^*_\pi|$ distribution, while the tracking and the kinematic fit uncertainties from 2009 are correlated with the corresponding components of the mass-dependent corrections in the pion channel of the new analysis.
Furthermore, the uncertainties related to the background subtraction are treated as correlated between the two analyses.
It is worth noting that treating all these sources of uncertainty as being pairwise $100\%$ correlated~(as implemented in this study)\footnote{Different sources of uncertainty in a given analysis are treated as independent between each other.} is a conservative choice for the evaluation of the significance of the differences between the (integrals of)~the two measurements. 
Indeed, the fact that different methodologies are employed in the two analyses is expected to induce a de-correlation among these uncertainty components.
Therefore, this scenario including correlations among the systematic uncertainties of the two analyses and the one employed for Fig.~\ref{fig:Significance_amu_09_26}, without such correlations, are expected to bracket the real level of systematic correlations among the analyses.
Furthermore, the difference between these two scenarios reflects the uncertainties in the systematic correlations.

Numerically, it was found that the correlation of the systematic uncertainties discussed above induces a correlation of about $17\%$ for the total uncertainty of $\mathrm{a}_{\mu}^{\pi\pi}$ computed on the full energy range of overlap of the measurements.
Furthermore, for all the significance values displayed in Fig.~\ref{fig:Significance_amu_09_26}, the same correlations induce an enhancement of at most~(generally less than) $0.2$ standard deviations.

These findings show that the statistical significances indicated in Fig.~\ref{fig:Significance_amu_09_26} are relatively stable with respect to both statistical and systematic correlations among the measurements.

\subsection{Pion form factor fits}
\label{sec:fffit}

\begin{figure}[bt]
    \centering
    \includegraphics[width=\linewidth]{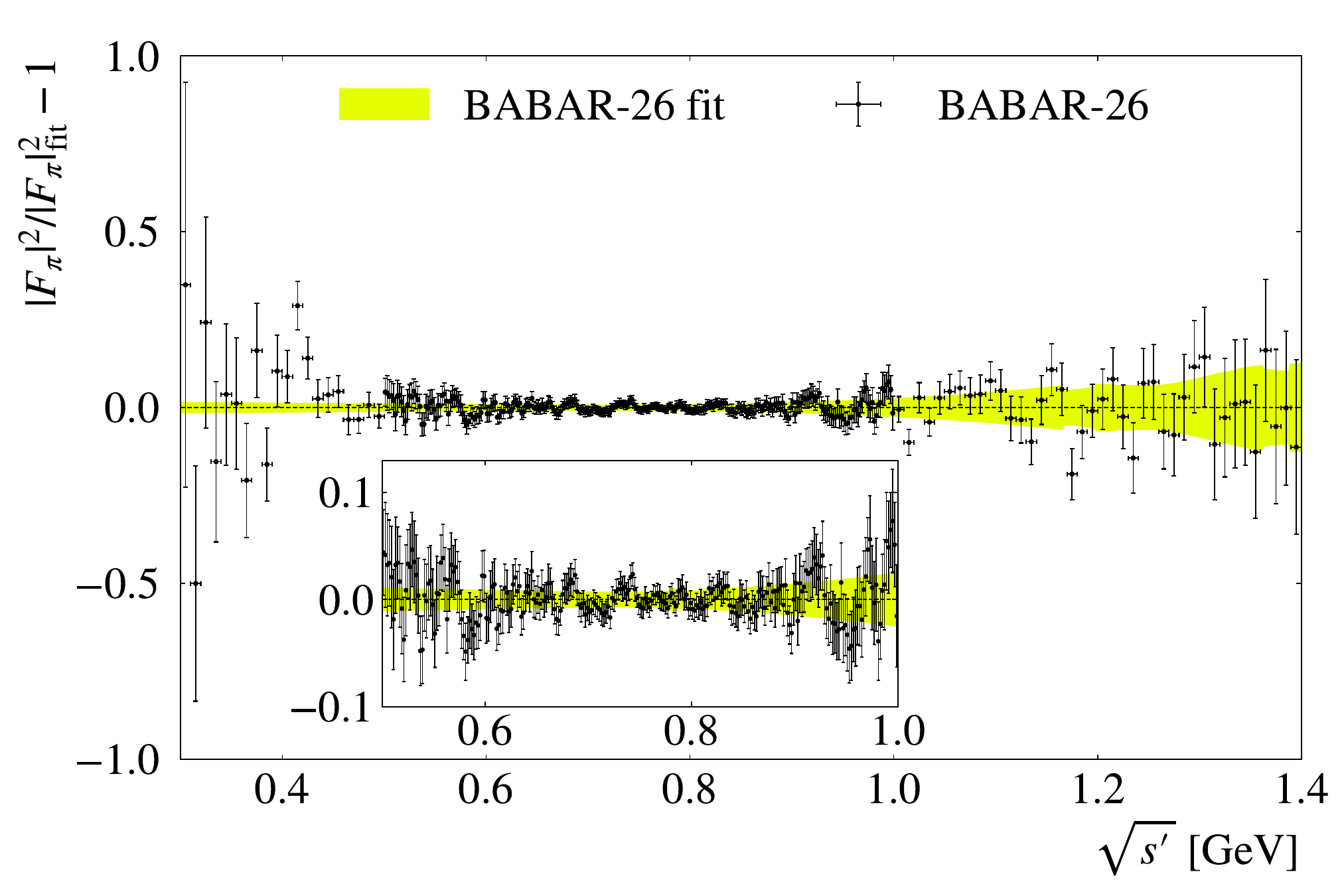}
    \caption{\small \justifying Relative differences of the pion form factors squared obtained via the fit using \babar\ 2026 results. The yellow band corresponds to the uncertainty in the fit.}
    \label{fig:fit_results_babar26}
\end{figure}

\begin{figure}
    \centering
    \includegraphics[width=\linewidth]{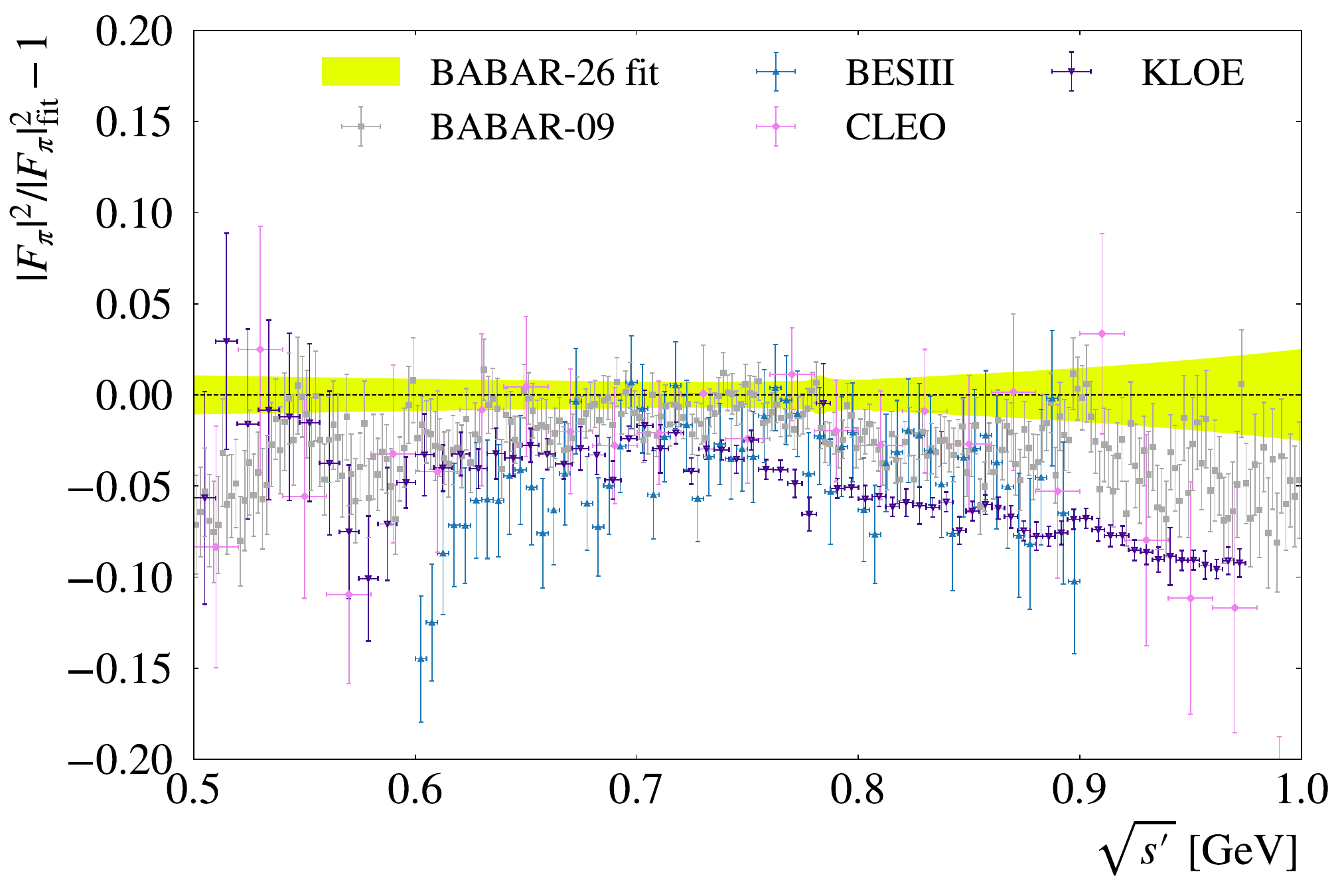}
    \includegraphics[width=\linewidth]{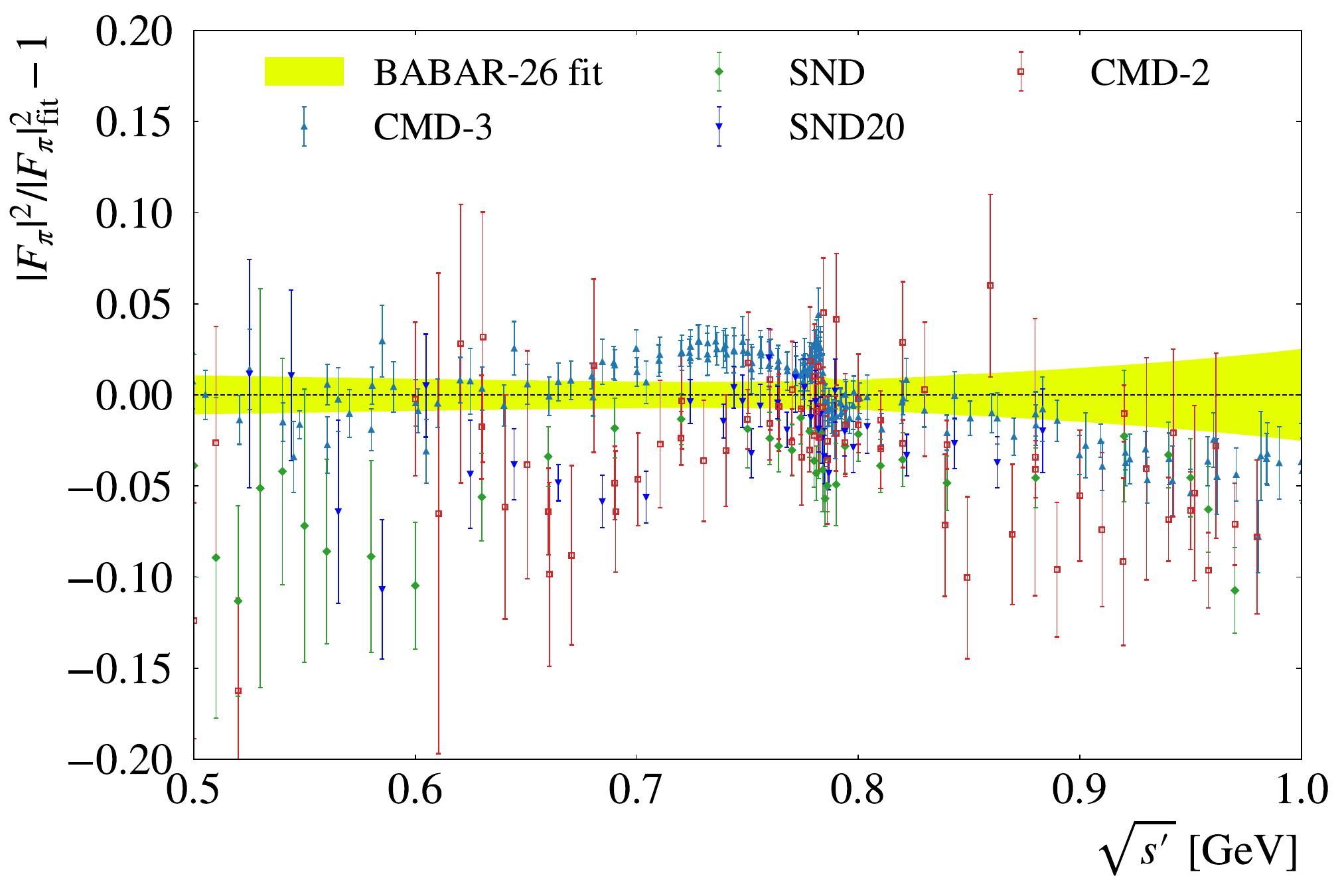}
    \caption{\small \justifying Relative differences of the pion form factors squared with respect to other ISR measurements (top) and beam-energy scan measurements (bottom). The yellow band corresponds to the uncertainty of the \babar-26 fit.}
    \label{fig:fit_comp}
\end{figure}

The measured $e^+e^-\rightarrow\pi^+\pi^-(\gamma)$ cross section can be expressed in terms of the pion electromagnetic form factor, which parametrizes the coupling of the virtual photon to the pion pair.
The pion form factor used to fit the cross section is defined in Ref.~\cite{BaBar:2012bdw}.
The fit contains eleven free parameters: the mass and width of $\rho$, $\omega$ and $\rho^\prime$, the relative amplitude and phase of the latter two with respect to the dominant $\rho$ resonance, and a global normalization factor. 
The other parameters, corresponding to high energy resonances, are fixed to those of the fit to the \babar\ 2009 data up to 3\gev.
The main purpose of the fit here is not really to determine the parameters of the model but to facilitate comparisons between different measurements. 
The resulting parameterization provides an excellent description of the measured form factor over the full energy range (Fig.~\ref{fig:fit_results_babar26}) that can be directly compared with previous measurements~\cite{BaBar:2009wpw,BaBar:2012bdw,BESIII:2015equ,Xiao:2017dqv,CMD-2:2003gqi,CMD-2:2005mvb,Aulchenko:2006dxz,CMD-2:2006gxt,CMD-3:2023alj,KLOE:2008fmq,KLOE:2010qei,KLOE:2012anl,Achasov:2006vp,SND:2020nwa} as shown in Fig.~\ref{fig:fit_comp}.
The \babar\ 2026 measurement is in general higher than the other measurements except for CMD-3 at around 0.73\gev where the latter is even higher by about 2.5\%. 

\section{Combination of \babar\ results}
\label{sec:combination}

\begin{figure}[tb]
    \centering
    \includegraphics[width=\linewidth]{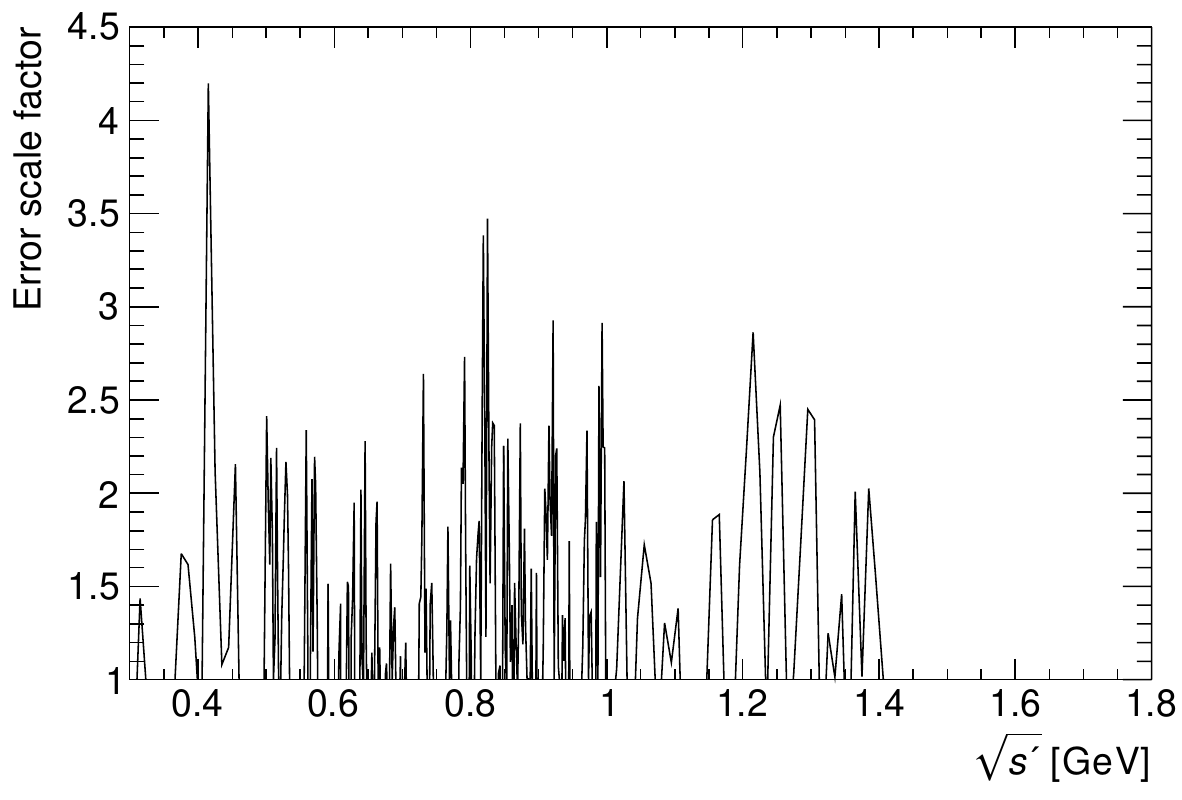}
    \includegraphics[width=\linewidth]{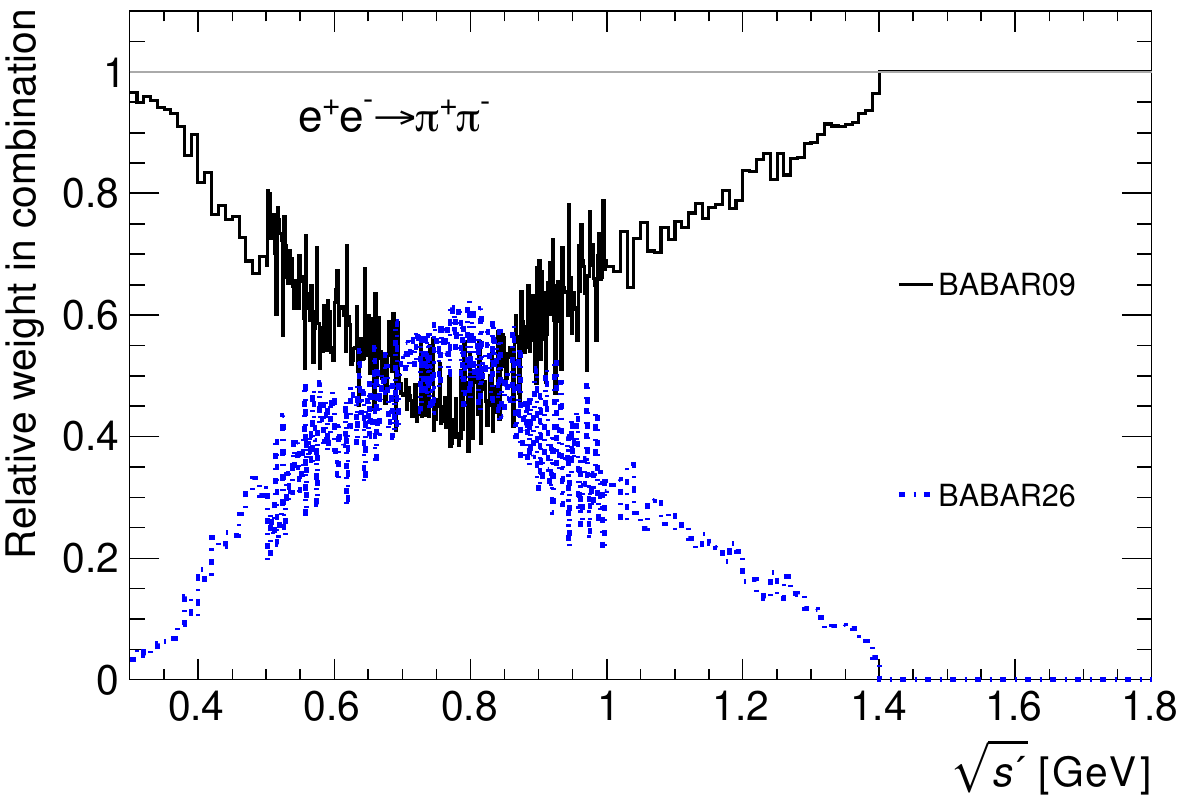}
    \caption{\small \justifying Value of the $\sqrt{\chi^2/{\rm ndf}}$ in the energy bins where this quantity is larger than unity~(top), and weights of the 2009 and new \babar\ measurements in the combination~(bottom), evaluated treating the two \babar\ measurements as systematically correlated.}
    \label{fig:LocalWeightsChi2}
\end{figure}

The combination of the new \babar\ measurements with those of 2009 in the $2\pi$ channel is performed employing a simplified version of the methodology developed in Refs.~\cite{Davier:2010fmf,Davier:2010rnx}.
The weights given to the input cross sections in the averages are derived on the basis of a $\chi^2$ minimization performed within each common bin of the measurements\footnote{This $\chi^2$ minimization can indeed be expressed as a weighted average of the input measurements.}, which allows quantifying their local level of agreement (see Fig.~\ref{fig:LocalWeightsChi2}).
The two measurements have similar weights on the $\rho$ peak region, while the 2009 measurement is more precise at lower or higher energies and has hence larger weights.
The relative weights of the two measurements have rather little sensitivity to the treatment of the systematic correlations among them (see discussion in Section~\ref{sec:comp09}).\footnote{On the basis of the methodology developed in Refs.~\cite{Davier:2010fmf,Davier:2010rnx}, we studied the option of having the average weights derived through a local $\chi^2$ minimization within the variable binning of the measurements (also used for the integrals and the combination), but performing an interpolation among larger intermediate bins of $0.1$~GeV filled on the basis of each measurement.
Such approach enables employing in the weight-derivation the information on intermediate-range correlations among the bins of every measurement (while avoiding shifts of systematic uncertainties that global $\chi^2$ fits would implicitly induce by over-exploiting the information of uncertainty correlations). 
While such change in the procedure impacts somewhat the determination of the weights (as expected), little impact was induced for the local $\chi^2/\mathrm{ndf}$, the integrals for the combined spectrum and their uncertainties.}

The statistical and systematic uncertainties, together with their correlations, are propagated from the input measurements to the combined spectrum, employing a set of $10^4$ pseudo-experiments.
They allow derivation of the covariance matrix of the combined spectrum.

\begin{figure}[tb]
    \centering
    \includegraphics[width=\linewidth]{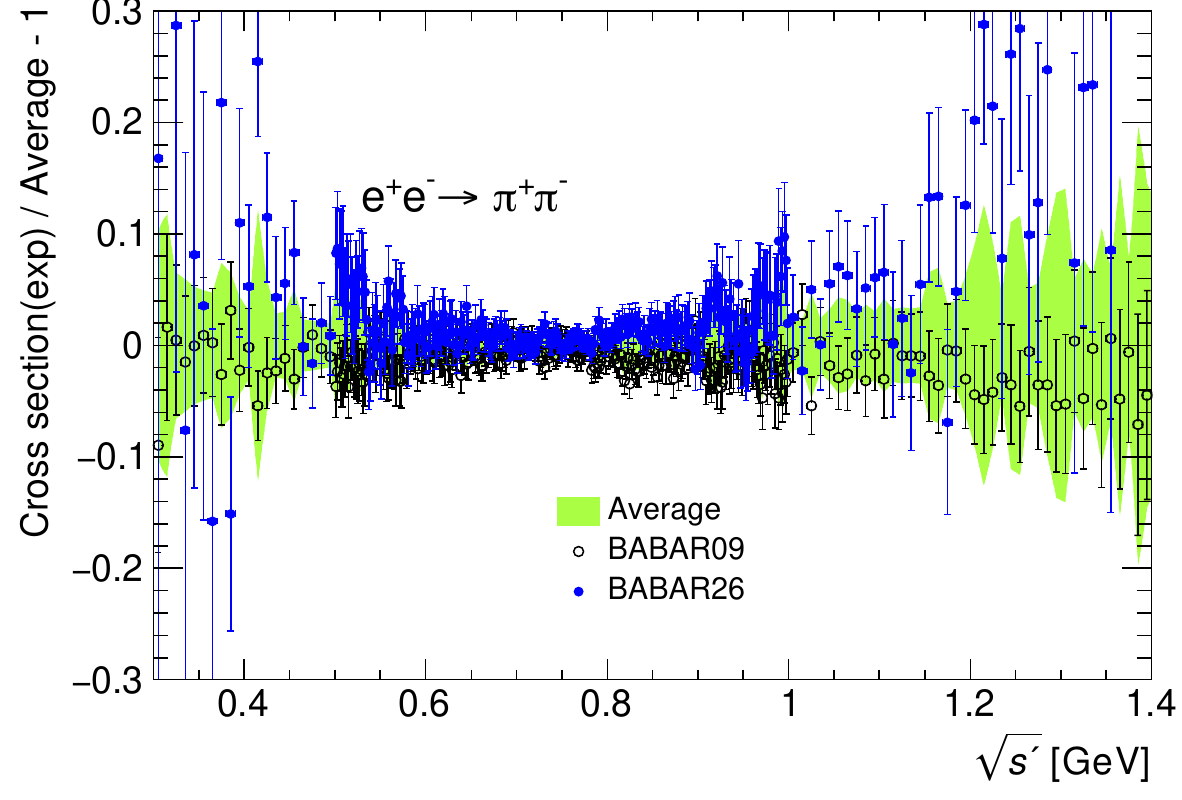}
    \includegraphics[width=\linewidth]{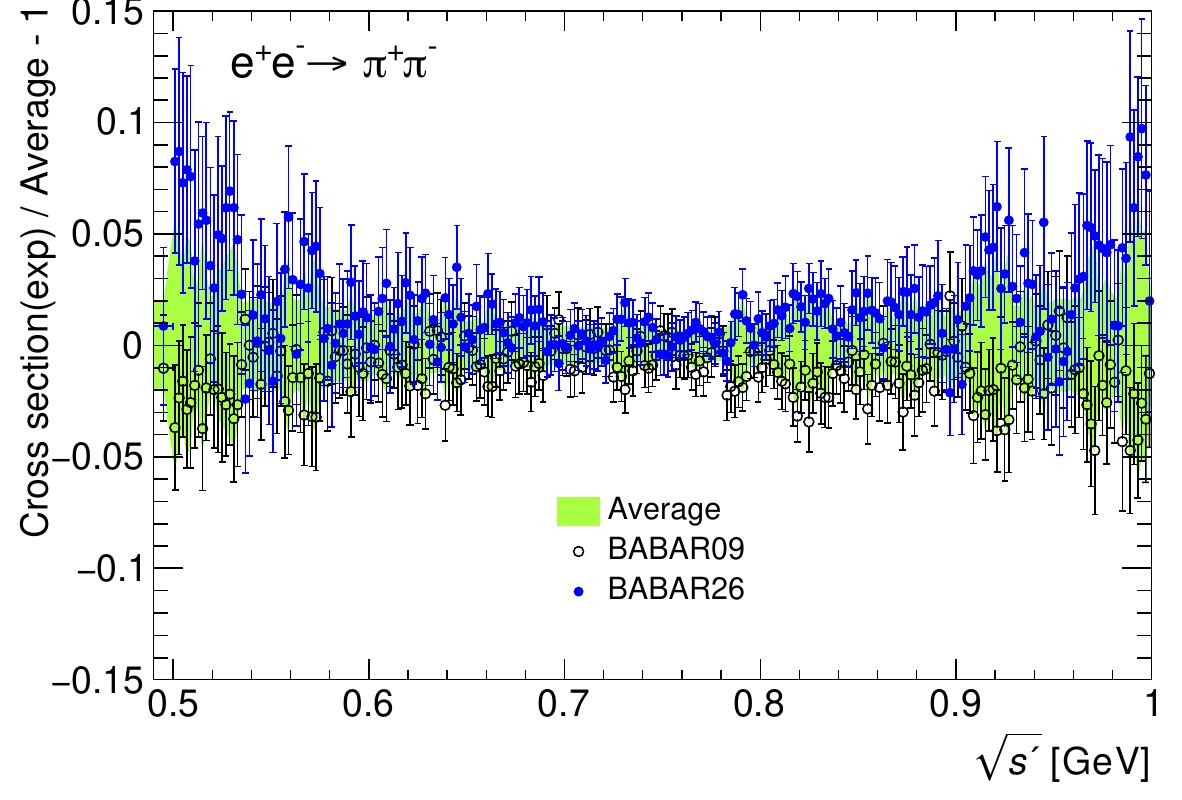}
    \caption{\small \justifying Relative difference among the 2009 and new \babar\ measurements and their combination derived treating the two measurements as systematically correlated, for the full energy range of overlap~(top) and a zoom on the $\rho$ peak region~(bottom).}
    \label{fig:RelativeDifferenceCombination}
\end{figure}

Figure~\ref{fig:RelativeDifferenceCombination} shows the relative difference between the 2009 and new \babar\ measurements and their combination, derived treating the two measurements as systematically correlated.
In the energy bins where the local $\sqrt{\chi^2/\mathrm{ndf}}$ is larger than unity~(see Fig.~\ref{fig:LocalWeightsChi2}), the uncertainty in the combination is enhanced by the corresponding factor, following the recommendation from the PDG~\cite{ParticleDataGroup:2026mpi}.
The same uncertainty enhancement is applied to the covariance matrix of the combined spectrum and hence the uncertainties of the dispersion integrals.

\section{The $\pi\pi$ contribution to the anomalous magnetic moment of the muon}
\label{sec:amu}

The leading-order HVP contribution of the $\pi\pi$ channel to the muon magnetic anomaly, $a_{\mu}^{\pi\pi}$, is evaluated using a dispersion integral over this 
\babar\ measurement of the $\pi\pi(\gamma)$ cross section.

For the full energy-range of the measurement, between $0.3$ and $1.4\gev$, the $a_{\mu}^{\pi\pi}$ integral has a statistical precision of $0.67\%$.
As expected, the dominant contribution to this statistical uncertainty originates from the low- and high-energy region where the overwhelming dominance of the $\mu\mu$ process renders the statistical separation of the $\pi\pi$ contribution very difficult.
Indeed, the integral between $0.3$ and $0.5$\gev (1.0--1.4\gev) has a statistical uncertainty of $2.7\%$ (1.5\%), while that between 0.5 and 1.0\gev is 0.57\%.
An anti-correlation between the statistical uncertainties of these energy regions is also induced by the migrations accounted for through the unfolding, while positive correlations are typically induced by the fit of the $\mumu$ data/\textsc{Phokhara} ratio employed for the determination of the luminosity.

The total systematic uncertainty amounts to $0.72\%$ for the full energy range and to $2.4\%$, 0.56\%, and 3.2\% in the 0.3--0.5\gev, 0.5--1.0\gev, and 1.0--1.4\gev energy regions, respectively. 
It includes the systematic uncertainty originating from the choice of the functional form used for the fit of the data/\textsc{Phokhara} ratio employed in the determination of the ISR luminosity, which amounts to $0.15$ (in units of $10^{-10}$) for the full energy range and to $-0.30$, $0.24$, and 0.08 in the 0.3--0.5\gev, 0.5--1.0, and 1.0--1.4\gev energy regions (with an anti-correlation among them).
The smoothing of the systematic uncertainties mentioned above, aiming to mitigate statistical fluctuations in their evaluations, is also applied for all the uncertainties employed here.

These uncertainties are to be compared with those of the 2009 analysis as shown in Table~\ref{tab:damu-comp}.
The two analyses employing different methods are thus complementary. 
It is recalled that in the 2009 analysis the $a_{\mu}^{\pi\pi}$ contributions were found to be 57.6, 445.9, and 9.7 in the corresponding regions.
These values are to be compared with the ones of the new 2026 analysis, amounting to 60.7, 454.6, and 10.6 in the same regions.
The combined 2009 and 2026 measurement, after taking into account their correlation and enhancing the uncertainty when the local $\chi^2$ value is greater than unity, gives 58.2, 449.7, and 9.9 in these energy regions.

\begin{table}[tbp]
\centering
\caption{\small \justifying Comparison of the value and uncertainties of $a_{\mu}^{\pi\pi}$~(units of $10^{-10}$) in this analysis with the previous \babar\ 2009 result. The first and second uncertainties are statistical and systematic. The latter includes a systematic uncertainty of 0.3, 0.24, and 0.08 respectively, for each of the energy intervals, corresponding to the difference between two types of fits employed for the evaluation of the luminosity~(see text).}
  \label{tab:damu-comp}
\resizebox{\columnwidth}{!}{%
\begin{tabular}{cccc}
\hline
Energy range  & Present  & BABAR & Combination \\
 (GeV) & result & 2009 & 2009+2026 \\
\hline
0.3--0.5 & $60.7(1.7)(1.4)$ & $57.6(0.63)(0.55)$ & $58.2(0.86)(0.78)$ \\
0.5--1.0 & $454.6(2.6)(2.5)$ & $445.9(2.1)(2.5)$ & $449.7(2.1)(2.5)$ \\
1.0--1.4 & $10.6(0.16)(0.34)$ & $9.7(0.10)(0.08)$ & $9.9(0.12)(0.14)$ \\\hline
Total & $525.9(3.5)(3.8)$ & $513.3(2.2)(3.1)$ & $517.8(2.4)(3.3)$ \\
\hline
\end{tabular}}
\end{table}

\begin{figure}
    \centering
    \includegraphics[width=\linewidth]{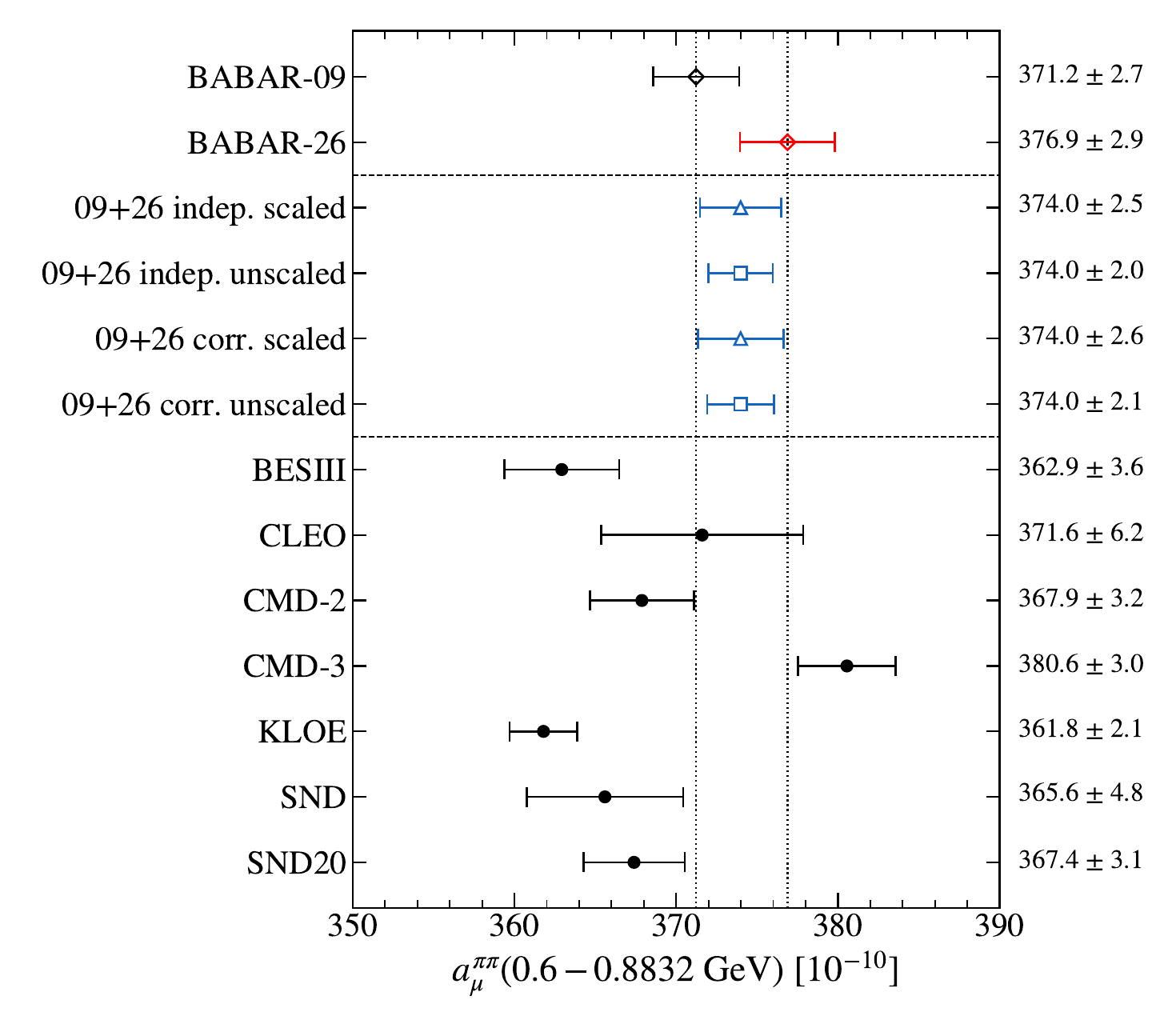}
    \caption{\small \justifying Comparison of the leading-order hadronic vacuum polarization contribution, $a_{\mu}^{\pi\pi}$, from $e^{+}e^{-}\!\to\!\pi^{+}\pi^{-}$ data in the $0.6$--$0.8832\gev$ energy range. The error bars denote the quoted experimental uncertainties.}
    \label{fig:amu_integrals}
\end{figure}

In the common region of measurement, which is from 0.6 to 0.8832\gev,
with the lower and upper limits determined by the BESIII~\cite{BESIII:2015equ} and SND20~\cite{SND:2020nwa} results, respectively, the new 2026 \babar\ measurement is also compared in Fig.~\ref{fig:amu_integrals} with other recent measurements~\cite{BaBar:2009wpw,BaBar:2012bdw,BESIII:2015equ,Xiao:2017dqv,CMD-2:2003gqi,CMD-2:2005mvb,Aulchenko:2006dxz,CMD-2:2006gxt,CMD-3:2023alj,KLOE:2008fmq,KLOE:2010qei,KLOE:2012anl,Achasov:2006vp,SND:2020nwa} and is found to lie between those of \babar\ 2009 and CMD-3 with a difference of $+1.4\sigma$ and $-0.9\sigma$, respectively.
In Fig.~\ref{fig:amu_integrals}, the combined 2009 and 2026 \babar\ measurements are also shown with different options including or excluding the correlation, and with or without enhancing the uncertainties. While there is essentially no difference in the central values, the uncertainties differ significantly from one option to another. The most conservative option with the correlation and uncertainty scaling is selected as our default choice and is shown in Table~\ref{tab:damu-comp}.
In the energy region of the CMD-3 measurement (0.32698--1.199168\gev)~\cite{CMD-3:2023alj}, the \babar\ 2026 and the default 2009+2026 combination measurements yield 520.7(3.3)(3.5) and 513.0(2.3)(3.3), respectively, to be compared with 520.3(0.52)(4.1) of CMD-3 in units of $10^{-10}$, where the first and second uncertainties are statistical and systematic.
\section{Conclusion}

The full \babar\ data sample on and just below the $\Upsilon(4S)$ resonance, corresponding to an integrated luminosity of 460\,fb$^{-1}$, is analyzed using the angular distributions, rather than particle identification (PID), to separate the $\pi\pi$ and $\mu\mu$ final states.
The analysis is thus complementary to the 2009, PID-based version. 
In both analyses, all relevant efficiencies are measured using data control samples. This new analysis features twice as much data, a lower momentum requirement on the charged particles, and improved treatment of higher-order photon radiation. 

The analysis was blinded using three different offset factors for each of the two processes. After many detailed checks, the $\mu\mu$ cross section was unblinded and found to be consistent with the QED prediction, with a fitted ratio of 
$0.9920\pm 0.0030_\mathrm{stat}\pm 0.0031_\mathrm{syst} \pm 0.0033_{\gamma\, \mathrm{ISR}}\pm 0.0043_{\mathrm{lumi}\, ee}$, 
where the uncertainties correspond to the statistical, systematic related to angular fits, photon efficiency correction, and luminosity uncertainties.
This measurement is converted into an equivalent ISR luminosity, used in the calculation of the $\pi\pi$ cross section.

Following the successful QED test for the $\mu\mu$ process, the unblinded energy spectrum of the $\pi\pi$ process in the energy range from threshold to 1.4\gev is obtained.
Using the new $\pi\pi(\gamma)$ cross section, the leading-order hadronic contribution to the muon $g-2$ in the energy range of 0.3--1.4\gev is obtained with the value of $525.8(3.5)(3.8)\times 10^{-10}$ where the first and second uncertainties are statistical and systematic, respectively. This is to be compared with the corresponding result of the 2009 \babar\ measurement of $513.3(2.2)(3.1)\times 10^{-10}$. The larger uncertainties are mainly due to the intrinsic limitation of the new method in the low and high energy regions where the $\pi\pi$ event yields are much lower than those of the $\mu\mu$ process.
The two measurements have been compared quantitatively, both in restricted energy ranges and more globally, on their full range of overlap.
The differences between the two measurements are found to be typically below $2.5$ standard deviations, with relatively little sensitivity to the uncertainties impacting the statistical and systematic correlations among the measurements. The two measurements are thus combined, giving $517.8(2.4)(3.3)\times 10^{-10}$ in the common energy range of 0.3--1.4\gev. 
\\

\section*{Acknowledgements}
We are grateful for the extraordinary contributions of our PEP-II colleagues in achieving the excellent luminosity and machine conditions that have made this work possible. The success of this project also relies critically on the expertise and dedication of the computing organizations that support BABAR, including GridKa, UVic HEP-RC, CC-IN2P3, and CERN. The collaborating institutions wish to thank SLAC for its support and the kind hospitality extended to them. We also wish to acknowledge the important contributions of J.~Dorfan and our deceased colleagues E.~Gabathuler, W.~Innes, D.W.G.S.~Leith, A.~Onuchin, G.~Piredda, and R. F.~Schwitters. This work is supported by ANR-22-CE31-0011.

\section*{Data Availability}

The data that support the findings of this article are not
publicly available. The data are available from the authors
upon reasonable request.

\newpage

\appendix
\section{Main changes since the preliminary 2025 results}
\label{app:changes}

In this appendix, the main changes with respect to the preliminary results presented in 2025 are presented.
The most significant methodological change concerns the analysis strategy, employing here $|\cos\theta_\pi^\ast|$ template fits performed only in $\pi\pi$ mass bins.
This avoids a large sensitivity on the resulting measured cross section to the shape of the pion form factor employed in the Monte Carlo simulation, which would have otherwise required a large enhancement of the uncertainties impacting the 2025 result.
On the more technical side, an important change was made regarding from the determination of the data/MC pion tracking efficiency corrections, where a bookkeeping problem in 2025 prevented the use of the full event statistics while at the same time introducing a bias. 
Because of these changes all the new studies in the analysis were carried out with new blinding factors, which were released only in the final analysis stage. It turned out that the updated results presented in this paper manifest significant differences with respect to the incorrect preliminary ones, which are therefore superseded.

The main changes are summarized in the following.
\begin{description}
    \item[Dressed pion form factor changed to bare one] This change is discussed in Section~\ref{subsec:vp-correction}. The impact of the correction on the $\pi\pi$ mass spectrum is negligible. The impact on the $\mu\mu$ spectrum is slightly larger and remains very small ($-0.5\%$) and localized at the $\rho$-$\omega$ interference mass. However, the small impact on the latter is achieved only when the template fit strategy is changed from using the templates under the muon mass hypothesis to performing the template fit under the pion mass hypothesis and then doing a mass transformation from the pion to muon mass to get the $\mu\mu$ mass spectrum. Had we not made this change, the impact on the $\mu\mu$ mass spectrum would have inflated to $+11\%$ and $-4\%$ around the $\rho$-$\omega$ interference region.
    \item[On the pion form factor parametrization] For the final 2026 results, we have studied potential uncertainties related to the choice of the form factor used in producing the pion \textsc{Phokhara} MC samples by varying the default choice based on the \babar\ 2009 result to a CMD-3-like form factor (Section~\ref{sec:CMD3-BaBar}). The resulting uncertainty in the $\pi\pi$ mass spectrum remains negligibly small and this is also true in the $\mu\mu$ mass spectrum. The uncertainty in the latter would have been over 2\% with the 2025 muon fitting strategy.
    \item[Changed muon fitting strategy] The necessity for changing the template fit strategy for obtaining the $\mu\mu$ mass spectrum has already been explained above (Section~\ref{subsec:mumu_strat}). Without such a change, the corresponding uncertainty would have been too large and would have invalidated the use of the template method for a new $\pi\pi$ cross-section measurement.
    \item[Pion tracking efficiency corrections] This is by far the largest technical fix, responsible for about 1.6\% of the 2\% increase of the two-pion cross-section result in the dominant $\rho$ peak region (Section~\ref{subsec:tracking_pions}). The change is due mainly to a bookkeeping error in 2025 in the handling of the final histograms. Also the selection for the pion tracking sample is loosened compared to 2025 in order to increase the statistical precision while
    at the same time removing a bias introduced by the 2025 tighter cuts. The changes concern two components: one is the shape correction of the pion template and the other the mass-dependent correction. The shape change is shown in Fig.~\ref{fig:track_pi_template}. It affects mainly the high $|\cos\theta_\pi^\ast|$ region where it is the most sensitive to the $\pi\pi$ and $\mu\mu$ separation. The impact of the shape change on the $\pi\pi$ and $\mu\mu$ mass spectra after the template fit is shown in Fig.~\ref{fig:impact_tracking_template}. It increases the $\pi\pi$ event yield by about 0.5\% fairly independently of the mass over the full range. On the other hand, it decreases the $\mu\mu$ event yield by about $-3\%$ at 0.7\gevcc. The latter in turn affects the data/MC ratio for the ISR luminosity determination such that a linear function fits the ratio best in 2026, compared to a quadratic function in 2025.
    The resulting comparison is shown in Fig.~\ref{fig:lumi_comp}. This new ISR luminosity determination alone increases the $\pi\pi$ cross section by $+0.8\%$ around 0.8\gev.
    The second component of the change is the pion mass-dependent correction. The comparison of the 2025 and 2026 tracking corrections is shown in 
     Fig.~\ref{fig:tracking_mass}. The impact is mass dependent and it increases the $\pi\pi$ event yield by about $+0.3\%$ in the $\rho$ peak region. 
    In total, fixing the erroneous tracking efficiency used in 2025 results in an increase of $+1.6\%$ in the two-pion cross section in the dominant $\rho$ peak region.
\begin{figure}[tb]
    \centering
    \includegraphics[width=\linewidth]{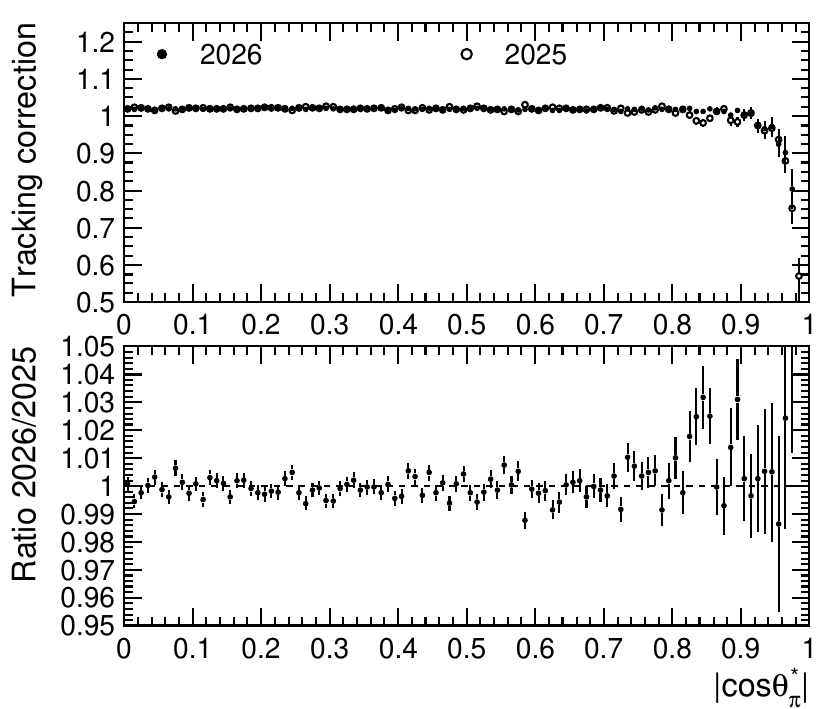}
    \caption{\small \justifying Shape comparison between the pion tracking template correction used for the preliminary 2025 results and the final 2026 one.}
    \label{fig:track_pi_template}
\end{figure}
\begin{figure}[tb]
    \centering
    \includegraphics[width=\linewidth]{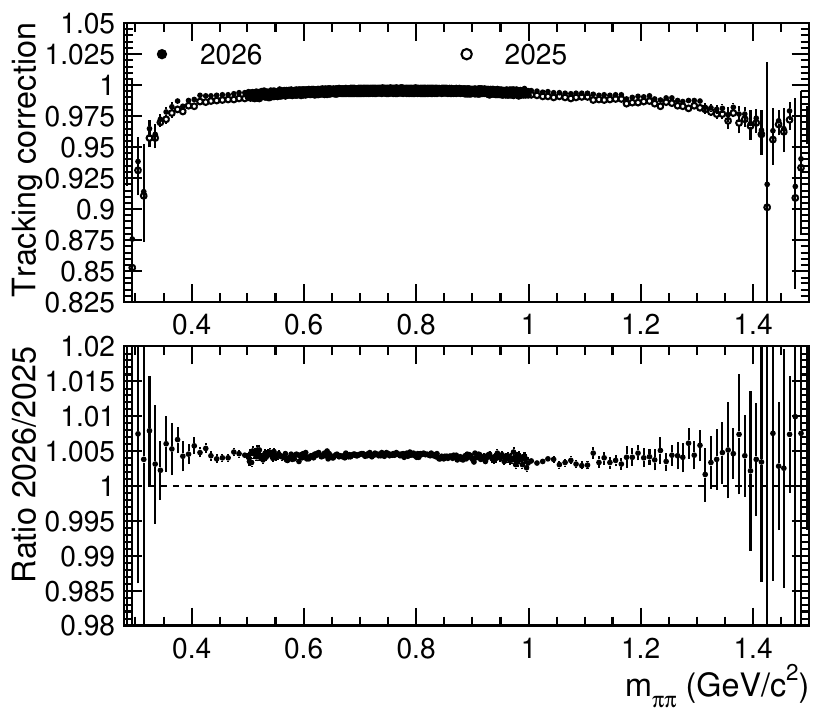}
    \includegraphics[width=\linewidth]{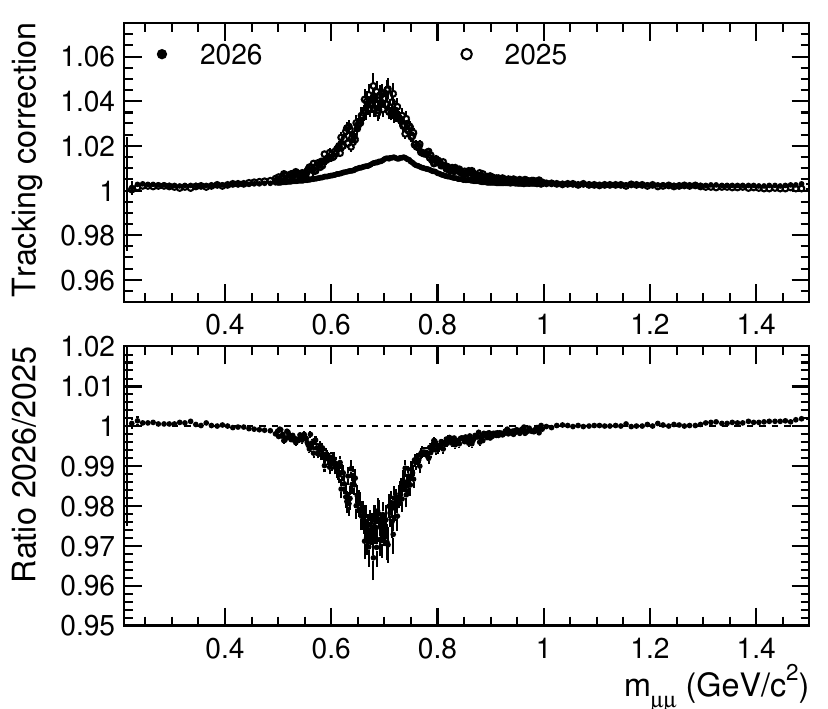}
    \caption{\small \justifying Comparison of the impact of the pion tracking template corrections (2025 vs 2026) on the fitted $\pi\pi$ (top) and $\mu\mu$ (bottom) mass spectra.}
    \label{fig:impact_tracking_template}
\end{figure}
\begin{figure}[tb]
    \centering
    \includegraphics[width=\linewidth]{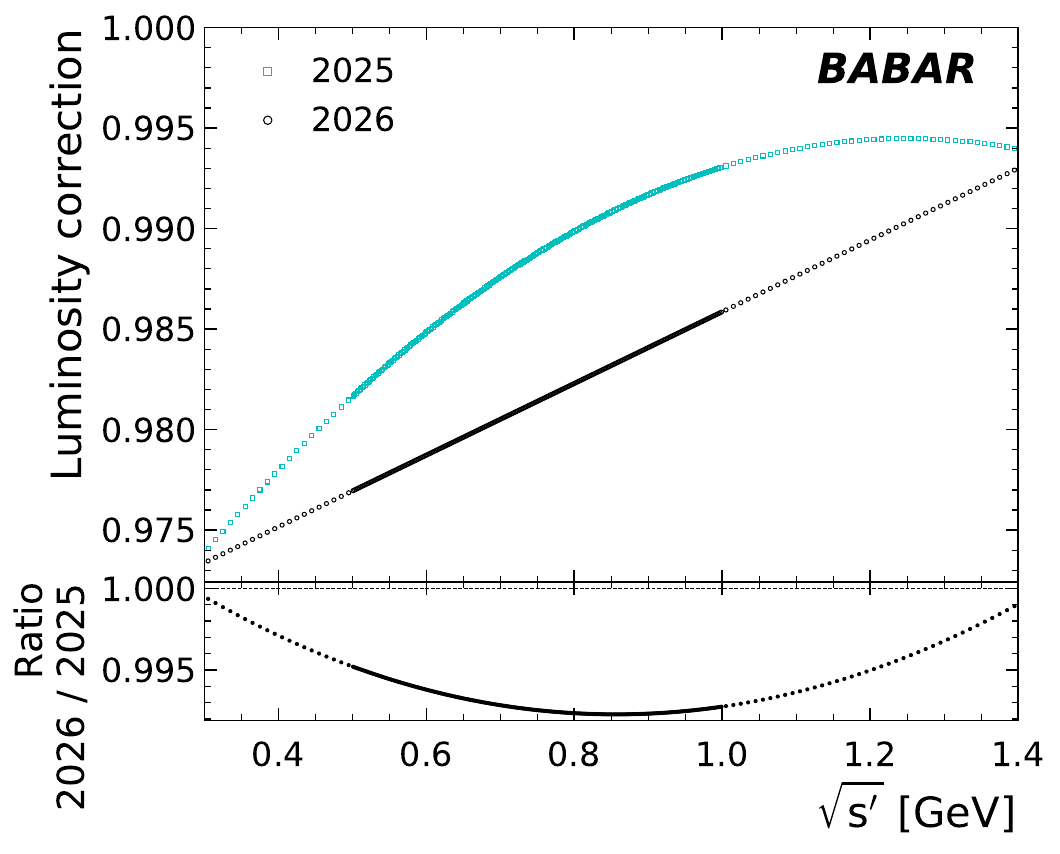}
    \caption{\small \justifying Impact of the modified pion template corrections on the choice of functional forms for the fitted data/\textsc{Phokhara} correction $f_{\mu\mu}(\sqrt{s'})$ to the effective ISR luminosity.}
    \label{fig:lumi_comp}
\end{figure}
\begin{figure}[tb]
    \centering
    \includegraphics[width=\linewidth]{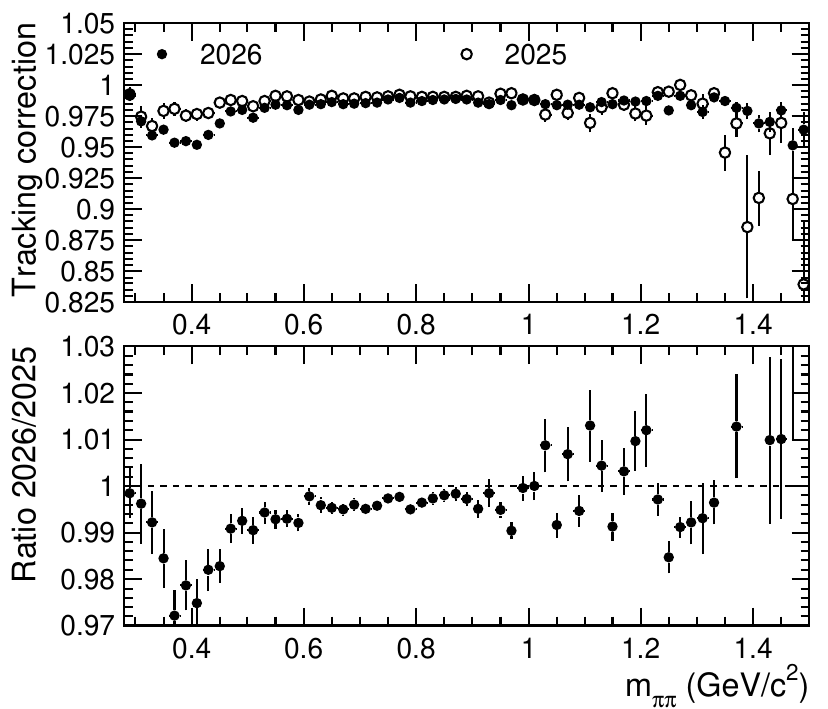}
    \caption{\small \justifying Comparison of the mass-dependent pion tracking 
    corrections (2025 vs 2026). 
    }
    \label{fig:tracking_mass}
\end{figure}

    \item[Pion BGFilter efficiencies] The corrections are improved to correct for a bias due to the correlated loss overlooked in 2025 (Section~\ref{subsubsec:BGF_pions}). 
    The impact of the change from the pion BGFilter template correction is however much smaller than that of the modified tracking template correction. On the other hand, the impact of the change from the BGFilter mass-dependent correction is comparable with that of the tracking correction (Fig.~\ref{fig:bgfilter_mass}).
    The total effect is an increase of about $+0.4\%$ in the $\pi\pi$ cross section in the $\rho$ peak region. 
    \item[Improved unfolding procedure] In the preliminary 2025 results, a significant slope in the data/MC ratio for the ISR luminosity was observed with respect to the flat data/MC ratio for the QED test. Further investigation shows that this is related to the single-step unfolding procedure used in 2025 for performing the transformation from a mass spectrum to the effective center-of-mass energy, which corrected
    only the detector resolution and FSR.
    The new three-step procedure, described in Section~\ref{sec:unfolding}, is needed to correct for some boundary event loss at the high energy tail.
\begin{figure}[tb]
    \centering
    \includegraphics[width=\linewidth]{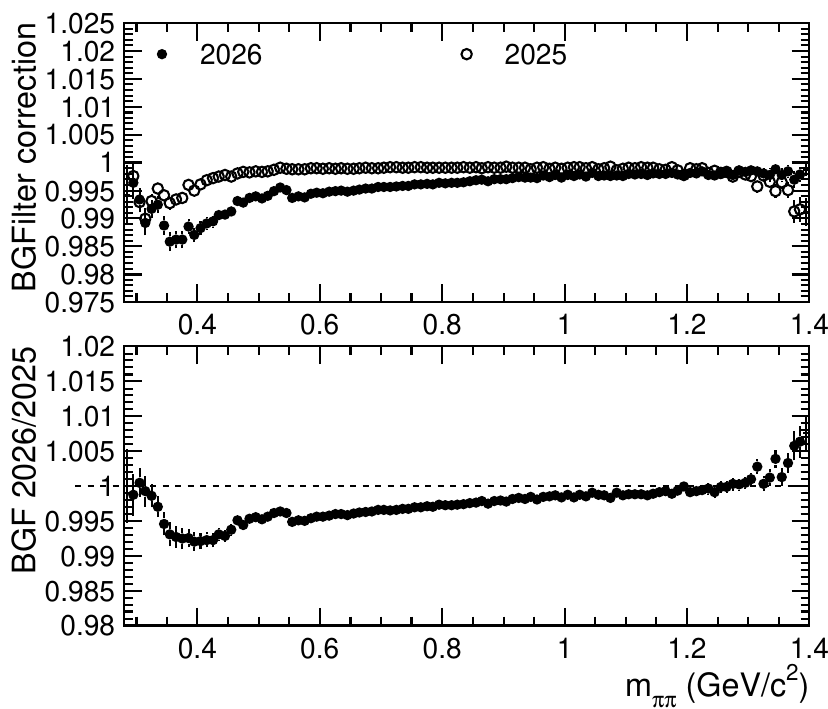}
    \caption{\small \justifying Comparison of the mass-dependent pion BGFilter 
    corrections (2025 vs 2026). 
    }
    \label{fig:bgfilter_mass}
\end{figure}    
    \item[Scaled kaon background prediction] The $KK$ contribution under the pion mass hypothesis is fitted together with the $\pi\pi$ and $\mu\mu$ contributions up to 0.4\gevcc. The remaining small tail contribution beyond this mass value in the 2025 analysis was subtracted using the MC prediction. In the final analysis, the MC prediction is scaled by $0.937(3)$, accounting for the data/MC acceptance correction for $KK\gamma$ events. This results in a local increase of the $\pi\pi$ event yield by up to $+15\%$ in the corresponding mass region.
    \item[Improved systematic error propagation] Some local large systematic variations for the 
    mass spectra obtained from the angular fits were observed for the preliminary 2025 results. This was investigated and found to be related to a small fit instability when performing the $ee\gamma$ background subtraction. This issue is fixed in the final results.  
    \item[Other minor improvements] Some of the systematic uncertainties have an important statistical component. A smoothing procedure is applied to reduce some of the fluctuation. Several minor systematic error sources are considered in the final analysis. One is related to a comparison of the 
    results between using the nominal MC migration matrix and an alternative one that takes into account small difference in the data/MC populations in the various zones of the 2d-$\chi^2$ plane. Another one concerns the variation of the ISR luminosity determination using the nominal linear fit or a constant one. The uncertainty in the subtracted backgrounds is also propagated to the final cross-section results. 
\end{description}

\clearpage
\bibliography{references}

@article{Boccaletti:2024guq,
    author = "Boccaletti, A. and others",
    title = "{Hybrid calculation of hadronic vacuum polarization in muon $g-2$ to 0.48{\%}}",
    eprint = "2407.10913",
    archivePrefix = "arXiv",
    primaryClass = "hep-lat",
    doi = "10.1038/s41586-026-10449-z",
    journal = "Nature",
    volume = "653",
    number = "8114",
    pages = "373--377",
    year = "2026"
}

@article{Bouchiat:1961lbg,
    author = "Bouchiat, Claude and Michel, Louis",
    title = "{La r{\'e}sonance dans la diffusion m{\'e}son $\pi$ -- m{\'e}son $\pi$ et le moment magn{\'e}tique anormal du m{\'e}son {\ensuremath{\mu}}}",
    doi = "10.1051/jphysrad:01961002202012101",
    journal = "J. Phys. Radium",
    volume = "22", 
    number = "2",
    pages = "121--121",
    year = "1961"
}

@article{SND:2020nwa,
    author = "Achasov, M. N. and others",
    collaboration = "SND",
    title = "{Measurement of the $e^+e^- \to\pi^+\pi^- $ process cross section with the SND detector at the VEPP-2000 collider in the energy region $0.525<\sqrt{s}<0.883$\,GeV}",
    eprint = "2004.00263",
    archivePrefix = "arXiv",
    primaryClass = "hep-ex",
    doi = "10.1007/JHEP01(2021)113",
    journal = "J.\ High Energy Phys.",
    volume = "01",
    pages = "113",
    year = "2021"
}

@article{Achasov:2006vp,
    author = "Achasov, M. N. and others",
    title = "{Update of the $e^+ e^- \to \pi^+ \pi^-$ cross-section measured by SND detector in the energy region 400\,MeV$<\sqrt{s}<$ 1000\,MeV}",
    eprint = "hep-ex/0605013",
    archivePrefix = "arXiv",
    doi = "10.1134/S106377610609007X",
    journal = "J. Exp. Theor. Phys.",
    volume = "103",
    pages = "380--384",
    year = "2006"
}

@article{KLOE:2012anl,
    author = "Babusci, D. and others",
    collaboration = "KLOE",
    title = "{Precision measurement of $\sigma(e^+e^-\rightarrow \pi^+\pi^-\gamma)/ \sigma(e^+e^-\rightarrow \mu^+\mu^-\gamma)$ and determination of the $\pi^+\pi^-$ contribution to the muon anomaly with the KLOE detector}",
    eprint = "1212.4524",
    archivePrefix = "arXiv",
    primaryClass = "hep-ex",
    doi = "10.1016/j.physletb.2013.02.029",
    journal = "Phys. Lett. B",
    volume = "720",
    pages = "336--343",
    year = "2013"
}

@article{KLOE:2010qei,
    author = "Ambrosino, F. and others",
    collaboration = "KLOE",
    title = "{Measurement of $\sigma(e^+ e^- \to \pi^+ \pi^-)$ from threshold to 0.85 GeV$^2$ using Initial State Radiation with the KLOE detector}",
    eprint = "1006.5313",
    archivePrefix = "arXiv",
    primaryClass = "hep-ex",
    doi = "10.1016/j.physletb.2011.04.055",
    journal = "Phys. Lett. B",
    volume = "700",
    pages = "102--110",
    year = "2011"
}

@article{KLOE:2008fmq,
    author = "Ambrosino, F. and others",
    collaboration = "KLOE",
    title = "{Measurement of $\sigma(e^+ e^- \to \pi^+ \pi^- \gamma(\gamma)$ and the dipion contribution to the muon anomaly with the KLOE detector}",
    eprint = "0809.3950",
    archivePrefix = "arXiv",
    primaryClass = "hep-ex",
    doi = "10.1016/j.physletb.2008.10.060",
    journal = "Phys. Lett. B",
    volume = "670",
    pages = "285--291",
    year = "2009"
}

@article{CMD-2:2006gxt,
    author = "Akhmetshin, R. R. and others",
    collaboration = "CMD-2",
    title = "{High-statistics measurement of the pion form factor in the rho-meson energy range with the CMD-2 detector}",
    eprint = "hep-ex/0610021",
    archivePrefix = "arXiv",
    doi = "10.1016/j.physletb.2007.01.073",
    journal = "Phys. Lett. B",
    volume = "648",
    pages = "28--38",
    year = "2007"
}

@article{Aulchenko:2006dxz,
    author = "Aul'chenko, V. M. and others",
    title = "{Measurement of the $e^+ e^- \to \pi^+ \pi^-$ cross section with the CMD-2 detector in the 370--520\,MeV c.m.\ energy range}",
    eprint = "hep-ex/0610016",
    archivePrefix = "arXiv",
    doi = "10.1134/S0021364006200021",
    journal = "JETP Lett.",
    volume = "84",
    pages = "413--417",
    year = "2006"
}

@article{CMD-2:2005mvb,
    author = "Aul'chenko, V. M. and others",
    collaboration = "CMD-2",
    title = "{Measurement of the pion form-factor in the range 1.04\,GeV to 1.38\,GeV with the CMD-2 detector}",
    eprint = "hep-ex/0603021",
    archivePrefix = "arXiv",
    doi = "10.1134/1.2175241",
    journal = "JETP Lett.",
    volume = "82",
    pages = "743--747",
    year = "2005"
}

@article{CMD-2:2003gqi,
    author = "Akhmetshin, R. R. and others",
    collaboration = "CMD-2",
    title = "{Reanalysis of hadronic cross-section measurements at CMD-2}",
    eprint = "hep-ex/0308008",
    archivePrefix = "arXiv",
    doi = "10.1016/j.physletb.2003.10.108",
    journal = "Phys. Lett. B",
    volume = "578",
    pages = "285--289",
    year = "2004"
}

@article{Xiao:2017dqv,
    author = "Xiao, T. and Dobbs, S. and Tomaradze, A. and Seth, Kamal K. and Bonvicini, G.",
    title = "{Precision Measurement of the Hadronic Contribution to the Muon Anomalous Magnetic Moment}",
    eprint = "1712.04530",
    archivePrefix = "arXiv",
    primaryClass = "hep-ex",
    doi = "10.1103/PhysRevD.97.032012",
    journal = "Phys. Rev. D",
    volume = "97",
    number = "3",
    pages = "032012",
    year = "2018"
}

@article{Castro:2024prg,
    author = "Castro, Gabriel L{\'o}pez and Miranda, Alejandro and Roig, Pablo",
    title = "{Isospin breaking corrections in $2\pi$ production in tau decays and $e^+e^-$ annihilation: Consequences for the muon $g-2$ and conserved vector current tests}",
    eprint = "2411.07696",
    archivePrefix = "arXiv",
    primaryClass = "hep-ph",
    doi = "10.1103/PhysRevD.111.073004",
    journal = "Phys. Rev. D",
    volume = "111",
    number = "7",
    pages = "073004",
    year = "2025"
}

@article{BESIII:2015equ,
    author = "Ablikim, M. and others",
    collaboration = "BESIII",
    title = "{Measurement of the $e^+ e^- \to \pi^+ \pi^-$ cross section between 600 and 900\,MeV using initial state radiation}",
    eprint = "1507.08188",
    archivePrefix = "arXiv",
    primaryClass = "hep-ex",
    doi = "10.1016/j.physletb.2015.11.043",
    journal = "Phys. Lett. B",
    volume = "753",
    pages = "629--638",
    year = "2016",
    note = "[Erratum: Phys.Lett.B 812, 135982 (2021)]"
}

@article{CMD-3:2023alj,
    author = "Ignatov, F. V. and others",
    collaboration = "CMD-3",
    title = "{Measurement of the $e^+e^-\to\pi^+\pi^-$ cross section from threshold to 1.2\,GeV with the CMD-3 detector}",
    eprint = "2302.08834",
    archivePrefix = "arXiv",
    primaryClass = "hep-ex",
    doi = "10.1103/PhysRevD.109.112002",
    journal = "Phys. Rev. D",
    volume = "109",
    number = "11",
    pages = "112002",
    year = "2024"
}

@book{Jegerlehner:2017gek,
    author = "Jegerlehner, Friedrich",
    title = "{The Anomalous Magnetic Moment of the Muon}",
    doi = "10.1007/978-3-319-63577-4",
    publisher = "Springer",
    series    = "Springer Tracts in Modern Physics",
    volume = "274",
    year = "2017",
    pages = "1--693",
}

@inproceedings{Polat:2026ysh,
    author = "Polat, L{\'e}onard",
    title = "{New precise measurement of the $e^+e^- \rightarrow \pi^+\pi^-(\gamma)$ cross section with \babar}",
    booktitle = "{32nd International Symposium on Lepton Photon Interactions at High Energies}: {Lepton-Photon 2025}",
    eprint = "2601.16587",
    archivePrefix = "arXiv",
    primaryClass = "hep-ex",
    month = "1",
    year = "2026"
}

@inproceedings{Zhang:2026jvb,
    author = "Zhang, Zhiqing",
    title = "{Review of hadronic vacuum polarization calculations via $e^+e^-$ measurements}",
    booktitle = "{32nd International Symposium on Lepton Photon Interactions at High Energies}: {Lepton-Photon 2025}",
    eprint = "2601.11966",
    archivePrefix = "arXiv",
    primaryClass = "hep-ex",
    month = "1",
    year = "2026"
}

@misc{davier-ti-2017,
  author = "Michel Davier",
  title = "{$e^+e^-$ results from \babar\ and implications for the muon $g-2$}",
  note = "{First Workshop of the Muon g-2 Theory Initiative, Q Center, 3-6 June 2017}",
  year ="2017",
  url = "https://indico.fnal.gov/event/13795/"
}

@article{Jadach:1999vf,
    author = "Jadach, S. and Ward, B. F. L. and Was, Z.",
    title = "{The Precision Monte Carlo event generator $KK$ for two fermion final states in $e^+ e^-$ collisions}",
    eprint = "hep-ph/9912214",
    archivePrefix = "arXiv",
    reportNumber = "DESY-99-106, CERN-TH-99-235, UTHEP-99-08-01",
    doi = "10.1016/S0010-4655(00)00048-5",
    journal = "Comput. Phys. Commun.",
    volume = "130",
    pages = "260",
    year = "2000"
}

@article{Aliberti:2025beg,
    author = "Aliberti, R. and others",
    title = "{The anomalous magnetic moment of the muon in the Standard Model: an update}",
    eprint = "2505.21476",
    archivePrefix = "arXiv",
    primaryClass = "hep-ph",
    reportNumber = "CERN-TH-2025-101, FERMILAB-PUB-25-0344-T, INT-PUB-25-015, IPARCOS-UCM-25-029, KEK Preprint 2025-22, LTH 1403, MITP-25-037, UWThPh 2025-15, UWThPh
    2025-15, ZU-TH 37/25, IPARCOS-UCM-25-029",
    doi = "10.1016/j.physrep.2025.08.002",
    journal = "Phys. Rept.",
    volume = "1143",
    pages = "1--158",
    year = "2025"
}

@article{Davier:2010fmf,
    author = "Davier, M. and Hoecker, A. and Lopez Castro, G. and Malaescu, B. and Mo, X. H. and Toledo Sanchez, G. and Wang, P. and Yuan, C. Z. and Zhang, Z.",
    title = "{The Discrepancy Between tau and $e^+e^-$ Spectral Functions Revisited and the Consequences for the Muon Magnetic Anomaly}",
    eprint = "0906.5443",
    archivePrefix = "arXiv",
    primaryClass = "hep-ph",
    reportNumber = "BIHEP-TH-09-01, CERN-OPEN-2009-007, LAL-09-50",
    doi = "10.1140/epjc/s10052-009-1219-4",
    journal = "Eur. Phys. J. C",
    volume = "66",
    pages = "127--136",
    year = "2010"
}

@article{Davier:2010rnx,
    author = "Davier, M. and Hoecker, A. and Malaescu, B. and Yuan, C. Z. and Zhang, Z.",
    title = "{Reevaluation of the hadronic contribution to the muon magnetic anomaly using new $e^+ e^- \to \pi^+ \pi^-$ cross section data from \babar}",
    eprint = "0908.4300",
    archivePrefix = "arXiv",
    primaryClass = "hep-ph",
    reportNumber = "BIHEP-TH-2009-004, CERN-OPEN-2009-010, LAL-09-115",
    doi = "10.1140/epjc/s10052-010-1246-1",
    journal = "Eur. Phys. J. C",
    volume = "66",
    pages = "1--9",
    year = "2010"
}

@article{Davier:2023fpl,
    author = "Davier, Michel and Hoecker, Andreas and Lutz, Anne-Marie and Malaescu, Bogdan and Zhang, Zhiqing",
    title = "{Tensions in $e^+e^-\rightarrow \pi ^+\pi ^-(\gamma )$ measurements: the new landscape of data-driven hadronic vacuum polarization predictions for the muon $g - 2$}",
    eprint = "2312.02053",
    archivePrefix = "arXiv",
    primaryClass = "hep-ph",
    doi = "10.1140/epjc/s10052-024-12964-7",
    journal = "Eur. Phys. J. C",
    volume = "84",
    number = "7",
    pages = "721",
    year = "2024"
}

@article{Binner:1999bt,
    author = "Binner, S. and Kühn, Johann H. and Melnikov, K.",
    title = "{Measuring $\sigma(e^+ e^- \to \textrm{hadrons})$ using tagged photon}",
    eprint = "hep-ph/9902399",
    archivePrefix = "arXiv",
    reportNumber = "TTP-99-09",
    doi = "10.1016/S0370-2693(99)00658-9",
    journal = "Phys. Lett. B",
    volume = "459",
    pages = "279--287",
    year = "1999"
}

@article{BaBar:2023xiy,
    author = "Lees, J. P. and others",
    collaboration = "\babar",
    title = "{Measurement of additional radiation in the initial-state-radiation processes $e^+e^-\to\mu^+\mu^-\gamma$ and $e^+e^-\to \pi^+\pi^-\gamma$ at \babar}",
    eprint = "2308.05233",
    archivePrefix = "arXiv",
    primaryClass = "hep-ex",
    reportNumber = "BABAR-PUB-23/005, SLAC-PUB-17736",
    doi = "10.1103/PhysRevD.108.L111103",
    journal = "Phys. Rev. D",
    volume = "108",
    number = "11",
    pages = "L111103",
    year = "2023"
}

@article{BaBar:2009wpw,
    author = "Aubert, B. and others",
    collaboration = "\babar",
    title = "{Precise measurement of the $e^+ e^- \to \pi^+ \pi^- (\gamma)$ cross section with the Initial State Radiation method at \babar}",
    eprint = "0908.3589",
    archivePrefix = "arXiv",
    primaryClass = "hep-ex",
    reportNumber = "SLAC-PUB-13754, BABAR-PUB-09-027",
    doi = "10.1103/PhysRevLett.103.231801",
    journal = "Phys. Rev. Lett.",
    volume = "103",
    pages = "231801",
    year = "2009"
}

@article{BaBar:2012bdw,
    author = "Lees, J. P. and others",
    collaboration = "\babar",
    title = "{Precise Measurement of the $e^+ e^- \to \pi^+\pi^- (\gamma)$ Cross Section with the Initial-State Radiation Method at \babar}",
    eprint = "1205.2228",
    archivePrefix = "arXiv",
    primaryClass = "hep-ex",
    reportNumber = "BABAR-PUB-12-003",
    doi = "10.1103/PhysRevD.86.032013",
    journal = "Phys. Rev. D",
    volume = "86",
    pages = "032013",
    year = "2012"
}

@article{BaBar:2001yhh,
    author = "Aubert, B. and others",
    collaboration = "\babar",
    title = "{The \babar\ detector}",
    eprint = "hep-ex/0105044",
    archivePrefix = "arXiv",
    reportNumber = "SLAC-PUB-8569, BABAR-PUB-01-08",
    doi = "10.1016/S0168-9002(01)02012-5",
    journal = "Nucl. Instrum. Meth. A",
    volume = "479",
    pages = "1--116",
    year = "2002"
}

@article{BaBar:2013byz,
    author = "Aubert, B. and others",
    collaboration = "\babar",
    title = "{The \babar\ Detector: Upgrades, Operation and Performance}",
    eprint = "1305.3560",
    archivePrefix = "arXiv",
    primaryClass = "physics.ins-det",
    reportNumber = "BABAR-PUB-13-009, SLAC-PUB-15456",
    doi = "10.1016/j.nima.2013.05.107",
    journal = "Nucl. Instrum. Meth. A",
    volume = "729",
    pages = "615--701",
    year = "2013"
}

@article{phokhara-v9.1,
    author = "Campanario, F. and others",
    title = "{Complete QED NLO contributions to the reaction $e^+ e^- \to \mu^+ \mu^-\gamma$ and their implementation in the event generator PHOKHARA.}",
    eprint = "hep-ph/1312.3610",
    archivePrefix = "arXiv",
    doi = "10.1007/JHEP02(2014)114",
    journal = "J.\ High Energy Phys.",
    volume = "02",
    pages = "114",
    year = "2014"
}

@article{Czyz:2003hc,
    author = "H.Czyż, A.Grzelińska, J.H.Kuehn and G.Rodrigo",  
    title = "{The radiative return at $\phi$- and $B$-factories: FSR at next-to-leading order}",
    eprint = "hep-ph/0308312",
    archivePrefix = "arXiv",
    doi = "10.1140/epjc/s2004-01605-0", 	 
    journal = "Eur.Phys.J.C",
    volume = "33",
    pages = "333-347",
    year = "2004"
}

@article{Czyz:2004rj,
    author = "Czyż, Henryk and Grzelińska, Agnieszka and Kühn, Johann H. and Rodrigo, German",
    title = "{The Radiative return at $\phi$ and $B$ factories: FSR for muon pair production at next-to-leading order}",
    eprint = "hep-ph/0404078",
    archivePrefix = "arXiv",
    reportNumber = "CERN-TH-2004-061, TTP-04-07",
    doi = "10.1140/epjc/s2004-02103-1",
    journal = "Eur. Phys. J. C",
    volume = "39",
    pages = "411--420",
    year = "2005"
}

@article{Czyz:2000wh,
    author = "Czyż, H. and Kühn, Johann H.",
    title = "{Four pion final states with tagged photons at electron positron colliders}",
    eprint = "hep-ph/0008262",
    archivePrefix = "arXiv",
    reportNumber = "KA-TTP-00-16",
    doi = "10.1007/s100520000553",
    journal = "Eur. Phys. J. C",
    volume = "18",
    pages = "497--509",
    year = "2001"
}

@article{Caffo:1997yy,
    author = "Caffo, Michele and Czyż, H. and Remiddi, E.",
    title = "{BHAGEN95: A Monte Carlo program for Bhabha scattering at LEP-1 / SLC and LEP-2 energies}",
    eprint = "hep-ph/9704443",
    archivePrefix = "arXiv",
    reportNumber = "DFUB-96-21, TP-USL-96-12",
    doi = "10.1007/BF03035898",
    journal = "Nuovo Cim. A",
    volume = "110",
    pages = "515--536",
    year = "1997"
}

@article{Barberio:1990ms,
    author = "Barberio, Elisabetta and van Eijk, Bob and Was, Zbigniew",
    title = "{PHOTOS: A Universal Monte Carlo for QED radiative corrections in decays}",
    reportNumber = "CERN-TH-5857-90",
    doi = "10.1016/0010-4655(91)90012-A",
    journal = "Comput. Phys. Commun.",
    volume = "66",
    pages = "115--128",
    year = "1991"
}

@article{Sjostrand:1993yb,
    author = "Torbjörn Sjöstrand",
    title = "{High-energy physics event generation with PYTHIA 5.7 and JETSET 7.4}",
    reportNumber = "CERN-TH-7111-93",
    doi = "10.1016/0010-4655(94)90132-5",
    journal = "Comput. Phys. Commun.",
    volume = "82",
    pages = "74--90",
    year = "1994"
}

@article{GEANT4:2002zbu,
    author = "Agostinelli, S. and others",
    collaboration = "GEANT4",
    title = "{GEANT4 - A Simulation Toolkit}",
    reportNumber = "SLAC-PUB-9350, FERMILAB-PUB-03-339, CERN-IT-2002-003",
    doi = "10.1016/S0168-9002(03)01368-8",
    journal = "Nucl. Instrum. Meth. A",
    volume = "506",
    pages = "250--303",
    year = "2003"
}

@article{Bystritskiy:2005ib,
    author = "Bystritskiy, Yu. M. and Kuraev, E. A. and Fedotovich, G. V. and Ignatov, F. V.",
    title = "{The cross sections of the muons and charged pions pairs production at electron-positron annihilation near the threshold}",
    eprint = "hep-ph/0505236",
    archivePrefix = "arXiv",
    doi = "10.1103/PhysRevD.72.114019",
    journal = "Phys. Rev. D",
    volume = "72",
    pages = "114019",
    year = "2005"
}

@article{Ji:2019yca,
    author = "Ji, Xiangpan and Gu, Wenqiang and Qian, Xin and Wei, Hanyu and Zhang, Chao",
    title = "{Combined Neyman\textendash{}Pearson chi-square: An improved approximation to the Poisson-likelihood chi-square}",
    eprint = "1903.07185",
    archivePrefix = "arXiv",
    primaryClass = "physics.data-an",
    doi = "10.1016/j.nima.2020.163677",
    journal = "Nucl. Instrum. Meth. A",
    volume = "961",
    pages = "163677",
    year = "2020"
}

@misc{Malaescu:2009dm,
    author = "Malaescu, Bogdan",
    title = "{An iterative, dynamically stabilized method of data unfolding}",
    eprint = "0907.3791",
    archivePrefix = "arXiv",
    primaryClass = "physics.data-an",
    reportNumber = "LAL-09-107",
    month = "7",
    year = "2009"
}

@inproceedings{Malaescu:2011yg,
    author = "Malaescu, Bogdan",
    title = "{An Iterative, Dynamically Stabilized (IDS) Method of Data Unfolding}",
    booktitle = "{PHYSTAT 2011}",
    eprint = "1106.3107",
    archivePrefix = "arXiv",
    primaryClass = "physics.data-an",
    reportNumber = "CERN-PH-EP-2011-111",
    doi = "10.5170/CERN-2011-006.271",
    pages = "271--275",
    year = "2011"
}

@article{Lees_2012,
   author={Lees, J. P. and others},
   collaboration = "\babar",
   title={Initial-state radiation measurement of the ${e}^{+}{e}^- \to \pi^+\pi^-\pi^+\pi^-$ cross section},
   volume={85},
   ISSN={1550-2368},
   url={http://dx.doi.org/10.1103/PhysRevD.85.112009},
   DOI={10.1103/physrevd.85.112009},
   number={11},
   journal={Physical Review D},
   publisher={American Physical Society (APS)},
   year={2012},
   month=jun }

@misc{hafner-thesis,
  author = {Andreas Hafner},
  title = "{Measurement of the $e^+e^- \to \pi^+\pi^-\pi^+\pi^-$ Cross Section Using Initial State Radiation at \babar}",
  note = "{PhD thesis, Johannes Gutenberg-Universität in Mainz}",
  year ="2012",
  url = "https://openscience.ub.uni-mainz.de/items/0d039d8d-f67f-4f65-9974-aa3c33847e30"
}

@article{BaBar:2015onb,
    author = "Lees, J. P. and others",
    collaboration = "\babar",
    title = "{Measurement of initial-state and final-state radiation interference in the processes $e^+e^- \to \mu^+\mu^-\gamma$ and $e^+e^- \to \pi^+\pi^-\gamma$}",
    eprint = "1508.04008",
    archivePrefix = "arXiv",
    primaryClass = "hep-ex",
    reportNumber = "SLAC-PUB-16361",
    doi = "10.1103/PhysRevD.92.072015",
    journal = "Phys. Rev. D",
    volume = "92",
    number = "7",
    pages = "072015",
    year = "2015"
}

@article{BaBarBelle2014,
    author = "A. J. Bevan and others",
    collaboration = "Belle, \babar",
    title = "{The Physics of the $B$ Factories}",
    eprint = "1406.6311",
    archivePrefix = "arXiv",
    primaryClass = "hep-ex",
    journal = "Eur. Phys. J. C",
    volume = "74",
    pages = "3026", 
    year = "2014"
}

@Inbook{Efron1992,
author="Efron, Bradley",
editor="Kotz, Samuel
and Johnson, Norman L.",
title="Bootstrap Methods: Another Look at the Jackknife",
bookTitle="Breakthroughs in Statistics: Methodology and Distribution",
year="1992",
publisher="Springer New York",
address="New York, NY",
pages="569--593",
isbn="978-1-4612-4380-9",
doi="10.1007/978-1-4612-4380-9_41",
url="https://doi.org/10.1007/978-1-4612-4380-9_41"
}

@article{HANLEY200657,
title = {Creating non-parametric bootstrap samples using Poisson frequencies},
journal = {Computer Methods and Programs in Biomedicine},
volume = {83},
number = {1},
pages = {57-62},
year = {2006},
issn = {0169-2607},
doi = {https://doi.org/10.1016/j.cmpb.2006.04.006},
url = {https://www.sciencedirect.com/science/article/pii/S0169260706000770},
author = {James A. Hanley and Brenda MacGibbon}
}

@book{Bohm:2005bu,
    author = "Bohm, G. and Zech, Gunter",
    title = "{Introduction to statistics and measurement analysis for physicists}",
    doi = "10.3204/PUBDB-2017-08987",
    isbn = "978-3-945931-13-4",
    year = "2005"
}

@misc{ATLAS:2021kho,
    author = "{ATLAS Collaboration}",
    title = "{Evaluating statistical uncertainties and correlations using the bootstrap method}",
    Note = "{ATL-PHYS-PUB-2021-011}",
    year = "2021"
}

@article{Alexe:2024smm,
    author = "Alexe, Cristina-Andreea and Bendavid, Joshua and Bianchini, Lorenzo and Bruschini, Davide",
    title = "{Under-coverage in high-statistics counting experiments with finite MC samples}",
    eprint = "2401.10542",
    archivePrefix = "arXiv",
    primaryClass = "physics.data-an",
    doi = "10.1016/j.nima.2026.171360",
    journal = "Nucl. Instrum. Meth. A",
    volume = "1086",
    pages = "171360",
    year = "2026"
}

@article{BaBar:2013agn,
    author = "Lees, J. P. and others",
    collaboration = "\babar",
    title = "{Time-Integrated Luminosity Recorded by the \babar\ Detector at the PEP-II $e^+ e^-$ Collider}",
    eprint = "1301.2703",
    archivePrefix = "arXiv",
    primaryClass = "hep-ex",
    reportNumber = "BABAR-PUB-12-031, SLAC-PUB-15347",
    doi = "10.1016/j.nima.2013.04.029",
    journal = "Nucl. Instrum. Meth. A",
    volume = "726",
    pages = "203--213",
    year = "2013"
}

@article{Czy__2010,
  title = {Narrow resonances studies with the radiative return method},
  author={Czyż, Henryk and Grzelińska, Agnieszka and Kühn, Johann H.},
  journal = {Phys. Rev. D},
  volume = {81},
  issue = {9},
  pages = {094014},
  numpages = {11},
  year = {2010},
  month = {May},
  publisher = {American Physical Society},
  doi = {10.1103/PhysRevD.81.094014},
  url = {https://link.aps.org/doi/10.1103/PhysRevD.81.094014}
}

@article{BaBar:2013ves,
author = "Lees, J. P. and others",
collaboration = "\babar",
title = "{Study of $e^+e^- \to p \bar{p}$ via initial-state radiation at \babar}",
eprint = "1302.0055",
archivePrefix = "arXiv",
primaryClass = "hep-ex",
reportNumber = "BABAR-PUB-12-030, SLAC-PUB-15324",
doi = "10.1103/PhysRevD.87.092005",
journal = "Phys. Rev. D",
volume = "87",
number = "9",
pages = "092005",
year = "2013"
}

@article{Hagiwara_2011,
   title={$(g-2)_\mu$ and $\alpha(M^2_Z)$ re-evaluated using new precise data},
   volume={38},
   ISSN={1361-6471},
   url={http://dx.doi.org/10.1088/0954-3899/38/8/085003},
   DOI={10.1088/0954-3899/38/8/085003},
   number={8},
   journal={Journal of Physics G: Nuclear and Particle Physics},
   publisher={IOP Publishing},
   author={Hagiwara, Kaoru and Liao, Ruofan and Martin, Alan D and Nomura, Daisuke and Teubner, Thomas},
   year={2011},
   month=June, pages={085003} }

@article{ParticleDataGroup:2026mpi,
    author = "Takahashi, F. and others",
    collaboration = "Particle Data Group",
    title = "{Review of Particle Physics}",
    doi = "10.1142/s0217751x26300115",
    journal = "Int. J. Mod. Phys. A",
    volume = "41",
    number = "22",
    pages = "2630011",
    year = "2026"
}

\end{document}